\pdfoutput=1

\documentclass[11pt,twoside,a4paper,cmspaper,final,collab]{cms-tdr}
\def\svnVersion{245642}\def\cmsCernNo{2013-220}\def\cmsCernDate{\today}\def\cmsMessage{Published in Physical Review D as \href{http://dx.doi.org/10.1103/PhysRevD.89.092007}{\doi{10.1103/PhysRevD.89.092007.}}}

\begin{document}\cmsNoteHeader{HIG-13-002}

\hyphenation{had-ron-i-za-tion}
\hyphenation{cal-or-i-me-ter}
\hyphenation{de-vices}

\RCS$Revision: 236345 $
\RCS$HeadURL: svn+ssh://svn.cern.ch/reps/tdr2/papers/HIG-13-002/trunk/HIG-13-002.tex $
\RCS$Id: HIG-13-002.tex 236345 2014-04-11 16:10:08Z emanuele $
\ifthenelse{\boolean{cms@external}}{\providecommand{\cmsLeft}{top}}{\providecommand{\cmsLeft}{left}}
\ifthenelse{\boolean{cms@external}}{\providecommand{\cmsRight}{bottom}}{\providecommand{\cmsRight}{right}}
\newlength\cmsFigWidth
\ifthenelse{\boolean{cms@external}}{\setlength\cmsFigWidth{0.48\textwidth}}{\setlength\cmsFigWidth{0.6\textwidth}}
\newlength\cmsFigWidthStd
\ifthenelse{\boolean{cms@external}}{\setlength\cmsFigWidthStd{0.45\textwidth}}{\setlength\cmsFigWidthStd{0.48\textwidth}}

\ifthenelse{\boolean{cms@external}}{\providecommand{\NA}{\ensuremath{\cdots}}}{\providecommand{\NA}{\text{---}}}
\ifthenelse{\boolean{cms@external}}{\providecommand{\CL}{\ensuremath{\mathrm{C.L.}}\xspace}}{\providecommand{\CL}{\ensuremath{\mathrm{CL}}\xspace}}
\newcommand{\Zee}{\ensuremath{\cPZ\to\Pep\Pem}\xspace}
\newcommand{\tauh}{\ensuremath{\Pgt_\mathrm{h}}}
\newcommand{\taue}{\ensuremath{\Pgt_\mathrm{e}}}
\newcommand{\taul}{\ensuremath{\Pgt_\mathrm{\ell}}}
\newcommand{\Hmu}{\ensuremath{\PH\to\cPZ\cPZ^{(\ast)}\to4\Pgm}}
\newcommand{\mmu}{\ensuremath{m_{4\Pgm}}}
\newcommand{\mH}{\ensuremath{m_{\PH}}}
\newcommand{\lnQ}{\ensuremath{\ln(Q)}}
\newcommand{\mlnQ}{\ensuremath{-2\,\ln(Q)}}
\newcommand{\LS}{\ensuremath{\mathcal{L}(S)}}
\newcommand{\LB}{\ensuremath{\mathcal{L}(B)}}
\newcommand{\LSB}{\ensuremath{\mathcal{L}(S+B)}}
\newcommand{\mZ}{\ensuremath{m_{\cPZ}}}
\newcommand{\Zbb}{\ensuremath{(\cPZ/\Pgg^*){\cPqb\cPaqb}}}
\newcommand{\ZZ}{\ensuremath{\cPZ/\Pgg^*\cPZ/\Pgg^*}}
\newcommand{\Zgam}{\ensuremath{\cPZ/\Pgg^*}}
\newcommand{\mumu}{\ensuremath{\Pgmp\Pgmm}}
\newcommand{\Wo}{\ensuremath{\PW}}%
\newcommand{\Wp}{\ensuremath{\PWp}}%
\newcommand{\Wm}{\ensuremath{\PWm}}%
\newcommand{\Zo}{\ensuremath{\cPZ}}%
\newcommand{\Ho}{\ensuremath{\PH}}%
\newcommand{\UpsNs}{\ensuremath{\PgU(\mathrm{nS})}}
\newcommand{\KD}{\ensuremath{\mathcal{D}^\text{kin}_\text{bkg}} }
\newcommand{\superKD}{\ensuremath{\mathcal{D}_\text{bkg}} }
\newcommand{\spinKD}{\ensuremath{\mathcal{D}_{J^P}} }
\newcommand{\VDj}{\mathrm{\mathcal{D}_\text{jet}} }
\newcommand{\VDu}{{\PT^{4\ell}} }
\newcommand{\MassD}{\mathcal{D}_\mathrm{m} }
\newcommand{\LikMu}{\mathcal{L}_{3D}^{\mu} }
\newcommand{\LikMuZOj}{\mathcal{L}_{3D}^{\mu,\,\text{0/1-jet}}(m_{4\ell},\KD,\VDu) }
\newcommand{\LikMuDj}{\mathcal{L}_{3D}^{\mu,\,\text{dijet}}(m_{4\ell},\KD,\VDj) }
\newcommand{\LikMuTwoD}{\mathcal{L}_{2D}^{\mu} }
\newcommand{\LikMuOneD}{\mathcal{L}_{1D}^{\mu} }
\newcommand{\LikMass}{\mathcal{L}_{3D}^{m,\Gamma} }
\newcommand{\LikMassTwoD}{\mathcal{L}_{2D}^{m,\Gamma} }
\newcommand{\LikMassOneD}{\mathcal{L}_{1D}^{m,\Gamma} }
\newcommand{\LikSpin}{\mathcal{L}_{2D}^{J^P} }
\newcommand{\X}{\ensuremath{\cmsSymbolFace{X}}\xspace}
\newcommand{\sip}{\ensuremath{\text{SIP}_\text{3D}} }
\newcommand{\fakerate}{\ensuremath{f(\ell,\PT^\ell,\vert\eta^\ell\vert)} }
\newcommand{\strength}{\ensuremath{\mu}}
\newcommand{\muV}{\ensuremath{\mu_{\mathrm{VBF},~\mathrm{V\PH}}} }
\newcommand{\muF}{\ensuremath{\mu_{\Pg\Pg\PH,\,\ttbar\PH}} }

\newcommand\T{\rule{0pt}{2.6ex}}       
\newcommand\B{\rule[-1.2ex]{0pt}{0pt}} 

\newcommand{\usedLumiA}{5.1\fbinv}
\newcommand{\usedLumiB}{19.7\fbinv}

\newcommand{\expSign}{6.7}
\newcommand{\obsSign}{6.8}
\newcommand{\expSignTwoD}{6.6}
\newcommand{\obsSignTwoD}{6.9}
\newcommand{\expSignOneD}{5.6}
\newcommand{\obsSignOneD}{5.0}
\newcommand{\valMuExp}{\ensuremath{1.00^{+0.31}_{-0.26}} }
\newcommand{\valMu}{\ensuremath{0.93^{+0.26}_{-0.23}\stat ^{+0.13}_{-0.09}\syst} }
\newcommand{\valMuTot}{\ensuremath{0.93^{+0.29}_{-0.24}} }
\newcommand{\valMuV}{\ensuremath{1.7^{+2.2}_{-2.1}} }
\newcommand{\valMuF}{\ensuremath{0.80^{+0.46}_{-0.36}} }
\newcommand{\valMuUntagged}{\ensuremath{0.83^{+0.31}_{-0.25}} }
\newcommand{\valMuDijet}{\ensuremath{1.45^{+0.89}_{-0.62}} }

\newcommand{\valMass}{\ensuremath{125.6 \pm 0.4\stat \pm 0.2\syst} }
\newcommand{\valMassFourMu}{\ensuremath{125.1^{+0.6}_{-0.9}} }
\newcommand{\valMassFourE}{\ensuremath{126.2^{+1.5}_{-1.8}} }
\newcommand{\valMassTwoETwoMu}{\ensuremath{126.3^{+0.9}_{-0.7}} }
\newcommand{\valWidth}{\ensuremath{0.0^{+1.3}_{-0.0}} }
\newcommand{\ulWidth}{3.4}
\newcommand{\expUlWidth}{2.8}

\newcommand{\ulFaThree}{0.47}
\newcommand{\valFaThree}{\ensuremath{0.00^{+0.15}_{-0.00}} }

\cmsNoteHeader{HIG-13-002} 
\title{Measurement of the properties of a Higgs boson in the four-lepton final state}

\date{\today}

\abstract{
The properties of a Higgs boson candidate are measured in the
$\PH\rightarrow\cPZ\cPZ\rightarrow 4\ell$ decay channel, with
$\ell=\Pe,\Pgm$, using data from $\Pp\Pp$ collisions corresponding to
an integrated luminosity of $\usedLumiA$ at the center-of-mass energy
of $\sqrt{s}=7$\TeV and $\usedLumiB$ at $\sqrt{s}= 8$\TeV, recorded
with the CMS detector at the LHC.  The new boson is observed as a
narrow resonance with a local significance of $\obsSign$ standard
deviations, a measured mass of $\valMass$\GeV, and a total width
$\le\ulWidth$\GeV at the 95\% confidence level.  The production cross
section of the new boson times its branching fraction to four leptons
is measured to be $\valMu$ times that predicted by the standard
model. Its spin-parity properties are found to be consistent with the
expectations for the standard-model Higgs boson.  The hypotheses of a
pseudoscalar and all tested spin-1 boson hypotheses are excluded at
the 99\% confidence level or higher. All tested spin-2 boson
hypotheses are excluded at the 95\% confidence level or higher.}

\hypersetup{%
pdfauthor={CMS Collaboration},
pdftitle={Measurement of the properties of a Higgs boson in the four-lepton final state},
pdfsubject={CMS},
pdfkeywords={CMS, physics, Higgs, multilepton, diboson}}

\maketitle 

\section{Introduction}
\label{sec:introduction}

The standard model (SM) of particle physics
~\cite{StandardModel67_1,StandardModel67_2,StandardModel67_3,PhysRevLett.30.1343,PhysRevLett.30.1346}
describes very successfully the electroweak and strong interactions of
elementary particles over a wide range of energies.  In the SM, the
massive mediators of the electroweak force, the $\PW$ and $\cPZ$
bosons, acquire mass through the mechanism of spontaneous symmetry
breaking~\cite{Englert:1964et,Higgs:1964ia,Higgs:1964pj,Guralnik:1964eu,Higgs:1966ev,Kibble:1967sv}.
This mechanism introduces a complex scalar field with four degrees of
freedom, three of which lead to the $\PW$ and $\cPZ$ bosons acquiring
mass while the fourth gives rise to a physical particle, the scalar
Higgs boson $\PH$.  The masses of the fermions arise through Yukawa
interactions between the fermions and the scalar
field~\cite{PhysRev.122.345,GellMann:1960np}.  The mass of the Higgs
boson $\mH$ is a free parameter of the model and has to be
determined experimentally.
General theoretical considerations on the unitarity of the
SM~\cite{Cornwall:1973tb,Cornwall:1974km,LlewellynSmith:1973ey,Lee:1977eg}
suggest that $\mH$ should be smaller than $\approx$1\TeV, while
precision electroweak measurements imply that $\mH < 152$\GeV at the
95\% confidence level (\CL)~\cite{EWKlimits}.
Using about 5\fbinv of data collected at $\sqrt{s}= 7$\TeV in
2011 and about 5\fbinv of additional data collected in the first
half of 2012 at $\sqrt{s} = 8$\TeV, the ATLAS and CMS experiments
have reported the discovery of a new boson at a mass around 125\GeV,
with properties compatible with those of the SM Higgs
boson~\cite{Aad:2012tfa, Chatrchyan:2012ufa, Chatrchyan:2013lba}.
Previously, direct searches for the Higgs boson have been carried out
at the LEP collider, leading to a lower bound of $\mH > 114.4$\GeV at
the 95\% \CL~\cite{Barate:2003sz}, and at the Tevatron
proton-antiproton collider, excluding the mass ranges 90--109\GeV and
149--182\GeV at the 95\% \CL and indicating a broad excess of events
in the range 120--135\GeV~\cite{PhysRevLett.109.071804,Tevatron2013}.

Searches for the SM Higgs boson in the
$\PH\rightarrow\cPZ\cPZ\rightarrow 4\ell$ ($\ell = \Pe, \Pgm$) channel
at the Large Hadron Collider (LHC) have been previously performed
using a sample corresponding to an integrated luminosity of about
$5\fbinv$ of 2011 data by the
ATLAS~\cite{ATLAS:2012ac,ATLAS:2012ae,atlas:20127tev} and Compact Muon
Solenoid
(CMS)~\cite{Chatrchyan:2012dg,Chatrchyan:2012hr,Chatrchyan:2012tx}
collaborations.
After the new boson discovery, the spin-parity properties have been
further studied by both experiments, using more data. The pseudoscalar
hypothesis is excluded by CMS~\cite{Chatrchyan:2012br} and ATLAS
experiments~\cite{Aad:2013wqa,Aad:2013xqa} at the 95\% \CL or
higher. ATLAS has also excluded at the 99\% \CL the hypotheses of
vector, pseudovector, and graviton-like spin-2 bosons, under certain
assumptions on their production mechanisms~\cite{Aad:2013xqa}.

In this paper, the analysis of the $\PH\rightarrow\cPZ\cPZ\rightarrow
4\ell$ channel is presented using the entire data set collected by the
CMS experiment during the 2011--2012 LHC running period. The data
correspond to an integrated luminosity of $\usedLumiA$ of $\Pp\Pp$
collisions at a center-of-mass energy of $\sqrt{s} = 7$\TeV, and
$\usedLumiB$ at $\sqrt{s} = 8$\TeV.  The search looks for a signal
consisting of two pairs of same-flavor, opposite-charge,
well-identified and isolated leptons, $\Pep\Pem$, $\Pgmp\Pgmm$,
compatible with a $\cPZ\cPZ$ system, where one or both the $\cPZ$
bosons can be off shell, appearing as a narrow resonance on top of a
smooth background in the four-lepton invariant mass distribution.
Improved calibrations and alignment constants with respect to those
used in Refs.~\cite{Chatrchyan:2012ufa,
Chatrchyan:2013lba,Chatrchyan:2012br}, based on the full data set, are
used in the reconstruction of the events considered for this paper.
The statistical significance of the observation of the new boson in
the four-lepton decay mode is reported, together with measurements of
the boson's mass and its cross section times its branching fraction
with respect to the SM prediction, an upper limit on the boson's
width, and the compatibility of the boson with nine alternative
spin-parity hypotheses.  The compatibility of the data with a mixed
scalar/pseudoscalar state is also assessed. A search is also conducted
for additional resonances compatible with the SM Higgs boson in the
$\PH\rightarrow\cPZ\cPZ\rightarrow 4\ell$ channel in the mass range
110--1000\GeV.

The paper is organized as follows: the apparatus, the data samples,
and the online selection are described in Secs.~\ref{sec:detector}
through~\ref{sec:triggers}.  Sections~\ref{sec:reconstruction}
through~\ref{sec:jets} describe the reconstruction and identification
algorithms used in this analysis for leptons, photons, and jets. The
event selection and categorization are discussed in
Sec.~\ref{sec:selection}.  The background estimation is described in
Sec.~\ref{sec:backgrounds}. Kinematic discriminants used to further
improve the separation between signal and background and to test the
spin and parity of the new boson are presented in
Sec.~\ref{sec:kd}. The event yields, kinematic distributions, and
measured properties are discussed in Secs.~\ref{sec:yields}
through~\ref{sec:results}.

\section{The CMS detector}
\label{sec:detector}

The central feature of the CMS apparatus is a superconducting solenoid
of 6\unit{m} internal diameter, providing a 3.8\unit{T} field. Within
the superconducting solenoid volume are a silicon pixel and strip
tracker, a lead tungstate crystal electromagnetic calorimeter (ECAL),
and a brass/scintillator hadron calorimeter (HCAL). Muons are detected
in gas-ionization detectors embedded in the iron flux return placed
outside the solenoid. Extensive forward calorimetry complements the
coverage provided by the barrel and end-cap detectors. The CMS detector
is described in detail in Ref.~\cite{cms:2008zzk}.

The CMS experiment uses a coordinate system with the origin at the
nominal interaction point, the $x$ axis pointing to the center of the
LHC ring, the $y$ axis pointing up (perpendicular to the LHC ring),
and the resulting $z$ axis along the beam direction using a
right-handed convention. The polar angle $\theta$ is measured from the
positive $z$ axis and the azimuthal angle $\phi$ is measured in the
$x$-$y$ plane in radians. The pseudorapidity is defined as $\eta =
-\ln[\tan(\theta/2)]$.

The inner tracker measures charged particle trajectories within the
range $\abs{\eta}< 2.5$. It consists of 1440 silicon pixel and 15\,148
silicon strip detector modules and is immersed in the magnetic field.
It provides an impact parameter resolution of ${\approx}15\mum$ and a
transverse momentum (\pt) resolution of about 1.5\% for 100\GeV
particles~\cite{Chatrchyan:2009ad,Chatrchyan:2009sr}.

The ECAL consists of 75\,848 lead tungstate crystals and provides
coverage of $\abs{ \eta }< 1.479 $ in the barrel region (EB), and
$1.479 <\abs{ \eta } < 3.0$ in the two end-cap regions (EE).  The EB
uses 23\unit{cm} long crystals with front-face cross sections of
around 2.2\unit{cm} $\times$ 2.2\unit{cm}, while the EE comprises
22\unit{cm} long crystals with front-face cross sections of
2.86\unit{cm} $\times$ 2.86\unit{cm}.  A preshower detector consisting
of two planes of silicon sensors interleaved with a total of 3
radiation lengths of lead is located in front of the EE. The ECAL
energy resolution for electrons with transverse energy $\ET \approx
45$\GeV from the $\cPZ\rightarrow\Pep\Pem$ decays is better than 2\%
in the central region of the EB $(\abs{\eta} < 0.8)$, and is between
2\% and 5\% elsewhere. For low-bremsstrahlung electrons that have 94\%
or more of their energy contained within a $3 \times 3$ array of
crystals, the energy resolution improves to 1.5\% for $\abs{\eta} <
0.8$~\cite{Chatrchyan:2013dga}. The Gaussian resolution of the
dielectron mass distribution for a $\cPZ$-boson sample, when both
electrons belong to this class, is $0.97 \pm 0.01$\GeV in $\sqrt{s} =
7$\TeV data.

The HCAL is a sampling calorimeter with brass as the passive material
and plastic scintillator tiles serving as active material, providing
coverage of $\abs{ \eta }< 2.9 $. The calorimeter cells are grouped in
projective towers of granularity $\Delta \eta \times \Delta \phi =
0.087 \times 0.087$ in the HB (covering $\abs{\eta} < 1.3$) and
$ \Delta \eta \times \Delta \phi \approx 0.17\times 0.17$ in the HE
(covering $1.3 < \abs{\eta} < 2.9$), the exact granularity depending
on $\abs{\eta}$. A hadron forward calorimeter extends the coverage up
to $\abs{ \eta }< 5.2$.

Muons are detected in the pseudorapidity range $\abs{\eta}< 2.4$, with
detection planes made using three technologies: drift tubes,
cathode-strip chambers, and resistive-plate chambers. The global fit
of the muon tracks matched to the tracks reconstructed in the silicon
tracker results in a transverse momentum resolution, averaged over
$\phi^\Pgm$ and $\eta^\Pgm$, from 1.8\% at $\PT^\Pgm$ = 30\GeV to
2.3\% at $\PT^\Pgm$ = 50\GeV~\cite{Chatrchyan:2009sr}.

\section{Simulated data samples}
\label{sec:datasets}

The Monte Carlo (MC) simulated samples, generated with programs based
on state-of-the-art theoretical calculations for both the SM Higgs
boson signal and relevant background processes, are used to optimize
the event selection and to evaluate the acceptance and systematic
uncertainties.
The samples of Higgs boson signal events produced in either gluon fusion ($\Pg\Pg \rightarrow \PH$)
or vector-boson fusion ($\Pq\Pq \rightarrow \Pq\Pq \PH$) processes are generated with the
\POWHEG~\cite{powheg,Bagnaschi:2011tu,Nason:2009ai}
generator at next-to-leading-order (NLO) QCD accuracy. The Higgs boson
decay is modeled with \textsc{jhugen}~3.1.8~\cite{Gao:2010qx,
Bolognesi:2012,Anderson:2013fba} and includes proper treatment of
interference effects associated with permutations of identical leptons
in the four-electron and four-muon final states. Alternative
spin-parity states are also modeled with
\textsc{jhugen}, where production of the spin-0 states is modeled in gluon fusion
with \POWHEG at NLO QCD accuracy. It is also found that NLO QCD
effects relevant for this analysis are approximated well with the
combination of leading-order (LO) QCD matrix elements and parton
showering. Therefore, simulation of spin-1 and spin-2 resonances
is performed in quark-antiquark and gluon fusion production at LO QCD
accuracy, followed by parton showering generated
with \PYTHIA~6.4.24~\cite{Sjostrand:2006za}.

For low-mass Higgs boson hypotheses ($\mH<400$\GeV), the Higgs boson
line shape is described with a Breit-Wigner (BW) distribution. At high
mass ($\mH \ge 400\GeV$), because of the very large Higgs boson width
($\Gamma_{\PH} > 70\GeV$), the line shape is described using the
complex pole scheme
(CPS)~\cite{Passarino:2010qk,Goria:2011wa,Kauer:2012hd}.  The
inclusive cross section for every $\mH$ is computed including
corrections due to the CPS~\cite{Heinemeyer:2013tqa}. The interference
between the Higgs boson signal produced by gluon fusion and the
background from $\Pg\Pg \rightarrow\cPZ\cPZ$ is taken into account, as
suggested in Ref.~\cite{Passarino:2012ri}.  The theoretical
uncertainty in the shape of the resonance due to missing NLO
corrections in the interference between background and signal is
considered, as well as the uncertainties due to electroweak
corrections~\cite{Goria:2011wa, Passarino:2012ri, Kauer:2012ma}.
Samples of \PW\PH, \cPZ\PH, and $\ttbar\PH$ events are generated with
\PYTHIA.  Higgs boson signal events for all the production mechanisms
are reweighted using the generator-level invariant mass, to include
contributions from gluon fusion up to next-to-next-to-leading order
(NNLO) and next-to-next-to-leading logarithm
(NNLL)~\cite{deFlorian:2012mx,Anastasiou:2008tj,deFlorian:2009hc,Baglio:2010ae,LHCHiggsCrossSectionWorkingGroup:2011ti,Djouadi:1991tka,Dawson:1990zj,Spira:1995rr,Harlander:2002wh,Anastasiou:2002yz,Ravindran:2003um,Catani:2003zt,Actis:2008ug},
and from the vector-boson fusion (VBF) contribution computed at NNLO
in
Refs.~\cite{LHCHiggsCrossSectionWorkingGroup:2011ti,Ciccolini:2007jr,Ciccolini:2007ec,Figy:2003nv,Arnold:2008rz,Bolzoni:2010xr}.

The dominant background to the Higgs signal in this channel is the SM
$\cPZ\cPZ$ or $\cPZ\gamma^\ast$ production via $\Pq\Paq$ annihilation
and gluon fusion, which is referred to as $\cPZ\cPZ$ in what follows.
Smaller contributions arise from $\cPZ$ + jets and $\ttbar$ production
where the final states contain two isolated leptons and two
heavy-flavor jets producing secondary leptons.  Additional backgrounds
arise from $\cPZ$ + jets, $\cPZ\gamma$ + jets, $\PW\PW$ + jets, and
$\PW\cPZ$ + jets events, where misidentified leptons can arise from
decays of heavy-flavor hadrons, in-flight decays of light mesons
within jets, and, in the case of electrons, overlaps of $\pi^0$ decays
with charged hadrons.  The $\cPZ\cPZ$ production via $\Pq\Paq$ is
generated at NLO with \POWHEG~\cite{Melia:2011tj}, while the $\PW\PW$,
$\PW\cPZ$ processes are generated with
\MADGRAPH~\cite{Alwall:2007st} and normalized to cross sections
computed at NLO.  The $\Pg\Pg \rightarrow\cPZ\cPZ$ contribution is
generated with \textsc{gg2zz}~\cite{Binoth:2008pr}.  The
$\cPZ\cPqb\cPaqb$, $\cPZ\cPqc\cPaqc$, $\cPZ\Pgg$, and
$\cPZ+\text{light jets}$ samples (referred to as $\cPZ$+ jets in the
following) are generated with \MADGRAPH, comprising inclusive $\cPZ$
production of up to four additional partons at the matrix-element level,
which is normalized to the cross section computed at NNLO.  The
$\ttbar$ events are generated at NLO with \POWHEG.  The event
generator takes into account the internal initial-state and
final-state radiation effects which can lead to the presence of
additional hard photons in an event.  In the case of LO generators,
the {CTEQ6L}~\cite{cteq66} set of parton distribution functions (PDFs)
is used, while the CT10~\cite{Lai:2010vv} set is used for the NLO and
higher-order generators.

All generated samples are processed with \PYTHIA for jet fragmentation
and showering. For the underlying event, the \PYTHIA 6.4.24
tunes {Z2} and Z2*, which rely on $\PT$-ordered
showers, are used for 7 and 8\TeV MC samples,
respectively~\cite{Chatrchyan:2011id}. Events are processed through
the detailed simulation of the CMS detector based on
\GEANTfour~\cite{Agostinelli:2002hh,GEANT} and are reconstructed with
the same algorithms as used for data. The simulations include
overlapping $\Pp\Pp$ interactions (pileup) matching the distribution
of the number of interactions per LHC beam crossing observed in data.
The average number of measured pileup interactions is approximatively
9 and 21 in the 7 and 8\TeV data sets, respectively.

\section{Online event selection}
\label{sec:triggers}

The first level (L1) of the CMS trigger system, composed of custom
hardware processors, uses information from the calorimeters and muon
detectors to select the most interesting events in a time
interval of less than 4\mus. The L1 trigger rate of 100\unit{kHz}
is further reduced by the high-level trigger (HLT) processor farm to
around 300\unit{Hz} before data storage.

Collision events analyzed in this paper are selected by the trigger
system, requiring the presence of two leptons: electrons or muons.
The minimal transverse momenta of the leading and subleading leptons
are 17 and 8\GeV, respectively, for both electrons and muons.  The
online selection includes double-electron, double-muon and mixed
electron-muon triggers. In the case of the $4\Pe$ final state, a
triple-electron trigger is added with thresholds of 15, 8, and 5\GeV
to increase the efficiency for low-$\PT$ electrons.  The trigger
efficiency for events within the geometrical acceptance of this
analysis is greater than 98\% for a Higgs boson signal with $\mH >
110\GeV$. The same trigger paths are applied on the 7 and 8\TeV data,
whereas different identification criteria are applied on the HLT
lepton candidates to account for the different LHC conditions.

In addition to the events selected to form the four-lepton sample,
dedicated triggers are used for lepton calibration and efficiency
measurements. In the case of dimuon events, the online trigger
algorithms used to select the signal events are sufficiently loose
that they can also be used to measure the selection efficiency with
the $\cPZ\to\Pgmp\Pgmm$ events. In order to measure the selection
efficiency of events with low-\pt leptons, low-mass resonances are
used. Events corresponding to these low-mass resonances are collected
in the dimuon case using dedicated triggers that require an
opposite-sign muon pair, with dedicated kinematic conditions on the
dimuon system.  In the case of electrons, low-mass resonances are
collected, with a smaller rate, with standard dielectron triggers.
Two specialized triggers are introduced to maximize the number of
$\cPZ\to\Pep\Pem$ events covering both high- and low-$\PT$ ranges.
The one having the most stringent (relaxed) identification and
isolation requirement on one electron requires the presence of a
cluster in the electromagnetic calorimeter with $\PT > 8\,$(17)\GeV,
forming an invariant mass with the other electron exceeding 50\GeV.

\section{Lepton reconstruction and selection}
\label{sec:reconstruction}

The analysis is performed by reconstructing a $\cPZ\cPZ$ system
composed of two pairs of same-flavor and opposite-charge isolated
leptons, $\Pep\Pem$ or $\Pgmp\Pgmm$.  The main background sources,
described in Sec.~\ref{sec:datasets}, are the SM $\cPZ\cPZ$
production, with smaller contributions from other diboson ($\PW\PW$,
$\PW\cPZ$) processes, single bosons with hadronic activity that can
mimic lepton signatures, and top-quark-pair events.  Given the very
low branching fraction of the $\PH\rightarrow\cPZ\cPZ\rightarrow
4\ell$ decay, of $\mathcal{O}(10^{-4})$ [$\mathcal{O}(10^{-3})$] for
$\mH = 125\,(200)\GeV $\cite{Dittmaier:2012vm}, it is important to
maintain a very high lepton selection efficiency in a wide range of
momenta, to maximize the sensitivity for a Higgs boson within the mass
range 110--1000\GeV.

The signal sensitivity also depends on the $4\ell$ invariant mass
resolution. The signal appears as a narrow resonance on top of a
smooth background, and therefore it is important to achieve the best
possible four-lepton mass resolution.  To obtain a precise measurement
of the mass of a resonance decaying into four leptons, it is crucial
to calibrate the individual lepton momentum scale and resolution to a
level such that the systematic uncertainty in the measured value of
$\mH$ is substantially smaller than the statistical uncertainty in the
current data set.  This section describes the techniques used in the
analysis to select electrons and muons in order to achieve the best
momentum resolution, measure the momentum scale, resolution, and
selection efficiency, and derive corrections based on dilepton
resonances.

The CMS particle flow (PF)
algorithm~\cite{CMS-PAS-PFT-09-001,CMS-PAS-PFT-10-001,CMS-PAS-PFT-10-002,CMS-PAS-PFT-10-003},
which combines information from all subdetectors, is used to provide
an event description in the form of reconstructed particle
candidates. The PF candidates are then used to build higher-level
objects, such as jets, missing transverse energy, and lepton isolation
quantities.

\subsection{Electron reconstruction and identification}
\label{sec:electrons}

Electron candidates are required to have a transverse momentum
$\PT^{\Pe} > 7\GeV$ and be within the geometrical acceptance, defined
by $\abs{\eta^{\Pe}} < 2.5$.  The electron reconstruction combines
information from the ECAL and the
tracker~\cite{Baffioni:2006cd,CMS-PAS-EGM-10-004,CMS_DPS_2011-003,CMS_DP_2013-003}.
Electron candidates are formed from arrays of energy clusters in the
ECAL (called superclusters) along the $\phi$ direction, which are
matched to tracks in the silicon tracker.  Superclusters, which
recover the energy of the bremsstrahlung photons emitted in the
tracker material and of some of the nearly collinear final-state
radiation (FSR) from the electron, are also used to identify hits in
the innermost tracker layers in order to initiate the reconstruction
of electron tracks.  This track seeding procedure is complemented by
an approach based on tracker seeds which improves the reconstruction
efficiency at low $\PT^\Pe$ and in the transition between the EB and
EE regions.  Trajectories, when initiated outside-in from the
ECAL superclusters as well as inside-out from the measurements in the
innermost tracker layers, are reconstructed using the Gaussian sum
filter (GSF) algorithm~\cite{Adam:2003kg}, which accounts for the
electron energy loss by bremsstrahlung.  Additional
requirements~\cite{Chatrchyan:2013dga} are applied in order to reject
electrons originating from photon conversions in the tracker
material. Electron candidates are selected using loose criteria on
track-supercluster matching observables that preserve the highest
possible efficiency while removing part of the QCD background.

Electron identification relies on a multivariate discriminant that
combines observables sensitive to the bremsstrahlung along the
electron trajectory, and the geometrical and momentum-energy matching
between the electron trajectory and the associated supercluster, as
well as ECAL shower-shape observables.  The multivariate discriminant
is trained using a sample of ${\approx}10^7$ simulated Drell-Yan
events for the signal (true electrons) and a high-purity $\PW$ + 1 jet
data sample for the background (misidentified electrons from
jets). The expected performance is validated using jets misidentified
as electrons in a $\cPZ(\to\Pgmp\Pgmm)$ and $\cPZ(\to\Pep\Pem)$ data
sample, with exactly one reconstructed electron not originated from
the $\cPZ$ boson decay. The sources of prompt electrons, such as
dibosons or $\ttbar$ decays are suppressed with appropriate selections
on the number of extra leptons and the presence of small missing
transverse energy in the event~\cite{CMS_DP_2013-003}. The selection
of the $\cPZ$ boson is the same as the one used in the analysis, so
the $\eta^\Pe$ and $\PT^\Pe$ spectrum is similar to the one for the
electrons characterizing the reducible background in the analysis. The
selection is optimized in six regions of the electron $\PT^\Pe$ and
$\abs{\eta^\Pe}$ to maximize the expected sensitivity for a low-mass
Higgs boson.  These regions correspond to two $\pt^\Pe$ ranges,
7--10\GeV and $>$10\GeV, and three pseudorapidity regions,
corresponding to two regions in the EB with different material in
front of the ECAL, the central barrel ($\abs{\eta^\Pe}<0.8$) and the
outer barrel ($0.800<\abs{\eta^\Pe}<1.479$), in addition to the EE,
$1.479<\abs{\eta^\Pe}<2.500$.

Several procedures are used to calibrate the energy response of
individual crystals~\cite{CMS_DPS_2012-015,Chatrchyan:2013dga}.  The
energy of the ECAL superclusters is corrected for the imperfect
containment of the clustering algorithm, the electron energy not
deposited in the ECAL, and leakage arising from showers near gaps
between crystals or between ECAL modules.  This is done using a
regression technique based on boosted decision trees (BDT)~\cite{tmva}
trained on a simulated dielectron sample with the pileup conditions
equivalent to the ones measured on data, covering a flat spectrum in
$\pt^\Pe$ from 5 to 100\GeV.  The variables include the electron
supercluster raw energy, $\eta$ and $\phi$ coordinates, several
shower-shape variables of the cluster with largest energy within the
supercluster (the seed cluster), the ratio of the energy in the HCAL
behind the seed cluster to the seed cluster energy, and the number of
clusters in the electron supercluster. In addition, the distance of
the seed crystal with respect to the gap between the ECAL modules, the
$\eta$ and $\phi$ coordinates of the seed cluster, and the energies of
the first three subleading clusters in the supercluster are used.  A
similar subset of variables is used depending on whether the electron
is detected in the EB or EE.  Using this multivariate technique, the
effective width and Gaussian resolution of the reconstructed invariant
mass are improved by 25\% and 30\%, respectively, for simulated
$\PH\to4\Pe$ decays compared to those obtained with a more traditional
approach based on ECAL-only energy measurements and corrections with a
parameterized energy response obtained from simulation.  The effective
width, $\sigma_\text{eff}$, is defined as the half-width of the
smallest interval that contains 68.3\% of the distribution.

\begin{figure}[!htb]
\begin{center}
\includegraphics[width=\cmsFigWidthStd]{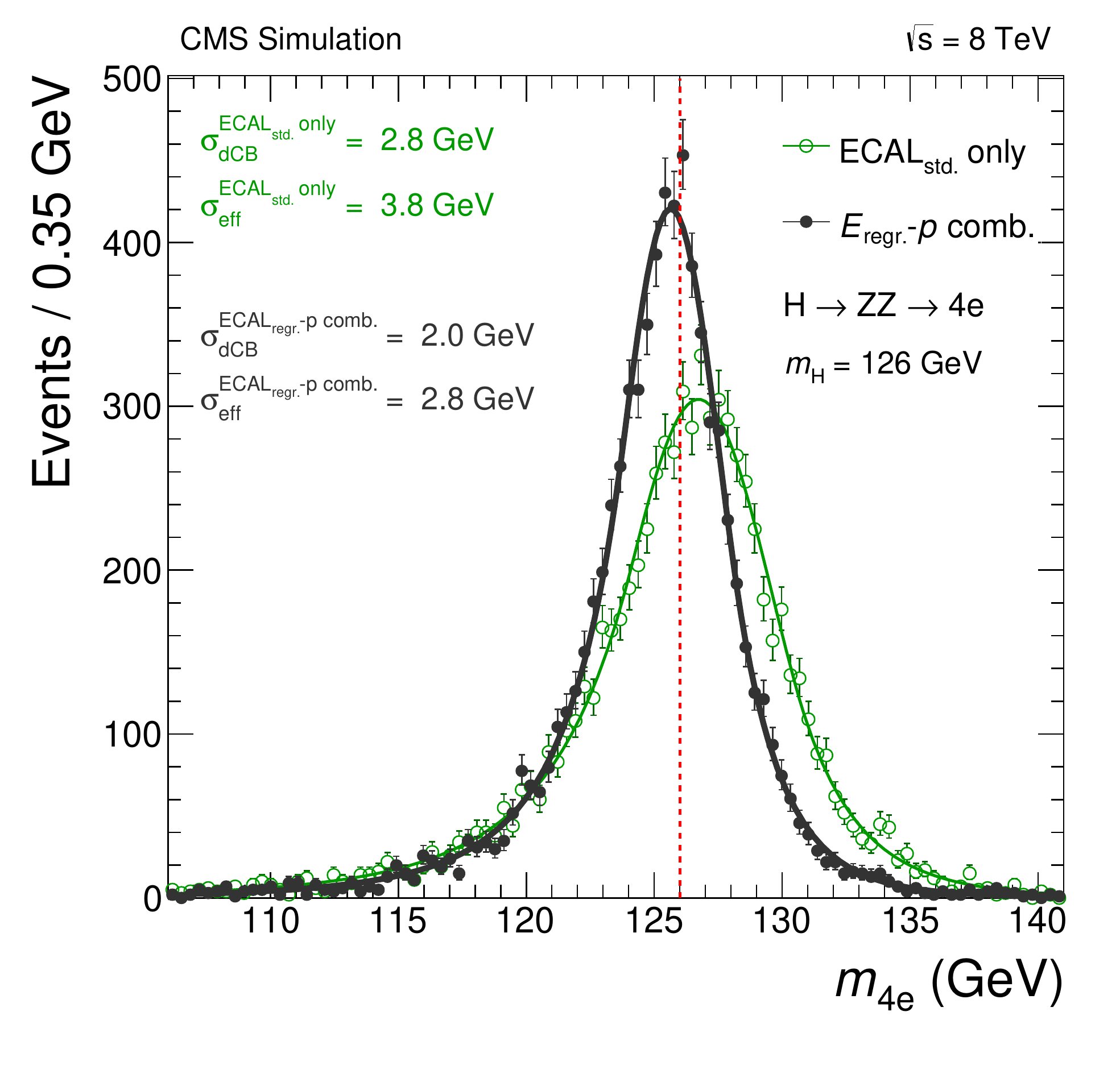}
\includegraphics[width=\cmsFigWidthStd]{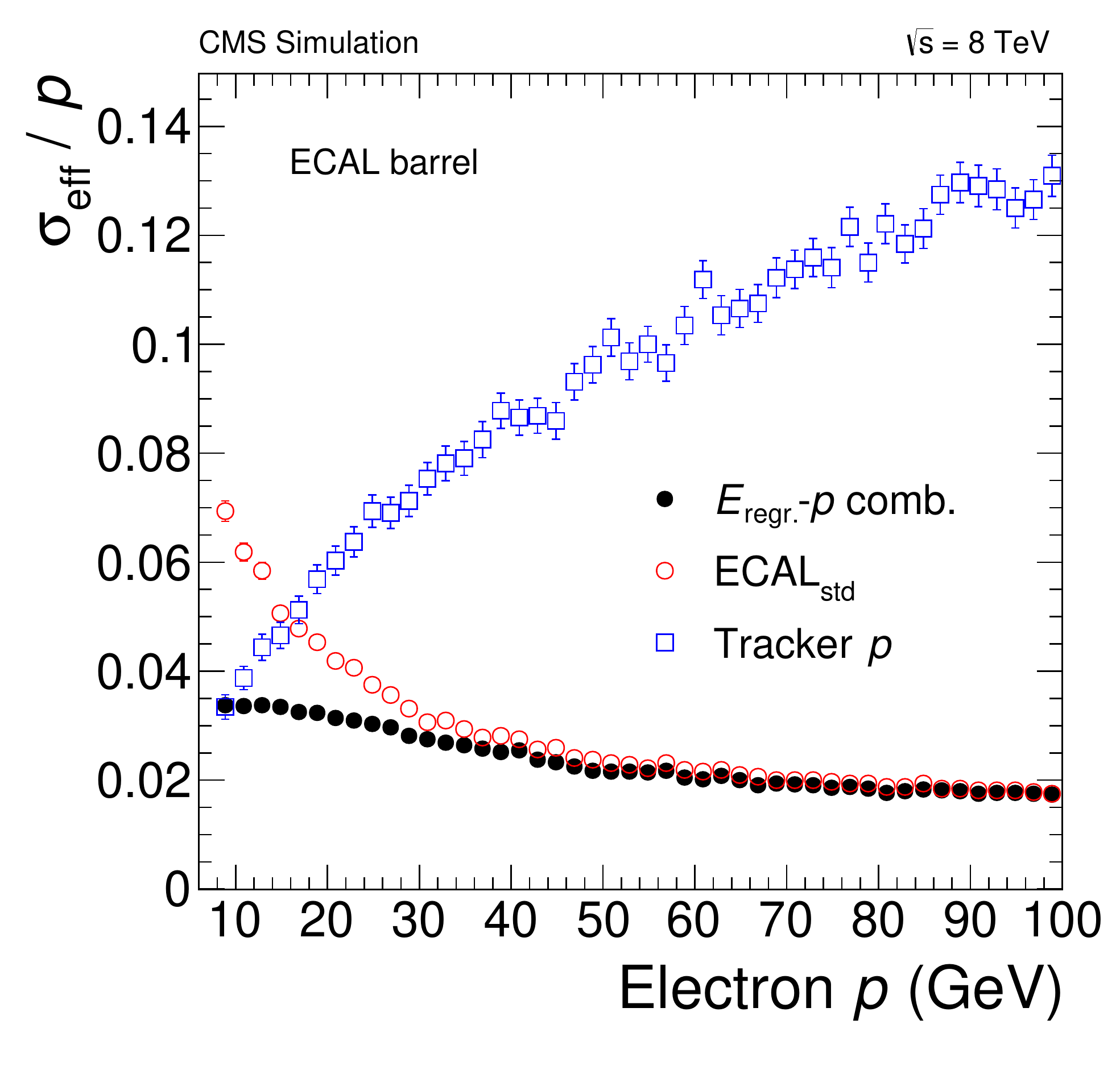}
\caption{(\cmsLeft) Expected four-lepton mass distribution for
  $\PH\to\cPZ\cPZ\to 4\Pe$ for $\mH = 126\GeV$ using ECAL-only
  electron momentum estimation (green open points:
  $\mathrm{ECAL}_\text{std.}$ only), and using the method employed in
  this analysis (black full points: $E_\text{regr}-p$ combination).
  The fitted standard deviation, $\sigma_\mathrm{dCB}$, of the
  double-sided Crystal-Ball~\cite{CrystalBall} function and effective
  width $\sigma_\text{eff}$ defined in the text are indicated.
  Electrons with $\PT^\Pe > 7$\GeV in the full $\eta^\Pe$ range are
  used.  (\cmsRight) Expected effective momentum resolution
  $\sigma_\text{eff}/p$ for electrons in the EB as a function of the
  momentum for the ECAL-only, the tracker-only, and the combined
  estimates. ~\label{fig:elereso}}
\end{center}
\end{figure}

The precision of the electron momentum measurement is dominated by the
ECAL at high energies, whereas for low-$\pt$ electrons the precision
is dominated by the tracker momentum determination.  Moreover, for
electrons near poorly instrumented regions, such as the crack between
the EB and the EE, the intermodule cracks~\cite{Chatrchyan:2009qm}, or
regions close to dead channels, the measurement accuracy and
resolution can also be improved by combining the ECAL energy with the
track momentum.  To account for biases arising from bremsstrahlung
losses in the tracker material, electron categories are defined based
on the cluster multiplicity inside the supercluster as well as on the
amount of bremsstrahlung as estimated from the GSF. The magnitude of
the electron momentum is then determined by combining the two
estimates with a multivariate regression function that takes as input
the corrected ECAL energy from the supercluster regression, the track
momentum estimate, their respective uncertainties, the ratio of the
corrected ECAL energy over the track momentum as obtained from the
track fit, the uncertainty in this ratio, and the electron category,
based on the amount of bremsstrahlung. The direction is taken from the
fitted track parameters at the point of closest approach to the
nominal beam spot position.  Figure~\ref{fig:elereso}~(\cmsLeft) shows
the reconstructed invariant mass for $\PH\to4\Pe$ decays, compared to
the traditional approach for the electron energy estimation. The
residual offset in the peak position [$<$0.2\%, black histogram in
Fig.~\ref{fig:elereso}~(\cmsLeft)] is irrelevant for the analysis,
because the absolute electron momentum scale is calibrated using known
resonances in data, as described in
Sec.~\ref{sec:calibrations}. Figure~\ref{fig:elereso}~(\cmsRight)
presents the expected effective resolution of the combined momentum
measurement as a function of the electron momentum at the vertex.  The
expected effective momentum resolution for the ECAL-only and
tracker-only estimates are also shown.

\subsection{Muon reconstruction and identification}
\label{sec:muons}

Muon candidates are required to have a transverse momentum $\PT^{\Pgm}
> 5\GeV$ and be within the geometrical acceptance, defined by
$\abs{\eta^{\Pgm}} < 2.4$.  The reconstruction combines information
from both the silicon tracker and the muon system.  The matching
between track segments is done either outside-in, starting from a
track in the muon system, or inside-out, starting from a track in the
silicon tracker. Both these candidates are referred to as global
muons.  Very low-\PT muons ($\PT^\Pgm\lesssim$ 5\GeV) may not have
sufficient energy to penetrate the entire muon system and leave track
segments in one or two stations of the muon system, where a station is
composed of multiple detection planes between two iron layers. Tracks
matched to such segments form so-called tracker muon objects. More
details on muon reconstruction in CMS can be found in
Ref.~\cite{Chatrchyan:2012xi}.  Both global and tracker muons are used
in this analysis.

The muons are selected among the reconstructed muon track candidates
by applying minimal requirements on the track segments in both the
muon system and inner tracker system and taking into account
compatibility with small energy deposits in the
calorimeters~\cite{Chatrchyan:2012xi,CMS-PAS-PFT-10-003}.

The $\PT$ resolution for muons in the momentum range relevant for this
analysis varies between 1.3\% and 2.0\% in the barrel, and up to 6\%
in the end caps.  The dominant effect determining this resolution is
the multiple scattering of muons in the tracker material.

The achieved statistical accuracy on the determination of the position
of the tracker modules is generally better than 10\mum, reaching a
level of $\leq$ 2\mum in the pixel tracker. Besides cosmic ray tracks,
the usage of resonance mass and vertex information in the alignment
procedure successfully constrains systematic deformations of the
geometry that could bias reconstructed track
parameters~\cite{Chatrchyan:2014wfa}.

The accuracy of the hit measurements in the muon chambers and the
overall alignment contribute to a lesser degree to the momentum
measurement.  This is achieved using several alignment procedures
using cosmic muons, optical surveys, a laser system, and, finally,
$\cPZ\to\Pgmp\Pgmm$ events.

\subsection{Lepton isolation and vertex compatibility}
\label{sec:isosip}

Lepton isolation is used to discriminate leptons originating from
high-$\PT$ boson decay, as in the case of the signal, from those
arising from hadronic processes, which are typically immersed in a jet
of other hadrons.

The isolation of individual leptons, measured relative to their
transverse momentum $\PT^{\ell}$, is defined by:

\ifthenelse{\boolean{cms@external}}{
\begin{equation}\begin{split}
\label{eqn:pfiso}
R_\text{Iso}^{\ell} \equiv& \Big( \sum \PT^\text{charged} +
                                 \max\big[ 0, \sum \PT^\text{neutral}
                                 +\\
                                  &\sum \PT^{\Pgg}
                                 - \PT^\mathrm{PU}(\ell) \big] \Big)
                                 / \PT^{\ell},
\end{split}
\end{equation}
}{
\begin{equation}
\label{eqn:pfiso}
R_\text{Iso}^{\ell} \equiv \Big( \sum \PT^\text{charged} +
                                 \max\big[ 0, \sum \PT^\text{neutral}
                                 +
                                  \sum \PT^{\Pgg}
                                 - \PT^\mathrm{PU}(\ell) \big] \Big)
                                 / \PT^{\ell},
\end{equation}
} where the sums are over charged and neutral PF candidates in a cone
$\Delta R = \sqrt{\smash[b]{(\Delta\eta)^{2} + (\Delta\phi)^{2}}} <
0.4$ around the lepton direction at the interaction vertex, where
$\Delta\eta=\eta^\ell-\eta^i$ and $\Delta\phi=\phi^\ell-\phi^i$
quantify the angular distance of the PF candidate $i$ from the lepton
$\ell$ in the $\eta$ and $\phi$ directions, respectively.  In
Eq.~(\ref{eqn:pfiso}), $\sum \PT^\text{charged}$ is the scalar sum of
the transverse momenta of charged hadrons originating from the chosen
primary vertex of the event. The primary vertex is selected to be the
one with the highest sum of $\PT^2$ of associated tracks.  The sums
$\sum \PT^\text{neutral}$ and $\sum \PT^{\Pgg}$ are the scalar sums of
the transverse momenta for neutral hadrons and photons, respectively.
The latter excludes photons that are candidates for final-state
radiation from the lepton, as defined in Sec.~\ref{sec:photons}.  The
contribution from pileup ($\PT^\text{PU}(\ell)$) in the isolation cone
is subtracted from $R_\text{Iso}^{\ell}$ with different techniques for
electrons and muons. For electrons, the \textsc{FastJet}
technique~\cite{Cacciari:2007fd,Cacciari:2008gn,Cacciari:2011ma} is
used, in which $\PT^\mathrm{PU}(\Pe) \equiv \rho \times A_\text{eff}$,
where the effective area, $ A_\text{eff}$, is the geometric area of
the isolation cone scaled by a factor that accounts for the residual
dependence of the average pileup deposition on the electron
$\eta^\Pe$.  The variable $\rho$ is defined as the median of the
energy-density distribution for the neutral particles within the area
of any jet in the event, reconstructed using the \kt clustering
algorithm~\cite{Catani:kt,Ellis:kt} with distance parameter $D=0.6$,
with $\PT^\text{jet} > 3$\GeV and $\abs{\eta}<2.5$. For muons,
$\PT^\mathrm{PU}(\Pgm) \equiv 0.5 \times \sum_i \PT^{\mathrm{PU}, i}$,
where $i$ runs over the momenta of the charged hadron PF candidates
not originating from the primary vertex. The factor 0.5 in the sum
corrects for the different fraction of charged and neutral particles
in the isolation cone.  The electrons or muons are considered isolated
if $ R_\text{Iso}^{\ell} < 0.4 $. The isolation requirement has been
optimized to maximize the discovery potential in the full $\mH$ range
of this analysis.

In order to suppress leptons originating from in-flight decays of
hadrons and muons from cosmic rays, all leptons are required to come
from the same primary vertex. This is achieved by requiring $\sip<4$,
where $\sip \equiv
{\mathrm{IP}_\mathrm{3D}}$/$\sigma_{\mathrm{IP}_\mathrm{3D}}$ is the
ratio of the impact parameter of the lepton track
(${\mathrm{IP}_\mathrm{3D}}$) in three dimensions (3D), with respect
to the chosen primary vertex position, and its uncertainty.
\subsection{Lepton momentum scale, resolution and selection efficiency}
\label{sec:calibrations}

The determination of the momentum differs for electrons and muons, and
it depends on the different CMS subdetectors involved in their
reconstruction. The CMS simulation used in this analysis is based on
the best knowledge of the detector conditions, as encoded in the ECAL
calibrations and tracker and muon system alignment. Nevertheless,
small discrepancies between data and simulation remain. In the case of
the electron momentum scale and resolution, the main sources of
discrepancy are the residual tracker misalignment and the imperfect
corrections at the crystal level of the transparency loss due to
irradiation, especially in the forward region.  The average measured
drop in energy response, before the crystal calibrations, is about
2\%--3\% in the barrel, rising to 20\% in the range 2.1
$\le\abs{\eta^\Pe}\le$ 2.5~\cite{Chatrchyan:2013dga}, and it is
reduced to a subpercent level after the calibrations.  In the case of
muons, the momentum determination is affected by the tracker and muon
system alignment geometry used for the reconstruction. The
misalignment of the tracker causes a dependence of the systematic
uncertainties in the reconstructed muon momentum on the $\eta^\Pgm$,
$\phi^\Pgm$, and charge measurements.

The momentum scale and resolution for electrons and muons are studied
using different data control samples for different $\PT^\ell$
ranges. In the range of interest for this analysis ($\PT^\ell<
100\GeV$), the dileptons from decays of the $\PJGy$, $\UpsNs$ and
$\cPZ$ resonances are used to calibrate or validate the momentum scale
and measure the momentum resolution. The $\PJGy$ and $\UpsNs$ decays
constitute a clean data source of low-$\PT$ electrons and muons and
are used to validate (calibrate) the electron (muon) momentum scale
for $\PT^\ell < 20\GeV$.  The $\cPZ\to\ell^+\ell^-$ decay mode is a
copious and pure source of leptons, with a wide momentum range
covering the full spectrum of leptons of interest to this analysis.
Table~\ref{tab:calibstat} provides the approximate number of dilepton
resonance decays reconstructed in the 7 and 8\TeV data used for the
calibration of the lepton momentum.

\begin{table}[!htbp]
  \begin{center} \topcaption{ Number of $\cPZ\to\ell^+\ell^-$,
    $\PJGy\to\ell^+\ell^-$ and $\UpsNs\to\ell^+\ell^-$ [sum of
        $\PgUa$, $\PgUb$ and $\PgUc$] used to calibrate or validate
      lepton momentum scale and resolution and to measure lepton
      efficiencies ($\cPZ\to\ell^+\ell^-$ only) in 7 and 8\TeV
      data. Low mass dimuon resonances are collected with specialized
      triggers.  \label{tab:calibstat}}
    \begin{scotch}{lccc}
    $\ell$   & $\cPZ\to\ell^+\ell^-$ & $\PJGy\to\ell^+\ell^-$ & $\UpsNs\to\ell^+\ell^-$ \T \\
    $\Pe$    & $10^7$                & $5\times10^3$          & $2.5\times10^4$ \T \\
    $\Pgm$   & $1.4\times10^7$       & $2.7\times10^7$        & $1.5\times10^7$ \T \B \\
    \end{scotch} \end{center}
\end{table}

For electrons, the calibration procedure consists of three steps.
First, a set of corrections for the momentum scale is obtained by
comparing the displacement of the peak position in the distributions
of the $\cPZ$-boson mass in the data and in the simulation in
different $\eta$ regions and in two categories depending on the amount
of bremsstrahlung. The corrections are derived as a function of time
in order to account for the time-dependent crystal transparency
loss~\cite{Chatrchyan:2013dga}.  Second, a linearity correction to the
momentum scale is applied to account for the $\PT$-dependent
differences between data and simulation by comparing the dielectron
mass distributions, binned in $\PT^\Pe$ of one of the two electrons,
in data and in simulated $\cPZ\to\Pep\Pem$ events. The
$\PJGy\to\Pep\Pem$ and $\PgUa\to\Pep\Pem$ events are used as
validation for electron $\pt^\Pe<20$\GeV. All the corrections on the
electron momentum scale from the first two steps are applied to
data. Third, the energies of single electrons in the simulation are
smeared by applying a random Gaussian multiplicative factor of mean 1
and width $\Delta \sigma$, in order to achieve the resolution observed
in the data $\cPZ$-boson sample.

For muons, an absolute measurement of momentum scale and resolution is
performed by using a reference model of the $\cPZ$ line shape
convolved with a Gaussian function.  The bias in the reconstructed
muon $\PT$ is determined from the position of the $\cPZ$ mass peak as
a function of muon kinematic variables, and a correction is derived
for the data according to the procedure of
Ref.~\cite{Chatrchyan:2012xi}.  A correction for the resolution is
also derived for the simulation from a fit to the $\cPZ\to\Pgmp\Pgmm$
mass spectrum. The large event sample based on low-mass dimuon
resonances provides an additional calibration source for the momentum
resolution in a similar manner.

After this calibration, the lepton momentum scale and resolution are
validated in data using dileptons from $\PJGy$, $\UpsNs$ and $\cPZ$
decays in several bins of lepton $\eta^\ell$ and $\PT^\ell$ in order
to cover the full momentum range relevant for the $\PH \rightarrow
\cPZ\cPZ \rightarrow 4\ell$ search. Electrons with $\PT^\Pe > 7$\GeV
and muons with $\PT^\mu > 5$\GeV are considered. For the selection of
$\cPZ \to \ell^+\ell^-$ events, all lepton selection criteria are
applied as in the $\PH \to \cPZ\cPZ \to 4\ell$ analysis.

\begin{figure}[!htb]
  \begin{center}
     \includegraphics[width=\cmsFigWidthStd]{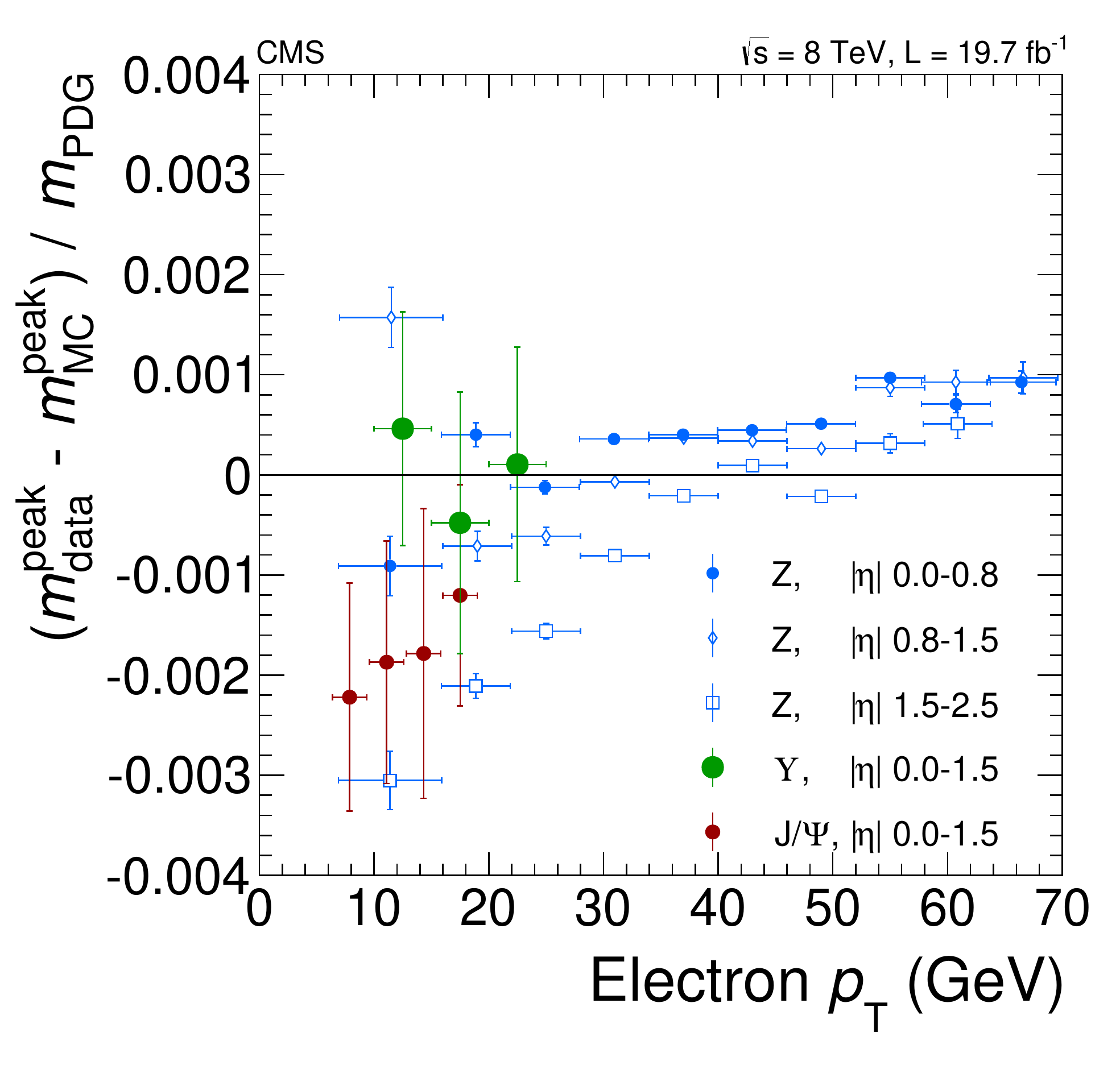}
     \includegraphics[width=\cmsFigWidthStd]{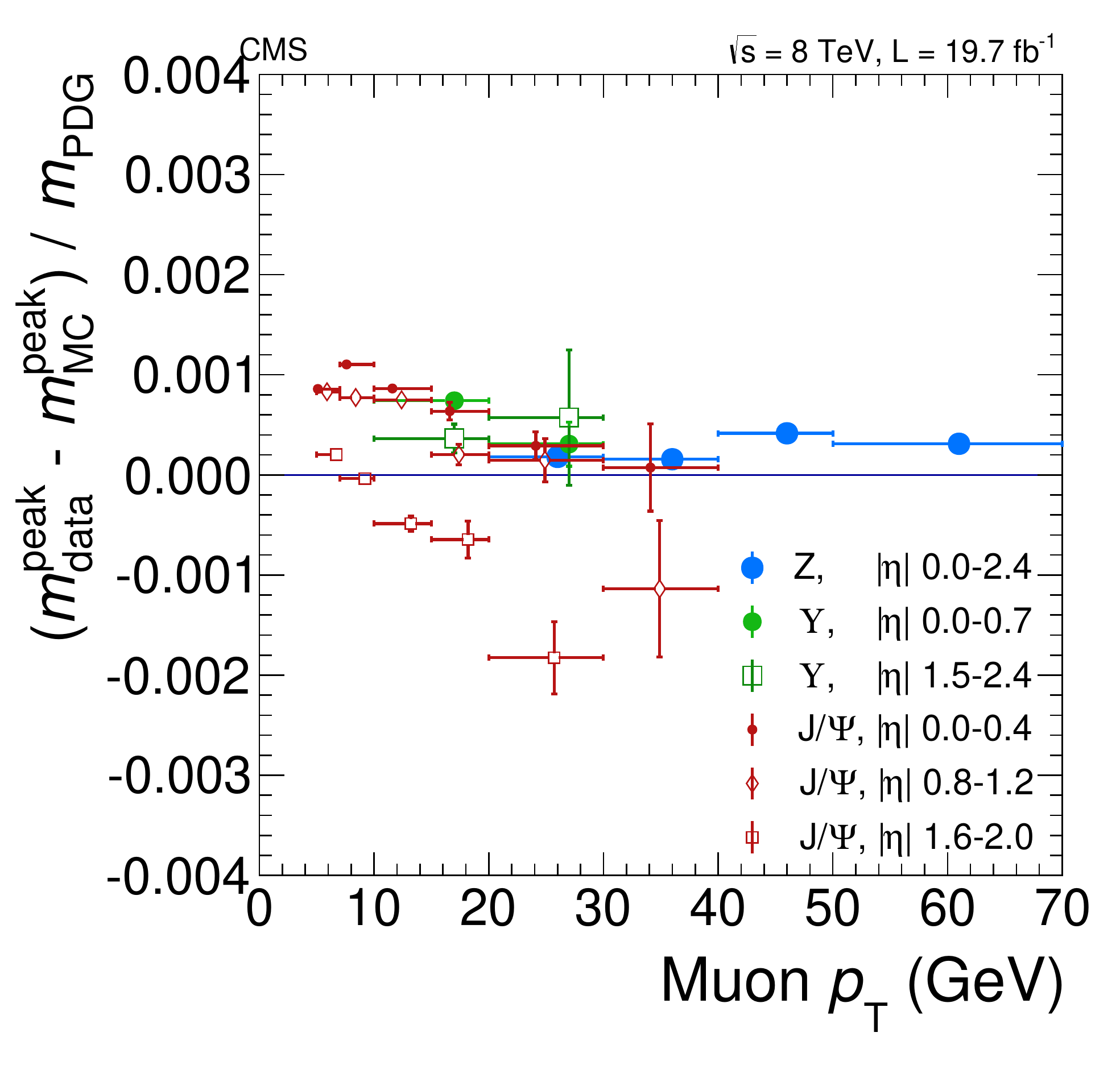}
    \caption{Relative difference between the dilepton mass peak
      positions in data and simulation as obtained from $\cPZ$,
      $\PJGy$ and $\UpsNs$ resonances as a function of (\cmsLeft) the
      transverse momentum of one of the electrons regardless of the
      second for dielectron events, and (\cmsRight) the average muon
      $\PT^\Pgm$ for dimuon events for the 8\TeV
      data. \label{fig:lepton_scale}}
  \end{center}
\end{figure}

The events are separated into categories according to the $\PT^\Pe$
and $\abs{\eta^\Pe}$ of one of the electrons, integrating over the
other, while for dimuons, the average $\PT^\Pgm$ and
$\vert\eta^\Pgm\vert$ are used. The dilepton mass distributions in
each category are fitted with a BW parameterization convolved with a
single-sided Crystal-Ball (CB) function~\cite{CrystalBall} [dimuon
resonances or dielectron $\PJGy$ and $\PgUa$] or with MC templates
($\cPZ\to\Pe\Pe$). From these fits, the offset in the measured peak
position in data with respect to the nominal $\cPZ$ mass, $\Delta
m_{\text{data}}=m^\text{peak}_\text{data}-m_{\cPZ}$, with respect to
that found in the simulation, $\Delta
m_{\mathrm{MC}}=m^\text{peak}_\mathrm{MC}-m_{\cPZ}$, is extracted.
Figure~\ref{fig:lepton_scale} shows the relative difference between
data and simulation of the dilepton mass scale.
After the electron calibration, the relative momentum scale between
data and simulation is consistent within 0.2\% in the central barrel
and up to $\approx$0.3\% in the forward part of the ECAL end caps.
The residual dependence at low momentum is due to the use of wide bins
in measured electron $\pt^\Pe$ in evaluating the $\cPZ$-peak mass
shift. The measured $\pt^\Pe$ dependence of the momentum scale before
the $\PT^\Pe$ linearity correction, up to 0.6\% in the central barrel
and up to 1.5\% in the end cap, is propagated to the reconstructed
four-lepton mass from simulated Higgs boson events. The resulting
shift of 0.3\% (0.1\%) for the $4\Pe$ ($2\Pe2\Pgm$) channel is
assigned as a systematic uncertainty in the signal mass scale. For
muons, the agreement between the observed and simulated mass scales is
within $0.1\%$ in the entire pseudorapidity range of interest. A
somewhat larger offset is seen for $\PJGy$ events with two
high-$\PT^\Pgm$ muons in the very forward region. However, for these
events, the muons are nearly collinear and such a kinematic
configuration is very atypical for the $\PH \to \cPZ\cPZ \to 4\ell$
events. Hence, the observed larger mass scale offset for such events
is irrelevant in the context of this analysis.

\begin{figure}[!htb]
  \centering
  \includegraphics[width=\cmsFigWidthStd]{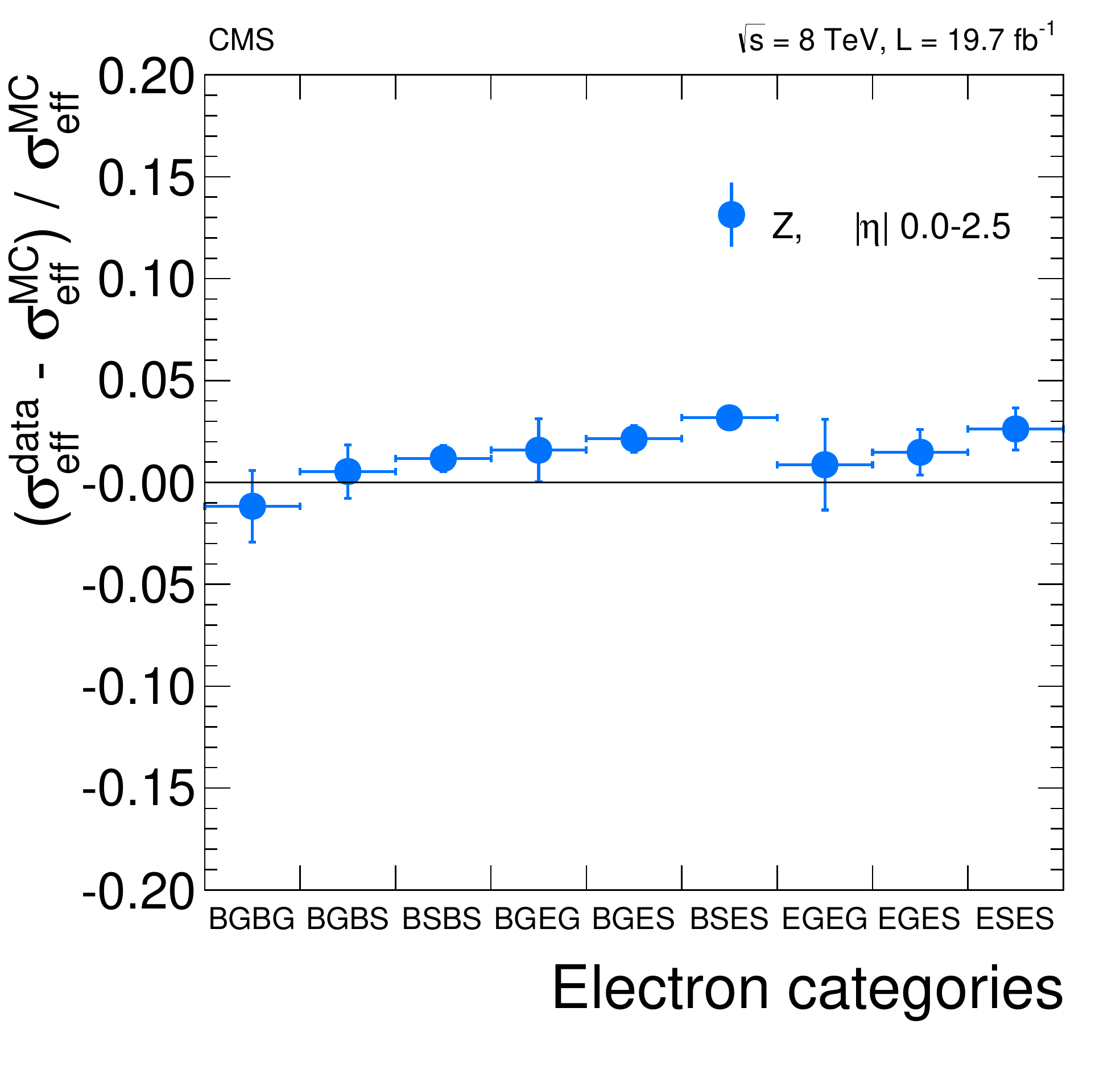}
  \includegraphics[width=\cmsFigWidthStd]{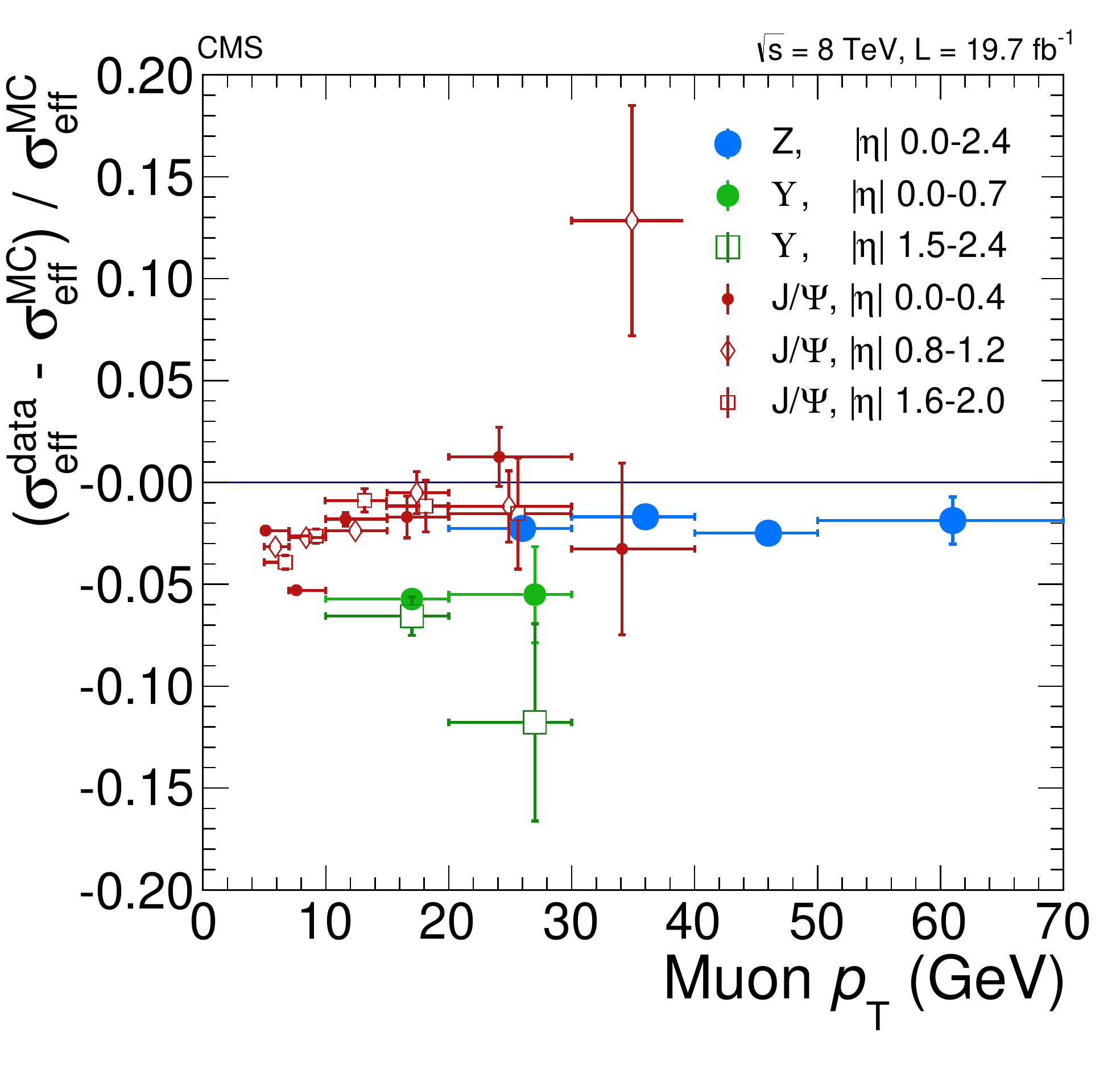}
  \caption{(\cmsLeft) Relative difference between the
  dielectron $\sigma_\text{eff}$ in data and simulation, as measured
  from $\cPZ\to\Pep\Pem$ events, where the electrons are classified
  into different categories (B: barrel, E: end caps, G: golden, S:
  showering). (\cmsRight) Relative difference between the dimuon mass
  resolutions in data and simulation as measured from~$\PJGy$,
  $\UpsNs$, and $\cPZ$ decays as functions of the average muon
  $\PT^\Pgm$.  The uncertainties shown are statistical only. Results
  are presented for data collected at $\sqrt{s}= 8$\TeV. \label{fig:lepton_reso}}

\end{figure}

Similarly, the widths of the peak due to instrumental resolution in
data, $\sigma_{\text{data}}$, and in the simulation,
$\sigma_{\mathrm{MC}}$, are compared. For electrons,
$\sigma_\text{eff}$ ranges from 1.2\% for the best category, which
consists of two central single-cluster electrons with a small amount
of bremsstrahlung [``barrel golden'' (BG)~\cite{Baffioni:2006sk}], to
4\% for the worst category, which consists of two electrons either
with multiple clusters or with a high amount of bremsstrahlung, one
central and one forward [``barrel showering'' (BS) and ``end cap
showering'' (ES)~\cite{Baffioni:2006sk}]. The amount of energy lost by
bremsstrahlung before the electron reaches the ECAL is estimated with
the GSF algorithm.  The relative difference in $\sigma_\text{eff}$
between data and simulation is less than 3\%, for different electron
categories [Fig.~\ref{fig:lepton_reso}~(\cmsLeft)]. For the muons, in
the whole kinematic range considered for this analysis, the
instrumental \cPZ-peak mass resolution observed in data is consistent
with that in the simulation within about 5\%, when not considering
$\PJGy$ events with two high-$\PT^\Pgm$, high-$\vert\eta^\Pgm\vert$
muons [Fig.~\ref{fig:lepton_reso} (\cmsRight)].

\begin{figure}[!htb]
  \begin{center}
      \includegraphics[width=\cmsFigWidthStd]{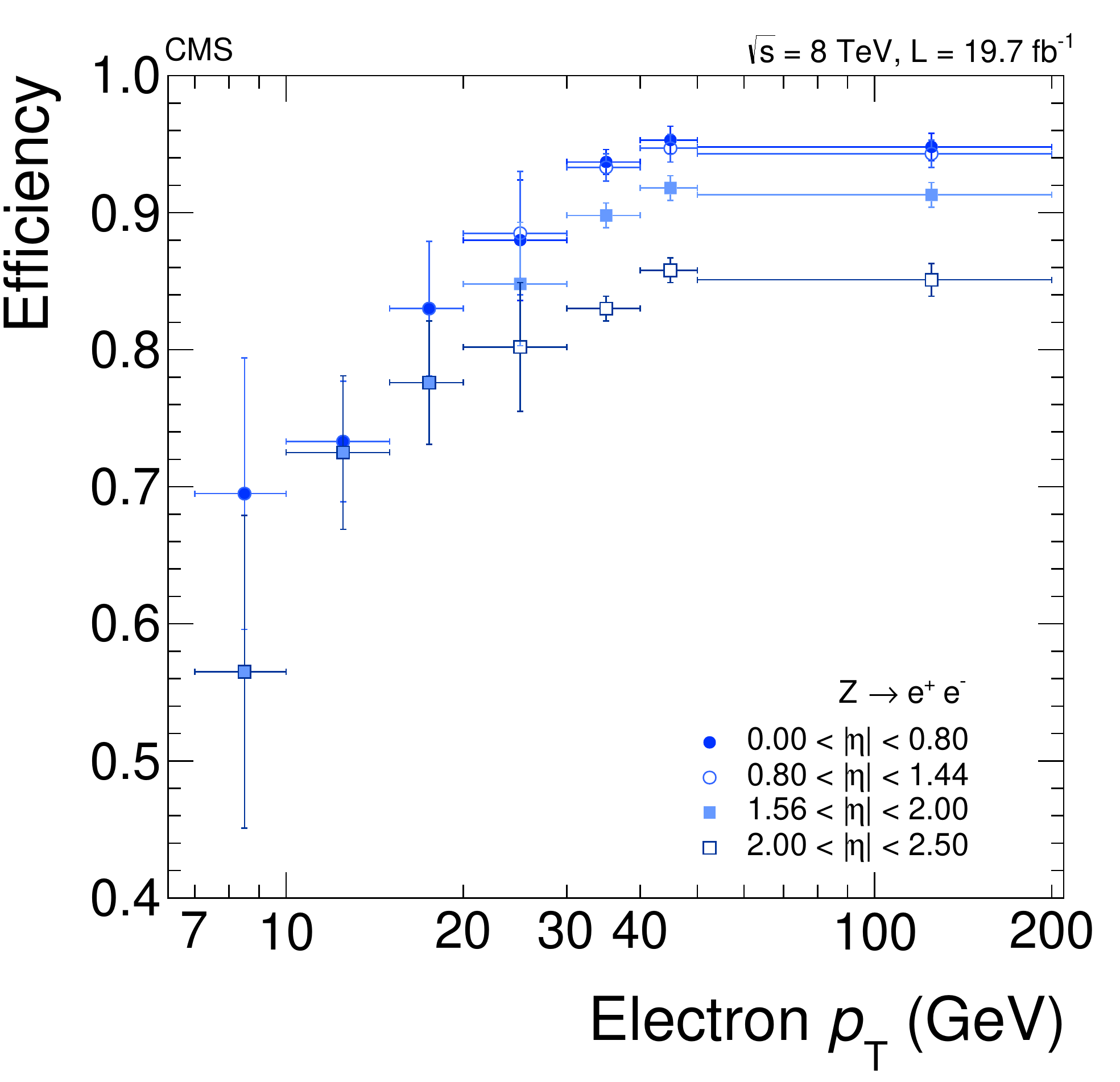}
      \includegraphics[width=\cmsFigWidthStd]{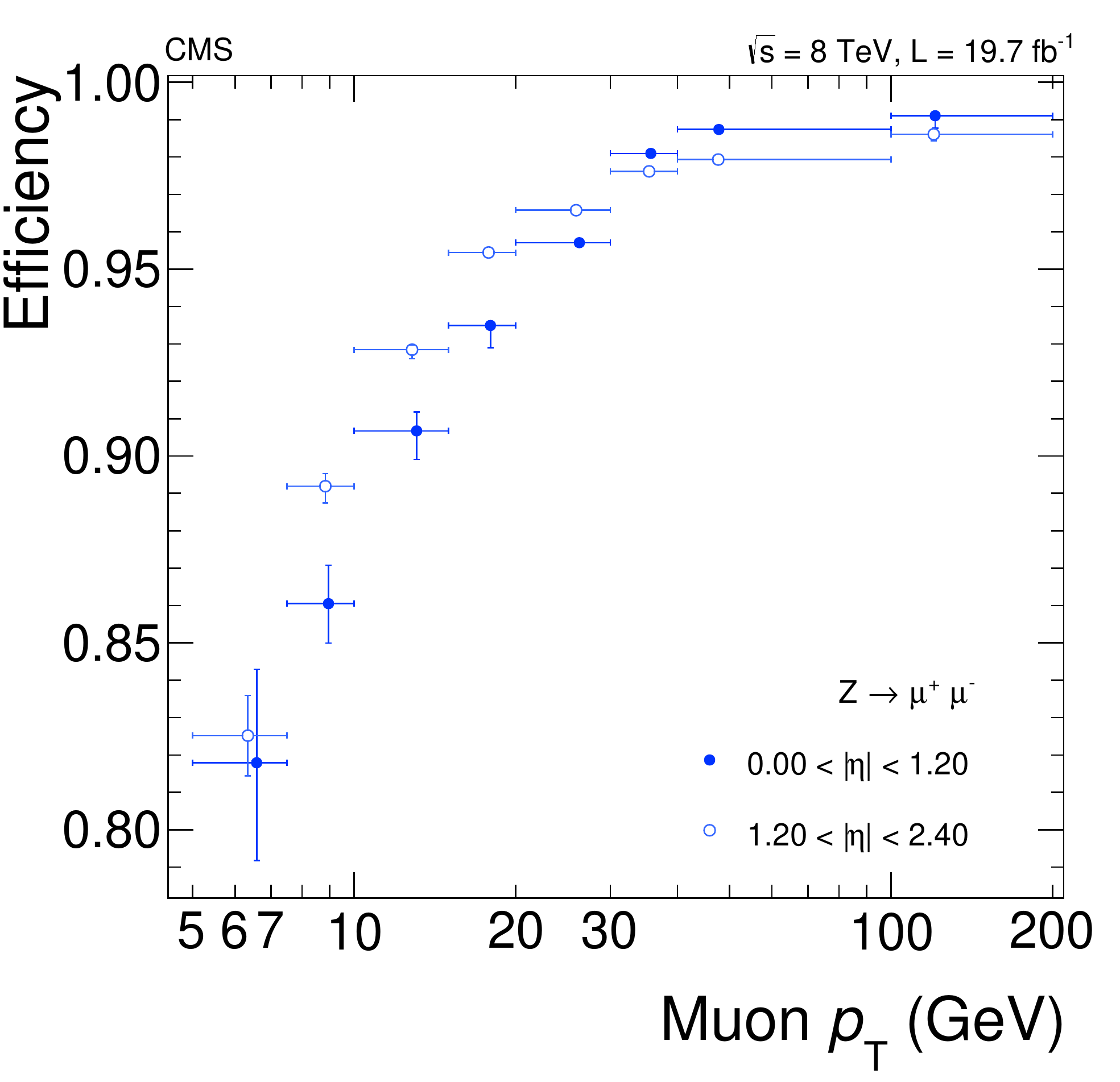}
    \caption{Efficiency, as a function of the lepton $\PT^\ell$, for
      reconstructing and selecting (\cmsLeft) electrons and
      (\cmsRight) muons, measured with a $\cPZ\to\ell\ell$ data sample
      by using a tag-and-probe
      method.  \label{fig:lepeff}} \end{center}
\end{figure}

The combined efficiency for the reconstruction, identification, and
isolation (and conversion rejection for electrons) of prompt electrons
or muons is measured in data using a ``tag and probe''
method~\cite{CMS:2011aa} based on an inclusive sample of \cPZ-boson
events, separately for 7 and 8\TeV data.  The efficiency is measured
from the $\cPZ\to\ell^+\ell^-$ yields obtained by fitting the $\cPZ$
line shape plus a background model to the dilepton mass distributions
in two samples, the first with the probe lepton satisfying the
selection criteria, and the second with the probe lepton failing
them. The same approach is used in both data and simulation, and the
ratio of the efficiency in the different $\PT^\ell$ and $\eta^\ell$
bins of the probed lepton is used in the analysis to rescale the
selection efficiency in the simulated samples. The efficiencies for
reconstructing and selecting electrons and muons in the full
$\PT^\ell$ and $\eta^\ell$ range exploited in this analysis are shown
in Fig.~\ref{fig:lepeff}. The deviation of the efficiency in
simulation relative to data, for the majority of the phase space of
the leptons, is less than 3\% for both electrons and muons. In the
case of electrons with $\pt^\Pe<15\GeV$, the deviation is larger,
5\%--9\%, but still consistent with unity, given the large statistical
uncertainty.  The dependency of the reconstruction and selection
efficiency on the number of reconstructed primary vertices in the
event is negligible for both the 7 and 8\TeV data samples. The
tracking efficiency decreases by about 0.4\% between 1 and 21 pileup
interactions, independent of the data-taking period.

\section{Final-state radiation recovery}
\label{sec:photons}

A $\cPZ$-boson decay into a lepton pair can be accompanied by
final-state radiation, in which case it is desirable to identify and
associate the radiated photon to the corresponding lepton to form the
$\cPZ$-boson candidate: $\cPZ \to \ell^+\ell^-\gamma$.  Photons
reconstructed within $\abs{\eta^{\cPgg}} < 2.4$ are possible FSR
candidates.  Low-energy photons are identified and reconstructed with
the PF reconstruction with a dedicated clustering algorithm,
efficient down to an energy of 230\MeV in the EB and 600\MeV in the
EE~\cite{CMS-PAS-PFT-10-002}.  The determination of the photon
energies and directions is monitored in the data with $\pi^0 \to
\gamma\gamma$ decays, and is in agreement with the predictions from
simulation.

Final-state radiated photons are mostly produced with a direction
nearly collinear with the parent lepton and have a harder spectrum
than background photons from initial-state radiation or pileup
interactions.  Therefore, to be identified as FSR, a reconstructed
photon must either have a transverse momentum $\pt^\cPgg > 2\GeV $ and
be found within a cone of size $ \Delta R < 0.07 $ from a selected
lepton candidate, or have $\PT^\cPgg > 4\GeV $ and be found isolated
from charged particles and energy deposits and within $ 0.07 < \Delta
R < 0.5 $ from a selected lepton candidate.

The photon isolation observable $R_\text{Iso}^{\cPgg}$ is the sum of
the transverse momenta of charged hadrons, other photons, and neutral
hadrons (including the ones originating from other vertices with
respect to the primary vertex of the event) identified by the PF
reconstruction within $\Delta R = 0.3$ around the candidate photon
direction, divided by the photon transverse momentum. Isolated photons
must satisfy $R_\text{Iso}^{\cPgg} < 1$.

If more than one FSR candidate is associated with a $\cPZ$ candidate,
the one with the highest $\pt^\cPgg$ is chosen, if there is at least
one with $\pt^\cPgg> 4\GeV$; otherwise, the one closest to any of the
individual daughter leptons of the $\cPZ$-boson candidate is
chosen. These criteria are chosen to maximize the efficiency of the
selection for photon emissions collinear with the lepton direction,
while keeping the contribution from background or pileup interactions
sufficiently low. The performance of the FSR recovery algorithm on the
simulation of signal events is described in Sec.~\ref{sec:selection}.

\section{Jet reconstruction and identification}
\label{sec:jets}

In the analysis the presence of jets is used as an indication of
vector-boson fusion (VBF) or associated production with a weak boson,
$V\PH$, with $V$ = $\PW$ or $\cPZ$, where the $V$ decays hadronically.
Jets are reconstructed using the anti-\kt clustering
algorithm~\cite{antikt} with distance parameter $D=0.5$, as
implemented in the \textsc{FastJet}
package~\cite{Cacciari:2011ma,Cacciari:fastjet2}, applied to the PF
candidates of the event.  Jet energy corrections are applied as a
function of the jet $\PT^\text{jet}$ and
$\eta^\text{jet}$~\cite{cmsJEC}. An offset correction is applied to
subtract the energy contribution not associated with the high-$\pt$
scattering, such as electronic noise and pileup, based on the jet-area
method~\cite{cmsJEC,Cacciari:2007fd,Cacciari:2008gn}. Jets are only
considered if they have $\PT^\text{jet}>30\GeV$ and
$\abs{\eta^\text{jet}}<4.7$. In addition, they are required to be
separated from the lepton candidates and from identified FSR photons by
$\Delta R > 0.5$.

Within the tracker acceptance, the jets are reconstructed with the
constraint that the charged particles are compatible with the primary
vertex.  In addition, in the entire acceptance, a multivariate
discriminator is used to separate jets arising from the primary
interaction from those reconstructed from energy deposits associated
with pileup interactions, especially due to neutral particles not
associated with the primary vertex of the event. The discrimination is
based on the differences in the jet shapes, the relative multiplicity
of charged and neutral components, and the fraction of transverse
momentum carried by the hardest components~\cite{jetIdPAS}.

\section{Selection and categorization of four-lepton candidates}
\label{sec:selection}

The event selection is designed to give a set of signal candidates in
the $\PH\rightarrow \cPZ\cPZ \rightarrow 4\ell$ final state in three
mutually exclusive subchannels: $4\Pe$, $2\Pe2\Pgm$, and $4\Pgm$.
Four well-identified and isolated leptons are required to originate
from the primary vertex to suppress the \cPZ + jet and $\ttbar$
backgrounds.

A $\cPZ$ candidate formed with a pair of leptons of the same flavor
and opposite charge ($\ell^+\ell^-$) is required.  When forming the
$\cPZ$-boson candidates, only FSR photon candidates that make the
lepton-pair mass closer to the nominal $\cPZ$-boson mass are
incorporated.  If the mass $m_{\ell\ell\Pgg}>100\GeV$, the photon is
not considered, to minimize the fraction of misidentified FSR
candidates. With the photon selection requirements described in
Sec.~\ref{sec:photons}, about 1.5\%, 4.6\%, and 9\% of the simulated
$\PH\to4\Pe$, $\PH\to2\Pe2\Pgm$, and $\PH\to4\Pgm$ decays,
respectively, are affected by the photon recovery procedure. As the
photon emission is most often collinear with one of the leptons,
measured electron energies, by construction, include the energy of a
large fraction of the emitted photons in the associated ECAL
supercluster, while measured muon momenta do not include the emitted
photons.  Therefore, without photon recovery, FSR is expected to
degrade the four-lepton mass resolution for Higgs boson candidates,
especially in the $4\mu$ and in the $2\Pe2\Pgm$ final states and, to a
lesser extent, in the 4$\Pe$ final state.  The performance of the FSR
recovery algorithm is estimated using simulated samples of
$\PH \to \cPZ\cPZ \to 4\ell$, and the rate is verified with inclusive
$\cPZ$ and $\cPZ\cPZ$ data events.  Genuine FSR photons within the
acceptance of the FSR selection are selected with an efficiency of
$\approx$50\% and with a mean purity of 80\%.  The FSR photons are
selected in 5\% of inclusive $\cPZ$ events with muon pairs, and in
0.5\% of single-\cPZ\ events with electron pairs.  A gain of
$\approx$3\% (2\%, 1\%) in efficiency is expected for the selection of
$\PH\rightarrow 4\mu$ (2\Pe2\Pgm, 4\Pe) events in this analysis. The
momentum of the selected FSR photon is added to the momentum of the
nearest lepton for the computation of every $4\ell$ kinematic
variable. Hereafter $\ell$ denotes a $\ell+\gamma$, in the case of a
recovered FSR photon.

Among all the possible opposite-charge lepton pairs in the event, the
one with an invariant mass closest to the nominal $\cPZ$-boson mass is
denoted $\cPZ_1$ and retained if its mass, $m_{\cPZ_1}$, satisfies 40
$< m_{\cPZ_1} < 120\GeV$.  Then, all remaining leptons are considered
and a second $\ell^+\ell^-$ pair is required ($\cPZ_2$), with the mass
denoted $m_{\cPZ_2}$.  If more than one ${\cPZ_2}$ candidate is
selected, the ambiguity is resolved by choosing the pair of leptons
with the highest scalar sum of $\PT$.  Simulation studies demonstrate
that this algorithm selects the true $\cPZ_2$ in the majority of cases
without sculpting the shape of the $\cPZ\cPZ$ background. The chosen
$\cPZ_2$ is required to satisfy 12 $< m_{\cPZ_2} <120\GeV$.  For the
mass range of $\mH<180\GeV$, at least one of the $\cPZ$ candidates is
off shell.  The lower bound for $m_{\cPZ_2}$ provides an optimal
sensitivity for a Higgs boson mass hypothesis in the range $110 < \mH
< 160\GeV$.

Among the four selected leptons forming the $\cPZ_1$ and the $\cPZ_2$,
at least one lepton is required to have $\PT^\ell > 20\GeV$, and
another one is required to have $\PT^\ell > 10\GeV$.  These
$\PT^\ell$ thresholds ensure that the selected events have leptons on
the efficiency plateau of the trigger.  To further remove events with
leptons originating from hadron decays produced by jet fragmentation
or from the decay of low-mass hadron resonances, it is required that
any opposite-charge pair of leptons chosen among the four selected
leptons (irrespective of flavor) satisfy $m_{\ell^+\ell^-} > 4\GeV$.
The phase space for the search of the SM Higgs boson is defined by
restricting the measured mass range to $m_{4\ell} > 100\GeV$.

\begin{figure}[!htb]
  \begin{center}
    \includegraphics[width=\cmsFigWidthStd]{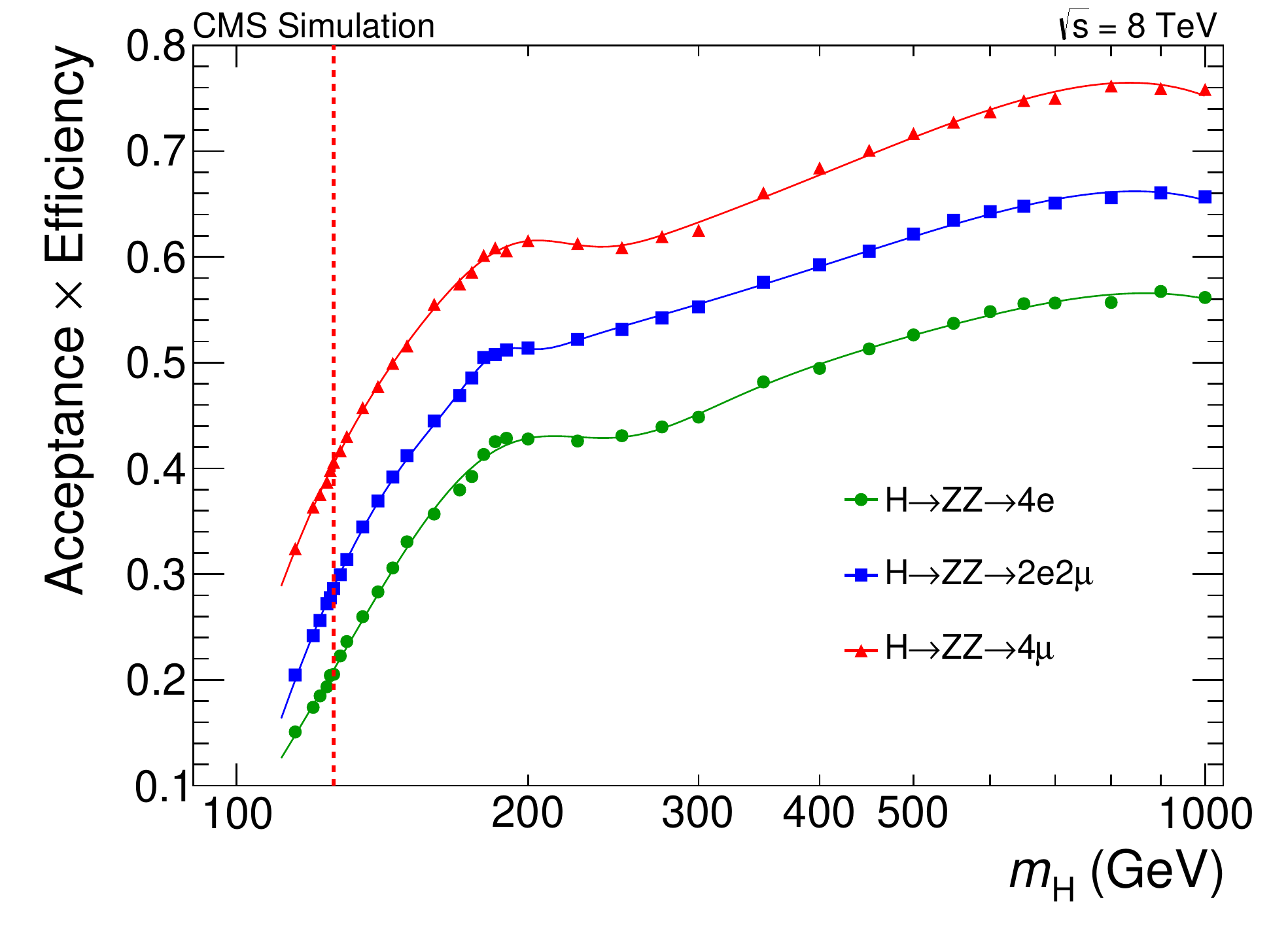}
    \caption{Geometrical acceptance times selection efficiency for the
      SM Higgs boson signal as a function of $\mH$ in the three final
      states for gluon fusion production. Points represent efficiency
      estimated from full CMS simulation; lines represent a smooth
      polynomial curve interpolating the points, used in the
      analysis. The vertical dashed line represents $\mH = 126\GeV$.
      \label{fig:effMH}} \end{center}
\end{figure}

The overall signal detection efficiencies, including geometrical
acceptance, for the $4\Pe$, $2\Pe2\Pgm$, and $4\Pgm$ channels increase
as a function of $\mH$ rapidly up to approximately 2$m_{\cPZ}$, where
both the $\cPZ$ bosons are on shell, and then flattens. The residual
rise for $\mH>300$\GeV is mostly due to the increased acceptance. The
efficiency versus $\mH$ is shown in Fig.~\ref{fig:effMH} for the gluon
fusion Higgs boson production mode, and it is very similar for other
production modes. The signal events are generated with
$\abs{\eta^\ell}<5$ and invariant mass of the dileptons from both the
$\cPZ_1$ and the $\cPZ_2$ boson decays $m_{\ell^+\ell^-}>1$\GeV.  The
efficiency within the geometrical acceptance is $\approx$30\%~(58\%),
43\%~(71\%), and 62\%~(87\%) for the three channels, respectively, for
$\mH = 126\,(200)\GeV$.

For a Higgs boson with $\mH = 126\GeV$, the resolution of the Gaussian
core of the mass distribution, estimated from simulated signal samples
with a double-sided Crystal-Ball function fit, is about 2.0, 1.6,
1.2\GeV for $4\Pe$, $2\Pe2\Pgm$, and $4\Pgm$, respectively.  The full
rms of the four-lepton mass distribution, including the asymmetric
tails, is estimated to be 2.9, 2.3, 1.7\GeV for the three channels,
respectively.  For a Higgs boson with $\mH = 600\GeV$, in which the
natural width of the resonance contributes most, the double-sided
Crystal-Ball-function core width parameter is about 75\GeV.

While in the dominant gluon fusion mechanism the Higgs boson is
produced only in association with jets from initial-state radiation of
the quarks, in the VBF production the two vector bosons are radiated
from the initial-state quarks to produce the Higgs boson.  The cross
section for VBF production is about 1 order of magnitude smaller than
that for the gluon fusion process.  In the vector-boson scattering
process, the two initial-state quarks deviate at a polar angle large
enough such that as final-state quarks they create measurable
additional jets in the event.  These two jets, being remnants of the
incoming proton beams, have typically a large separation in $\eta$ and
high momentum.  These characteristics are used to distinguish gluon
fusion from VBF Higgs boson production in the analysis. Jets in the
final state also come from $\ttbar\PH$ and V$\PH$ production, where
the $V$ decays hadronically.

In order to improve the sensitivity to the Higgs boson production
mechanisms, the event sample is split into two categories based on the
jet multiplicity, where a jet is defined as in Sec.~\ref{sec:jets}.
These categories are defined as the 0/1-jet category, containing
events with fewer than two jets, and the dijet category, containing
events with at least two jets.  In the 0/1-jet category, the transverse
momentum of the four-lepton system ($\VDu$) is used to distinguish VBF
production and associated production with a weak boson, $V\PH$, from
gluon fusion. In the dijet category, a linear discriminant ($\VDj$) is
formed combining two VBF-sensitive variables, the absolute difference
in pseudorapidity ($\vert\Delta
\eta_{jj}\vert$) and the invariant mass of the two leading jets
($m_{jj}$). The discriminant maximizes the separation between
vector-boson and gluon fusion processes. In the 0/1-jet (dijet)
category, about 5\% (20\%) of the signal events are expected to come
from the VBF production mechanism, as estimated from simulation. The
expected signal yield, split by category and by production mode, is
reported in Table~\ref{tab:PreFitYieldsSigRegionCat}.

\subsection{Per-event mass uncertainties}
\label{sec:masserrors}

For the Higgs boson mass and width measurement, the uncertainty in the
four-lepton mass, which can be estimated on a per-event basis, is
relevant because it varies considerably over the small number of
selected events.

Uncertainties in the measured lepton momentum arise from imperfect
calibration of the ECAL supercluster and uncertainty in the GSF track
fit due to possible high-bremsstrahlung emissions in the case of the
electrons, and from the uncertainty in the muon track fit due to the
multiple scattering of the muons in the material of the inner tracker.
These uncertainties depend on and are evaluated from the lepton's
direction and transverse momentum, as well as from possible
mismeasurements specific to each lepton.  In the case of electrons,
the momentum uncertainties are assessed from the combination of the
quality of the ECAL supercluster and the GSF track fit, through a
similar multivariate regression as the one used to refine the estimate
of the electron momentum, described in Sec.~\ref{sec:electrons}. In
the case of muons, the momentum uncertainties are assessed from the
properties of hits in the tracker and in the muon system, and the
quality of the muon candidate fit. If FSR photons are identified and
associated with the event, their uncertainty, assessed by the quality
of the ECAL clusters, is also accounted for in the event mass
uncertainty.

The momentum uncertainties for each of the four leptons in an event
are then propagated into a relative uncertainty
$\MassD\equiv\sigma_{m_{4\ell}}/m_{4\ell}$ in the four-lepton mass.
The per-event mass uncertainty is given as the sum in quadrature of
the individual mass uncertainty contributions from each lepton and any
identified FSR photon candidate.  A calibration of the per-lepton
uncertainties is derived using large $\PJGy\to\Pgmp\Pgmm$,
$\cPZ\to\Pgmp\Pgmm$, and $\cPZ\to\Pep\Pem$ event samples, both in data
(Table~\ref{tab:calibstat}) and in simulation. The line shape of these
resonances is modeled, as for the SM Higgs boson, with a BW convolved
with a double-sided CB function, where the resolution is estimated as
$\lambda \times \sigma(m_{4\ell})$. In this procedure,
$\sigma(m_{4\ell})$ is fixed to the value computed using the
uncertainties in the individual momenta of the leptons, and $\lambda$,
defined as the calibration constant, is a floating parameter.  The
latter is derived for electrons and muons in several bins of the
average $\PT^\ell$ and $\eta^\ell$ of the lepton: $\PJGy\to\Pgmp\Pgmm$
is used for muons with $\PT^\Pgm < $ 20\GeV, while, for lack of
a sufficiently large sample of $\PJGy\to\Pep\Pem$, $\cPZ\to\Pep\Pem$
events are used in the entire $\PT^\Pe$ range. The value of $\lambda$
obtained from the fit is approximately 1.2 for electrons and 1.1 for
muons, in the entire kinematic range of the leptons used in this
analysis.

\begin{figure}[bh]
  \begin{center}
    \includegraphics[width=\cmsFigWidthStd]{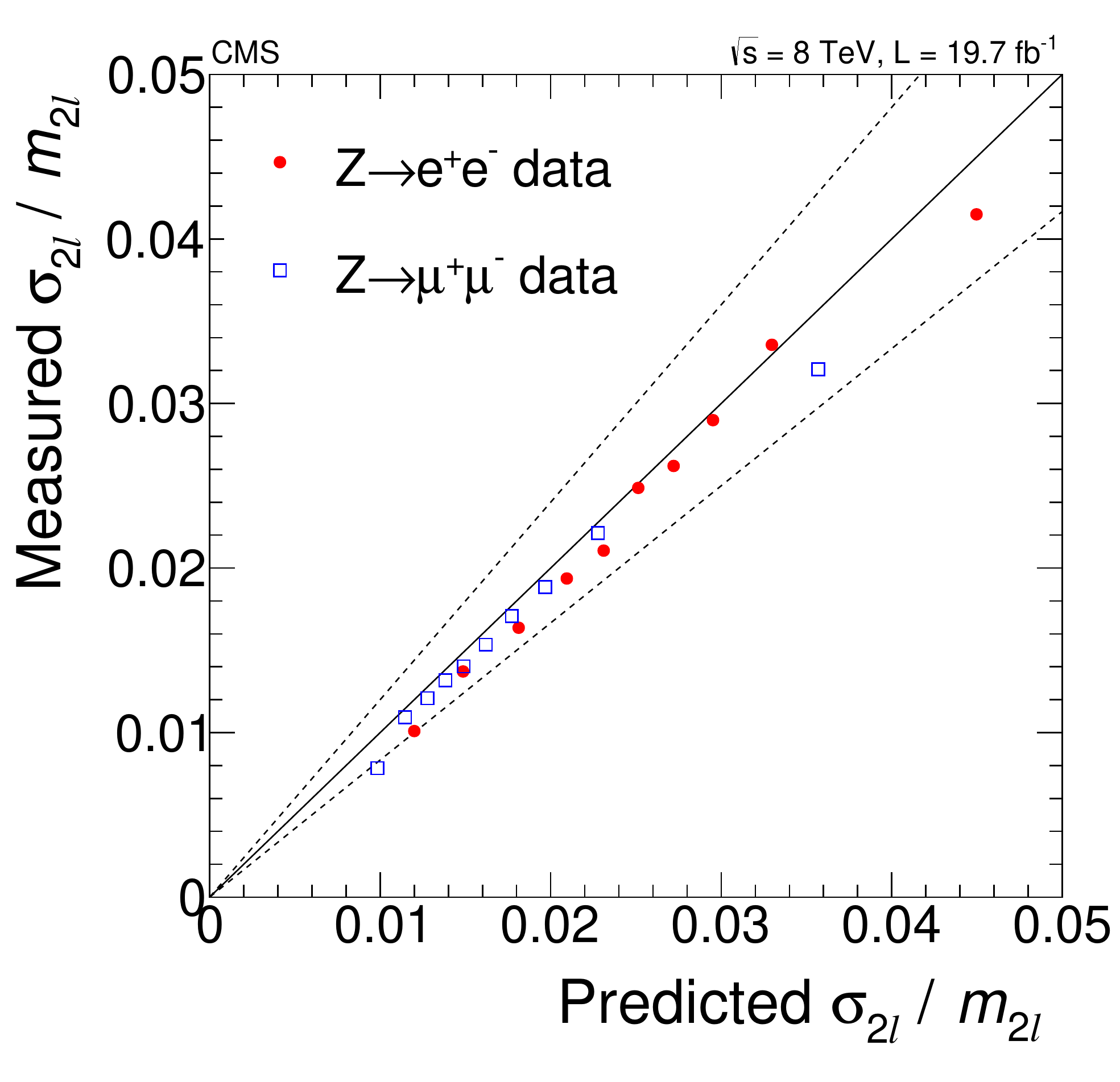}
    \includegraphics[width=\cmsFigWidthStd]{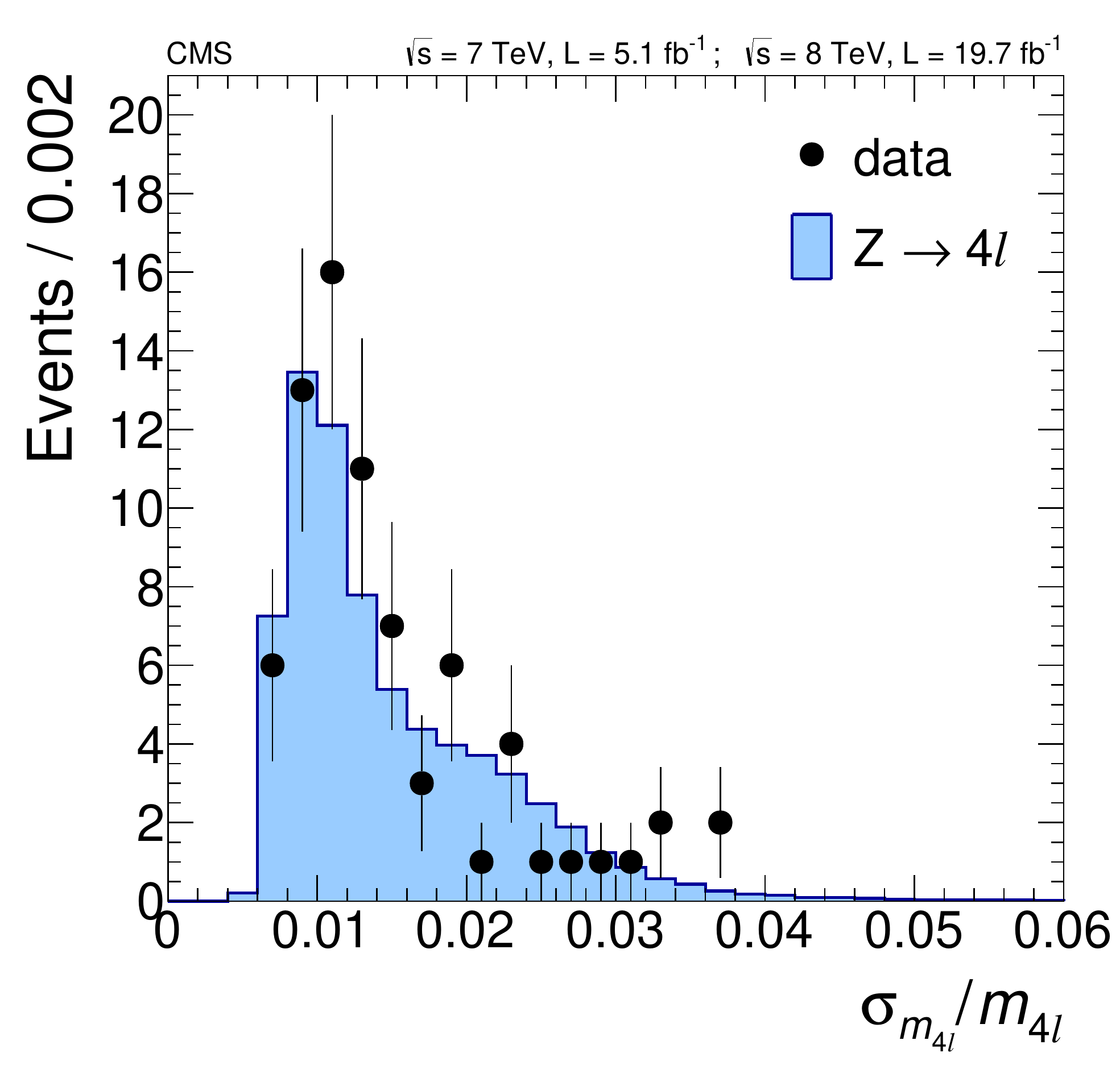}
    \caption{(\cmsLeft) Measured versus predicted relative mass uncertainties
    for $\cPZ\to\Pep\Pem$ and $\cPZ\to\Pgmp\Pgmm$ events in data. The
    dashed lines represent the $\pm$20\% envelope, used as systematic
    uncertainty in the resolution. (\cmsRight) Relative mass uncertainty
    distribution for data and simulation in the $\cPZ\to4\ell$ mass
    region of $80 <m_{4\ell}< 100\GeV$. \label{fig:EbE_validation}}
    \end{center}
\end{figure}

As a closure test, the $\cPZ \to \ell\ell$ events are grouped into
subsets based on their per-event predicted dilepton mass resolution
and fit to the $\cPZ$ line shape in each subset as described above.  A
systematic uncertainty of $\pm$20\% is assigned to the per-event mass
uncertainty for both electrons and muons based on the agreement
between per-event computed and observed mass resolutions as shown in
Fig.~\ref{fig:EbE_validation}~(\cmsLeft).  In
Fig.~\ref{fig:EbE_validation}~(\cmsRight), the comparison between data
and simulation of the $\MassD$ observable in the $\cPZ\to4\ell$ mass
region is shown.

\section{Background estimation}
\label{sec:backgrounds}

The dominant background contribution in the $\PH\to\cPZ\cPZ\to4\ell$
search is irreducible and is due to direct $\cPZ\cPZ$ production via
$\Pq\Paq$ annihilation and gluon fusion. The remaining subleading
contributions arise from reducible multilepton sources, $\cPZ+\text{jets}$, $\ttbar$, and $\PW\cPZ+\text{jets}$.

\subsection{Irreducible background}
\label{sec:irreducible_bkg}

The expected yield and shape of the $\cPZ\cPZ$ background is evaluated
by simulation.  The NLO cross section for $\Pq\Paq\to\cPZ\cPZ$
production and the LO cross section for $\Pg\Pg \to \cPZ\cPZ$
production are calculated with
\MCFM~\cite{MCFM,Campbell:1999ah,Campbell:2011bn}.  The relative
contribution of LO $\Pg\Pg \to \cPZ\cPZ$ with respect to NLO
$\Pq\Paq\to\cPZ\cPZ$ is about 2\% at four-lepton mass $m_{4\ell} =
126\GeV$ and about 6\% at 1\TeV.  The expected contribution of the
$\cPZ\cPZ$ processes to the total background, in the region $100 <
m_{4\ell} < 1000$ ($121.5 < m_{4\ell} < 130.5$)\GeV, is approximately
91\%, 94\%, and 97\% (58\%, 71\%, and 86\%) in the $4\Pe$,
$2\Pe2\Pgm$, and $4\Pgm$ channels, respectively. The shape
uncertainties arising from imperfect simulation of the $\PT^\ell$ and
$\eta^\ell$ dependence of the efficiency and other experimental
sources are completely overshadowed by the uncertainties from the
normalization systematics, such that shape variations have negligible
effects compared to the normalization variations.

The irreducible four-lepton background arising from double-parton
interactions (DPI), $\cPZ$ + Drell-Yan (DY), is evaluated
using \PYTHIA~6.4.24 with the overall cross section calculated as
$\sigma_\mathrm{DPI} = \sigma_{\cPZ} \cdot \sigma_\mathrm{DY}
/ \sigma_\text{pheno}$, where the phenomenological effective cross
section, measured at $\sqrt{s} = 7$\TeV, is $\sigma_\text{pheno}$ =
15\unit{mb}~\cite{Aad:2013bjm}, and the cross sections $\sigma_{\cPZ}$
and $\sigma_\mathrm{DY}$ are taken from simulation.  The DPI $\cPZ$+DY
background is much smaller than normalization uncertainties on either
$\Pq\Paq\to\cPZ\cPZ$, $\Pg\Pg \to\cPZ\cPZ$ or a reducible background;
hence, the DPI $\cPZ$ + DY background is neglected in the analysis.

\subsection{Reducible background}
\label{sec:reducible_bkg}

Two independent methods, using dedicated control regions in data, are
considered to estimate the reducible background, denoted as \cPZ + \X in
the following paragraphs because it is dominated by the
$\cPZ+\text{jets}$ process.  The control regions are defined by a
dilepton pair satisfying all the requirements of a $\cPZ_1$ candidate
and two additional leptons, opposite sign (OS) or same sign (SS),
satisfying certain relaxed identification requirements when compared
to those used in the analysis. The invariant mass of the additional
dilepton pair is required to be larger than 12\GeV, in order to be
consistent with the criteria imposed on the $\cPZ_2$ candidate in the
signal selection.

In both methods, the extrapolation from the control region to the
signal region is performed using the lepton misidentification
probability, $\fakerate$, which is defined as the fraction of
nonsignal leptons identified with the analysis selection criteria,
estimated in an enriched sample of nongenuine electrons and
muons. This sample is composed of $\cPZ_1 + 1 \ell_\text{loose}$
events in data consisting of a pair of leptons, both passing the
selection requirements used in the analysis, and exactly one
additional lepton passing the relaxed selection.  The mass of the
$\cPZ_1$ candidate is required to satisfy $\vert m_{\ell\ell} -
m_{\cPZ}\vert < 10$\GeV for the OS leptons method.  Such a stringent
requirement suppresses from the $\fakerate$ calculation the
contribution of events with FSR where the photon converts and one of
the conversion products is not reconstructed.  For the SS leptons
method, a requirement of $\abs{ m_{\ell\ell} - m_{\cPZ}} < 40$\GeV is
imposed.  In order to suppress the contribution from $\PW\cPZ$ and
$\ttbar$ processes, which have a third lepton, the missing transverse
energy ($\ETslash$) is required to be less than 25\GeV. The $\ETslash$
is defined as the modulus of the vector sum of the transverse momenta
of all reconstructed PF candidates (charged or neutral) in the event.
The invariant mass of the loose lepton and the opposite-sign lepton
from the $\cPZ_1$ candidate, if they have the same flavor, is required
to be greater than 4\GeV to reject contributions from low-mass
resonances, such as $\PJGy$. As a result of these requirements, the
control sample largely consists of events with a $\cPZ$ boson and a
misidentified additional lepton. Hence, the fraction of these events
in which the additional lepton passes the analysis identification and
isolation requirements gives a rate $\fakerate$ that ranges from
1\%--15\% (5\%--10\%) depending on the $\pt^\ell$ and $\eta^\ell$ of
the electron (muon).

\subsubsection{Method using opposite-sign (OS) leptons}
\label{sec:xzos}
In this method, the control region consists of events with a $\cPZ_1$
candidate and two additional leptons with the same flavor and opposite
charge. Two categories of events are considered in this method.

The category 2P2F is composed of events in which two leptons pass (P)
the selection requirements of the analysis and two fail (F), but pass
the loose selection. It is used to estimate the contribution from
backgrounds that intrinsically have only two prompt leptons
($\cPZ+\text{jets}$, $\ttbar$).  To estimate the contribution of these
background processes in the signal region, each 2P2F event $i$ is
weighted by a factor $\frac{f^i_{3}}{1-f^i_{3}}
\frac{f^i_{4}}{1-f^i_{4}}$, where $f^i_{3}$ and $f^i_{4}$ are the
$\fakerate$ for the third and fourth leptons.  Analogously, the 3P1F
category consists of events where exactly one of the two additional
leptons passes the analysis selection.  It is used to estimate the
contribution from backgrounds with three prompt leptons and one
misidentified lepton ($\PW\cPZ+\text{jets}$ and
$\cPZ\gamma+\text{jets}$ with the photon converting to an $\Pep\Pem$
pair).  Each event $j$ in the 3P1F control region is weighted by a
factor $\frac{f^j_a}{1-f^j_a}$, where $f^j_a$ is the $\fakerate$ for
the third or fourth lepton to fail the analysis selection.  This
control region also has contributions from $\cPZ\cPZ$ events where one
of the four prompt leptons fails the analysis selection, and from the
processes with only two prompt leptons (2P2F type), where one of the
two nonprompt leptons passes the selection requirements. The
contribution from $\cPZ\cPZ$ events, $n^{\cPZ\cPZ}_\mathrm{3P1F}$, is
estimated from simulation, and the background estimate is reduced by a
factor of $1 - n^{\cPZ\cPZ}_\mathrm{3P1F}/N_\mathrm{3P1F}$, where
$N_\mathrm{3P1F}$ is the number of events of the 3P1F control
region. The contribution from 2P2F-type processes in the 3P1F region
is estimated as $\sum_i ( \frac{f^j_3}{1-f^i_3}
+ \frac{f^j_4}{1-f^i_4})$. It contributes to the final weighted sum of
the 3P1F events with the component $\sum_i
(2 \frac{f^i_3}{1-f^i_3} \frac{f^i_4}{1-f^i_4})$, which has to be
subtracted from the background estimate. Therefore, in this method,
the expected yield for the reducible background in the signal region,
$N^\text{reducible}_\mathrm{SR}$, becomes
\ifthenelse{\boolean{cms@external}}{
\begin{equation}\begin{split}
  \label{eq:PredictionSR}
  N^\text{reducible}_\mathrm{SR} =
  \left( 1 - \frac{n^{\cPZ\cPZ}_\mathrm{3P1F}}{N_\mathrm{3P1F}} \right)
  \sum_j^{N_\mathrm{3P1F}} \frac{f^j_a}{1-f^j_a}-\\
  \sum_i^{N_\mathrm{2P2F}} \frac{f^i_3}{1-f^i_3} \frac{f^i_4}{1-f^i_4}.
\end{split}\end{equation}
}{
\begin{equation}
  \label{eq:PredictionSR}
  N^\text{reducible}_\mathrm{SR} =
  \left( 1 - \frac{n^{\cPZ\cPZ}_\mathrm{3P1F}}{N_\mathrm{3P1F}} \right)
  \sum_j^{N_\mathrm{3P1F}} \frac{f^j_a}{1-f^j_a}-
  \sum_i^{N_\mathrm{2P2F}} \frac{f^i_3}{1-f^i_3} \frac{f^i_4}{1-f^i_4}.
\end{equation}
}
\subsubsection{Method using same-sign (SS) leptons}
\label{sec:zxss}
In this method, the control region consists of events with a $\cPZ_1$
candidate and two additional leptons with the same flavor and same
charge.  The $\fakerate$ is measured using a $\cPZ_1 + 1 \ell_\text{loose}$ sample, which is similar to that used for the {\it OS} control
region, but with the invariant mass of the $\cPZ_1$ candidate, $\abs{
m_{\ell\ell} - m_\cPZ} <  40\GeV$, consistent with the
requirement on the $\cPZ_1$ candidate used in the analysis.  Here, the
contribution from FSR photons to the electron misidentification
probability is much larger and needs to be taken into account.  This
is done by exploiting the observed linear dependence of the
$f(\Pe,\PT^\Pe, \abs{\eta^\Pe})$ on the fraction of loose electrons with
tracks having one missing hit in the pixel detector, $r_\text{miss}(\PT^\Pe, \abs{\eta^\Pe})$, which is indicative of a possible
conversion.  The fraction $r_\text{miss}(\PT^\Pe, \abs{\eta^\Pe})$ is
estimated using samples with different FSR contributions obtained by
varying the requirements on $\abs{m_{\ell\ell} - m_\cPZ}$ and
$\abs{m_{\ell\ell \Pe_\text{loose}} - m_\cPZ}$.  The corrected
$\tilde f(\Pe,\PT^\Pe, \abs{\eta^\Pe})$ is then computed using the value
$r_\text{miss}(\PT^\Pe, \abs{\eta^\Pe})$ measured in the control sample
where the method is applied.

The expected number of reducible background events in the signal
region is obtained as:

\begin{equation}
  N^\text{reducible}_\mathrm{SR} =
  r_\mathrm{OS/SS} \, \cdot \,
  \sum_i^{N_\mathrm{2P2L_{SS}}}  \tilde f^i_3  \cdot \tilde f^i_4,
\end{equation}

where $N_\mathrm{2P2L_{SS}}$ is the number of observed events in the
region $\mathrm{2P2L_{SS}}$, in which both the additional leptons
fulfill the loose selection requirements for leptons, having the same
flavor and charge.  The ratio $r_\mathrm{OS/SS}$ between the number of
events in the $\mathrm{2P2L_{OS}}$ and $\mathrm{2P2L_{SS}}$ control
regions is obtained from simulation.

\subsubsection{Combination of the two methods}
\label{sec:zxcomb}
The predicted yields of the \cPZ + \X background from the two methods
are in agreement within their statistical uncertainties.  The dominant
sources of these uncertainties are the limited number of events in the
3P1F, 2P2F, and $\mathrm{2P2L_{SS}}$ control regions, as well as in
the region where the correction factor for $\tilde f(\Pe,\PT^\Pe,
\abs{\eta^\Pe})$ is computed.  Since they are mutually independent,
results of the two methods are combined.

The shape of the $m_{4\ell}$ distribution for the reducible background
is obtained from the OS method by fitting the $m_{4\ell}$
distributions of 2P2F and 3P1F events separately with empirical
functional forms built from Landau~\cite{Landau:1944if} and
exponential distributions. The systematic uncertainty in the
$m_{4\ell}$ shape is determined by the envelope that covers
alternative functional forms or alternative binning for the fit used
to determine its parameters. The additional discriminating variables
for this background are described by binned templates, as discussed in
Sec.~\ref{sec:statisticalanalysis}.

\begin{figure}[thbp]
 \begin{center}
   \includegraphics[width=\cmsFigWidthStd]{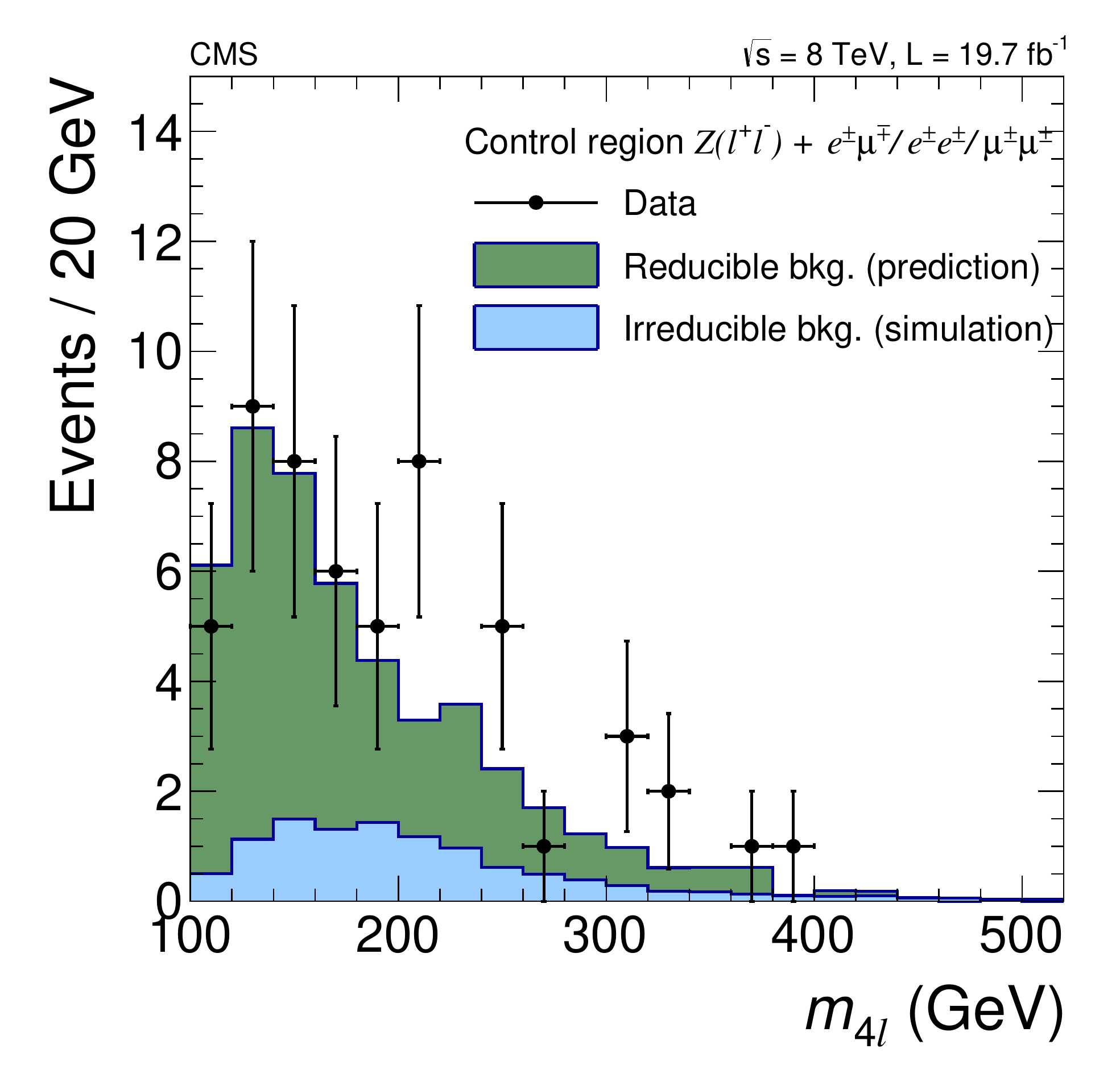}
   \includegraphics[width=\cmsFigWidthStd]{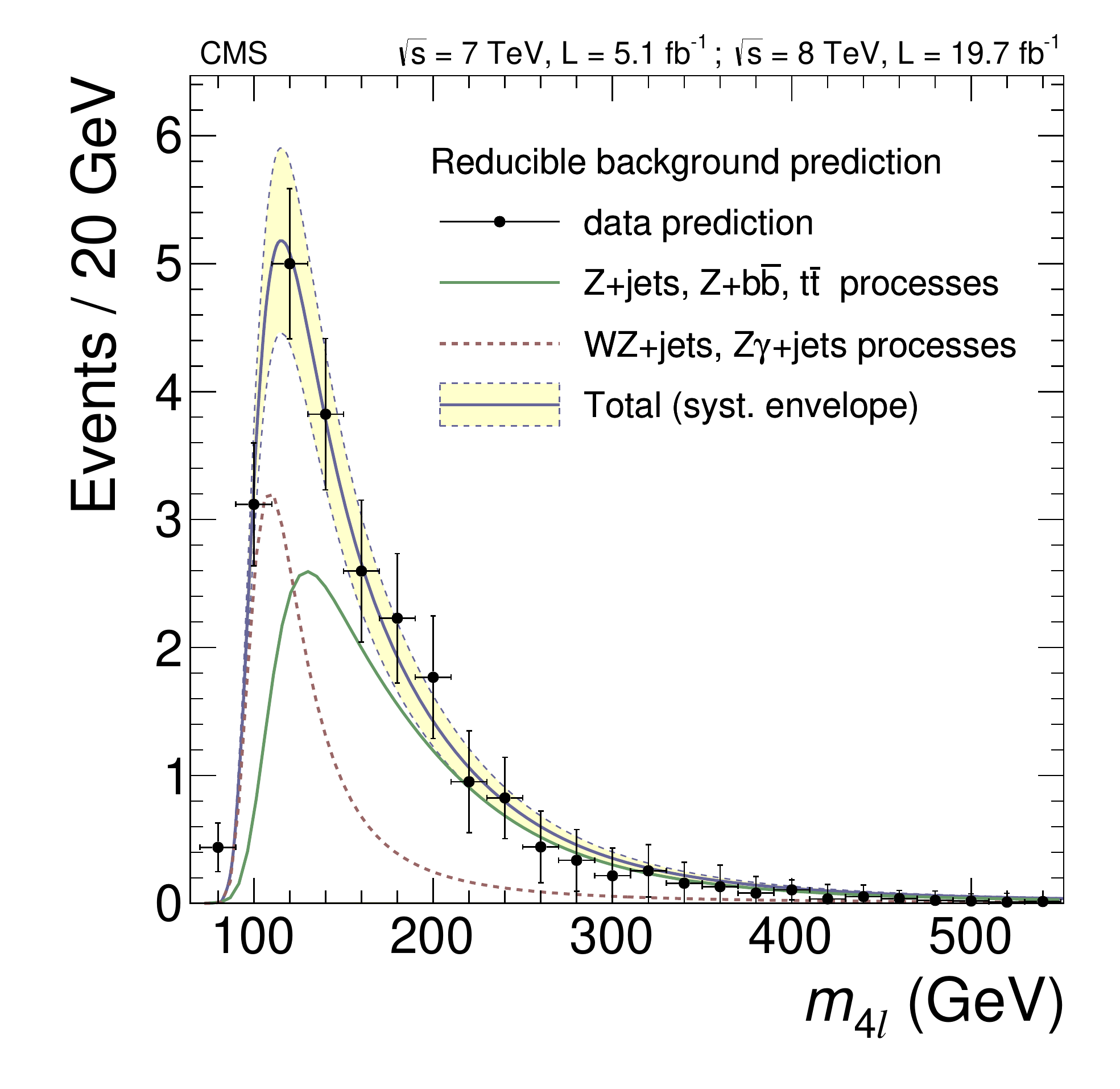}
   \caption{(\cmsLeft) Validation of the method using the SS control
     sample. The observed $m_{4\ell}$ distribution (black dots),
     prediction of the reducible background (dark green area), and
     expected contributions from $\cPZ\cPZ$ (light blue area) are
     shown. (\cmsRight) Prediction for the reducible background in all three
     channels together (black dots) fitted using an empirical shape
     (blue curve) with indicated total uncertainty (yellow band). The
     contributions from the 2P2F-like (solid green) and 3P1F-like
     (dashed red) processes are fitted separately.
 \label{fig:zx}}
 \end{center}
 \end{figure}

The total systematic uncertainties assigned to the \cPZ + \X background
estimate take into account the uncertainty in the $m_{4\ell}$
shape. They also account for the difference in the composition of the
$\cPZ_1 + 1 \ell_\text{loose}$ sample used to compute $\fakerate$ and
the control regions in the two methods used to estimate the \cPZ + \X
background, in particular the contribution of the heavy flavor jets
and photon conversions.  The systematic uncertainty is estimated to be
20\%, 25\%, and 40\% for the $4\Pe$, $2\Pe2\Pgm$, and $4\Pgm$ decay
channels, respectively. The two methods have been further validated
using events that pass the analysis selection with the exception that
the $\cPZ_2$ candidate is formed out of a lepton pair with the wrong
combination of flavors or charges (control region $\cPZ_1 +
\Pe^{\pm}\Pe^{\pm}/\Pe^{\pm}\mu^{\mp}/\mu^{\pm}\mu^{\pm}$).  The predicted
contribution of the reducible background in this control region is in
agreement with the observed number of events within the
uncertainties. Figure~\ref{fig:zx}~(\cmsLeft) shows the validation of the OS
method.

The prediction for the \cPZ + \X background yields with combined
statistical and systematic uncertainties is given in
Sec.~\ref{sec:yields} and also shown in
Fig.~\ref{fig:zx}~(\cmsRight). The expected yields of the \cPZ + \X
background in the signal region from the 2P2F-like and 3P1F-like
sources are estimated separately. The weighted events of the two
control regions are also fitted independently and then added together
to give the total \cPZ + \X $m_{4\ell}$ probability density function
used in the fit. The relative contribution of the reducible background
to the total background in the region $100 < m_{4\ell} < 1000$ $(121.5
< m_{4\ell} < 130.5)\GeV$ depends on the final state, being
approximately 9\% (42\%), 6\% (28\%), and 3\% (14\%) in the $4\Pe$,
$2\Pe2\Pgm$, and $4\Pgm$ channels, respectively.  The estimated yields
of this background are reported in Sec.~\ref{sec:yields}.

\section{Kinematic discriminants}
\label{sec:kd}

The four-lepton decay mode has the advantage that the kinematics of
the Higgs boson and its decay products are all visible in the
detector, providing many independent observables that can be used for
different purposes.  First, in addition to their invariant mass, the
angular distributions of the four leptons and the dilepton pairs
invariant masses can be used to further discriminate signal from
background and thus increase the signal sensitivity and reduce the
statistical uncertainty in measurements, including the cross section,
the mass, and the width of the resonance.  Second, this extra
information on angular correlations can be used to experimentally
establish the consistency of the spin and parity quantum numbers with
respect to the SM. This section describes how the full kinematic
information from the production and decay can be encoded in a
kinematic discriminant optimized for the separation of two processes,
be it signal from background or between different signal hypotheses.

\begin{figure}[thbp]
\begin{center}
\centering
\includegraphics[width=\cmsFigWidthStd]{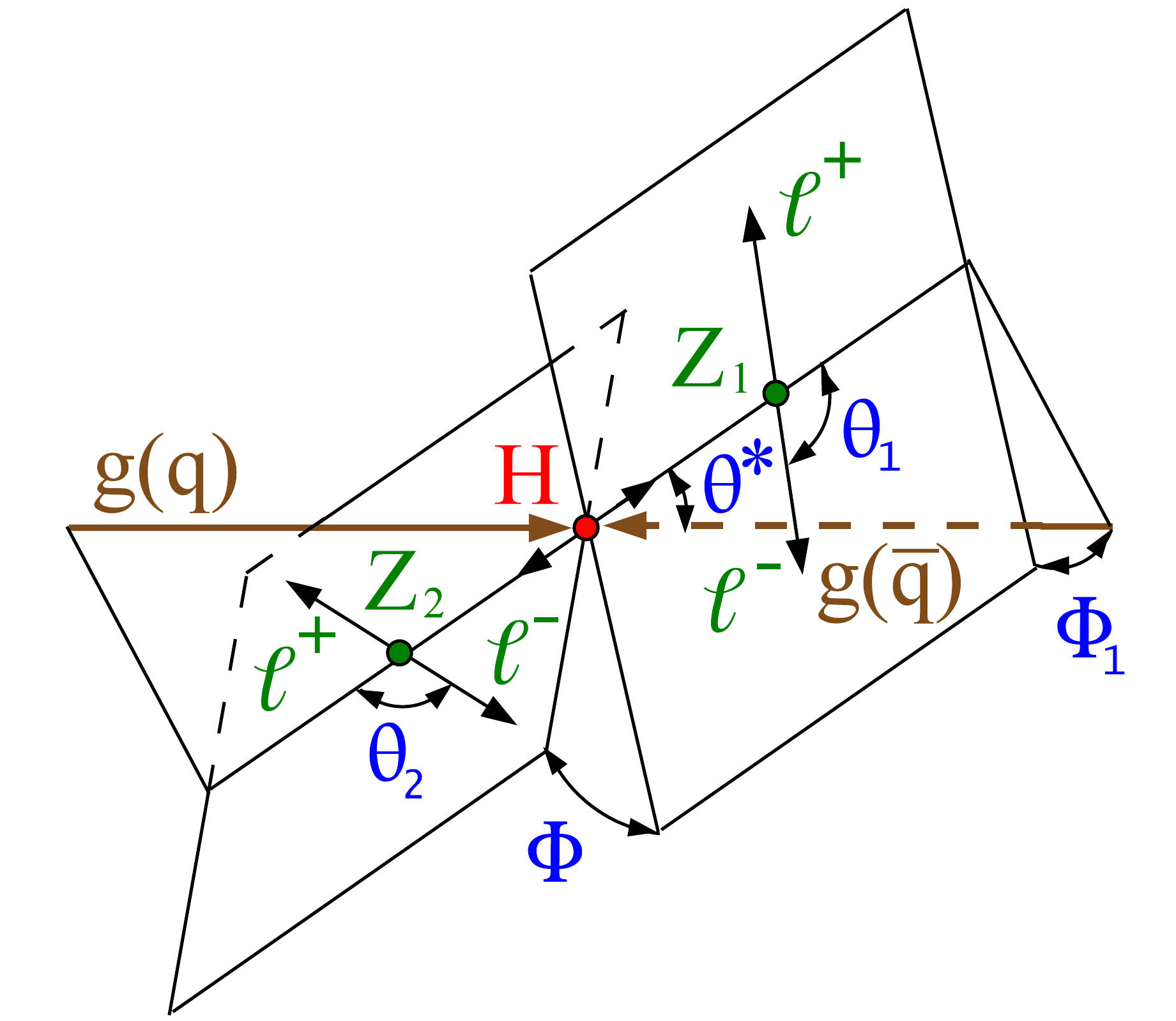}
\caption{ Illustration of the production and decay of a particle
  $\PH$, $\cPg\cPg(\cPq\cPaq)\to \PH\rightarrow\cPZ\cPZ\rightarrow
  4\ell$, with the two production angles $\theta^*$ and $\Phi_1$ shown
  in the $\PH$ rest frame and three decay angles $\theta_1$,
  $\theta_2$, and $\Phi$ shown in the $\cPZ_1$, $\cPZ_2$, and $\PH$
  rest frames, respectively.
\label{fig:decay}}
\end{center}
\end{figure}

The kinematic properties of the SM Higgs boson or any non-SM exotic
boson decay to the four-lepton final state has been extensively
studied in
Refs.~\cite{Choi:2002jk,Soni:1993jc,Barger:1993wt,Allanach:2002gn,
  Buszello:2002uu,Godbole:2007cn,Keung:2008ve,Antipin:2008hj,Hagiwara:2009wt,
  Gao:2010qx,DeRujula:2010ys,Gainer:2011xz,Bolognesi:2012,Chen:2012jy,Avery:2012um,Artoisenet:2013puc,Anderson:2013fba}.
Five angles $\vec\Omega\equiv(\theta^*, \Phi_1, \theta_1, \theta_2, \Phi)$
defined in Fig.~\ref{fig:decay}~\cite{Gao:2010qx,PhysRev.168.1926} and
the invariant masses of the lepton pairs, $m_{\cPZ_1}$ and
$m_{\cPZ_2}$, fully describe the kinematic configuration of a
four-lepton system in its center-of-mass frame, up to an arbitrary
rotation around the beam axis.  These observables provide significant
discriminating power between signal and background, as well as between
alternative signal models.  A matrix-element likelihood approach is
used to construct kinematic discriminants related to the decay
observables~\cite{Chatrchyan:2012ufa,Chatrchyan:2012br}.

In addition to the four-lepton center-of-mass-frame observables, the
four-lepton transverse momentum and rapidity are needed to completely
define the system in the lab frame. The transverse momentum of the
four-lepton system is used in the analysis as an independent
observable because it is sensitive to the production mechanism of the
Higgs boson, but it is not used in the spin-parity analysis.  The
four-lepton rapidity is not used because the discrimination power of
this observable for events within the experimental acceptance is
limited.

Kinematic discriminants are defined based on the event probabilities
depending on the background ($\mathcal{P}_\text{bkg}$) or signal
spin-parity ($J^P$) hypotheses under consideration ($\mathcal{P}_{
J^P}$):
\begin{align}
\label{eq:kd-prob-sm}
\mathcal{P}_\text{bkg} &= \mathcal{P}^\text{kin}_\text{bkg} (m_{\cPZ_1}, m_{\cPZ_2}, \vec\Omega|m_{4\ell}) \times \mathcal{P}^\text{mass}_\text{bkg} (m_{4\ell}), \\
\label{eq:kd-prob-jp}
\mathcal{P}_{J^P} &= \mathcal{P}^\text{kin}_{J^P} (m_{\cPZ_1}, m_{\cPZ_2}, \vec\Omega|m_{4\ell}) \times \mathcal{P}^\text{mass}_{\rm sig} (m_{4\ell}|\mH),
\end{align}
where $\mathcal{P}^\text{kin}$ is the probability distribution of
angular and mass observables $(\vec\Omega, m_{\cPZ_1}, m_{\cPZ_2})$
computed from the LO matrix element squared for signal and $\cPZ\cPZ$
processes, and $\mathcal{P}^\text{mass}$ is the probability
distribution of $m_{4\ell}$ and is calculated using the
parameterization described in Sec.~\ref{sec:combinedfit}.  Matrix
elements for the signals are calculated with the assumption that $\mH =
m_{4\ell}$.  The probability distributions for spin-0 resonances
are independent of an assumed production mechanism.  Only the dominant
$\cPq\cPaq \to
\cPZ\cPZ$ background is considered in the probability
parameterization. For the reducible backgrounds, empirical templates
derived from the data control samples defined in
Sec.~\ref{sec:reducible_bkg} are used to model the probability density
functions of the kinematic discriminants, as described in
Sec.~\ref{sec:statisticalanalysis}.

For the alternative signal hypotheses, nine models have been tested,
following the notations from Refs.~\cite{Gao:2010qx,Bolognesi:2012}.
The most general decay amplitude for a spin-0 boson decaying to two
vector bosons can be defined as:
\ifthenelse{\boolean{cms@external}}{
\begin{equation}\begin{split}
A(\PH \to \cPZ\cPZ) =& v^{-1} \Big (
  a_1 m_{Z}^2 \epsilon_1^* \epsilon_2^*
+ a_2 f_{\mu \nu}^{*(1)}f^{*(2),\mu \nu}
+ \\
&a_3  f^{*(1)}_{\mu \nu} {\tilde f}^{*(2),\mu  \nu}
\Big ),
\label{eq:fullampl-spin0}
\end{split}\end{equation}
}{
\begin{equation}
A(\PH \to \cPZ\cPZ) = v^{-1} \Big (
  a_1 m_{Z}^2 \epsilon_1^* \epsilon_2^*
+ a_2 f_{\mu \nu}^{*(1)}f^{*(2),\mu \nu}
+
a_3  f^{*(1)}_{\mu \nu} {\tilde f}^{*(2),\mu  \nu}
\Big ),
\label{eq:fullampl-spin0}
\end{equation}
}
where $f^{(i),{\mu \nu}} = \epsilon_i^{\mu}q_i^{\nu} -
\epsilon_{i}^\nu q_i^{\mu} $ is the field-strength tensor of a gauge
boson with momentum $q_i$ and polarization vector $\epsilon_i$,
${\tilde
f}^{(i)}_{\mu \nu}=1/2\epsilon_{\mu\nu\alpha\beta}f^{(i),\alpha\beta}=\epsilon_{\mu\nu\alpha\beta}\epsilon_i^\alpha
q_i^\beta$ is the conjugate field strength tensor, $f^{*}$ denotes the
complex conjugate field strength tensor, and $v$ is the vacuum
expectation value of the SM Higgs field.
$\epsilon_{\mu\nu\alpha\beta}$ is the Levi-Civita completely
antisymmetric tensor. The $a_i$ coefficients generally depend on
$q^2_i$.  In this analysis, we consider the lowest-dimension operators
in the effective Lagrangian corresponding to each of the three unique
Lorentz structures, therefore taking $a_i$ to be constant for the
relevant range $q^2_i=m_{\cPZ_i}^2<m_H^2$.  The SM Higgs boson decay
is dominated by the tree-level coupling $a_1$.  The $0^-$ model
corresponds to a pseudoscalar (dominated by the $a_3$ coupling), while
$0^+_{\mathrm{h}}$ is a scalar (dominated by the $a_2$ coupling) not
participating in the electroweak symmetry breaking, where {h} refers
to higher-dimensional operators in Eq.~(\ref{eq:fullampl-spin0}) with
respect to the SM Higgs boson.  The spin-0 signal models are simulated
for the gluon fusion production process, and their kinematics in the
boson center-of-mass frame is independent of the production mechanism.

The $1^-$ and $1^+$ hypotheses represent a vector and a pseudovector
decaying to two $\cPZ$ bosons.  The spin-1 resonance models are
simulated via the quark-antiquark production mechanism, as the gluon
fusion production of such resonances is expected to be strongly
suppressed. The spin-1 hypotheses are considered under the
assumption that the resonance decaying into $4\ell$ is not necessarily
the same resonance observed in the $\PH\to\Pgg\Pgg$
channel~\cite{Aad:2012tfa,Chatrchyan:2012ufa}, as $J=1$ in the latter
case is prohibited by the Landau-Yang
theorem~\cite{Landau:1948xy,Yang:1950rg}.  This also provides a test
of the spin-1 hypothesis in an independent way.

The spin-2 model with minimal couplings, $2^+_{\mathrm{m}}$,
represents a massive graviton-like boson $\X$ suggested, for example,
in models with warped extra dimensions
(ED)~\cite{Randall:1999vf,Randall:1999ee}, where gluon fusion is the
dominant process.  For completeness, 100\% quark-antiquark
annihilation is also considered, which provides a projection of the
spin of the resonance on the parton collision axis equal to 1, instead
of 2, as in the case of the gluon's fusion with minimal couplings. A
modified minimal coupling model $2^+_{\mathrm{b}}$ is also considered,
where the SM fields are allowed to propagate in the bulk of the
ED~\cite{Agashe:2007zd}, corresponding to $g_1 \ll g_5$ in the
$\X\cPZ\cPZ$ coupling for the $2^+_{\mathrm{m}}$ model, where the
$g_i$'s are the couplings in the effective Lagrangian of
Ref.~\cite{Bolognesi:2012}. Finally, two spin-2 models with
higher-dimension operators are considered with both positive and
negative parity, $2_{\mathrm{h}}^+$ and $2_{\mathrm{h}}^-$,
corresponding to the $g_4$ and $g_8$ couplings.  The
$2^+_{\mathrm{b}}$, $2_{\mathrm{h}}^+$, and $2_{\mathrm{h}}^-$
resonances are assumed to be produced in gluon fusion.  The above list
of the spin-2 models does not exhaust all possible scenarios, nor does
it cover possible mixed states. However, it does provide a
representative sample of spin-2 alternatives to the $J^P=0^+$
hypothesis.

For discrimination between the SM Higgs boson ($J^P=0^+$) and the SM
backgrounds (nonresonant $\cPZ\cPZ$ and reducible backgrounds), an
observable is created from the probability distributions in
Eqs. (\ref{eq:kd-prob-sm}) and (\ref{eq:kd-prob-jp}):
\begin{equation}
  \label{eq:kd-mela}
  \KD = \frac{\mathcal{P}^\text{kin}_{0^+} }{\mathcal{P}^\text{kin}_{0^+} +\mathcal{P}^\text{kin}_\text{bkg} }=
  \left[1+\frac{\mathcal{P}^\text{kin}_\text{bkg}(m_{\cPZ_1}, m_{\cPZ_2}, \vec\Omega | m_{4\ell})}
    {\mathcal{P}^\text{kin}_{0^+} (m_{\cPZ_1}, m_{\cPZ_2}, \vec\Omega | m_{4\ell})}  \right]^{-1}. \\
\end{equation}
The discriminant defined this way does not carry direct discrimination
power based on the four-lepton mass $m_{4\ell}$ between the signal and
the background.  Hence, it can be used as a second discriminating
observable in addition to the $m_{4\ell}$ distribution.  The
$\mathcal{P}_i$ are normalized with additional constant factors for a
given value of $m_{4\ell}$, such that the ratio of probabilities is
scaled by a constant factor leading to probabilities
$P(\mathcal{D}>0.5 \, | \,
\PH)=P(\mathcal{D} < 0.5 \, | \, \text{bkg})$.

In this analysis, the SM Higgs boson signal is distinguished
simultaneously from the background and from alternative signal
hypotheses. The former is separated with $\superKD$, and the latter
with $\spinKD$ observables constructed from the background, signal,
and the probability of the alternative hypotheses defined in
Eqs.~(\ref{eq:kd-prob-sm}) and (\ref{eq:kd-prob-jp}). The $\superKD$
observable extends $\KD$ defined in Eq.~(\ref{eq:kd-mela}) with the
four-lepton mass probability for separation at a fixed value of the
mass $m_{0^+}$:
\begin{equation}
  \label{eq:kd-supermela}
        \mathcal{D}_\text{bkg} =
        \left[1+\frac{\mathcal{P}^\text{kin}_\text{bkg} (m_{\cPZ_1}, m_{\cPZ_2}, \vec\Omega | m_{4\ell})\times \mathcal{P}^\text{mass}_\text{bkg} (m_{4\ell})  }
          {\mathcal{P}^\text{kin}_{0^+} (m_{\cPZ_1}, m_{\cPZ_2}, \vec\Omega | m_{4\ell}) \times \mathcal{P}^\text{mass}_{\rm sig} (m_{4\ell}|m_{0^+}) } \right]^{-1}.
\end{equation}

The other observable discriminates between the SM Higgs boson and the alternative signal hypothesis:
\begin{equation}
  \label{eq:kd-spinmela}
        \mathcal{D}_{J^P} =
        \left[1+\frac{\mathcal{P}^\text{kin}_{J^P} (m_{\cPZ_1}, m_{\cPZ_2}, \vec\Omega | m_{4\ell}) }
          {\mathcal{P}^\text{kin}_{0^+} (m_{\cPZ_1}, m_{\cPZ_2}, \vec\Omega | m_{4\ell}) } \right]^{-1}.
\end{equation}

The spin-0 discriminants $\mathcal{D}_{0^-}$ and
$\mathcal{D}_{0^+_\mathrm{h}}$ are independent of any production
mechanism, since in the production of a spin-0 particle the angular
decay variables are independent of production mechanism.  This is not
the case for the spin-1 and spin-2 signal hypotheses.  Therefore, it
is desirable to test the spin-1 and spin-2 hypotheses in a way that
does not depend on assumptions about the production mechanism.  This
is achieved by either averaging over the spin degrees of freedom of
the produced boson or, equivalently, integrating the matrix elements
squared over the production angles $\cos\theta^*$ and
$\Phi_1$~\cite{Heinemeyer:2013tqa}. With the latter the discriminants
are defined as
\ifthenelse{\boolean{cms@external}}{
\begin{align}
\label{eq:melaSigProd1}
\begin{split}
&\mathcal{D}^\text{dec}_\text{bkg} =
\Bigg[1+\\
  &\frac{ \frac{1}{4\pi}\int \rd\Phi_1  \rd\cos\theta^{*}
    \mathcal{P}^\text{kin}_\text{bkg} (m_{\cPZ_1}, m_{\cPZ_2}, \vec\Omega | m_{4\ell})\times \mathcal{P}^\text{mass}_\text{bkg} (m_{4\ell})  }
       {\mathcal{P}^\text{kin}_{0^+} (m_{\cPZ_1}, m_{\cPZ_2}, \vec\Omega | m_{4\ell}) \times \mathcal{P}^\text{mass}_\text{sig} (m_{4\ell}|m_{0^+}) }
\Bigg]^{-1},\\
\end{split}
\end{align}
}{
\begin{equation}
\label{eq:melaSigProd1}
\mathcal{D}^\text{dec}_\text{bkg} =
\Bigg[1+
  \frac{ \frac{1}{4\pi}\int \rd\Phi_1  \rd\cos\theta^{*}
    \mathcal{P}^\text{kin}_\text{bkg} (m_{\cPZ_1}, m_{\cPZ_2}, \vec\Omega | m_{4\ell})\times \mathcal{P}^\text{mass}_\text{bkg} (m_{4\ell})  }
       {\mathcal{P}^\text{kin}_{0^+} (m_{\cPZ_1}, m_{\cPZ_2}, \vec\Omega | m_{4\ell}) \times \mathcal{P}^\text{mass}_\text{sig} (m_{4\ell}|m_{0^+}) }
\Bigg]^{-1},
\end{equation}

}
\begin{equation}
\label{eq:melaSigProd2}
\mathcal{D}^\text{dec}_{J^P} =
\left[1+
  \frac{ \frac{1}{4\pi}\int \rd\Phi_1\,  \rd\cos\theta^{*}
    \mathcal{P}^\text{kin}_{J^P} (m_{\cPZ_1}, m_{\cPZ_2}, \vec\Omega | m_{4\ell}) }
       {\mathcal{P}^\text{kin}_{0^+} (m_{\cPZ_1}, m_{\cPZ_2}, \vec\Omega | m_{4\ell}) }
\right]^{-1}.
\end{equation}
The superscript ``dec'' indicates that these discriminants use
decay-only information.  The probabilities for spin-0 resonances
are already independent of the production mechanism, however, their
distributions, for all the $J^P$ hypotheses, do carry some production
dependence due to detector and analysis acceptance effects.  Such
production-dependent variations in the discriminant distribution
shapes are found to be small and are treated as systematic
uncertainties.

Table~\ref{tab:kdlist} summarizes all kinematic observables used in
this analysis, for different purposes.  To make an optimal use of the
available information, the distribution of these observables is used
without any selection in a fit.

\begin{table*}[thb]
\centering
\topcaption{List of observables and kinematic discriminants used for
  signal versus background separation and studies of the properties of
  the observed resonance. The alternative hypotheses for $J=0$ are
  independent of the production mechanism without the need of
  integrating out the production angles $\cos\theta^*$ and
  $\Phi_1$. \label{tab:kdlist} }
\begin{scotch}{cl}
Discriminant & Note \\
\hline
\multicolumn{2}{c}{ Observables used for the signal strength measurement } \\
\hline
\vspace{-0.2cm} & \\
$m_{4\ell}$ & Four-lepton invariant mass, main background discrimination. \\
\vspace{-0.2cm} & \\
$\mathcal{D}^\text{kin}_\text{bkg}$ &  Discriminate SM Higgs boson against $ZZ$ background. \\
\vspace{-0.2cm} & \\
$\mathcal{D}_\text{jet}$ & Linear discriminant, uses jet information to identify VBF topology.  \\
\vspace{-0.2cm} & \\
$\VDu$ & $\PT$ of the $4\ell$ system, discriminates between production mechanisms. \\
 & \\
\hline
\multicolumn{2}{c}{ Observables used in the spin-parity hypothesis testing } \\
\hline
\vspace{-0.2cm} & \\
$\mathcal{D}_\text{bkg}$ & Discriminates SM Higgs boson against $\cPZ\cPZ$ background, includes $m_{4\ell}$.  \\
\vspace{-0.2cm} & \\
$\mathcal{D}_{1^-}$ &  Exotic vector  ($1^-$), $\Pq\Paq$ annihilation. \\
\vspace{-0.2cm} & \\
$\mathcal{D}_{1^+}$ &  Exotic pseudovector ($1^+$), $\Pq\Paq$ annihilation. \\
\vspace{-0.2cm} & \\
$\mathcal{D}^{\Pg\Pg}_{2_\mathrm{m}^+}$ &  Graviton-like with minimal couplings ($2_\mathrm{m}^+$), gluon fusion. \\
\vspace{-0.2cm} & \\
$\mathcal{D}^{\Pq\Paq}_{2_\mathrm{m}^+}$ &  Graviton-like with minimal couplings ($2_\mathrm{m}^+$), $\Pq\Paq$ annihilation.  \\
\vspace{-0.2cm} & \\
$\mathcal{D}^{\Pg\Pg}_{2_\mathrm{b}^+}$ &  Graviton-like with SM in the bulk ($2_\mathrm{b}^+$), gluon fusion. \\
\vspace{-0.2cm} & \\
$\mathcal{D}^{\Pg\Pg}_{2_\mathrm{h}^+}$ &  Tensor with higher-dimension operators ($2_\mathrm{h}^+$), gluon fusion. \\
\vspace{-0.2cm} & \\
$\mathcal{D}^{\Pg\Pg}_{2_\mathrm{h}^-}$ & Pseudotensor with higher-dimension operators  ($2_\mathrm{h}^-$), gluon fusion. \\
 & \\
\hline
\multicolumn{2}{c}{ Production-independent observables used in the spin-parity hypothesis testing } \\
\hline
\vspace{-0.2cm} & \\
$\mathcal{D}_{0^-}$ & Pseudoscalar (${0^-}$), discriminates against SM Higgs boson. \\
\vspace{-0.2cm} & \\
$\mathcal{D}_{0_\mathrm{h}^+}$ & Non-SM scalar with higher-dimension operators (${0_\mathrm{h}^+}$).  \\
\vspace{-0.2cm} & \\
$\mathcal{D}^\text{dec}_\text{bkg}$ & Discriminates against $\cPZ\cPZ$ background, includes $m_{4\ell}$, excludes  $\cos\theta^*$, $\Phi_1$.  \\
\vspace{-0.2cm} & \\
$\mathcal{D}^\text{dec}_{1^-}$ &  Exotic vector ($1^-$), decay-only information. \\
\vspace{-0.2cm} & \\
$\mathcal{D}^\text{dec}_{1^+}$ &  Exotic pseudovector ($1^+$), decay-only information. \\
\vspace{-0.2cm} & \\
$\mathcal{D}^\text{dec}_{2_\mathrm{m}^+}$ & Graviton-like with minimal couplings ($2_\mathrm{m}^+$), decay-only information. \\
\vspace{-0.2cm} & \\
\end{scotch}
\end{table*}

This analysis uses the matrix-element likelihood approach ({MELA})
framework~\cite{Chatrchyan:2012ufa, Bolognesi:2012, Anderson:2013fba},
with the matrix-elements for different signal models taken
from \textsc{jhugen}~\cite{Gao:2010qx,Bolognesi:2012,Anderson:2013fba}
and the matrix element for the $\cPq\cPaq\to\cPZ\cPZ$ background taken
from
\MCFM~\cite{MCFM,Campbell:1999ah,Campbell:2011bn}.  Within the {MELA} framework, an analytical parameterization of matrix elements
for signal~\cite{Gao:2010qx,Bolognesi:2012} and
background~\cite{Chen:2012jy} was adopted in the previous analyses of
CMS data with results reported in
Refs.~\cite{Chatrchyan:2012ufa,Chatrchyan:2012br}.  The above
matrix-element calculations are validated against each other and also
tested with the matrix-element kinematic discriminant (MEKD)
framework~\cite{Avery:2012um}, based on \MADGRAPH~\cite{Alwall:2007st}
and \textsc{FeynRules}~\cite{Christensen:2008py}, and with a
stand-alone framework implementation of \MADGRAPH.  The inclusion of
the lepton interference in the kinematic discriminant parameterization
is a small improvement in the expected separation significance of
$\sim$3\% for spin-0 models with respect to earlier published
results~\cite{Chatrchyan:2012ufa,Chatrchyan:2012br}, as indicated by
cross-checks with generator-based matrix-element calculations
performed in the MELA and MEKD frameworks within studies reported in
Ref.~\cite{Chatrchyan:2012br}.

Detector acceptance effects approximately cancel in the probability
ratios, such as those in Eq.~(\ref{eq:kd-mela}).  In principle, the
kinematic discriminants could be modified to account for detector
resolution effects. However, the matrix-element approach with detector
transfer functions modeling detector resolution effects showed nearly
identical performance. This is not unexpected for leptons, as their
resolutions are of $\mathcal{O}(1\%)$ and are therefore negligible.

In order to provide additional validation of the kinematic
discriminants, machine-learning techniques have been used to construct
discriminants.  Two techniques have been used: the Bayesian neural
networks (BNN) framework~\cite{BNN1,BNN2} and the BDT
framework~\cite{bdt1,bdt2,tmva}.  In the BNN framework, a Bayesian
procedure is used to create a posterior probability density over the
space of neural network parameters. This probability density is then
used to calculate a BNN.  In both frameworks, a discriminant is built
using the four-lepton angular and mass variables, and the output is
used in the same way as the $\KD$ in the analysis described above.
The BNN and BDT discriminants are trained using simulated samples to
discriminate signatures for signal events from those for background
events or to discriminate between different signal hypotheses.  The MC
samples generated for training are based on the same matrix elements
for signal and background as used in the analysis and include the
effects of the full detector simulation.  The machine-trained
discriminants are found to give similar performance to the
matrix-element approaches described above.

\section{Yields and kinematic distributions}
\label{sec:yields}

The signal and background yields are extracted from a fit to the
invariant mass and other kinematic properties, characterizing the
decay of the Higgs boson candidate and its production mechanism.  The
expected distributions of signal and background components are used as
probability density functions in the likelihood function.  Simulation
and control samples from data are used to estimate the initial fit
values for the signal and background yields.

\begin{figure}[!htb]
\centering

\includegraphics[width=\cmsFigWidth]{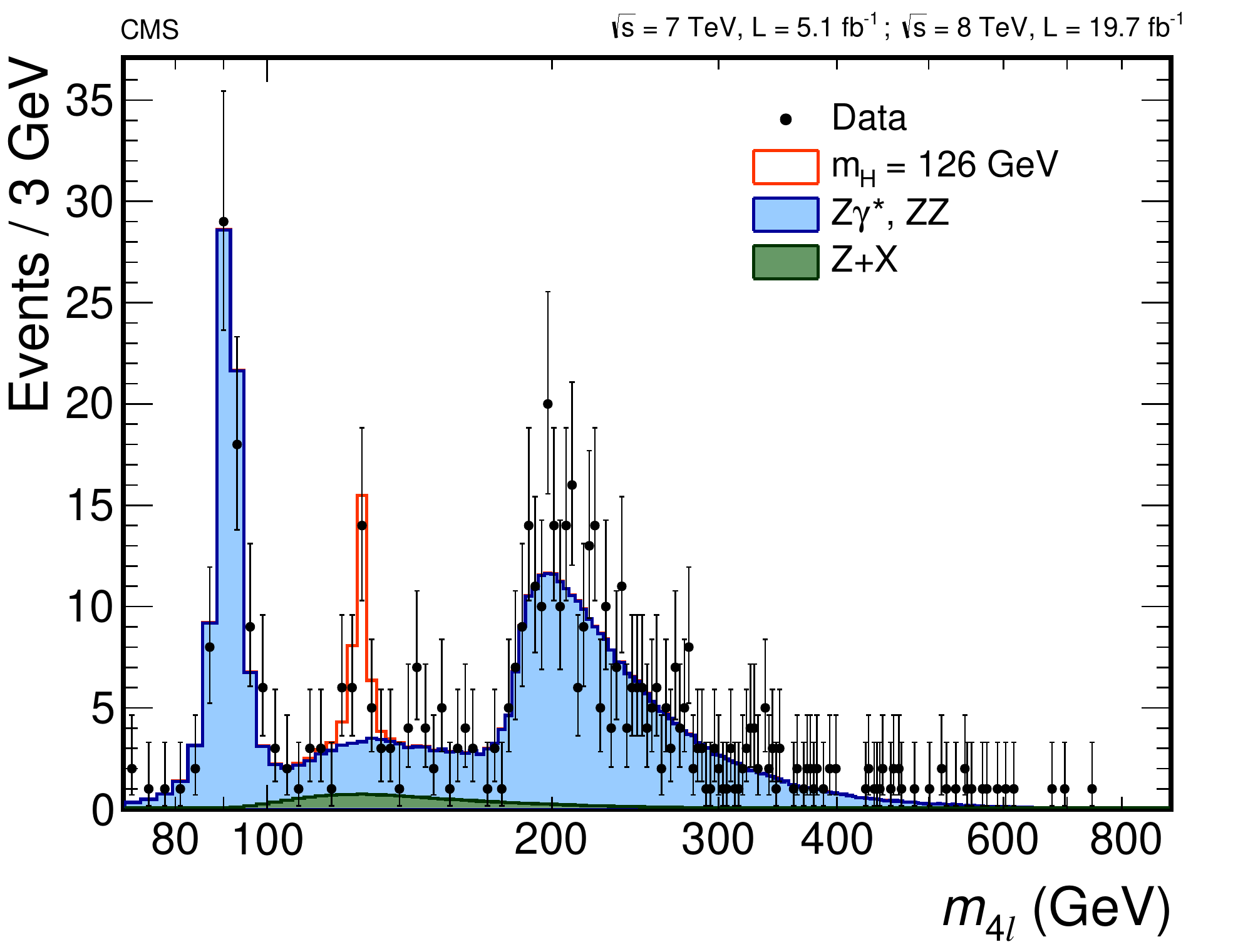} \\
\caption{ Distribution of the four-lepton reconstructed mass in the
  full mass range $70<m_{4\ell}<1000$\GeV for the sum of the $4\Pe$,
  $2\Pe2\Pgm$, and $4\Pgm$ channels.  Points with error bars represent
  the data, shaded histograms represent the backgrounds, and the
  unshaded histogram represents the signal expectation for a mass
  hypothesis of $\mH=126$\GeV. Signal and the $\cPZ\cPZ$ background are
  normalized to the SM expectation; the $\cPZ+\X$ background to the
  estimation from data.  The expected distributions are presented as
  stacked histograms. No events are observed with
  $m_{4\ell}>800\GeV$.  \label{fig:Mass4l}}

\end{figure}

The background from $\cPZ\cPZ$ and $\cPZ+\X$ processes dominates after
the event selection.  The reconstructed four-lepton invariant mass
distribution for the combined $4\Pe$, $2\Pe2\Pgm$, and $4\Pgm$
channels is shown in Fig.~\ref{fig:Mass4l} and compared with the
expectations from background processes. Here, and in the other figures
of this section, the normalization and shape of the $\cPZ\cPZ$
background and the signal ($\mH=126$\GeV) are obtained from
simulation, while the normalization and shape of the reducible
background is estimated from control samples in data, as described in
Sec.~\ref{sec:reducible_bkg}. The error bars on data points are
asymmetric Poisson uncertainties that cover the 68\% probability
interval around the central value~\cite{Cousins:1994yw}.  A clear peak
around $m_{4\ell} = 126$\GeV is seen, not expected from background
processes, confirming with a larger data sample the results reported
in
Refs.~\cite{Aad:2012tfa,Chatrchyan:2012ufa,Chatrchyan:2013lba,Chatrchyan:2012br}. The
observed distribution is in good agreement with the expected
backgrounds and a narrow resonance compatible with the SM Higgs boson
with $\mH$ around 126\GeV.  The $\cPZ\rightarrow 4\ell$ resonance peak
at $m_{4\ell}=m_{\cPZ}$ is observed in agreement with simulation.  The
measured distribution at masses greater than $2m_{\cPZ}$ is dominated
by the irreducible $\cPZ\cPZ$ background, where the two $\cPZ$ bosons
are produced on shell.

\begin{figure}[!htb]
  \begin{center} \includegraphics[width=\cmsFigWidthStd]{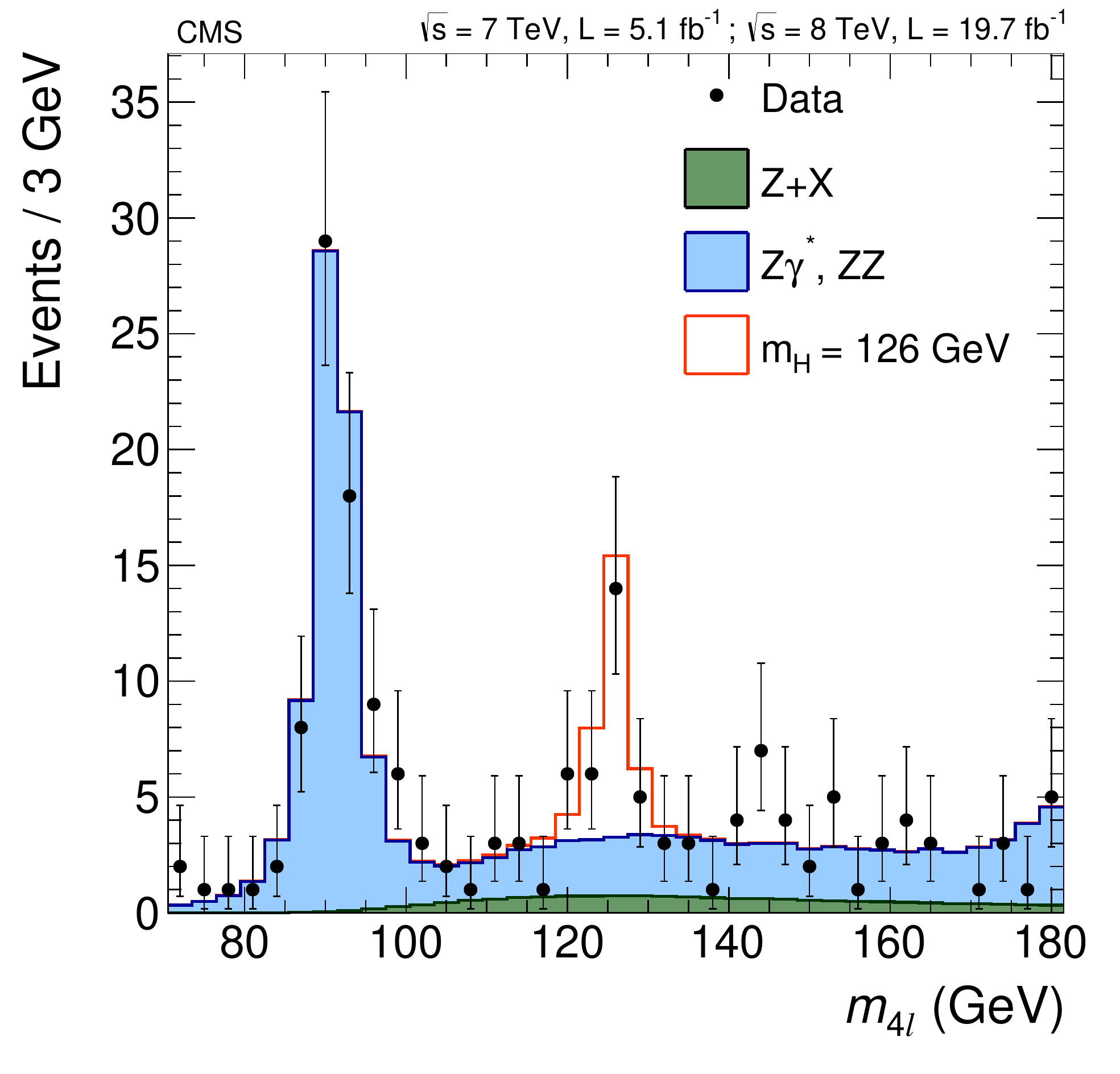}
   \caption{Distribution of the four-lepton reconstructed mass for the
    sum of the $4\Pe$, $2\Pe2\Pgm$, and $4\Pgm$ channels for the mass
    region $70<m_{4\ell}<180$\GeV.  Points with error bars represent
    the data, shaded histograms represent the backgrounds, and the
    unshaded histogram represents the signal expectation for a mass hypothesis of
    $\mH=126$\GeV. Signal and the $\cPZ\cPZ$ background are normalized to
    the SM expectation, the $\cPZ+\X$ background to the estimation from
    data.  \label{fig:Mass4lLow}} \end{center}
\end{figure}

\begin{figure*}[thb!]
\begin{center}
  \includegraphics[width=0.32\linewidth]{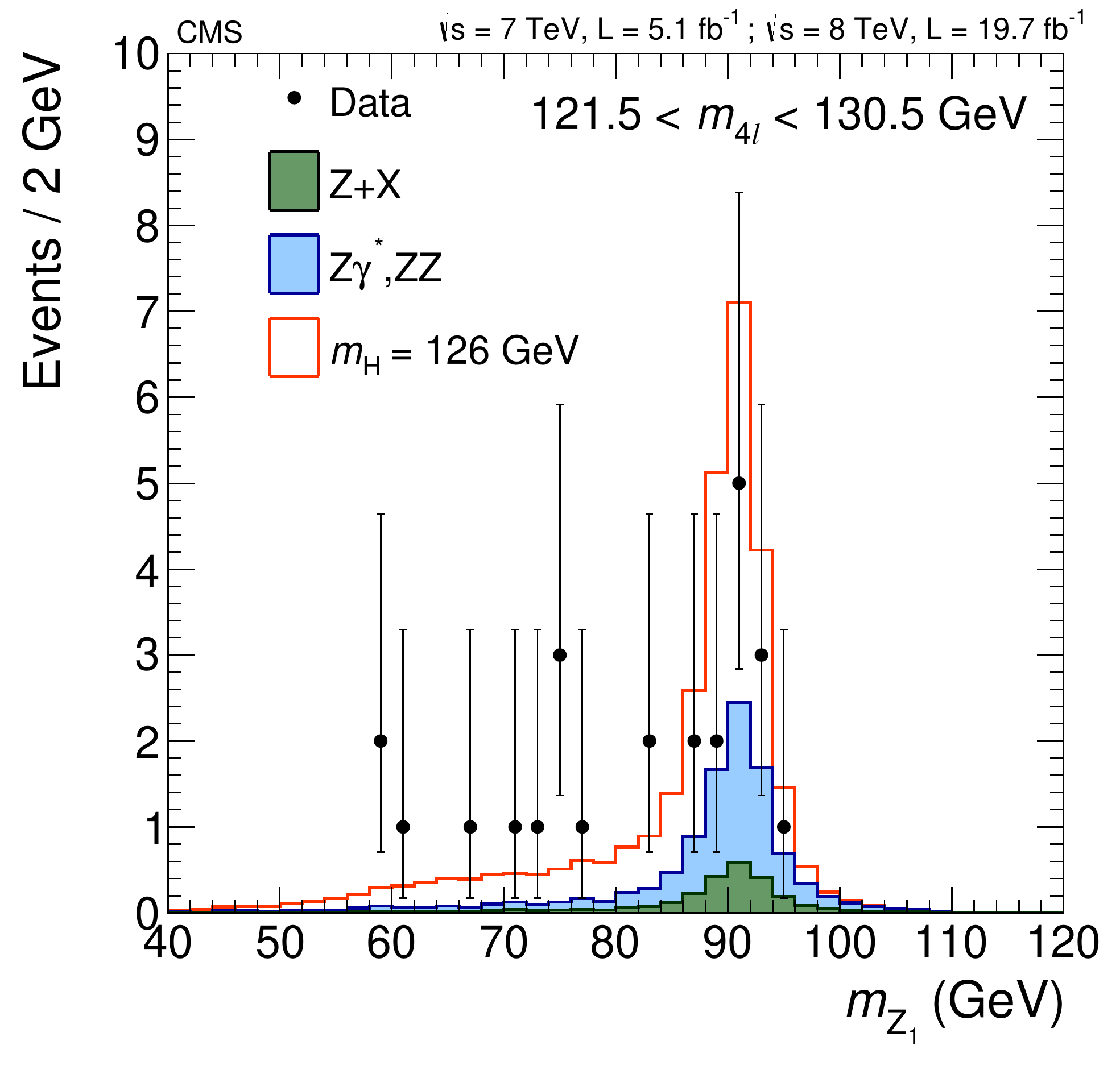}
  \includegraphics[width=0.32\linewidth]{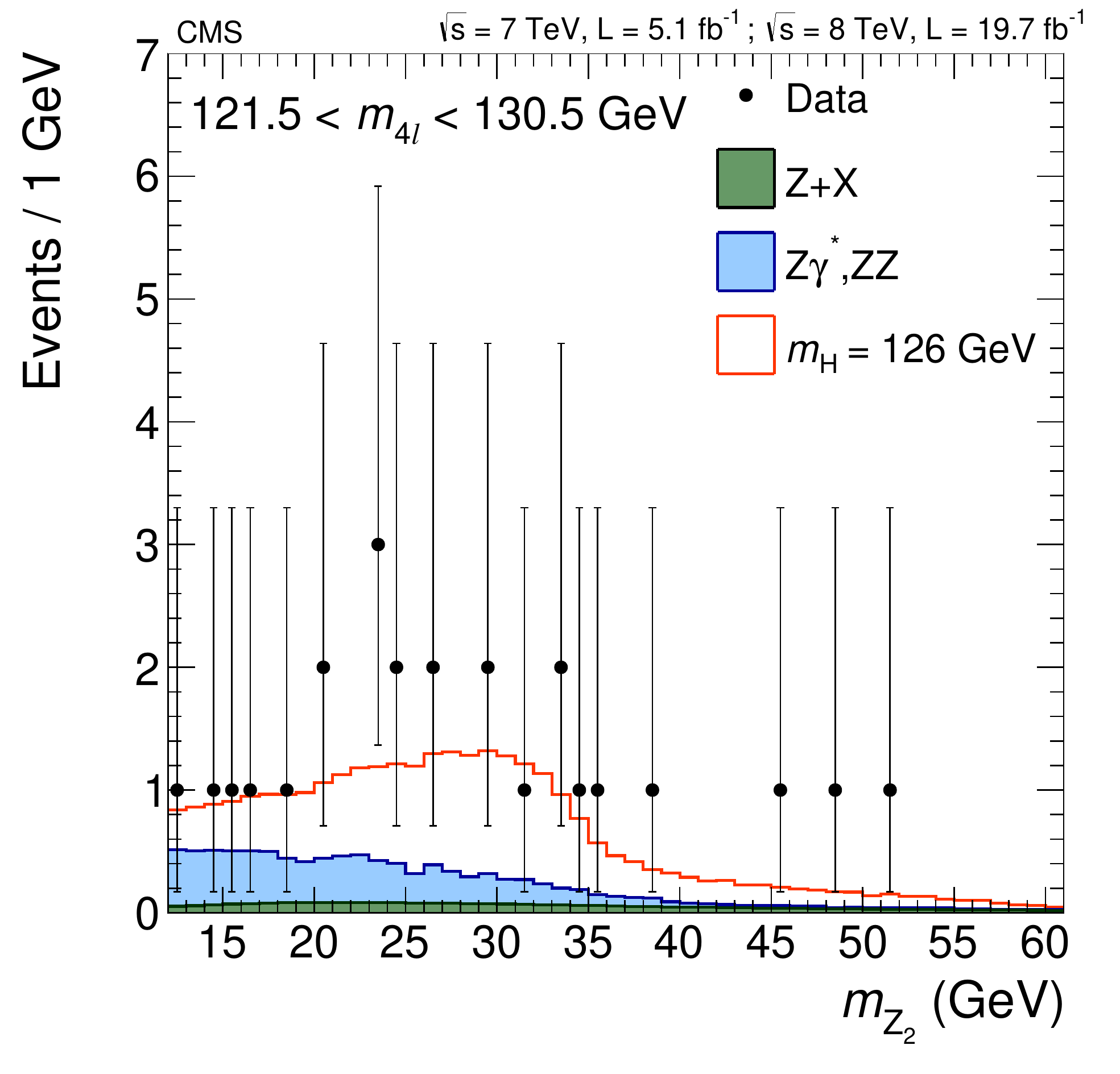}
  \includegraphics[width=0.32\linewidth]{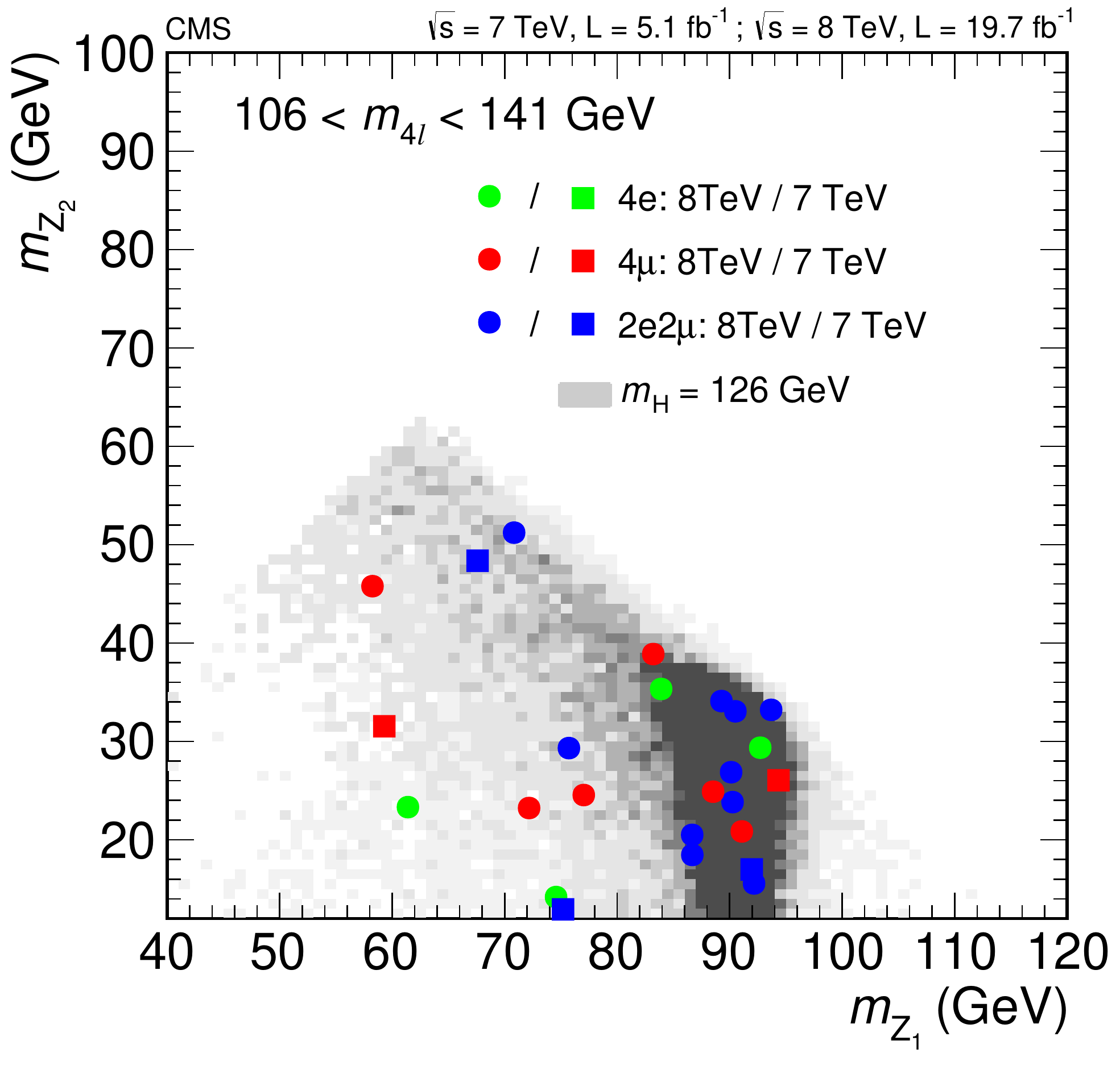}
  \caption{Distribution of (left) the ${\cPZ_1}$ and (center)
    the ${\cPZ_2}$ reconstructed invariant masses, in the mass region
    $121.5 < m_{4\ell} < 130.5\GeV$, for the sum of the four-lepton
    channels. Points represent the data, and shaded histograms
    represent the background. The signal expectation at $\mH =
    126\GeV$ is shown as the unshaded histogram. Signal and background
    histograms are stacked. (right) Two-dimensional distribution of the two
    variables in the mass region $106 < m_{4\ell} < 141\GeV$,
    corresponding to the range used in the signal extraction for $\mH = 126\GeV$,
    for the sum of the $4\ell$ channels. The signal expectation at
    $\mH = 126\GeV$ is shown as the grey scale.  \label{fig:mZ12}}
\end{center}
\end{figure*}

The number of candidates observed in data as well as the expected
yields for background and several SM Higgs boson mass hypotheses are
reported in Table~\ref{tab:PreFitYields}, for $m_{4\ell} > 100
\GeV$. The observed event rates for the various channels are
compatible with SM background expectations in the $m_{4\ell}$ region
above 2$m_\cPZ$, while a deviation is observed in the lower region.
Given that the excess of events observed in the $4\ell$ mass spectrum
is localized in a narrow region in the vicinity of 126\GeV, the events
expected in a narrower range, $121.5 < m_{4\ell} < 130.5\GeV$, are
reported in
Table~\ref{tab:PreFitYieldsSigRegion}. Table~\ref{tab:PreFitYieldsSigRegionCat}
reports the breakdown of the events observed in data and the expected
background yields in the same $m_{4\ell}$ region in the two analysis
categories, together with the expected yield for a SM Higgs boson with
$\mH=126$\GeV, split by production mechanism.  The $m_{4\ell}$
distribution for the sum of the $4\Pe$, $2\Pe2\Pgm$, and $4\Pgm$
channels, in the mass region $70<m_{4\ell}<180$\GeV, is shown in
Fig.~\ref{fig:Mass4lLow}. Figure~\ref{fig:mZ12} shows the
reconstructed invariant masses of the $\cPZ_1$ and $\cPZ_2$ in a
$m_{4\ell}$ range between 121.5 and 130.5\GeV.

\begin{table}[!hbt]
  \begin{center}
    \topcaption{The number of observed candidate events compared to the
      mean expected background and signal rates for each final state.
      Uncertainties include statistical and systematic sources.  The results
      are given integrated over the full mass measurement range $m_{4\ell} >
      100\GeV$ and for 7 and 8\TeV data combined.
      \label{tab:PreFitYields}}
    \begin{scotch}{lcccc}

      Channel         & $4\Pe$ & $2\Pe2\Pgm$ & $4\Pgm$ & $4\ell$  \\
      \hline
      $\cPZ\cPZ$ background  & 77  $\pm$ 10    &  191  $\pm$  25  & 119  $\pm$  15     &  387  $\pm$ 31\\
      $\cPZ + \X$  background & 7.4 $\pm$ 1.5   & 11.5  $\pm$ 2.9  & 3.6  $\pm$ 1.5     &  22.6 $\pm$ 3.6  \\
      \hline
      All backgrounds        & 85 $\pm$ 11     & 202  $\pm$ 25    &  123  $\pm$ 15     &  410 $\pm$ 31 \\
      \hline
      $\mH$ =  500\GeV &  5.2  $\pm$  0.6  & 12.2  $\pm$  1.4 &   7.1  $\pm$  0.8  &  24.5 $\pm$ 1.7  \\
      $\mH$ =  800\GeV &  0.7  $\pm$  0.1  &  1.6  $\pm$  0.2 &   0.9  $\pm$  0.1  &  3.1  $\pm$ 0.2 \\
      \hline
      Observed  & 89 & 247 & 134 & 470\\
    \end{scotch}
  \end{center}
\end{table}

\begin{table}[!hbt]
  \begin{center}
    \topcaption{ The number of observed candidate events compared to the
      mean expected background and signal rates for each final state.
      Uncertainties include statistical and systematic sources.  The results
      are integrated over the mass range from 121.5 to 130.5\GeV and for 7
      and 8\TeV data combined.
      \label{tab:PreFitYieldsSigRegion}}
    \begin{scotch}{lcccc}
      Channel        & $4\Pe$ & $2\Pe2\Pgm$ & $4\Pgm$  & $4\ell$ \\
      \hline
      $\cPZ\cPZ$ background &  1.1  $\pm$  0.1  &  3.2  $\pm$  0.2  &  2.5  $\pm$  0.2  &  6.8 $\pm$ 0.3  \\
      $\cPZ + \X$ background &  0.8  $\pm$  0.2  &  1.3  $\pm$  0.3  &  0.4  $\pm$  0.2  &  2.6 $\pm$ 0.4 \\
      \hline
      All backgrounds            &  1.9  $\pm$  0.2   &  4.6  $\pm$ 0.4  & 2.9  $\pm$ 0.2  & 9.4  $\pm$ 0.5\\
      \hline
      $\mH$ =  125\GeV &  3.0  $\pm$  0.4  &  7.9  $\pm$  1.0  &  6.4  $\pm$  0.7  & 17.3  $\pm$ 1.3 \\
      $\mH$ =  126\GeV &  3.4  $\pm$  0.5  &  9.0  $\pm$  1.1  &  7.2  $\pm$  0.8  & 19.6  $\pm$ 1.5 \\
      \hline
      Observed  & 4 & 13 & 8 & 25 \\
    \end{scotch}
  \end{center}
\end{table}

\begin{table}[!hbt]
  \begin{center}
    \topcaption{ The number of observed candidate events compared to the
      mean expected background and signal rates for the sum of the three
      final states for each of the two analysis categories.
      Uncertainties include statistical and systematic sources.  The
      results are integrated over the mass range from 121.5 to 130.5\GeV
      and for 7 and 8\TeV data combined.  The expected signal yield for a
      SM Higgs boson with $\mH=126$\GeV is reported, broken down by the
      production mechanism. \label{tab:PreFitYieldsSigRegionCat}}
    \begin{scotch}{lcc}
      Category       & 0/1-jet  &   Dijet  \\
      \hline
      $\cPZ\cPZ$ background &  6.4  $\pm$  0.3  &  0.38 $\pm$  0.02  \\
      $\cPZ + \X$ background &  2.0  $\pm$  0.3  &  0.5  $\pm$  0.1  \\
      \hline
      All backgrounds       &  8.5  $\pm$  0.5  &  0.9  $\pm$ 0.1 \\
      \hline
      $\Pg\Pg \PH$               &  15.4 $\pm$  1.2   &   1.6  $\pm$  0.3 \\
      $\ttbar\PH$           &         \NA          &  0.08  $\pm$  0.01 \\
      VBF                   &  0.70 $\pm$  0.03  &  0.87  $\pm$  0.07 \\
      $\PW\PH$              &  0.28 $\pm$  0.01  &  0.21  $\pm$  0.01 \\
      $\cPZ\PH$             &  0.21 $\pm$  0.01  &  0.16  $\pm$  0.01 \\
      \hline
      All signal, $\mH=126$\GeV            &  16.6 $\pm$  1.3  &  3.0  $\pm$  0.4 \\
      \hline
      Observed              & 20 & 5 \\
    \end{scotch}
  \end{center}
\end{table}

\begin{figure}[!htb]
  \begin{center}
    \includegraphics[width=\cmsFigWidthStd]{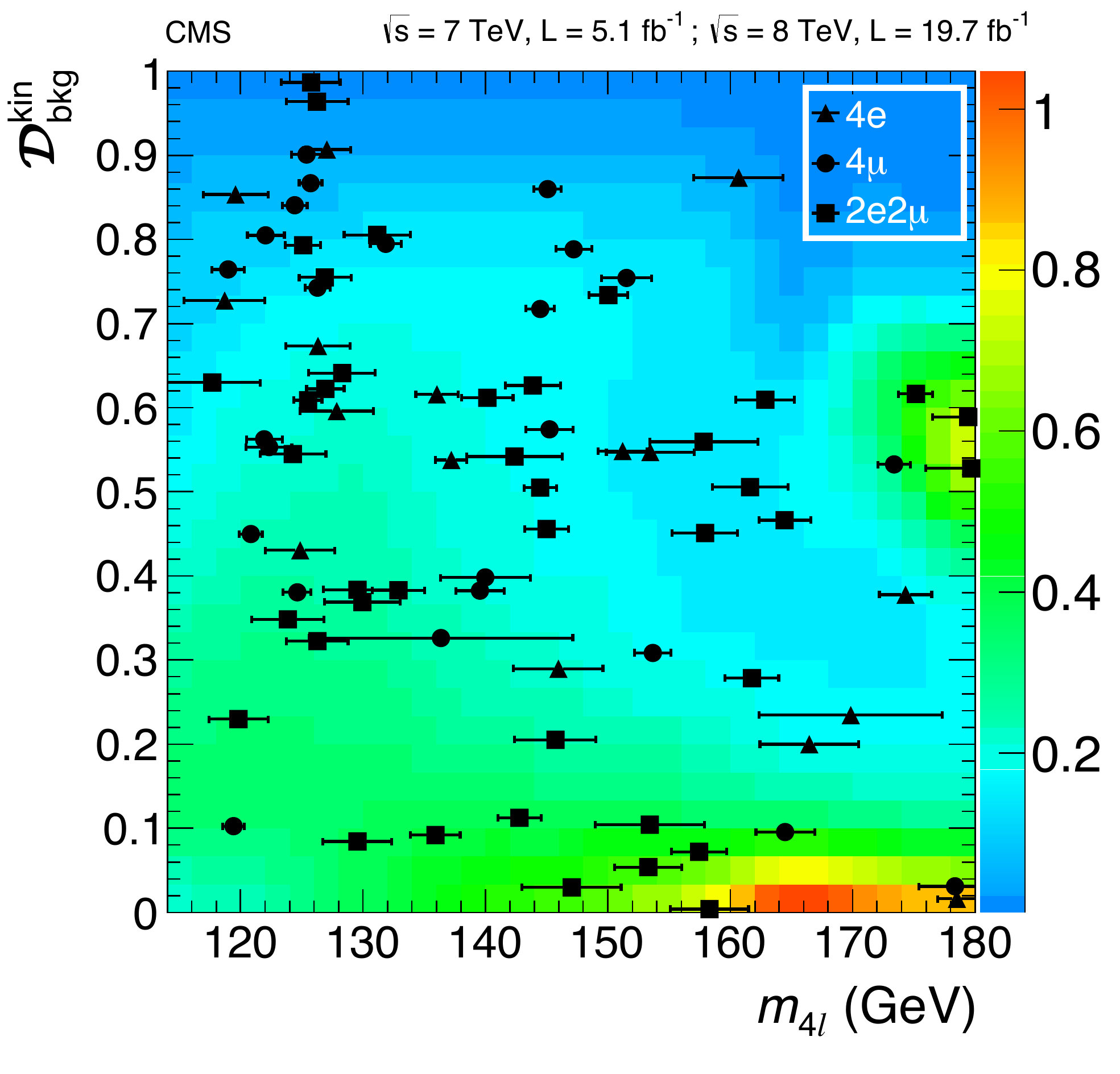}
    \includegraphics[width=\cmsFigWidthStd]{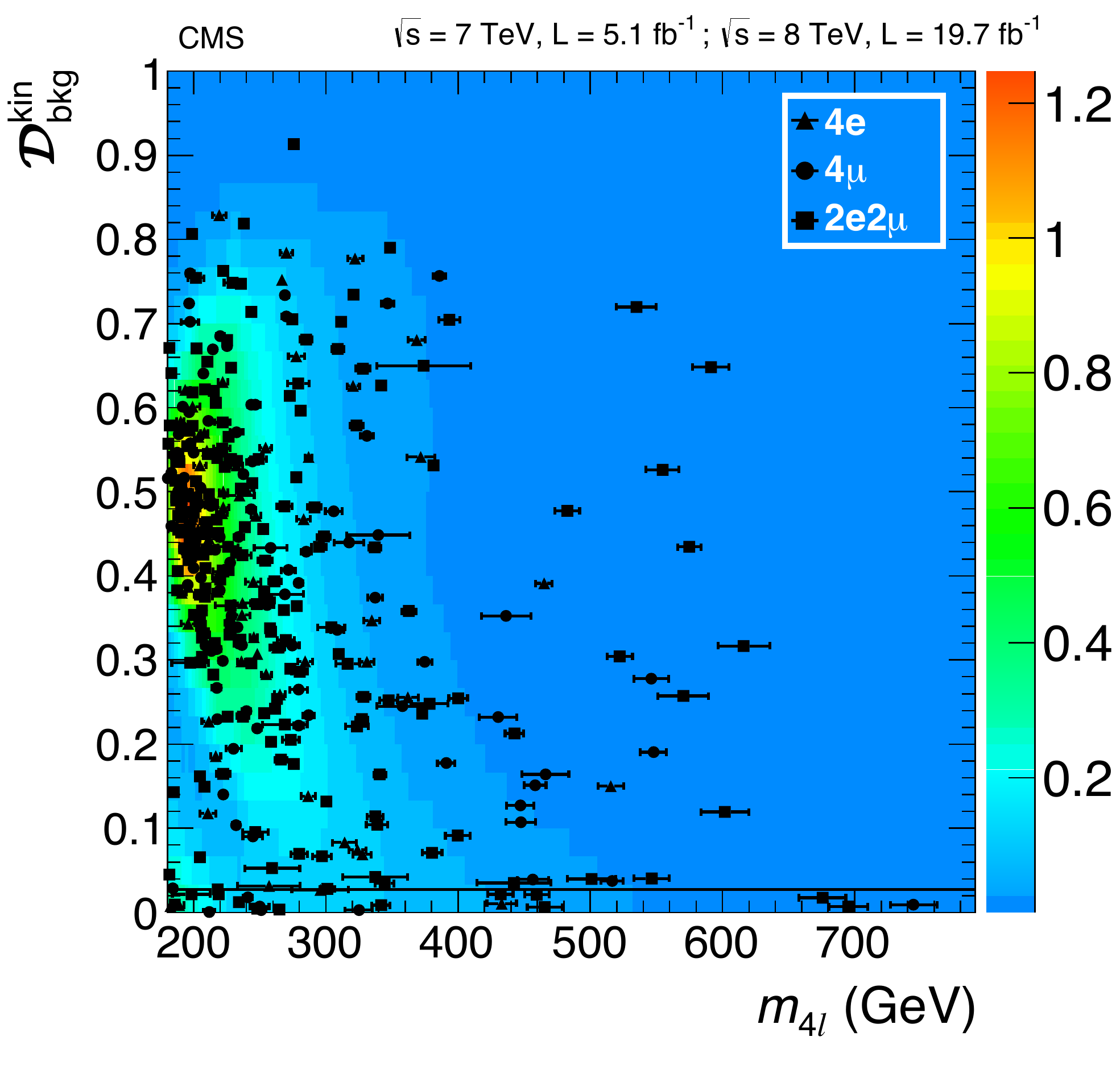}
    \caption{Distribution of the kinematic discriminant $\KD$ versus
      the four-lepton reconstructed mass $m_{4\ell}$ in the (\cmsLeft)
      low-mass and (\cmsRight) high-mass regions.  The color scale represents
      the expected relative density in linear scale (in arbitrary
      units) of background events.  The points show the data and the
      measured per-event invariant mass uncertainties as horizontal
      bars. One $2\Pe2\Pgm$ event with $m_{4\ell}\approx 220\GeV$ and
      small $\KD$ has a huge mass uncertainty, and it is displayed as the
      horizontal line. No events are observed for $m_{4\ell}>800$\GeV.
      \label{fig:KDvsM4lFullMass}}
  \end{center}
\end{figure}

\begin{figure}[!htb]
  \begin{center}
    \includegraphics[width=\cmsFigWidthStd]{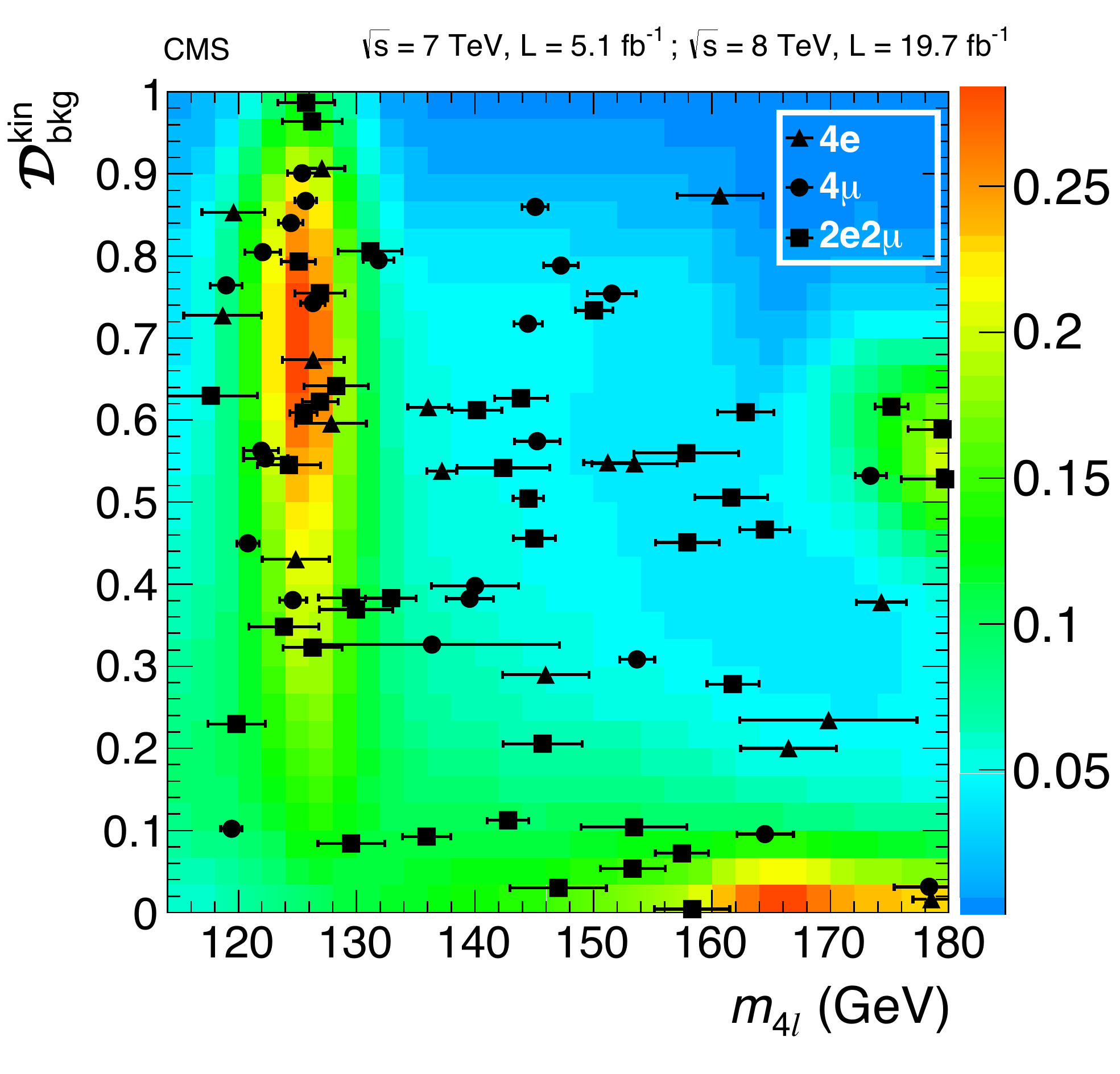}
    \includegraphics[width=\cmsFigWidthStd]{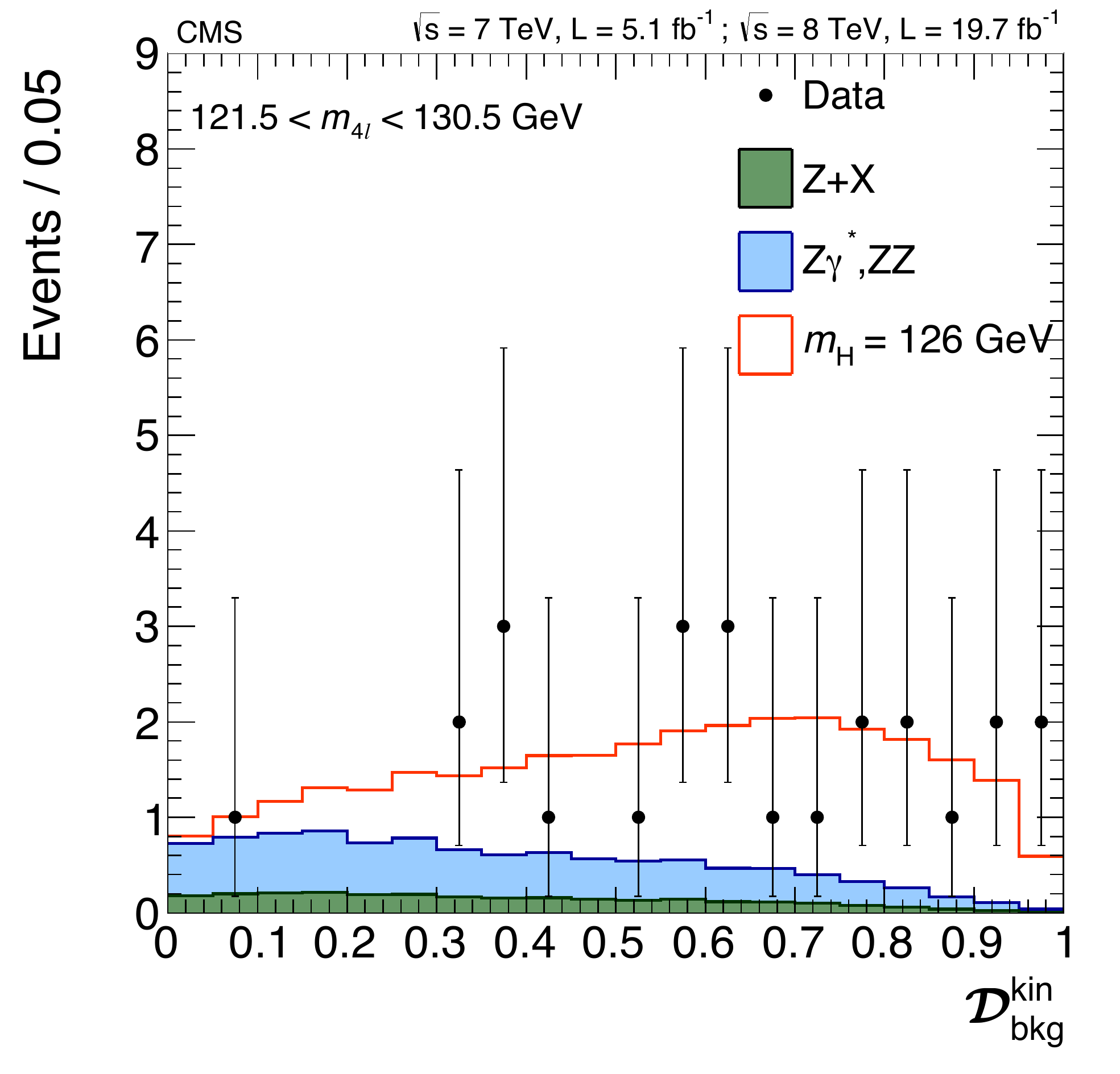}
        \caption{(\cmsLeft) Distribution of $\KD$ versus $m_{4\ell}$ in
      the low-mass range with colors shown for the expected relative
      density in linear scale (in arbitrary units) of background plus
      the Higgs boson signal for $m_H=126$\GeV. The points show the data,
      and horizontal bars represent the measured mass
      uncertainties. (\cmsRight) Distribution of the kinematic
      discriminant $\KD$ for events in the mass region $121.5 <
      m_{4\ell} < 130.5$\GeV.  Points with error bars represent the
      data, shaded histograms represent the backgrounds, and the
      unshaded histogram the signal expectation.  Signal and
      background histograms are
      stacked.  \label{fig:KDLow}} \end{center}
\end{figure}

The distributions of the $\KD$ versus $m_{4\ell}$ are shown for the
selected events and compared to the SM background expectation in
Fig.~\ref{fig:KDvsM4lFullMass}. The distribution of events in the
$(m_{4\ell}, \KD)$ plane agrees well with the SM background
expectation in the high-mass range
[Fig.~\ref{fig:KDvsM4lFullMass}~(\cmsRight)], while discrepancies in
the two-dimensional plane are observed in the low-mass range
$110<m_{4\ell}<180$\GeV [Fig.~\ref{fig:KDvsM4lFullMass}~(\cmsLeft)],
indicative of the presence of a signal.
Figure~\ref{fig:KDLow}~(\cmsLeft) shows the same data points as in
Fig.~\ref{fig:KDvsM4lFullMass}~(\cmsLeft), but compared with the expected
distribution from SM backgrounds plus the contribution of a Higgs
boson with $\mH = 126\GeV$. A signal-like clustering of events is
apparent at high values of $\KD$ and for $m_{4\ell} \approx 126
\GeV$.  Figure~\ref{fig:KDLow}~(\cmsRight) shows the distribution of the
kinematic discriminant $\KD$ in the mass region $121.5 < m_{4\ell} <
130.5$\GeV.

\begin{figure}[!htb]
  \begin{center}
    \includegraphics[width=\cmsFigWidthStd]{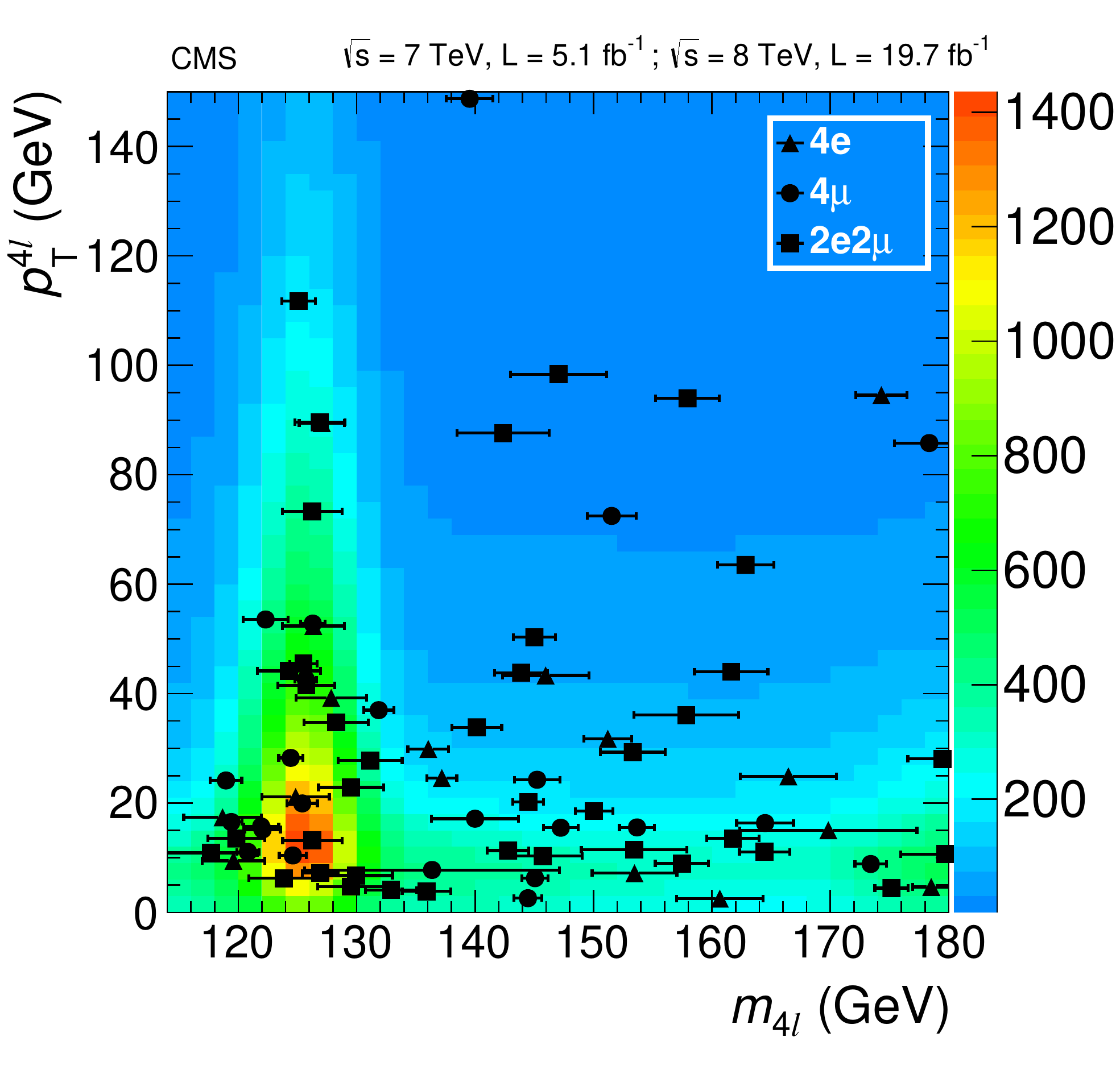}
    \includegraphics[width=\cmsFigWidthStd]{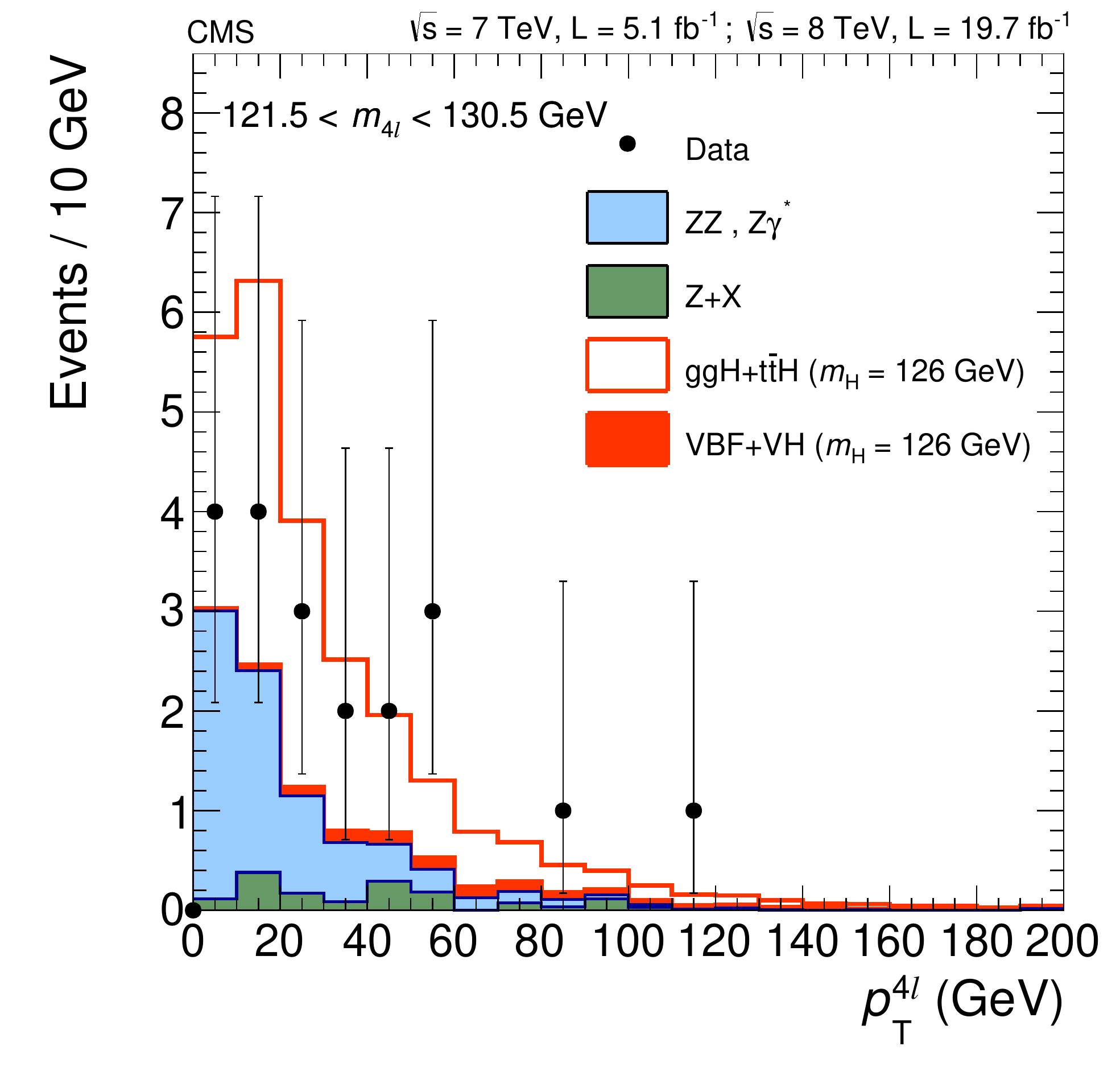}
    \caption{(\cmsLeft) Distribution of $\VDu$ versus $m_{4\ell}$ in the
      low-mass-range 0/1-jet category with colors shown for the
      expected relative density in linear scale (in arbitrary units)
      of background plus the Higgs boson signal for $m_H=126$\GeV. No
      events are observed for $\PT>150$\GeV. The points show the data,
      and horizontal bars represent the measured mass
      uncertainties. (\cmsRight) Distribution of $\VDu$ in the 0/1-jet
      category for events in the mass region $121.5 < m_{4\ell} <
      130.5$\GeV.  Points with error bars represent the data, shaded
      histograms represent the backgrounds, and the red histograms represent the
      signal expectation, broken down by production mechanism. Signal
      and background histograms are stacked.
 \label{fig:DpTLow}}
  \end{center}
\end{figure}

\begin{figure}[!htb]
\begin{center}
  \includegraphics[width=\cmsFigWidthStd]{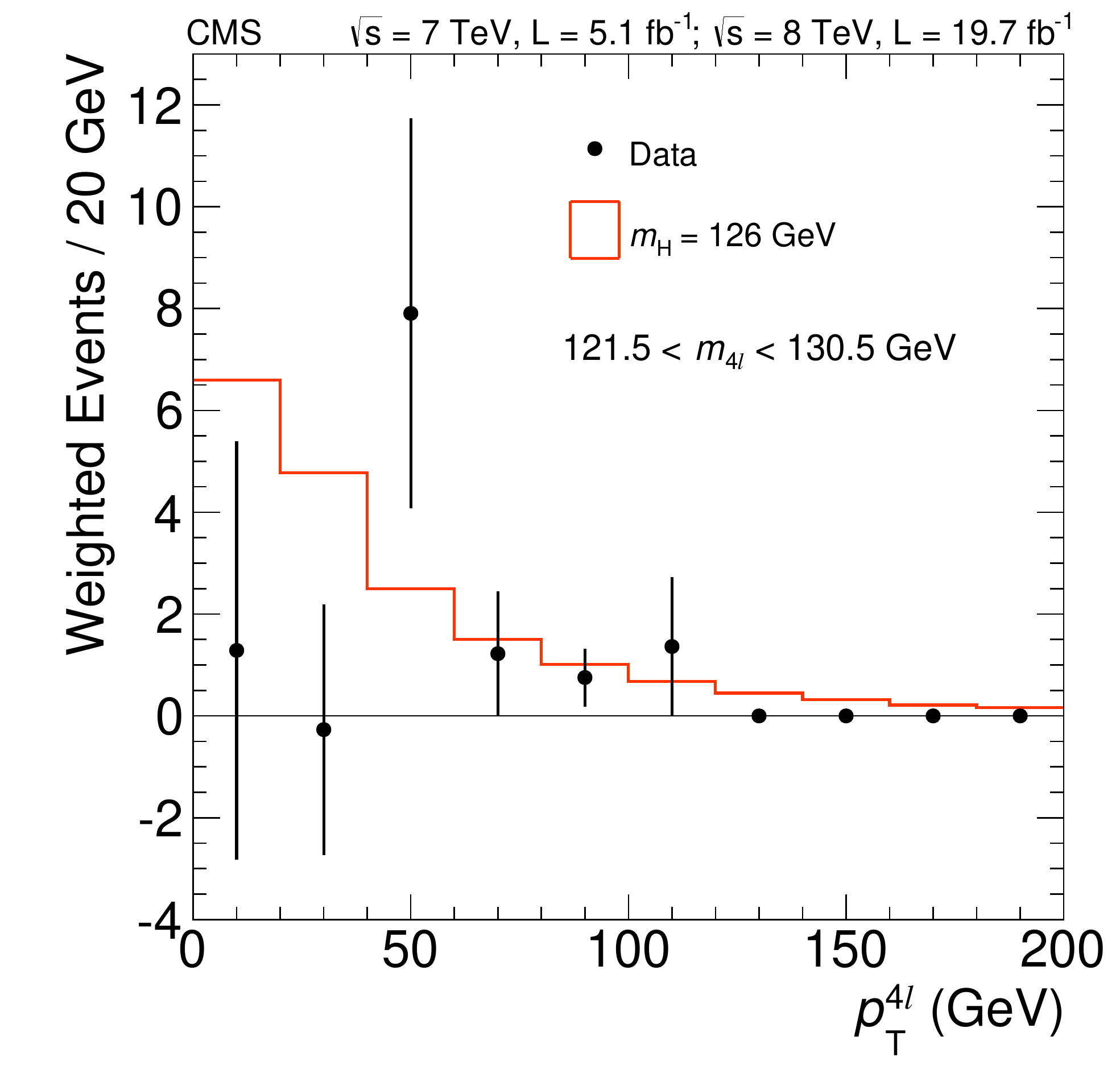}
  \caption{${}_s\mathcal{P}lot$ signal-weighted distribution of the
    four-lepton system $\VDu$ for all the selected events in the mass
    region $121.5 < m_{4\ell} < 130.5$\GeV.  The red solid line
    represents the expectation from a SM Higgs
    boson. \label{fig:ptsplot}}
\end{center}
\end{figure}

The distribution of the transverse momentum of the $4\ell$ system in
 the 0/1-jet category and its joint distribution with $m_{4\ell}$ are
 shown in Fig.~\ref{fig:DpTLow}. The $\PT$ spectrum shows good
 agreement with a SM Higgs boson hypothesis with $\mH=126$\GeV in the
 0/1-jet category with few events having $\PT>60$\GeV, where VBF and
 V$\PH$ production are relatively more relevant. In order to compare
 the $\PT$ spectrum in data with the SM Higgs boson distribution more
 quantitatively, a background subtraction using the
 ${}_s\mathcal{P}lot$ weighting technique~\cite{Pivk:2004ty} is
 performed. The event weights, related to the probability for each
 event to be signal-like or background-like, are computed according to
 the one-dimensional likelihood based on the $m_{4\ell}$ distribution,
 which shows a small correlation with the four-lepton $\VDu$.  The
 weighted distribution has the property that it corresponds to the
 signal-only distribution and is normalized to the fitted signal
 yield.  The background-subtracted weighted $\VDu$ distribution is
 shown in Fig.~\ref{fig:ptsplot}.

\begin{figure}[!htb]
  \begin{center}
    \includegraphics[width=\cmsFigWidthStd]{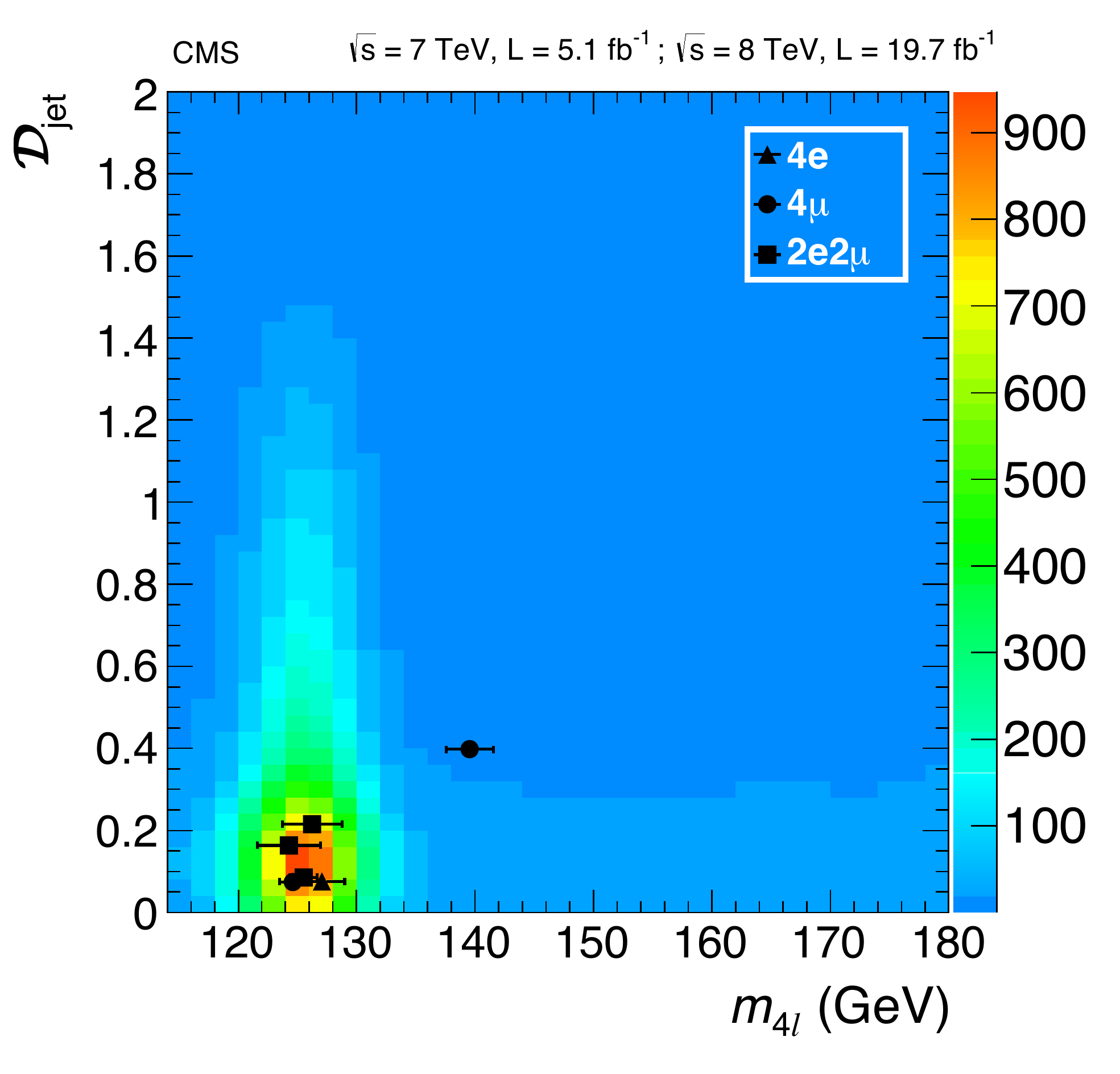}
    \includegraphics[width=\cmsFigWidthStd]{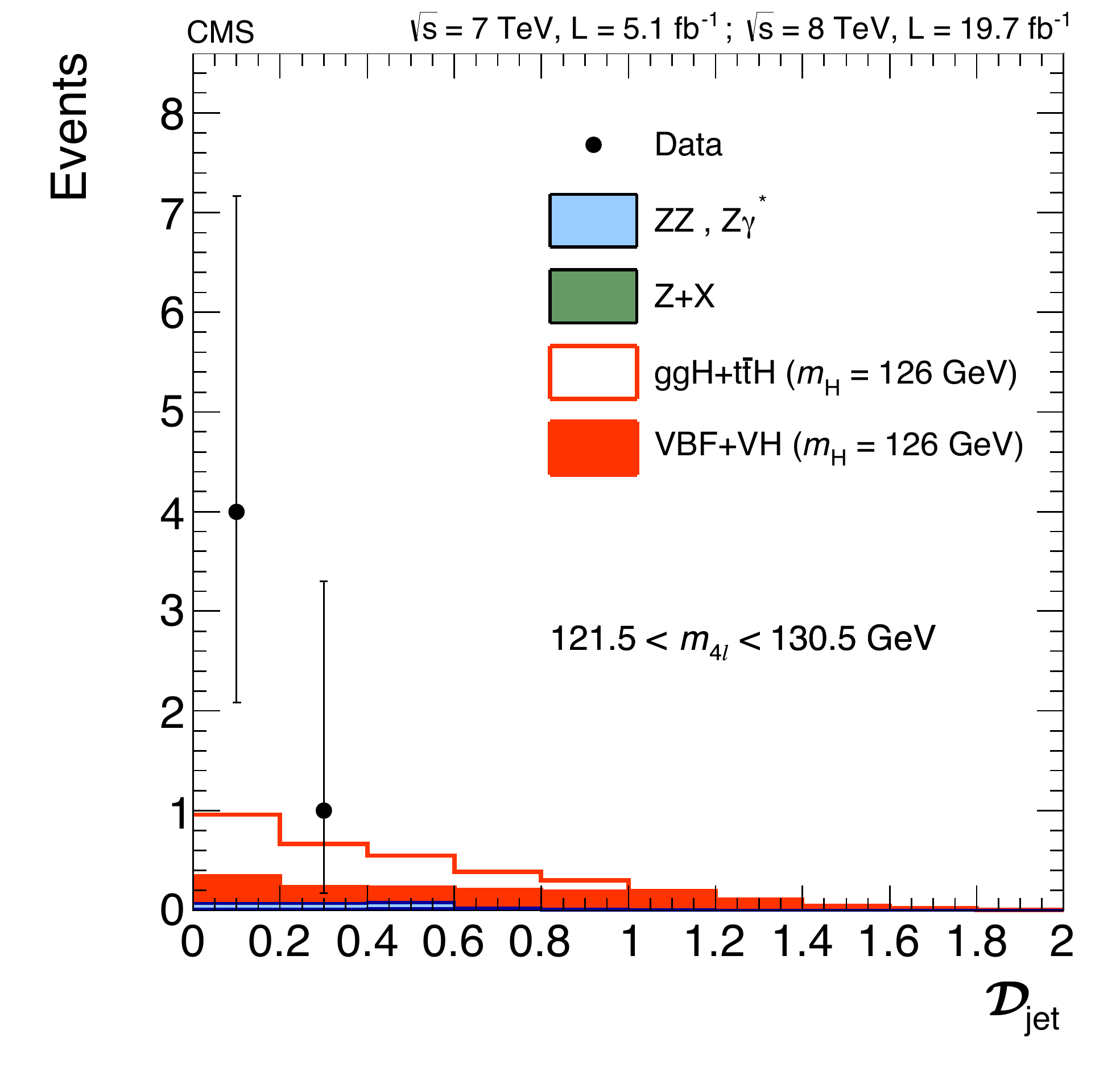}
    \caption{(\cmsLeft) Distribution of $\VDj$ versus $m_{4\ell}$ in the
      low-mass-range dijet category with colors shown for the expected
      relative density in linear scale (in arbitrary units) of
      background plus the Higgs boson signal for $m_H=126$\GeV.  The
      points show the data and horizontal bars represent the measured
      mass uncertainties. (\cmsRight) Distribution of $\VDj$ in the dijet
      category for events in the mass region $121.5 < m_{4\ell} <
      130.5$\GeV.  Points with error bars represent the data, shaded
      histograms represent the backgrounds, and the red histograms represent the
      signal expectation, broken down by production mechanism. Signal
      and background histograms are stacked. \label{fig:FisherLow}}
  \end{center}
\end{figure}

The distribution of the production mechanism discriminant in the dijet
category and its joint distribution with $m_{4\ell}$ are shown in
Fig.~\ref{fig:FisherLow}. Good agreement is found with the expectation
from simulation, which predicts a negligible background and a fraction
of 42\% of the signal events arising from vector-boson-induced
production (VBF and V$\PH$). No events with a high rank of the $\VDj$
($\VDj>0.5$) discriminant are observed.

\section{Higgs boson properties measurement}
\label{sec:statisticalanalysis}

In this section the fit models used to perform the measurements in the
$\PH\to\cPZ\cPZ\to4\ell$ channel, based on the observables defined in
the previous sections, are presented. Then, the systematic
uncertainties effects considered in the fits for both assessing the
presence of a signal and performing the measurement of different
properties are described.

\subsection{Multidimensional likelihoods}
\label{sec:combinedfit}

\begin{figure*}[htb]
\centering
     \includegraphics[width=0.32\linewidth]{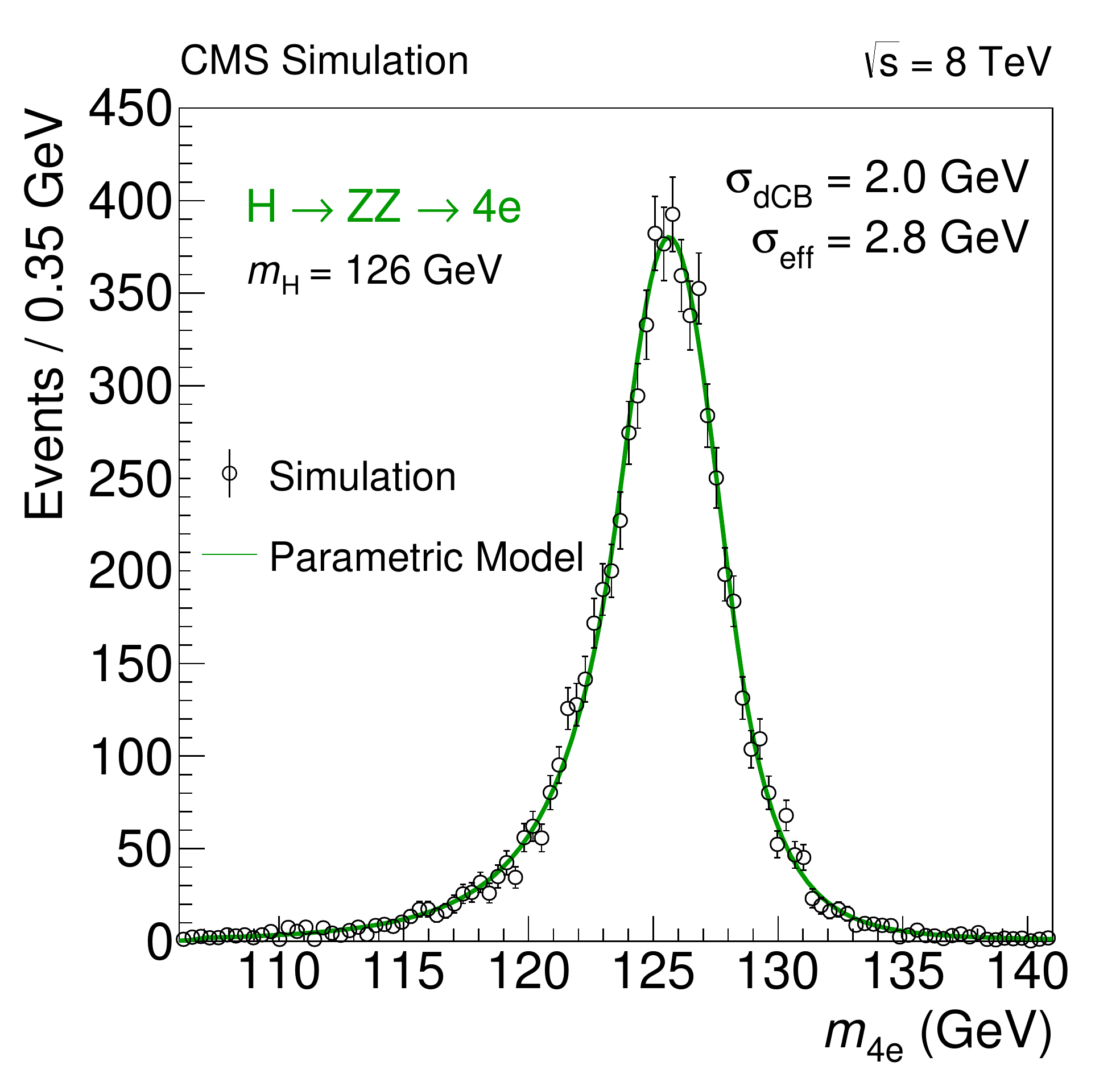}
     \includegraphics[width=0.32\linewidth]{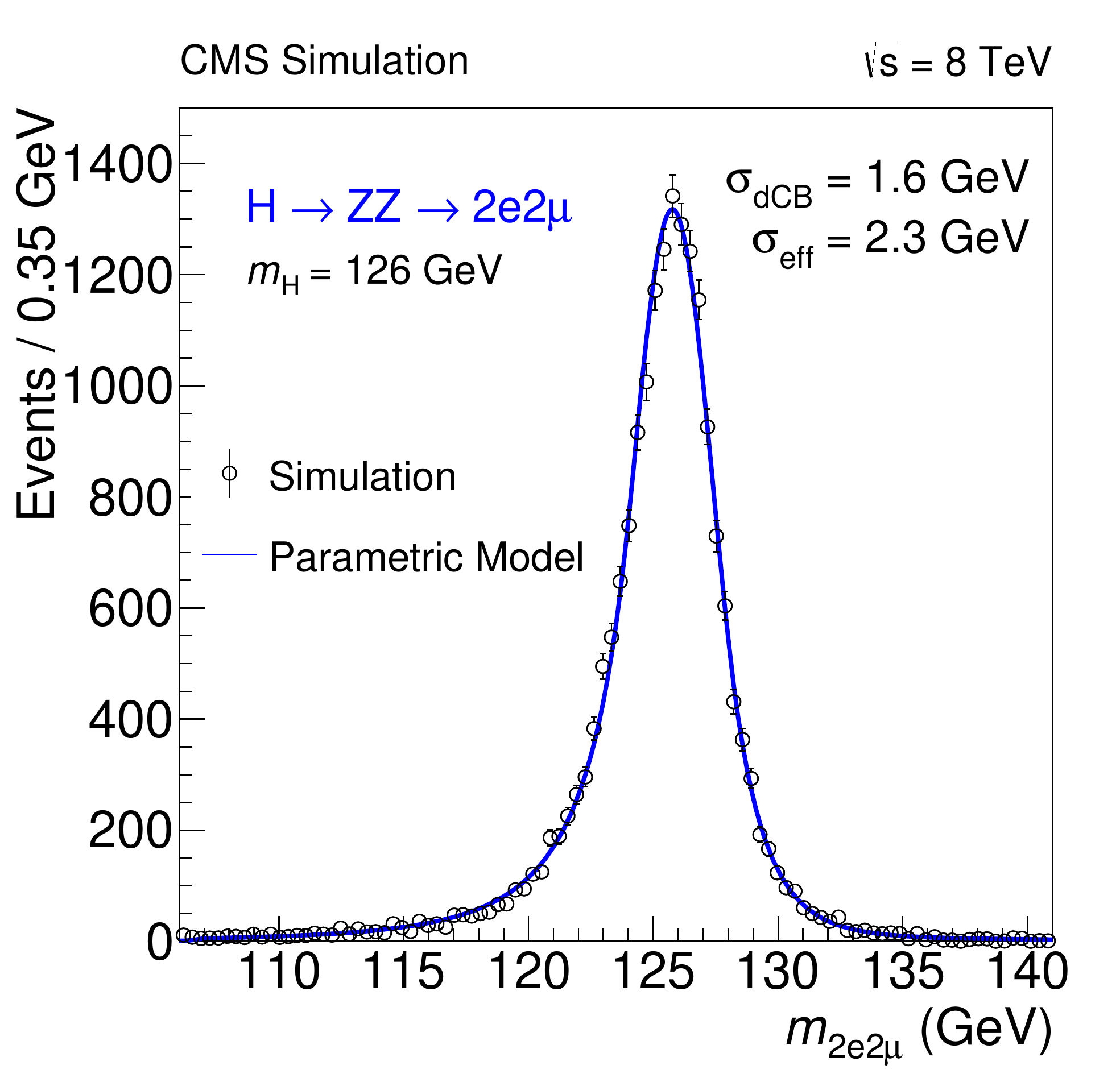}
     \includegraphics[width=0.32\linewidth]{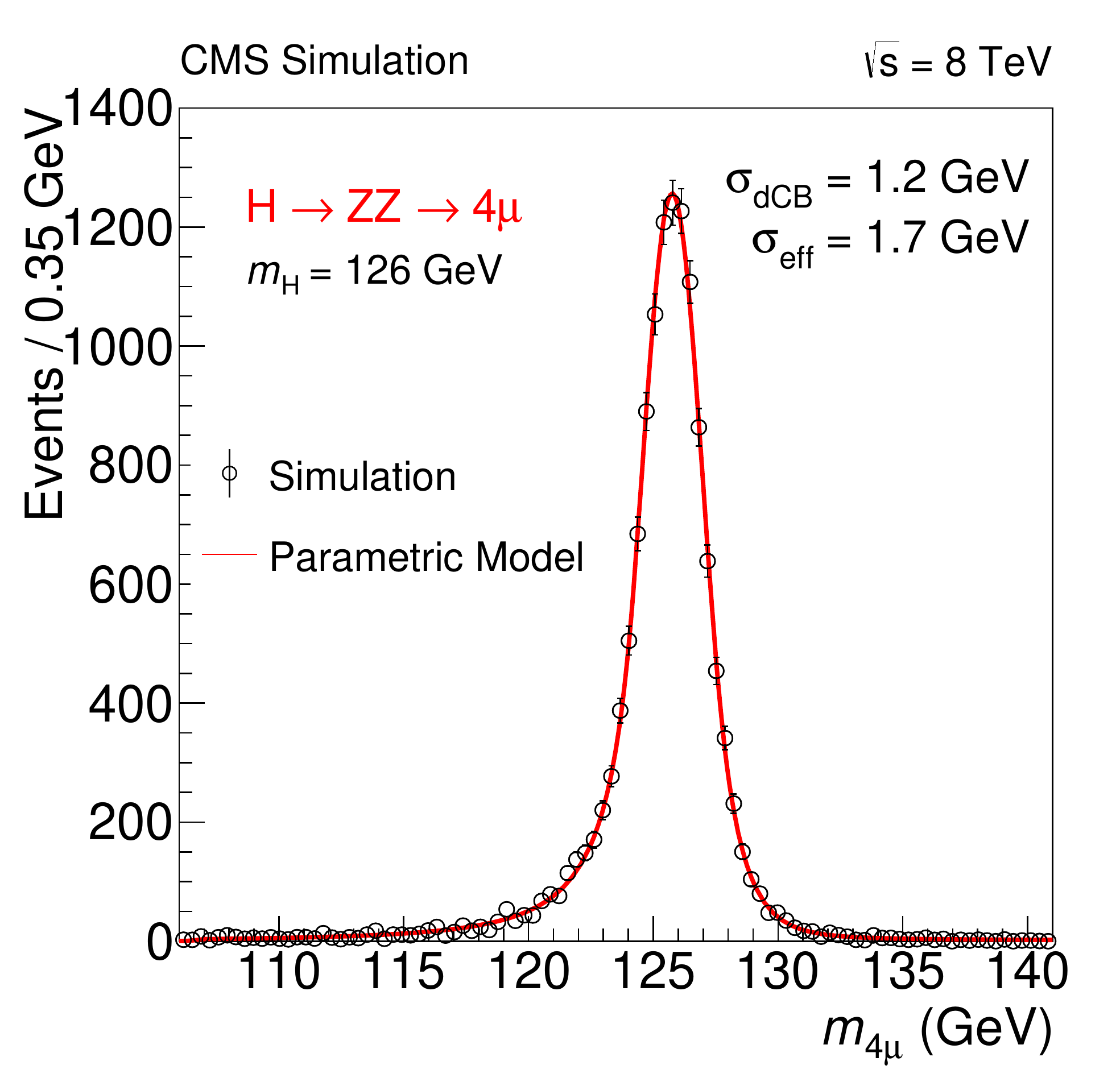}
\caption{ The $\PH\rightarrow\cPZ\cPZ\rightarrow 4\ell$ invariant mass
  distribution for $\mH = 126\GeV$ in the (left) $4\Pe$, (center) $2\Pe2\Pgm$,
  and (right) $4\Pgm$ channels. The distributions are fitted with a
  double-sided CB function and the fitted values of the CB width
  $\sigma_{\mathrm{dCB}}$ are indicated. The values of effective resolution,
  defined as half the smallest width that contains 68.3\% of the
  distribution, are also indicated. The distributions are arbitrarily
  normalized.
\label{fig:4lmassmc}}
\end{figure*}

The properties of interest to be measured in this analysis, such as
the signal and background yields, the mass and width of the resonance,
and the spin-parity quantum numbers, are determined with unbinned
maximum-likelihood fits performed to the selected events. The fits
include probability density functions for five signal components
(gluon fusion, VBF, $\PW\PH$, $\cPZ\PH$, and $\ttbar\PH$ productions)
and three background processes ($\Pq\Paq\to\cPZ\cPZ$,
$\Pg\Pg \to\cPZ\cPZ$, and $\cPZ+\X$). The normalizations of these
components and systematic uncertainties are introduced in the fits as
nuisance parameters, assuming log-normal {\it a priori} probability
distributions, and are profiled during the minimization. The shapes of
the probability density functions for the event observables are also
varied within alternative ones, according to the effect induced by
experimental or theoretical systematic
uncertainties~\cite{LHC-HCG,Chatrchyan:2012tx}. Depending on the
specific result to be extracted, different multidimensional models,
using different sets of discriminating variables, are used.  The
dimension refers to the number of input variables used in the
likelihood function. In the cases where one of the discriminants
listed in Table~\ref{tab:kdlist} is used, this observable typically
combines more than one discriminating variable.  Each of these models
is outlined below:
\begin{enumerate}
\item For the assessment of exclusion limits as a function
  of $\mH$, the signal significance, and the measurement of the signal
  strength, $\mu\equiv\sigma/\sigma_{SM}$, defined as the measured
  cross section times the branching fraction into $\cPZ\cPZ$, relative
  to the expectation for the SM Higgs boson, the following 3D
  likelihood functions are used:
\ifthenelse{\boolean{cms@external}}{
  \begin{align}\begin{split}
    \label{eqn:likmu_01j}
    \LikMu \equiv& \LikMuZOj \\
           =&
            \mathcal{P}(m_{4\ell}|\mH,\Gamma)
            \mathcal{P}(\KD|m_{4\ell})\times\\
            &\mathcal{P}(\VDu|m_{4\ell}),
    \end{split}\end{align}\begin{align}\begin{split}
    \label{eqn:likmu_2j}
    \LikMu \equiv& \LikMuDj \\
            =&
            \mathcal{P}(m_{4\ell}|\mH,\Gamma)
            \mathcal{P}(\KD|m_{4\ell})\times\\
            &\mathcal{P}(\VDj|m_{4\ell}).
  \end{split}\end{align}
}{
 \begin{align}
    \label{eqn:likmu_01j}
    \LikMu \equiv& \LikMuZOj
           =&
            \mathcal{P}(m_{4\ell}|\mH,\Gamma)
            \mathcal{P}(\KD|m_{4\ell})\times
            \mathcal{P}(\VDu|m_{4\ell}),\\
    \label{eqn:likmu_2j}
    \LikMu \equiv& \LikMuDj
            =&
            \mathcal{P}(m_{4\ell}|\mH,\Gamma)
            \mathcal{P}(\KD|m_{4\ell})\times
            \mathcal{P}(\VDj|m_{4\ell}),\\
 \end{align}
}
  where $\mH$ and $\Gamma$ are the mass and the width of the
  SM Higgs boson.  The likelihood $\LikMu$ includes the kinematic
  discriminant to differentiate the Higgs boson signal from the $\cPZ\cPZ$
  background, defined in Eq.~(\ref{eq:kd-mela}). As the third
  dimension of the fit, depending on the category, the production-mode-sensitive 
  discriminant $\VDu$ of Eq.~(\ref{eqn:likmu_01j}) (0/1-jet
  category) or the $\VDj$ of Eq.~(\ref{eqn:likmu_2j}) (dijet category) is
  used. These discriminants are defined in
  Sec.~\ref{sec:selection}.  The template distributions used as
  probability density functions for $\mathcal{P}(\VDu|m_{4\ell})$ and
  $\mathcal{P}(\VDj|m_{4\ell})$ are derived in the same way as for the
  $\mathcal{P}(\KD|m_{4\ell})$, which is discussed later in this
  section.

\item For the measurement of the mass and width of the resonance we
  use the following 3D likelihood:
\ifthenelse{\boolean{cms@external}}{
  \begin{equation}\begin{split}
    \label{eqn:likmass}
    \LikMass \equiv& \mathcal{L}_{3D}^{m,\Gamma}(m_{4\ell},\MassD,\KD)\\
             =&
             \mathcal{P}(m_{4\ell}|\mH,\Gamma,\MassD)
             \mathcal{P}(\MassD|m_{4\ell})\times\\
             &\mathcal{P}(\KD|m_{4\ell}).
  \end{split}\end{equation}
  }{
  \begin{equation}
    \label{eqn:likmass}
    \LikMass \equiv \mathcal{L}_{3D}^{m,\Gamma}(m_{4\ell},\MassD,\KD)
             =
             \mathcal{P}(m_{4\ell}|\mH,\Gamma,\MassD)
             \mathcal{P}(\MassD|m_{4\ell})\times
             \mathcal{P}(\KD|m_{4\ell}).
    \end{equation}
  }

  In this case, the information about the per-event mass uncertainty,
  $\MassD$, based on the estimated resolution of the single leptons,
  as described in Sec.~\ref{sec:masserrors}, is used. The
  probability density function $\mathcal{P}(\MassD|\mH)$ is used for the
  simulated signal, while $\mathcal{P}(\MassD|m_{4\ell})$ is used for
  backgrounds. The parameterization of the $\mathcal{P}(\MassD|\mH)$ and
  $\mathcal{P}(\MassD|m_{4\ell})$ probability density functions is
  discussed later in Sec.~\ref{sec:mass}.

\item For the spin-parity hypothesis tests, the following two-dimensional (2D) likelihood is
  used:

  \begin{equation}
    \label{eqn:likjcp}
    \LikSpin \equiv \mathcal{L}_{2D}^{J^P}(\superKD,\spinKD).
  \end{equation}

  In this case, as described in Sec.~\ref{sec:kd}, the four-lepton
  invariant mass and the separation of the Higgs boson signal from the
  $\cPZ\cPZ$ background using angular variables are condensed in a
  single discriminant, $\superKD$, defined in
  Eq.~(\ref{eq:kd-supermela}). The second dimension of the likelihood
  provides discrimination between the SM Higgs boson ($0^+$) and the
  alternative $J^P$ hypothesis. The discriminant $\spinKD$ is defined
  in Eq.~(\ref{eq:kd-spinmela}). In the case of production-independent
  hypothesis tests, $\superKD^\text{dec}$ and $\spinKD^\text{dec}$ are
  used.

\end{enumerate}
As mentioned in Sec.~\ref{sec:datasets}, the theoretical line shape
is described by the functional form of a relativistic BW function
centered at $\mH$ and with the expected natural width for the SM Higgs
boson, $\Gamma_\PH$, in the mass region $\mH<400\GeV$.  The BW
function is convolved with a double-sided CB function (to account for
the core and for the asymmetric non-Gaussian tails of the experimental
resolution) to parameterize the reconstructed signal $m_{4\ell}$
distributions, $\mathcal{P}(m_{4\ell}|\mH,\Gamma)$.  The expected
four-lepton mass distributions with their parameterizations
superimposed for the three final states are shown in
Fig.~\ref{fig:4lmassmc} for the SM Higgs boson with $\mH = 126\GeV$.
For a SM Higgs boson with mass $\mH \geq 400$\GeV, the total width is much
larger than the experimental four-lepton mass resolution, as described
in Sec.~\ref{sec:datasets}. Given the $m_{4\ell}$ distribution of the
signal in the high-mass (HM) range, the functional form of the theoretical
line shape has to be modified as follows:
\begin{equation}
\label{eqn:relHighMassBW}
f_\mathrm{BW}^\mathrm{HM}(m_{4\ell}|\mH)
\propto \frac{m_{4\ell}}{(m_{4\ell}^2-\mH^2)^2+m_{4\ell}^2\cdot\Gamma^2_\mathrm{HM}},
\end{equation}
where the $\Gamma_\mathrm{HM}$ parameter is left floating in the fit
used to determine the signal parameterization.  This modified BW
function is convolved with a double-sided CB function to account for
the experimental resolution as in the low-mass case. In the fit used
to determine the $m_{4\ell}$ parameterization for $\mH \ge $ 400\GeV,
the constraint that the experimental resolution parameter,
$\sigma_{\mathrm{dCB}}$, must be much smaller than the natural Higgs
boson width is imposed.  Systematics on the line shape are
incorporated by varying the signal weights for the interference
effects, as a function of the generated Higgs boson mass, by
${\pm}1\sigma$.

The probability distribution $\mathcal{P}(m_{4\ell})$ for the
background is parameterized with empirical functions using
simulation for the $\Zo\Zo$ background and data control regions for the
$\cPZ+\X$ background.

The correlated three-dimensional likelihood $\LikMu$, defined in
Eqs.~(\ref{eqn:likmu_01j}) and~(\ref{eqn:likmu_2j}) for the 0/1-jet
and dijet categories, respectively, is described by the
one-dimensional (1D) parametric probability distribution
$\mathcal{P}(m_{4\ell})$ multiplied by a two-dimensional template
distribution of $(m_{4\ell}, \KD)$, and a two-dimensional
$(m_{4\ell},\VDu)$ or $(m_{4\ell},\VDj)$ template distribution, where
$\VDu$ is used in the 0/1-jet category and $\VDj$ is used in the dijet
category.  The $\mathcal{P}(m_{4\ell}, \KD)$,
$\mathcal{P}(m_{4\ell},\VDu)$, and $\mathcal{P}(m_{4\ell},\VDj)$
probabilities are normalized to 1 in the second dimension for each
bin of $m_{4\ell}$.

For the signal and background, the 2D probability density functions
$\mathcal{P}(\KD|m_{4\ell})$ are obtained from simulation, for each of
the four-lepton final states and two center-of-mass energies.  The
effect of instrumental uncertainties (lepton reconstruction efficiency
and momentum resolution) on the shapes of this parameterization is
incorporated using alternative distributions or Gaussian nuisance
parameters in the likelihood and is small.  For the reducible
background, the probability density function is built using the
control regions. The reducible background templates are found to be
similar to the ones of the $\Pq\Paq\to\cPZ\cPZ$ background.  The
difference in shapes is taken as a systematic uncertainty in the
reducible background templates. The binning used for
$\mathcal{P}(\KD|m_{4\ell})$ is shown in
Figs.~\ref{fig:KDvsM4lFullMass}~(\cmsLeft) and
~\ref{fig:KDvsM4lFullMass}~(\cmsRight) for the low- and high-mass
regions, respectively.

The template distributions for $\mathcal{P}(\VDu|m_{4\ell})$ are
derived from simulation for both the signal and SM $\cPZ\cPZ$
processes and from control regions for the $\cPZ+\X$ background.  The
Higgs boson $\PT^\PH$ spectrum for gluon fusion production is obtained
by tuning the \POWHEG simulation to include contributions up to NNLO
and NNLL expectations, including effects from
resummation~\cite{Bozzi:2005wk, deFlorian:2011xf, Bozzi:2003jy}.  For
the $\PT^\PH$ spectra for VBF production and the $\cPZ\cPZ$
background, \POWHEG is used. Several uncertainties are taken into
account for the probability density function
$\mathcal{P}(\VDu|m_{4\ell})$: using alternative PDF sets and varying
the fixed-order QCD scales produces systematic uncertainties for all
the samples. For gluon fusion Higgs boson production, variations of
the default scale for NNLL resummation, and of the quark mass effects
are also considered.  For the associated production process, the LO
spectrum predicted by \PYTHIA is used, and the difference due to NLO
effects is considered as a systematic uncertainty.  For the
$\Pq\Paq\to\cPZ\cPZ$ process, a systematic uncertainty is extracted
comparing the $\PT^\cPZ$ distribution of the inclusive $\cPZ$-boson
production in events simulated with \POWHEG and in the data. The
binning used for the $\mathcal{P}(\VDu|m_{4\ell})$ template is shown
in Fig.~\ref{fig:DpTLow}~(\cmsLeft) for the low-mass region.

The template distributions for $\mathcal{P}(\VDj|m_{4\ell})$ are taken
from \POWHEG simulations for both the signal and SM $\cPZ\cPZ$
processes and from control regions for the $\cPZ+\X$ background.
Alternative shapes are introduced to account for statistical and
systematic uncertainties in these observables. In the dijet category,
alternative shapes of $\VDj$ arise from the comparison with different
generators and underlying event tunes. The change in the $\VDj$ shape
with variations of the jet energy scale is negligible. The binning
used for the $\mathcal{P}(\VDj|m_{4\ell})$ template is shown in
Fig.~\ref{fig:FisherLow}~(\cmsLeft) for the low-mass region.

\subsection{Systematic uncertainties}
\label{sec:systematics}

Experimental systematic uncertainties in the normalization of the
signal and the irreducible background processes are evaluated from
data for the trigger, which contributes 1.5\%, and for the combined
lepton reconstruction, identification, and isolation efficiencies,
which vary from 5.5\% to 11\% in the $4\Pe$ channel, and from 2.9\% to
4.3\% in the $4\Pgm$ channel, depending on the considered $\mH$.  The
theoretical uncertainties in the irreducible background are computed
as functions of $m_{4\ell}$, varying both the renormalization and
factorization scales and the PDF set following the PDF4LHC
recommendations~\cite{Alekhin:2011sk,Botje:2011sn,Lai:2010vv,Martin:2009iq,Ball:2011mu}.
Depending on the four-lepton mass range, the theoretical uncertainties
for $\Pq\Paq\to\cPZ\cPZ$ and $\Pg\Pg \to \cPZ\cPZ$ are 4\%--14\% and
25\%--50\%, respectively.

Samples of $\cPZ \to \ell^+\ell^-$, $\UpsNs \to \ell^+\ell^-$, and
$\PJGy \to \ell^+\ell^-$ events are used to set and validate the
absolute momentum scale and resolution. For electrons, a $\PT^\Pe$
dependence of the momentum scale is observed, but it only marginally
affects the four-lepton mass, and the per-electron uncertainty is
propagated, accounting for the correlations, to the $4\Pe$ and
$2\Pe2\Pgm$ channels. This dependence is corrected for, but the
observed deviation is conservatively used as a systematic uncertainty,
resulting in effects of 0.3\% and 0.1\% on the mass scales of the two
channels, respectively.  The systematic uncertainty in the muon
momentum scale translates into a 0.1\% uncertainty in the 4$\Pgm$ mass
scale.  The effect of the energy resolution uncertainties is taken
into account by introducing a 20\% uncertainty in the simulated width
of the signal mass peak, according to the maximum deviation between
data and simulation observed in the $\cPZ\to\ell^+\ell^-$ events, as
shown in Fig.~\ref{fig:lepton_reso}.

Additional systematic uncertainties arise from the limited statistical
precision in the reducible background control regions as well as from
the difference in background composition between the control regions
and the sample from which the lepton misidentification probability is
derived. As described in Sec.~\ref{sec:backgrounds}, systematic
uncertainties of 20\%, 25\%, and 40\% are assigned to the
normalization of the reducible background for the $4\Pe$, $2\Pe2\Pgm$,
and $4\Pgm$ final states, respectively. All reducible background
sources are derived from control regions and the comparison of data
with the background expectation in the signal region is independent of
the uncertainty in the LHC integrated luminosity of the data sample.
The uncertainty in the luminosity measurement (2.2\% at 7\TeV and
2.6\% at 8\TeV)~\cite{lumiPAS,CMS:2013gfa} enters the evaluation of
the \cPZ\cPZ\ background and the calculation of the cross-section
limit through the normalization of the signal.

Systematic uncertainties in the Higgs boson cross section and
branching fraction are taken from
Refs.~\cite{LHCHiggsCrossSectionWorkingGroup:2011ti,
  Denner:2011mq}. In the 0/1-jet category, an additional systematic
uncertainty in the $\cPZ\cPZ$ background normalization comes from the
comparison of \POWHEG and \MADGRAPH. In the dijet category, a $30\%$
normalization uncertainty is taken into account for the $\Pg\Pg\rightarrow
\PH+2$ jets signal cross section, while 10\% is retained for the VBF
production cross section. Table~\ref{tab:systematics} shows the
summary of the systematic uncertainties in the normalization of the
signal and background processes.

\begin{table*}[!hbt]
  \begin{center}
    \topcaption{Effect of systematic uncertainties on the yields of
      signal ($\mH = 126\GeV$) and background processes for the 8\TeV
      data set and 0/1-jet category. Uncertainties appearing on the same line are 100\% correlated, with two exceptions: those related to the missing higher orders are not correlated, and those from the $\alpha_S$ + PDF (gg) in $\ttbar\PH$ are 100\% anticorrelated. 
 Uncertainties for the 7\TeV data set are similar.
      \label{tab:systematics}}
    \begin{scotch}{lccccccc}
      Source         & \multicolumn{4}{c}{Signal ($\mH = 126\GeV$)}  &  \multicolumn{3}{c}{Backgrounds} \\
      \hline
      & $\Pg\Pg \PH$       & VBF     &  $V\PH$   &  $\ttbar\PH$ &  $\Pq\Paq\to\cPZ\cPZ$  &  $\Pg\Pg \to\cPZ\cPZ$  & $\cPZ+\X$ \\
      \hline
      $\alpha_S$ + PDF (gg)           & 7.2\%         &  \NA      &   \NA                   &  7.8\%       &  \NA                     &  7.2\%            & \NA   \\
      $\alpha_S$ + PDF ($\Pq\Paq$)    & \NA             &  2.7\%  &   3.5\%               &  \NA           &  3.4\%                 &  \NA                & \NA   \\
      Missing higher orders           & 7.5\%         &  0.2\%  &  0.4\%, 1.6\%         & 6.6\%        &  2.9\%                 &  24\%             & \NA   \\
      Signal acceptance               & \multicolumn{4}{c}{ 2\% }                                      &  \NA                     & \NA                 & \NA   \\
      BR($\PH\to\cPZ\cPZ$)              & \multicolumn{4}{c}{ 2\% }                                      &  \NA                     & \NA                 & \NA   \\
      Luminosity                      & \multicolumn{6}{c}{ 2.6\% }                                                                                 & \NA   \\
      Electron efficiency             & \multicolumn{6}{c}{ 10\% (4$\Pe$),   4.3\% (2$\Pe$2$\Pgm$)}                                                 & \NA   \\
      Muon efficiency                 & \multicolumn{6}{c}{ 4.3\% (4$\Pgm$), 2.1\% (2$\Pe$2$\Pgm$)}                                                 & \NA   \\
      Control region                  & \NA             & \NA       & \NA                     & \NA            & \NA                      & \NA                 & 40\% \\
    \end{scotch}
  \end{center}
\end{table*}

Shape uncertainties for both categories are considered, accounting for
the lepton scale and resolution variations on the $m_{4\ell}$ line
shape, theoretical uncertainties in the $\VDu$ signal and background
models, and theoretical and experimental uncertainties (such as the
variations on the jet energy scale and resolution) in the $\VDj$
distribution.

\section{Results and interpretation}
\label{sec:results}

The results of the search for a signal consistent with a SM Higgs
boson in the $\mH$ range 110--1000\GeV are described along with the
estimation of the significance of the excess observed in the low-mass
region.  Then, the measurement of the mass of the new boson in the
hypothesis of a narrow resonance and limits on its width are reported.
For this resonance, the compatibility of the cross section measurement
with the SM Higgs boson calculation is given together with constraints
on the production mechanisms.  Finally, the spin and parity of the
boson are tested to check the compatibility with the hypothesis of a
$0^+$ resonance as compared with the alternatives, and the measurement
of the fraction of a CP-odd contribution to the decay amplitude is
reported.

\subsection{Signal significance and exclusion limits}
\label{sec:limits_significance}

The selected events are split into twelve subcategories based on the
three final states, two data-taking periods (7 and 8\TeV), and two jet
categories. These events are examined for 187 hypothetical SM-like
Higgs boson masses in a range between 110 and 1000\GeV, where the mass
steps are optimized to account for the expected width and
resolution~\cite{LHC-HCG}.  A 3D model, $\LikMuZOj$ and $\LikMuDj$,
defined, respectively, in Eqs.~(\ref{eqn:likmu_01j})
and~(\ref{eqn:likmu_2j}) for the 0/1-jet category and for the dijet
category, is used.  The statistical approach discussed in
Ref.~\cite{LHC-HCG} is followed to set exclusion limits and to
establish the significance of an excess. The modified frequentist
construction $\mathrm{CL_s}$~\cite{Junk,Read:2002hq,LHC-HCG} is
adopted as the primary method for reporting limits. As a complementary
method to the frequentist construction, a Bayesian approach~\cite{PDG}
yields consistent results.

\begin{figure}[th!]
  \begin{center}
    \includegraphics[width=\cmsFigWidthStd]{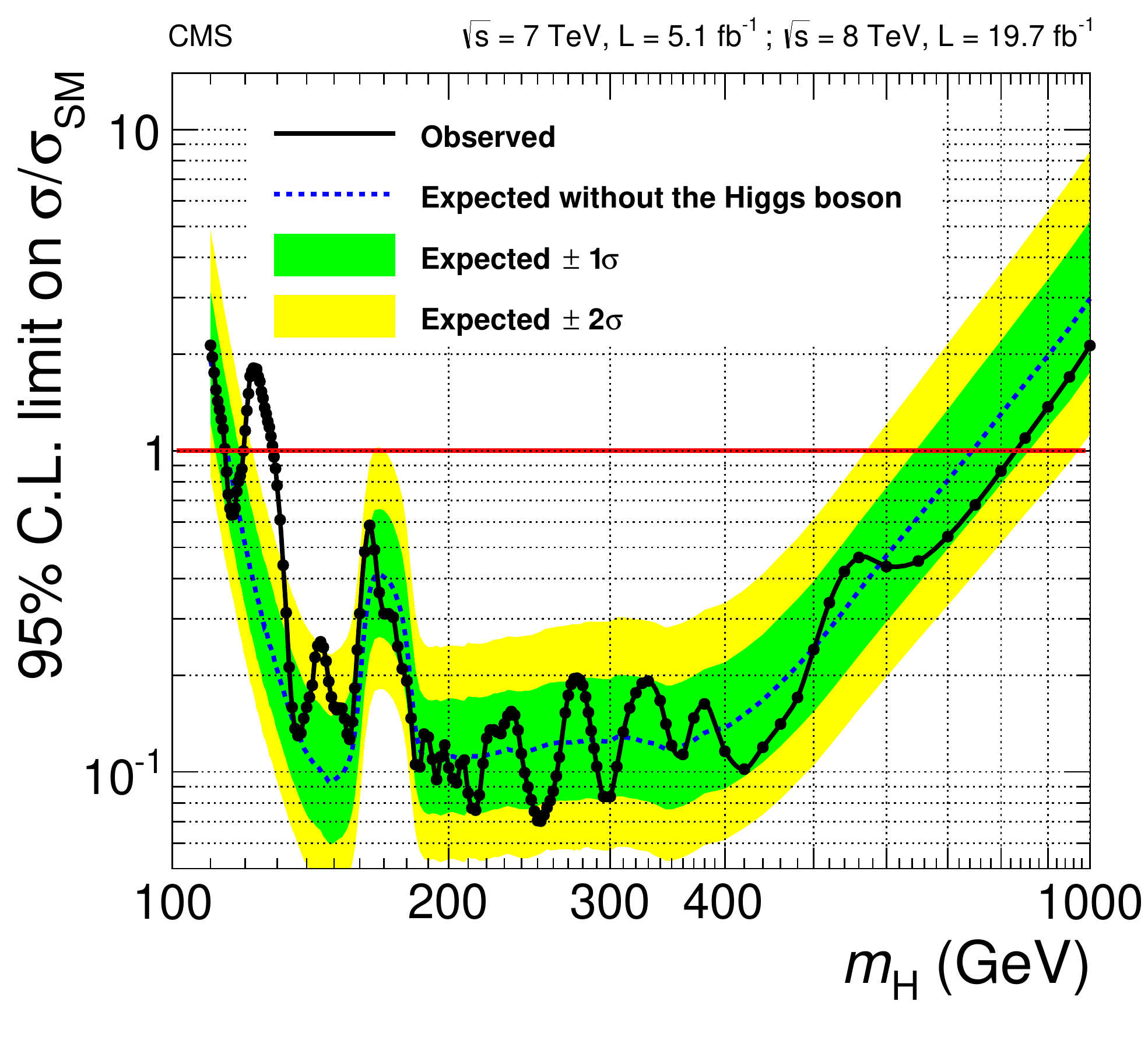}
    \includegraphics[width=\cmsFigWidthStd]{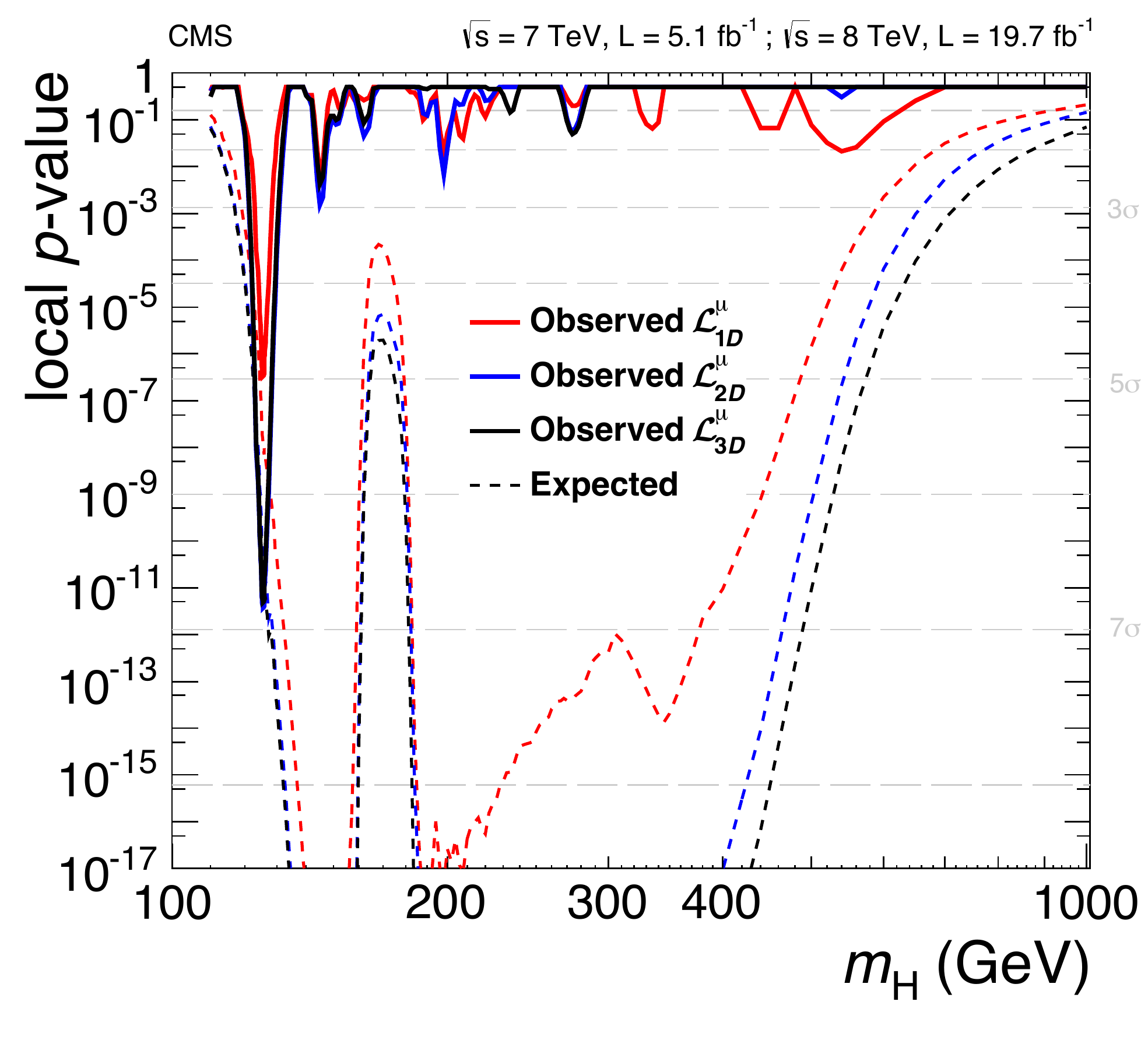}
    \caption{(\cmsLeft) Observed and expected 95\% \CL upper limit on the
      ratio of the production cross section to the SM expectation.
      The expected 1$\sigma$ and 2$\sigma$ ranges of expectation for
      the background-only model are also shown with green and yellow
      bands, respectively. (\cmsRight) Significance of the local excess with
      respect to the SM background expectation as a function of the
      Higgs boson mass in the full mass range 110--1000\GeV. Results
      are shown for the 1D fit ($\LikMuOneD$), the 2D fit ($\LikMuTwoD$),
      and the reference 3D fit
      ($\LikMu$).  \label{fig:UpperLimit_ASCLS-H}}
    \end{center}
\end{figure}

\begin{figure}[th!]
  \begin{center}
    \includegraphics[width=\cmsFigWidthStd]{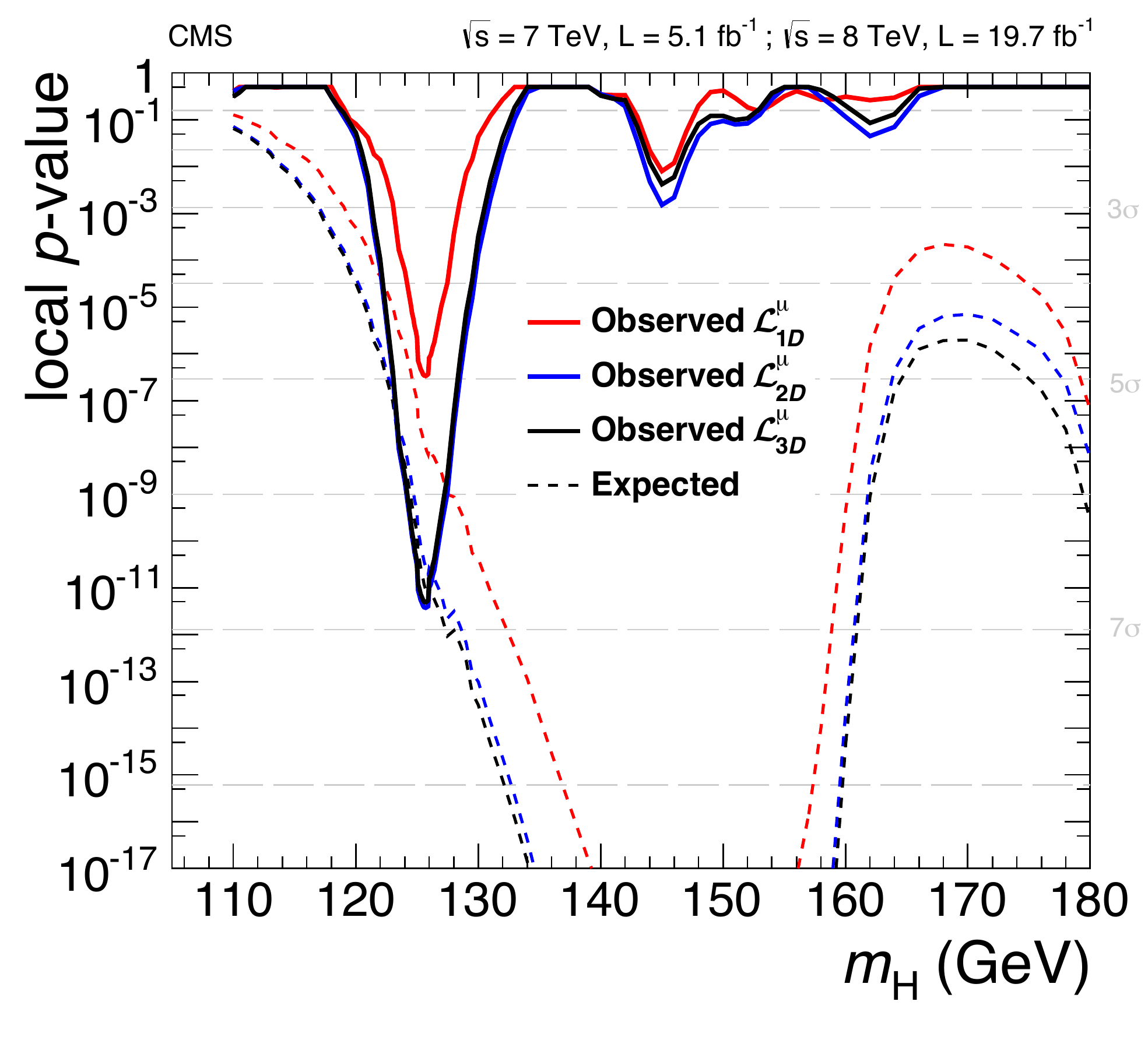}
    \caption{ Significance of the local excess with respect to the SM
    background expectation as a function of the Higgs boson mass for
    the 1D fit ($\LikMuOneD$), the 2D fit ($\LikMuTwoD$), and the
    reference 3D fit ($\LikMu$).  Results are shown for the full data
    sample in the low-mass region
    only.  \label{fig:UpperLimit_ASCLS-L}} \end{center}
\end{figure}

Upper limits on the ratio of the production cross section to the SM
expectation are shown in Fig.~\ref{fig:UpperLimit_ASCLS-H}~(\cmsLeft). The
results presented in this section make use of asymptotic formulas from
Ref.~\cite{Cowan:2010st}. The SM-like Higgs boson is excluded by the
four-lepton channels at the 95\% \CL in the mass ranges 114.5--119.0\GeV
and 129.5--832.0\GeV, for an expected exclusion range of 115--740\GeV.
The local $p$values, representing the significance of a local excess
relative to the background expectation, are shown for the full mass
range as a function of $\mH$ in
Fig.~\ref{fig:UpperLimit_ASCLS-H}~(\cmsRight). The minimum of the
local $p$value is reached around $m_{4\ell}$ = 125.7\GeV, near the
mass of the new boson, confirming the result in
Ref.~\cite{Chatrchyan:2012ufa}, and corresponds to a local
significance of $\obsSign\sigma$, consistent with the expected
sensitivity of $\expSign\sigma$.
As a cross-check, 1D
[$\LikMuOneD \equiv \mathcal{L}_{1D}^{\mu}(m_{4\ell})$] and 2D
[$\LikMuTwoD \equiv \mathcal{L}_{2D}^{\mu}(m_{4\ell},\KD)$] models are
also studied, as shown in
Figs.~\ref{fig:UpperLimit_ASCLS-H}~(\cmsRight)
and~\ref{fig:UpperLimit_ASCLS-L}, resulting in an observed local
significance of $\obsSignOneD$$\sigma$ and $\obsSignTwoD$$\sigma$, for
an expectation of $\expSignOneD$$\sigma$ and $\expSignTwoD$$\sigma$,
respectively. These results are consistent with the 3D model; however,
with a systematically lower expected sensitivity to the signal. No
other significant deviations with respect to the expectations is found
in the mass range 110--1000\GeV. The second most significant $p$-value
minimum is reached around $m_{4\ell}$=146\GeV, with a local
significance of $2.7\sigma$. This computation does not take into
account the look-elsewhere effect~\cite{Gross:2010qma}.

\subsection{Mass and width}
\label{sec:mass}

In order to measure the mass and width of the new boson precisely and
to correctly assign the uncertainties in these measurements, the
four-lepton mass uncertainties estimated on a per-event basis, as
described in Sec.~\ref{sec:masserrors}, are incorporated into the
likelihood.  This approach has the largest impact in a context of a
low number of events and a wide spread of per-event uncertainties,
both of which are present in the $\PH \to \cPZ\cPZ \to 4\ell$
analysis. Tests on simulation indicate that, with this approach, the
uncertainties in the measured mass and the upper limit on the width of
the Higgs boson are expected to improve by about 8\% and 10\%,
respectively, with respect to using the average resolution.

The experimental resolution parameter of the double-sided CB function,
used to model the $m_{4\ell}$ line shape, is substituted with the
per-event estimation of the mass uncertainty $\MassD$. The parameters
describing the tail of the double-sided CB from simulation are also
corrected on a per-event basis.

The likelihood used for the mass and width measurements is defined in
Eq.~(\ref{eqn:likmass}). By construction, this likelihood neglects
potential correlations between $\KD$ and $\MassD$. Simulated Higgs
boson and $\cPq\cPaq \to \cPZ\cPZ$ events show no evident correlations
between these two observables.  The probability density functions
$\mathcal{P}(\MassD \, | \, \mH)$ of the per-event uncertainty
distributions for the signal are obtained from simulation.  The
probability density functions $\mathcal{P}(\MassD \, | \, m_{4\ell})$
for the $\cPZ\cPZ$ background are obtained from simulation and are
cross-checked with data in control regions dominated by the $\cPZ\cPZ$
background events ($m_{4\ell} > 180$\GeV) and $\cPZ \to 4\ell$ events
($80 <m_{4\ell}< 100\GeV$)~\cite{CMS:2012bw}, as shown in
Fig.~\ref{fig:EbE_validation}~(\cmsRight).  The $\mathcal{P}(\MassD \,
| \, m_{4\ell})$ for the reducible background is obtained from the
control regions in the data with the same technique used to derive the
$m_{4\ell}$ line shapes. The $\mathcal{P}(\MassD \, | \, m_{4\ell})$
is a conditional probability distribution function, where for all the
channels and both signal and background components the probability
density functions $\mathcal{P}(\MassD)$ are parameterized as a sum of
a Landau~\cite{Landau:1944if} and a Gaussian function.

\begin{figure}[!htb]
  \centering
   \includegraphics[width=\cmsFigWidthStd]{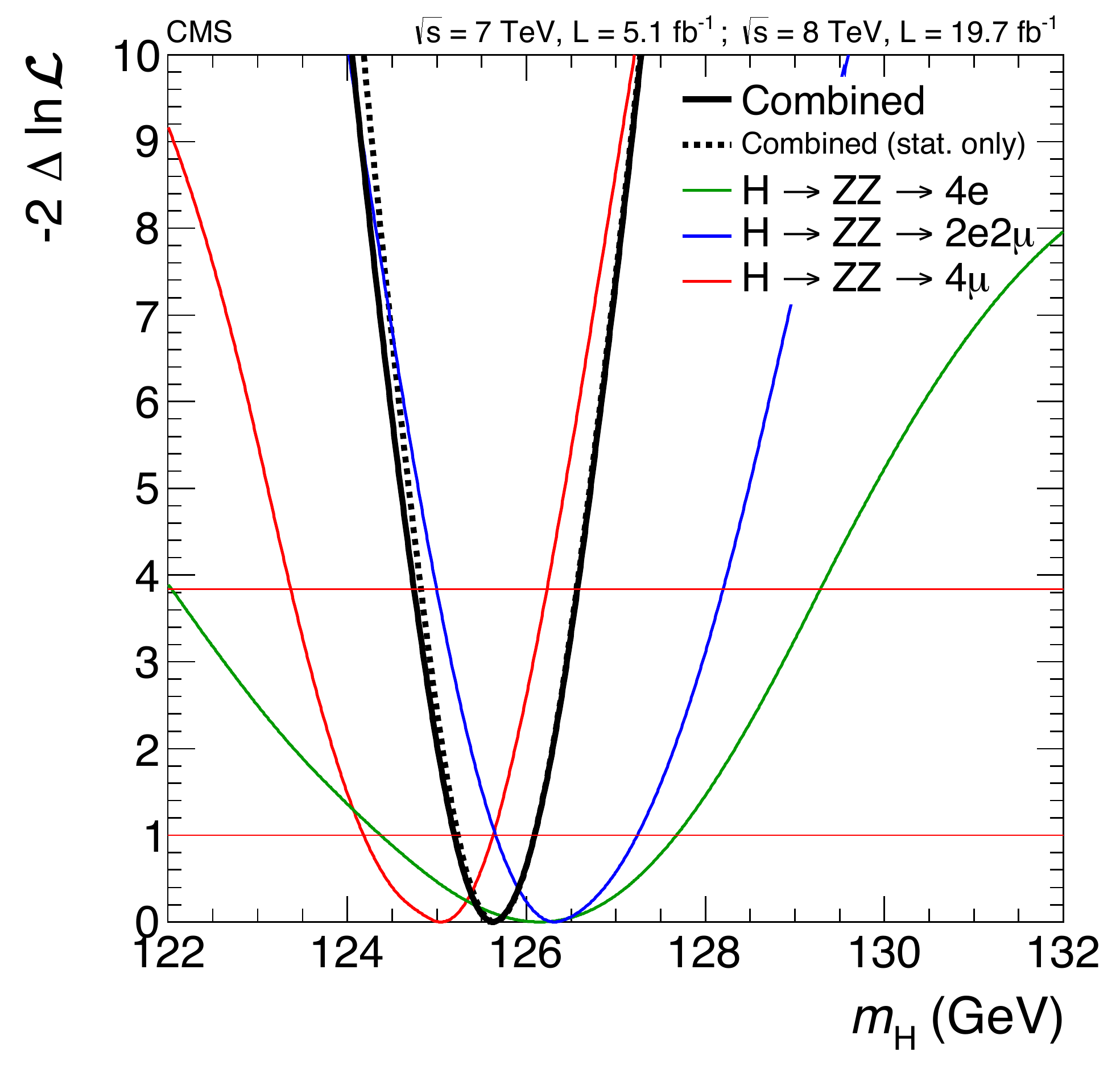} \includegraphics[width=\cmsFigWidthStd]{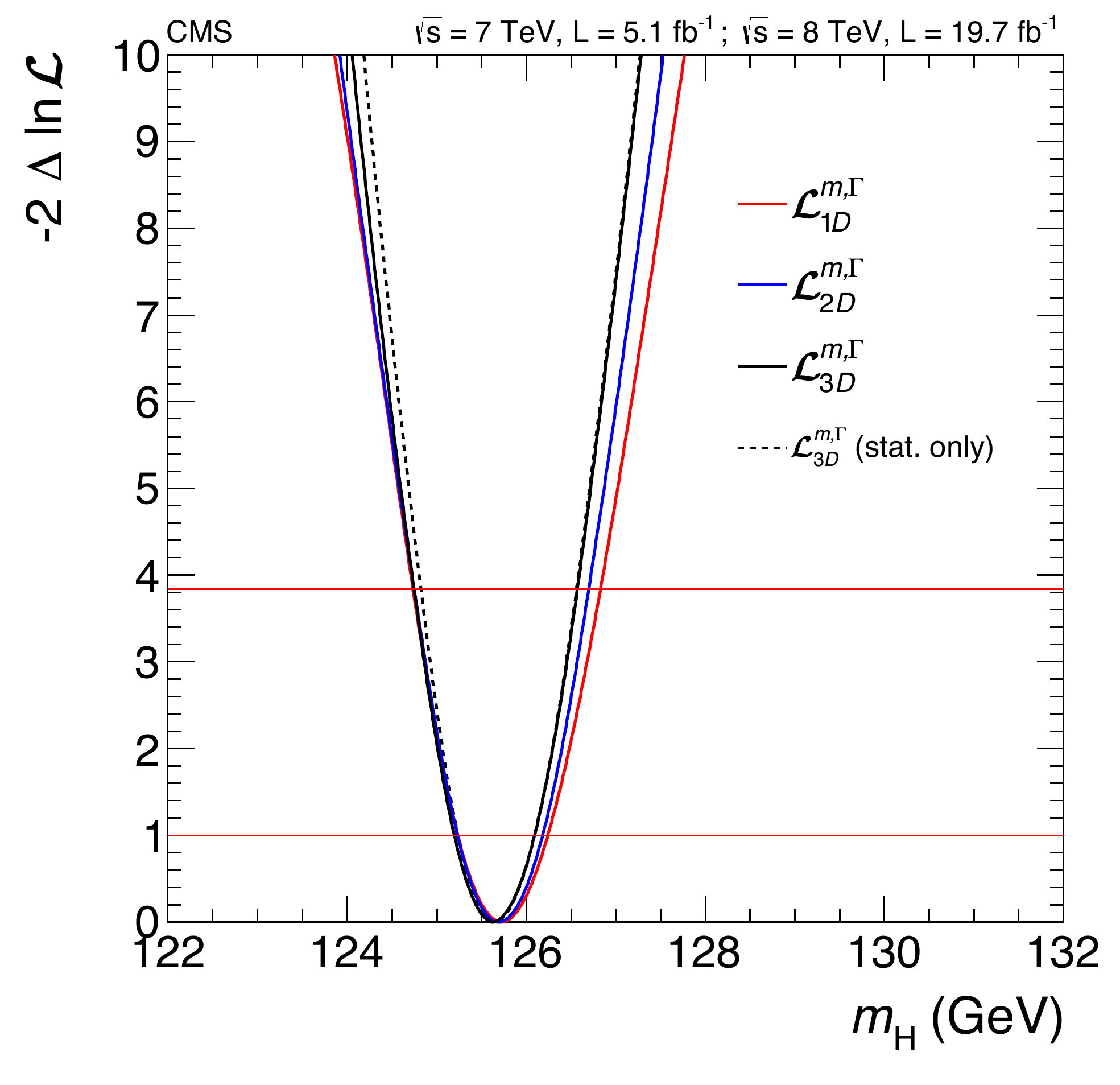}
    \caption{(\cmsLeft) Scan of the negative log likelihood
      $-2\Delta \ln \mathcal{L}$ versus the SM Higgs boson mass $\mH$,
      for each of the three channels separately and the combination of
      the three, where the dashed line represents the scan including
      only statistical uncertainties when using the 3D model.
      (\cmsRight) Scan of $-2\Delta \ln \mathcal{L}$ versus $\mH$ for
      the combination of the three channels, and using the 1D fit
      ($\LikMassOneD)$, 2D fit ($\LikMassTwoD$), and 3D fit
      ($\LikMass$). The horizontal lines at $-2\Delta \ln \mathcal{L}
      = 1$ and 3.84 represent the 68\% and 95\% \CL's,
      respectively. \label{fig:mass}}
\end{figure}

Figure~\ref{fig:mass}~(\cmsLeft) shows the profile likelihood scan
versus the SM Higgs boson mass performed under the assumption that its
width is much smaller than the detector resolution, for the single
channels, combining 7 and 8\TeV data, and for the combination of all
the channels. The Higgs boson cross section is left floating in the
fit.  To decompose the total mass uncertainty into statistical and
systematic components, a fit with all nuisance parameters fixed at
their best-fit values is performed.  The mass uncertainty obtained in
this way is purely statistical.  The systematic uncertainties account
for an effect on the mass scale of the lepton momentum scale and
resolution, shape systematics in the $\mathcal{P}(\KD \, | \,
m_{4\ell})$ probability density functions used as signal and
background models, and normalization systematics due to acceptance and
efficiency uncertainty.  The measured mass is $\mH = \valMass\GeV$.

Figure~\ref{fig:mass}~(\cmsLeft) also shows likelihood scans
separately for the $4\Pe$, $2\Pe2\Pgm$ and $4\Pgm$ final states when
using the 3D model $\LikMass$ of Eq.~(\ref{eqn:likmass}). The
measurements in the three final states are statistically compatible.
The best-fit values for each subchannel are also shown in
Table~\ref{tab:mass}.  The dominant contribution to the systematic
uncertainty is the limited knowledge of the lepton momentum scale.

Two more mass measurements are performed with a reduced level of
information, by dropping the $\mathcal{P}(\KD \, | \, m_{4\ell})$ term of
the likelihood in Eq.~(\ref{eqn:likmass}), resulting in a 2D model,
$\LikMassTwoD\equiv\mathcal{L}_{2D}^{m,\Gamma}(m_{4\ell},\MassD)$, or by
performing only a mass line shape fit and assuming the average mass
resolution is applicable for each channel, resulting in a 1D model,
$\LikMassOneD\equiv\mathcal{L}_{1D}^{m,\Gamma}(m_{4\ell})$.  The measured
central value is the same in all three cases, with an increasing
uncertainty, due to the reduced information available to the fit in
the case of 2D or 1D models. Figure~\ref{fig:mass}~(\cmsRight) shows the
likelihood scans for the combination of all the final states
separately for the $\LikMassOneD$, $\LikMassTwoD$, and $\LikMass$
models.

\begin{table}[!hb]
  \begin{center}
    \topcaption{Best fit values for the mass of the Higgs boson candidate,
      measured in the $4\ell$, $\ell=\Pe,\Pgm$ final states using
      $\LikMass$ model. For the combination of all the final states
      $\PH\to4\ell$, the separate contribution of the statistical and
      systematic uncertainty to the total one is given.
      \label{tab:mass}}
    \begin{scotch}{lc}
      Channel & Measured mass (\GeVns{}) \\
      \hline
      $4\Pe$          &    $\valMassFourE$  \T \B   \\
      $2\Pe2\Pgm$     &    $\valMassTwoETwoMu$  \T \B  \\
      $4\Pgm$         &    $\valMassFourMu$ \T \B  \\
      \hline
      $4\ell$         &    $\valMass$ \T \B \\
    \end{scotch}
  \end{center}
\end{table}

The mass distribution for the $\cPZ \to 4\ell$ decay exhibits a
pronounced resonant peak at $m_{4\ell} = m_{\cPZ}$ close to the new
boson ($80 < m_{4\ell}< 100\GeV$). Hence, the $\cPZ \to 4\ell$ peak
can be used as validation of the measurement of the mass of the new
boson using the same techniques as for the Higgs boson.  The mass of
the reconstructed $\cPZ$ boson in $\cPZ \to 4\ell$ decays, with the
assumption of the Particle Data Group (PDG)~\cite{PDG} value for the
$\cPZ$-boson natural width, is consistent in each subchannel.  The
measured value for the combination of all the $\cPZ \to 4\ell$ final
states is $m_{\cPZ}=91.1$\GeV, compatible with the PDG value ($91.1876
\pm 0.0021\GeV$) within the total estimated uncertainty of
0.4\GeV~\cite{PDG}.

\begin{figure}[!htb]
  \begin{center}
    \includegraphics[width=\cmsFigWidthStd]{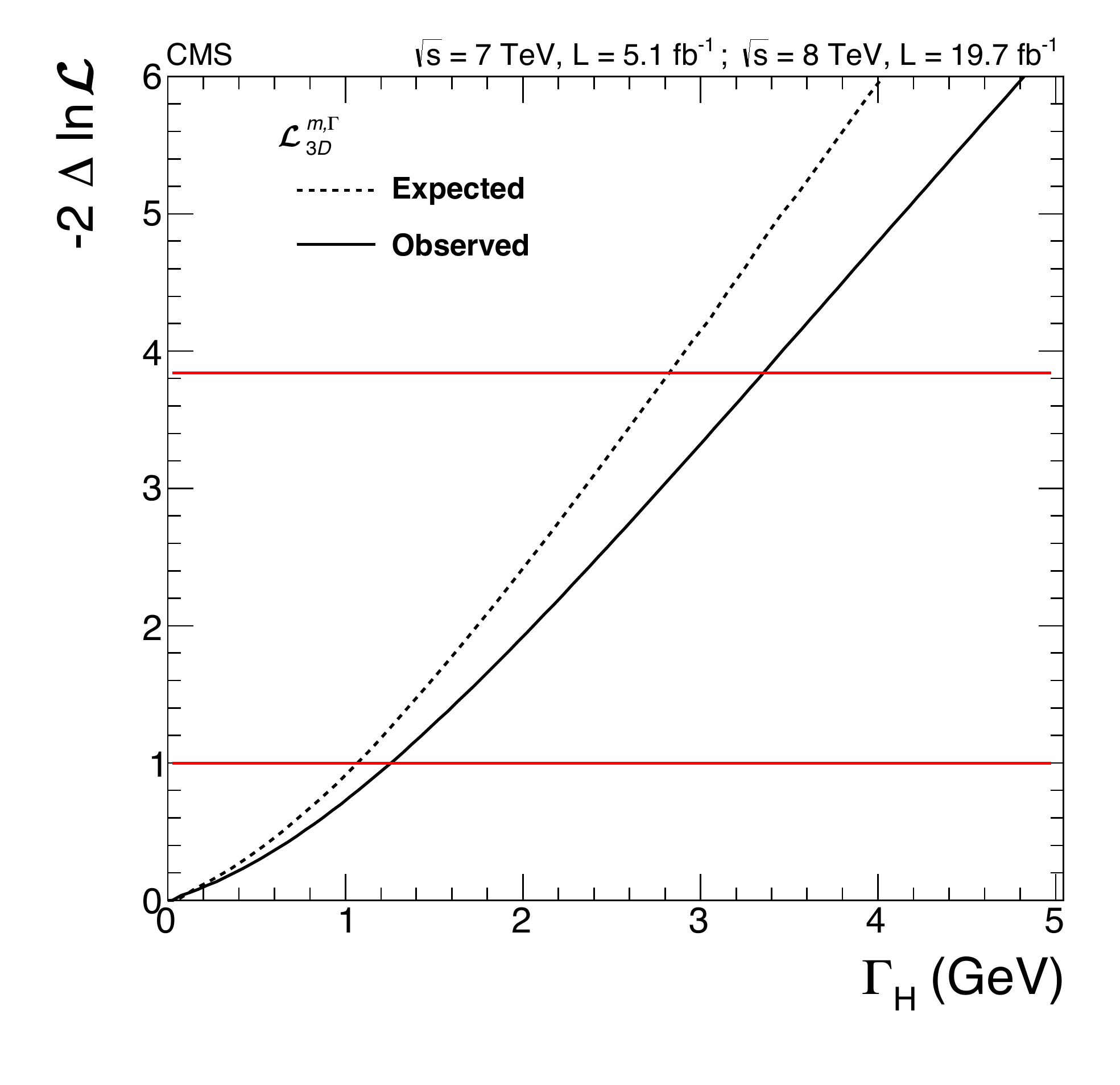}
    \caption{ Scan of the average expected and observed negative log
      likelihood $-2\Delta \ln \mathcal{L}$ versus the tested SM Higgs
      boson width~$\Gamma_{\PH}$ obtained with the 3D fit
      ($\LikMass$). The horizontal lines at $-2\Delta \ln \mathcal{L}$
      = 1 and 3.84 represent the 68\% and 95\% \CL's,
      respectively.  \label{fig:width}} \end{center}
\end{figure}

Figure~\ref{fig:width} shows the scan of the 3D likelihood versus the
width of the SM-like Higgs boson with an arbitrary width. In this
scan, the mass and the signal strength $\strength$ are profiled, as
all other nuisance parameters. This shows that the data are compatible
with a narrow-width resonance. The measured width is
$\Gamma_\PH=\valWidth$\GeV, and the upper limit on the width
is \ulWidth\GeV at the 95\% \CL The expected upper limit
is \expUlWidth\GeV.

\subsection{Signal strength}
\label{sec:strength}

\begin{figure}[htb]
  \begin{center}
    \includegraphics[width=\cmsFigWidthStd]{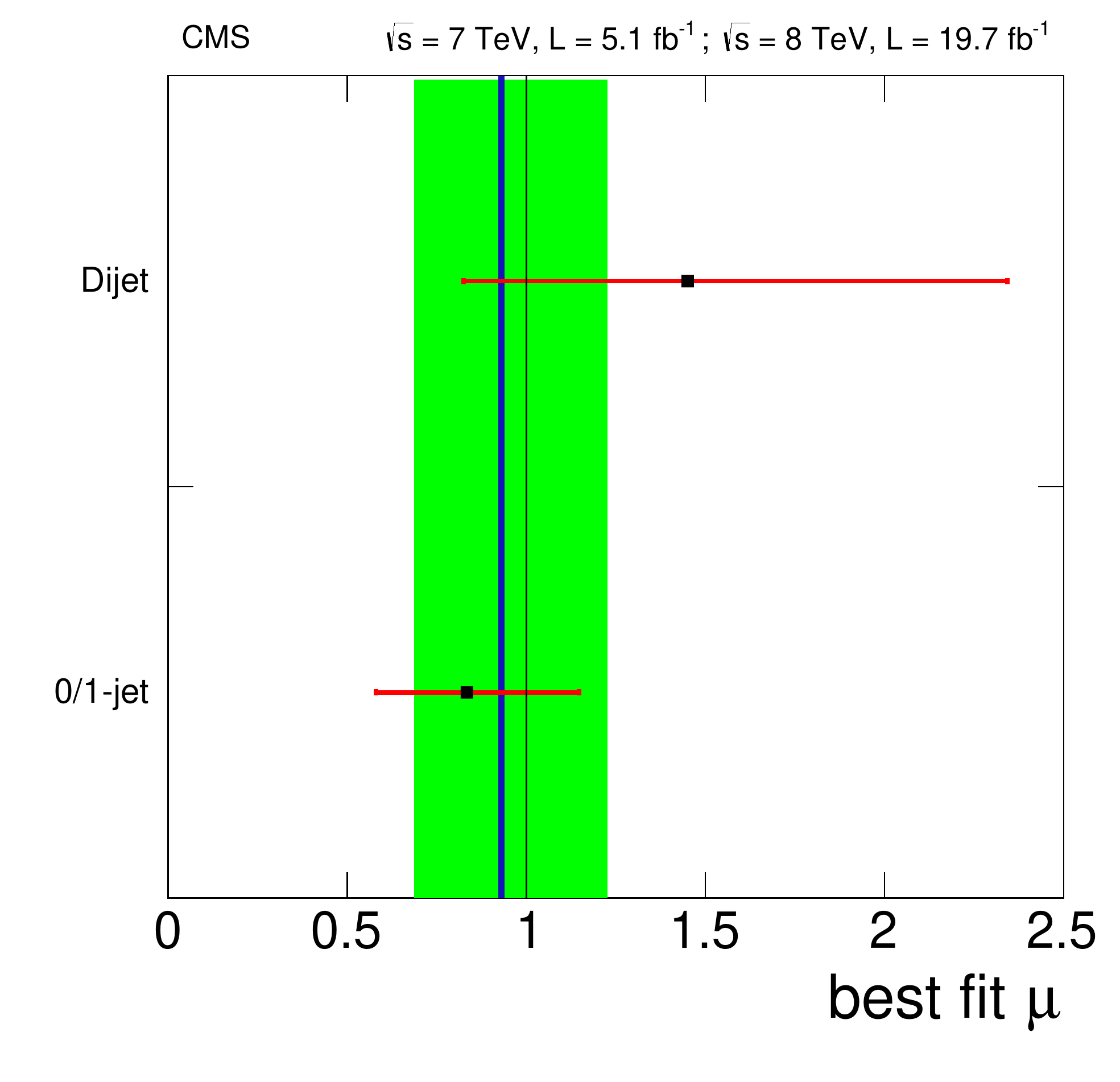}
    \includegraphics[width=\cmsFigWidthStd]{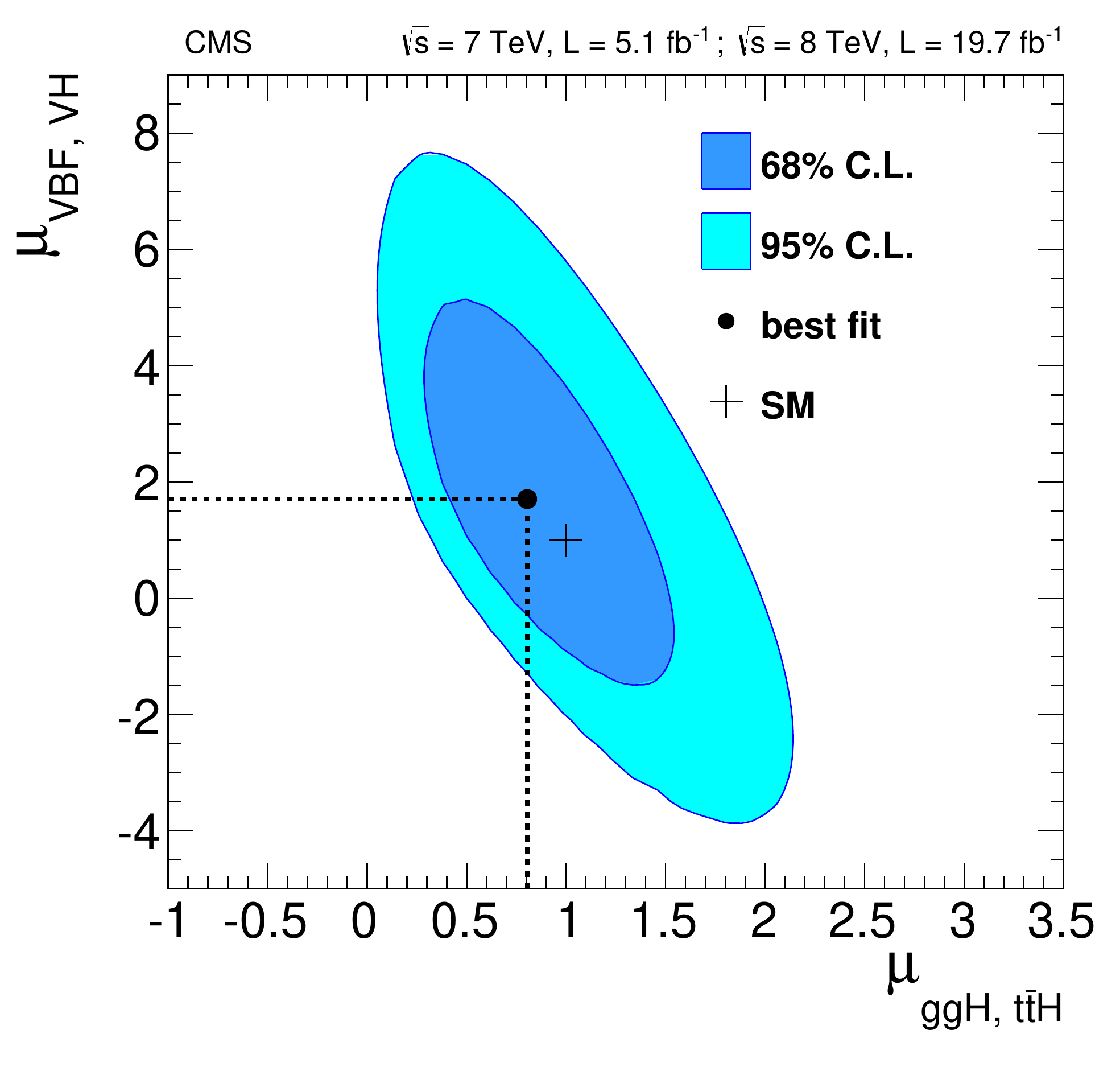}
    \caption{(\cmsLeft) Values of $\strength$ for the two
    categories. The vertical line shows the combined $\strength$
    together with its associated ${\pm}1\sigma$ uncertainties, shown
    as a green band. The horizontal bars indicate the ${\pm}1\sigma$
    uncertainties in $\strength$ for the different categories. The
    uncertainties include both statistical and systematic sources of
    uncertainty.  (\cmsRight) Likelihood contours on the
    signal-strength modifiers associated with fermions ($\muF$) and
    vector bosons ($\muV$) shown at a 68\% and
    95\% \CL \label{fig:mucat}} \end{center}
\end{figure}

The measured signal strength is $\strength=\sigma/\sigma_{SM}=\valMu$
at the best-fit mass ($\mH = 125.6\GeV$) with the models of
Eqs.~(\ref{eqn:likmu_01j}) and~(\ref{eqn:likmu_2j}) for the 0/1-jet
category and the dijet category, respectively. The median expected
signal strength is $\strength=\valMuExp$, for which the total
uncertainty agrees with the observed one.  The result is
$\valMuUntagged$ in the 0/1-jet category and $\valMuDijet$ in the
dijet category. The best-fit values are shown in
Fig.~\ref{fig:mucat}~(\cmsLeft). For each category, the signal
strength is consistent with SM expectations within the uncertainties,
which are dominated by the statistical ones with the current data set.

The categorization according to jet multiplicity and the inclusion of
VBF-sensitive variables in the likelihood, like $\VDu$ and $\VDj$,
used to measure the cross section in the inclusive category, are also
used to disentangle the production mechanisms of the observed new
state.  The production mechanisms are split into two families
depending on whether the production is through couplings to fermions
(gluon fusion, $\ttbar\PH$) or vector bosons (VBF, $V\PH$).  For $\mH
= 126$\GeV, about 55\% of the VBF events are expected to be included
in the dijet category, while only 8\% of the gluon fusion events are
included in the dijet category. As shown in
Table~\ref{tab:PreFitYieldsSigRegionCat}, a fraction of 43\% of
$\PW\PH$ and $\cPZ\PH$ production contributes to the dijet
category. Events that contribute are those in which the vector boson
decays hadronically.

Two signal-strength modifiers ($\muF$ and $\muV$) are introduced as
scale factors for the fermion and vector-boson induced contribution to
the expected SM cross section. A two-dimensional fit is performed for
the two signal-strength modifiers assuming a mass hypothesis of $\mH =
125.6\GeV$. The likelihood is profiled for all nuisance parameters and
68\% and 95\% \CL contours in the ($\muF,\muV$) plane are obtained.
Figure~\ref{fig:mucat}~(\cmsRight) shows the result of the fit leading
to the measurements of $\muF=\valMuF$ and $\muV=\valMuV$.  The
measured values are consistent with the expectations for the SM Higgs
boson, $(\muF,\muV)=(1,1)$.  With the current limited statistics, we
cannot establish yet the presence of VBF and $V\PH$ production, since
$\muV=0$ is also compatible with the data.  Since the decay (into
$\cPZ\cPZ$) is vector-boson mediated, it is necessary that such a
coupling must exist in the production side and that the SM VBF and SM
$V\PH$ production mechanisms must be present. The fitted value of
$\muV$ larger than 1 is driven partly by the hard $\VDu$ spectrum of
the events observed in data when compared to the expectation from the
production of the SM Higgs boson (Fig.~\ref{fig:DpTLow}).

\subsection{Spin and parity}
\label{sec:jcp}

To measure the spin and parity properties of the new boson, the
methodology discussed in Sec.~\ref{sec:kd} is followed.  In addition
to the models tested in Ref.~\cite{Chatrchyan:2012br} ($0^-$ and
$\Pg\Pg \to 2^+_{\mathrm{m}}$), seven additional models are examined:
$0^+_{\mathrm{h}}$, $\Pq\Paq \to 1^-$, $\Pq\Paq \to 1^+$, $\Pq\Paq \to
2^+_{\mathrm{m}}$, $\Pg\Pg \to 2^+_{\mathrm{h}}$, $\Pg\Pg \to
2^-_{\mathrm{h}}$, $\Pg\Pg \to 2^+_{\mathrm{b}}$. The discrimination
is based on 2D probability density functions $(\superKD, \spinKD)$,
where the kinematic discriminants $\superKD$ and $\spinKD$ are defined
by Eqs.~(\ref{eq:kd-supermela}) and~(\ref{eq:kd-spinmela}).  The
$1^{\pm}$ and $2^+_{\mathrm{m}}$ signal hypotheses are also tested by
relying only on their decay information, i.e. in a
production-independent way, using pairs of kinematic discriminants
$( \mathcal{D}_\text{bkg}^{\text{dec}} \,, \mathcal{D}_{J^P}^{\text{dec}}
)$, defined by Eqs.~(\ref{eq:melaSigProd1}) and
(\ref{eq:melaSigProd2}).  All models and discriminants, discussed in
Section~\ref{sec:kd}, are listed in Table~\ref{tab:kdlist}.

For spin and parity studies, the event categorization based on jets is
not used in order to reduce the dependence on the production
mechanisms.  Consequently, the VBF discriminants, $\VDu$ and $\VDj$,
are not used, resulting in the $\LikSpin$ model defined in
Eq.~(\ref{eqn:likjcp}).  Events in the mass range $106< m_{4\ell} <
141\GeV$ are used to perform these studies. The Higgs boson mass is
assumed to be $m_{0^+}$ = 125.6\GeV. The 2D probability density
functions for signal and background, $\mathcal{P}(\spinKD, \superKD)$
in Eq.~(\ref{eqn:likjcp}), are obtained as 2D templates from
simulation for the signal and irreducible background, and from control
regions for the reducible backgrounds.

\begin{figure}[th!]
  \begin{center}
    \includegraphics[width=\cmsFigWidthStd]{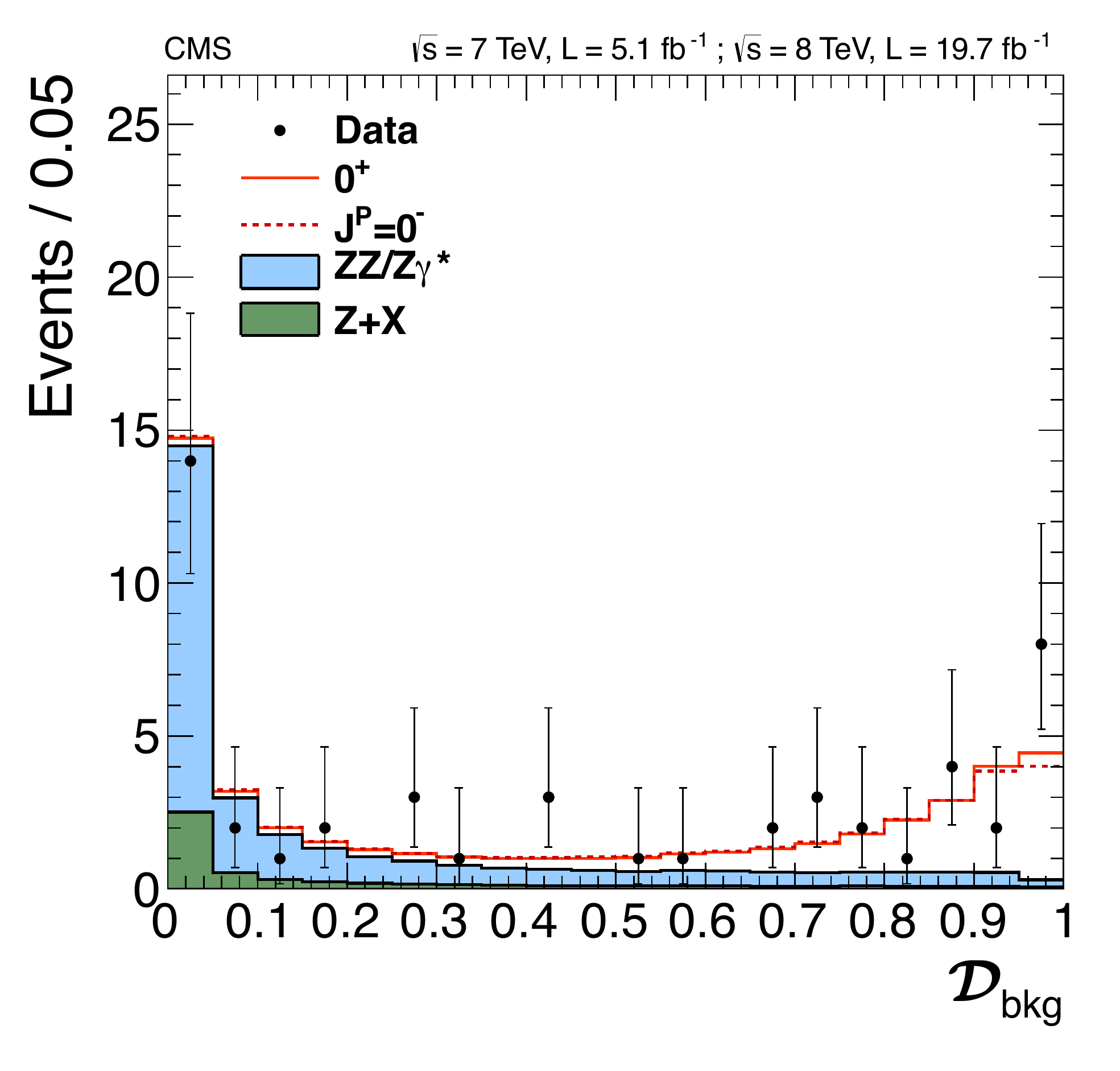}
    \includegraphics[width=\cmsFigWidthStd]{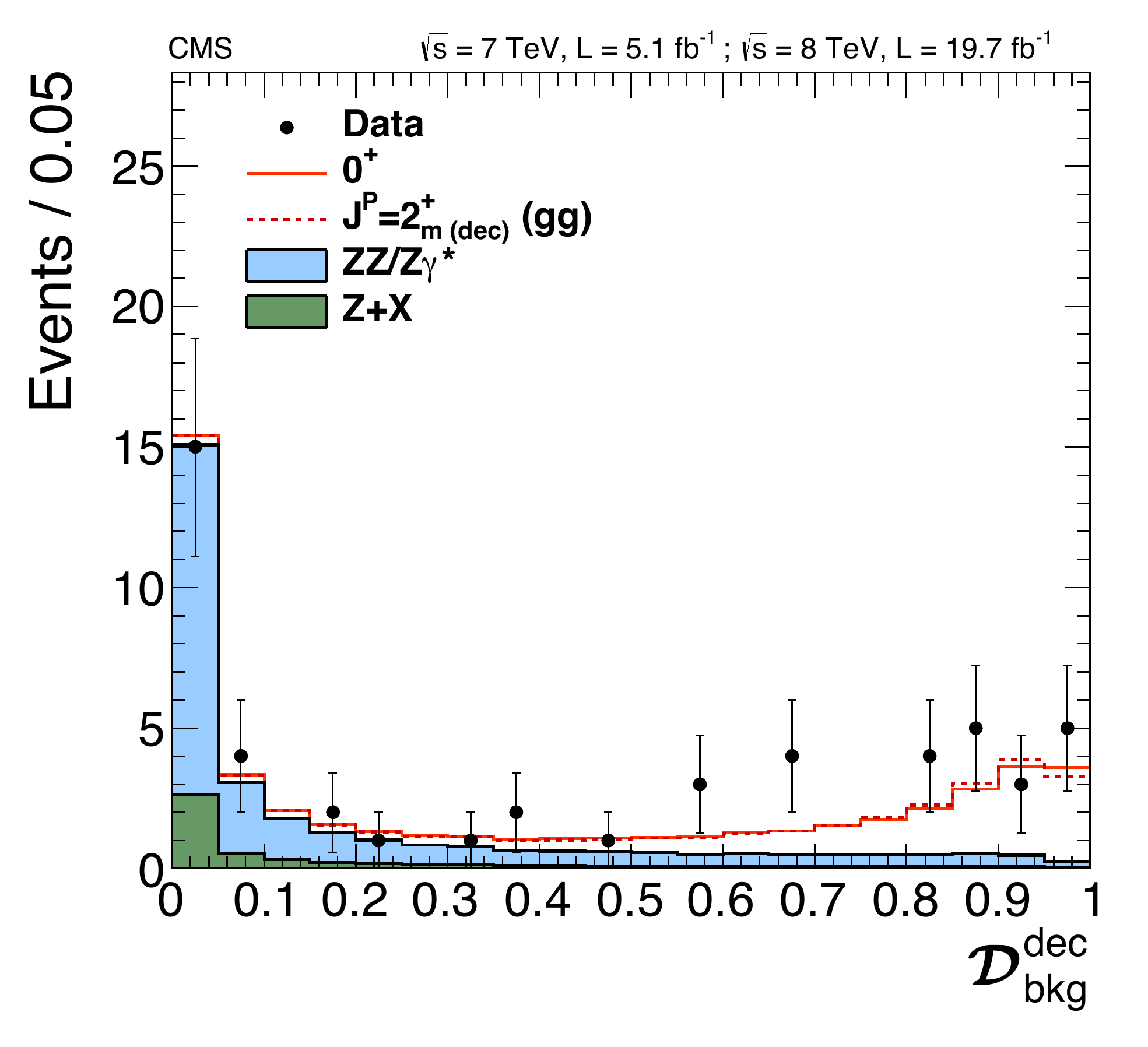}
    \caption{Distribution of $\superKD$ (\cmsLeft) and $\mathcal{D}_\text{bkg}^\text{dec}$ for the production-independent scenario (\cmsRight) in
    data and MC expectations for the background and for a signal
    resonance consistent with the SM Higgs boson with $m_{0^+}$ = 125.6\GeV. \label{fig:supermela}} \end{center}
\end{figure}

\begin{figure*}[th!]
  \begin{center}
    \includegraphics[width=0.32\linewidth]{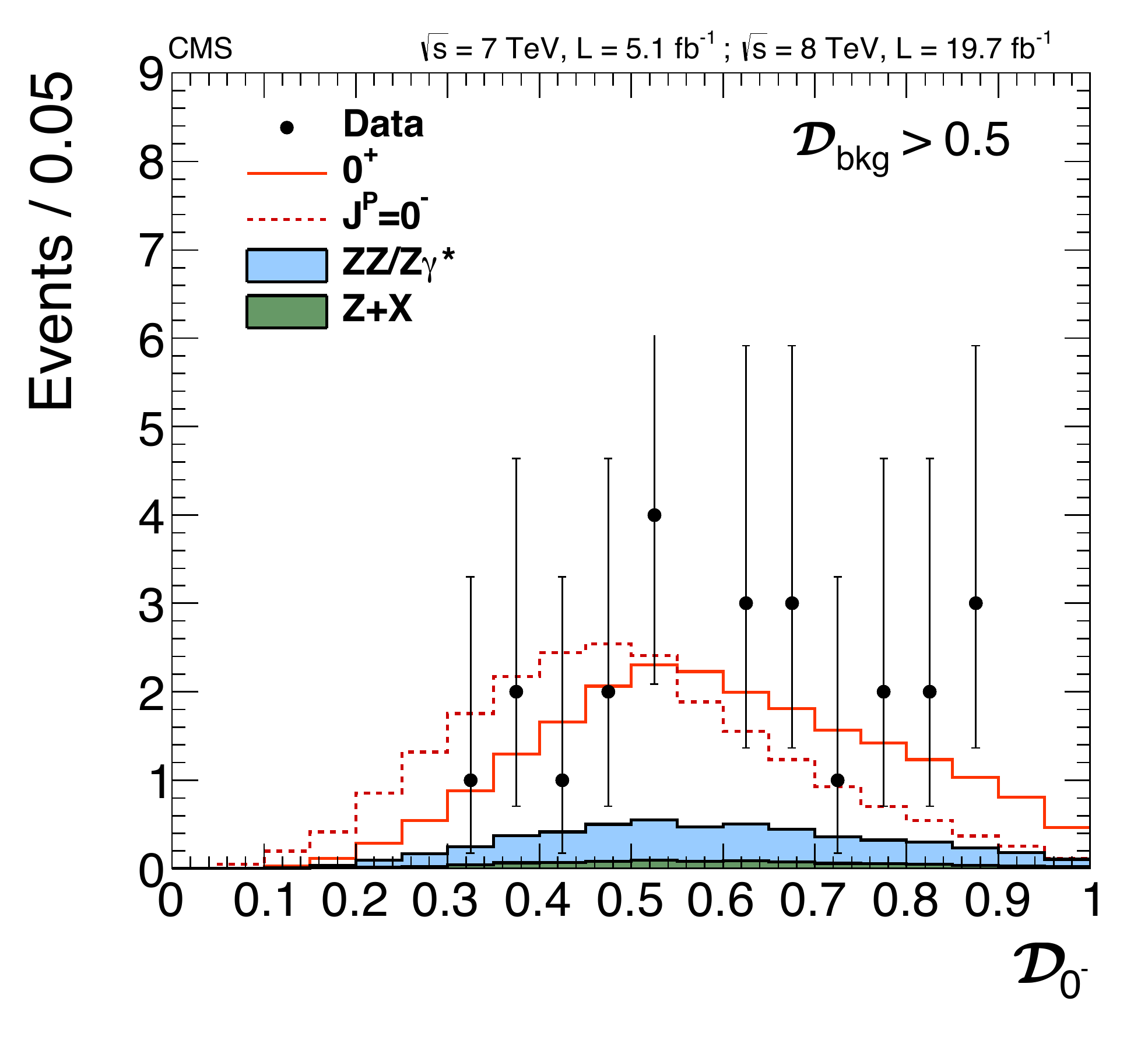} \includegraphics[width=0.32\linewidth]{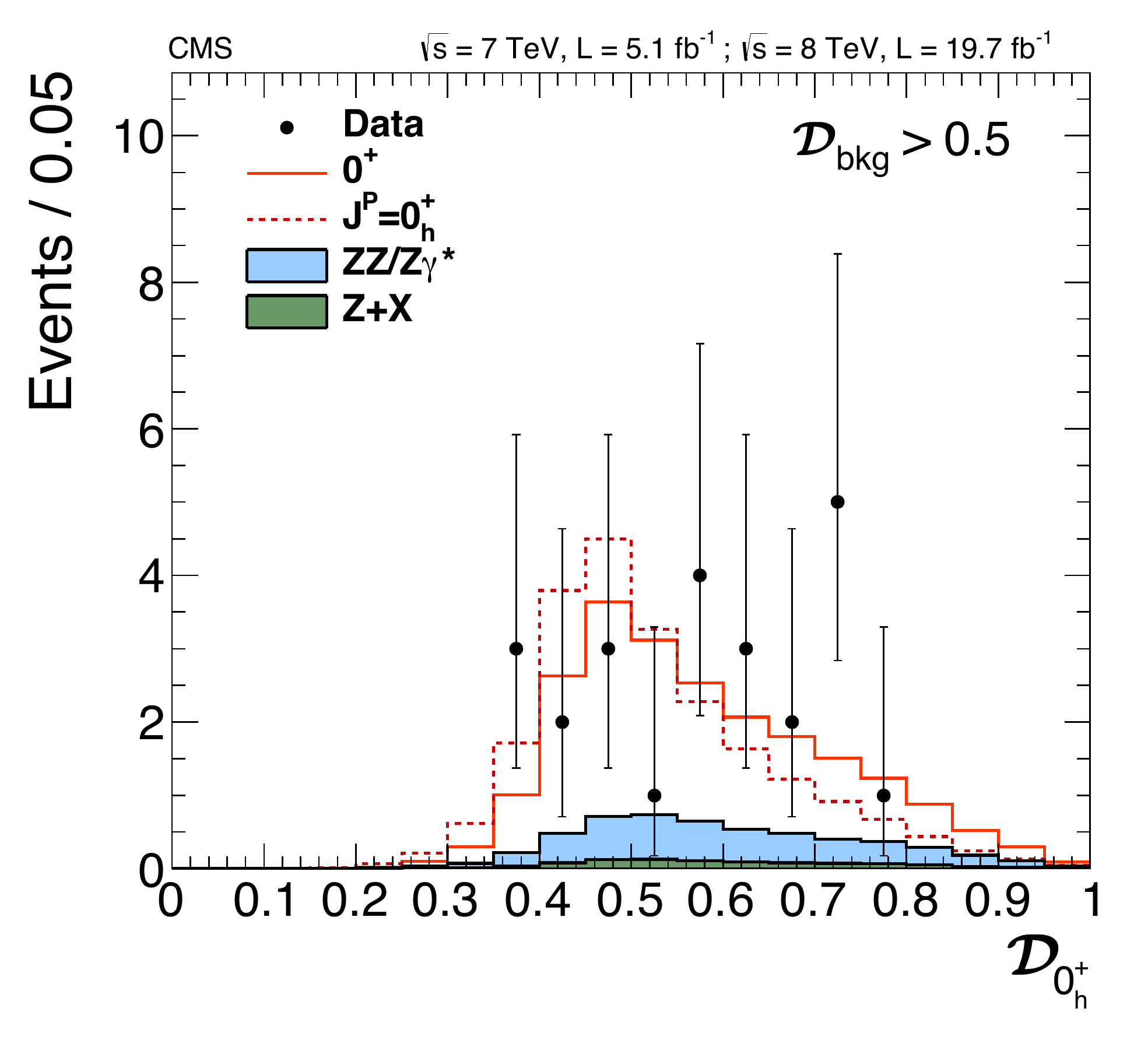}
    \includegraphics[width=0.32\linewidth]{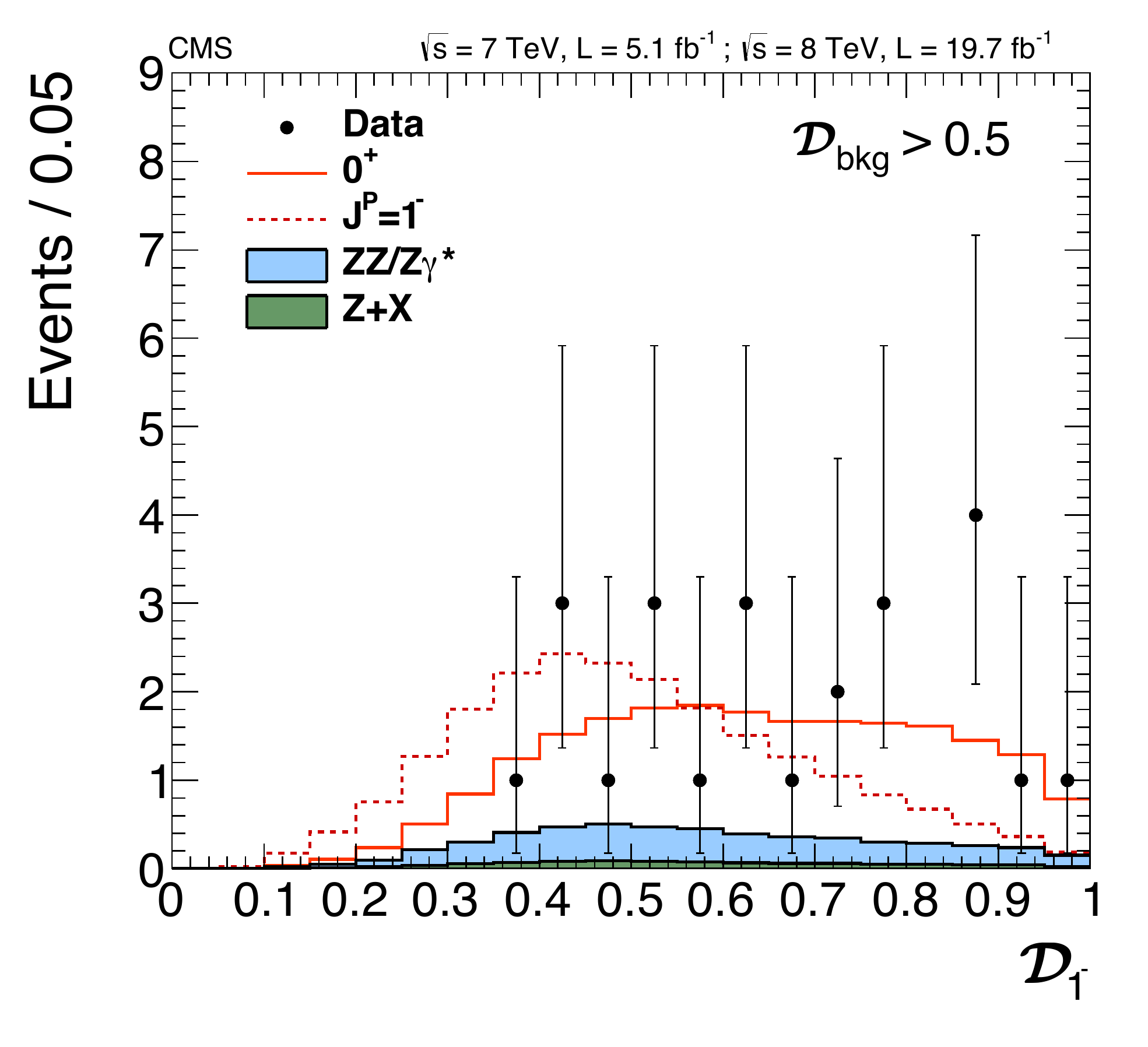}
    \includegraphics[width=0.32\linewidth]{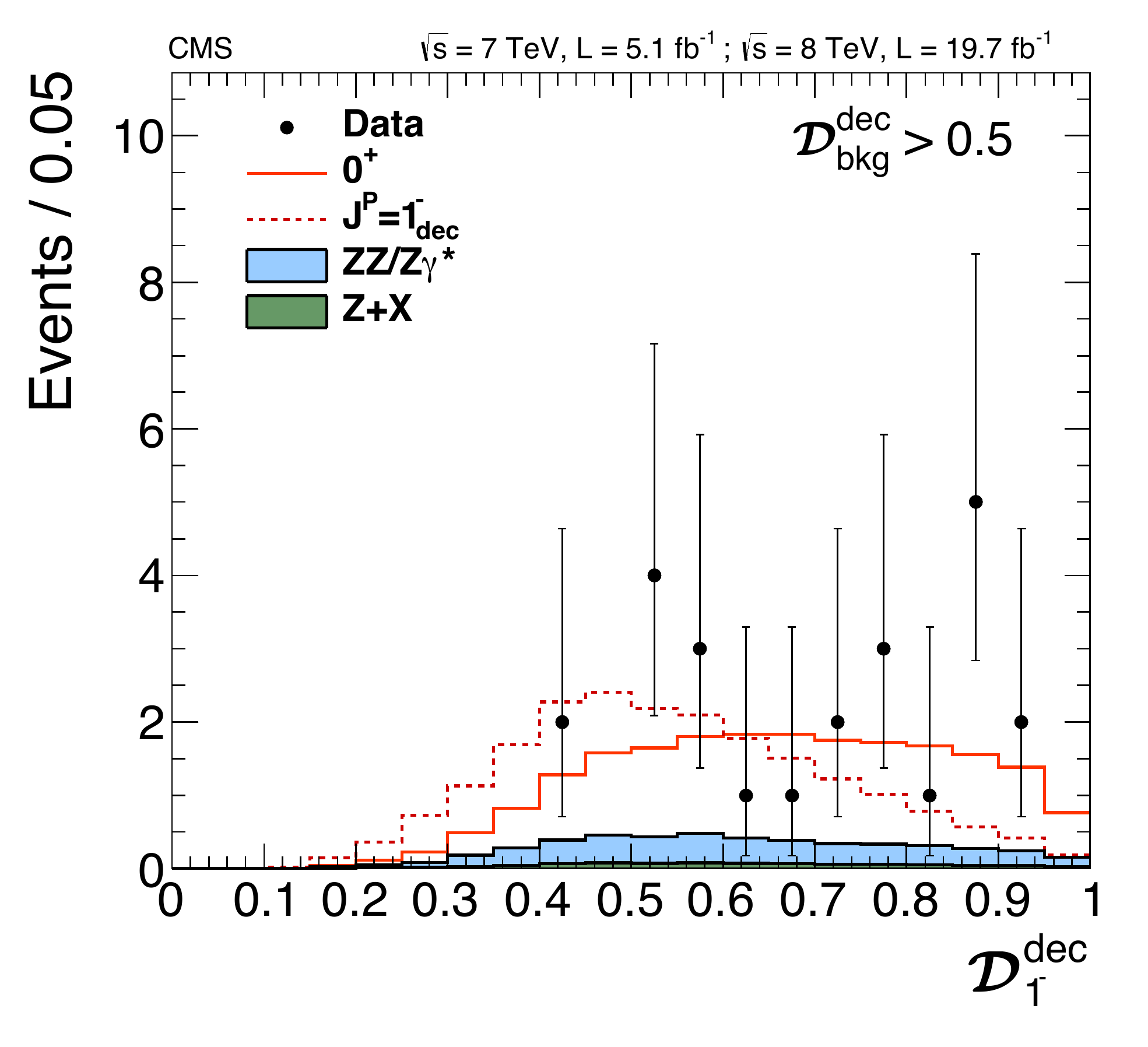}
    \includegraphics[width=0.32\linewidth]{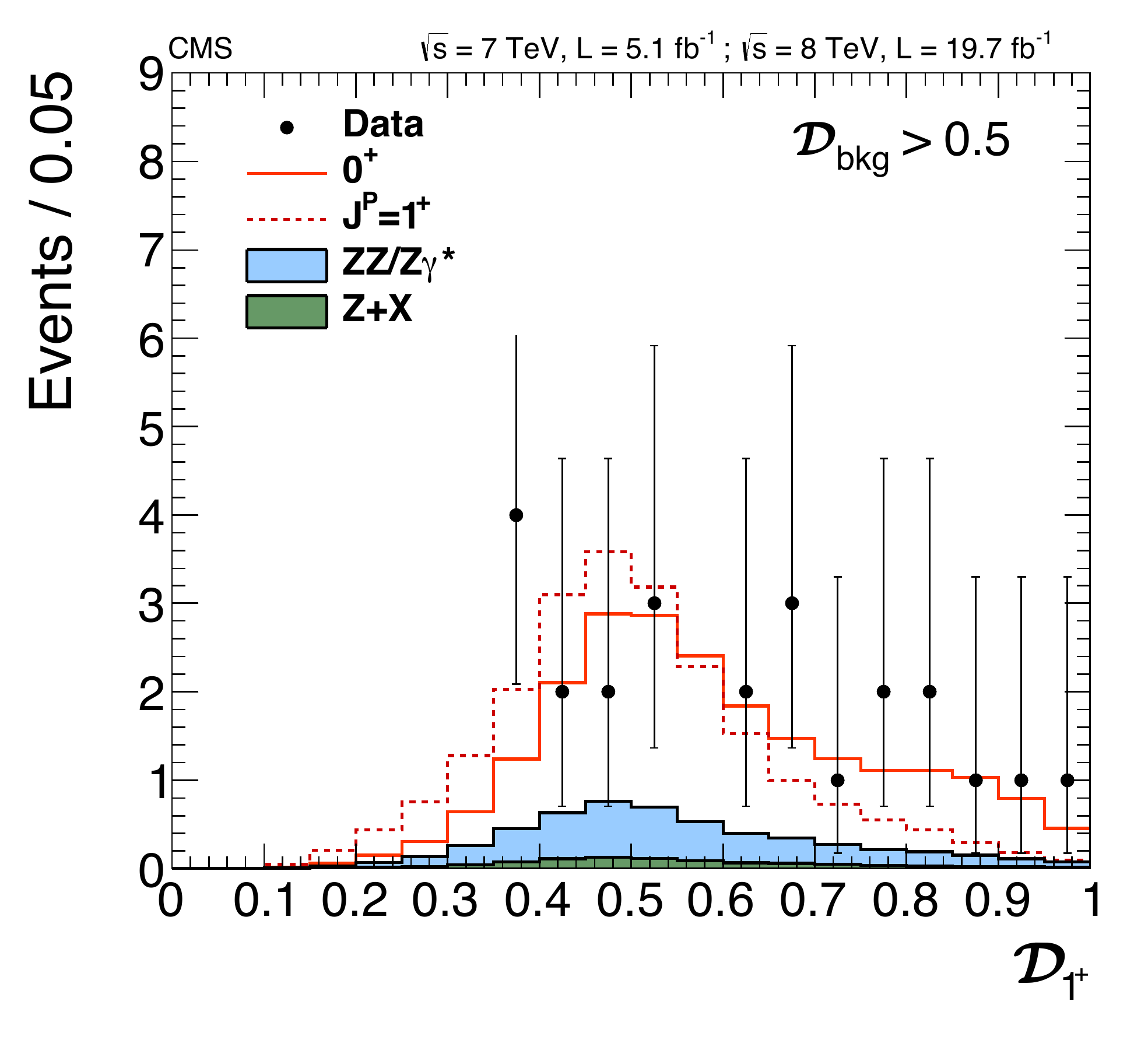} \includegraphics[width=0.32\linewidth]{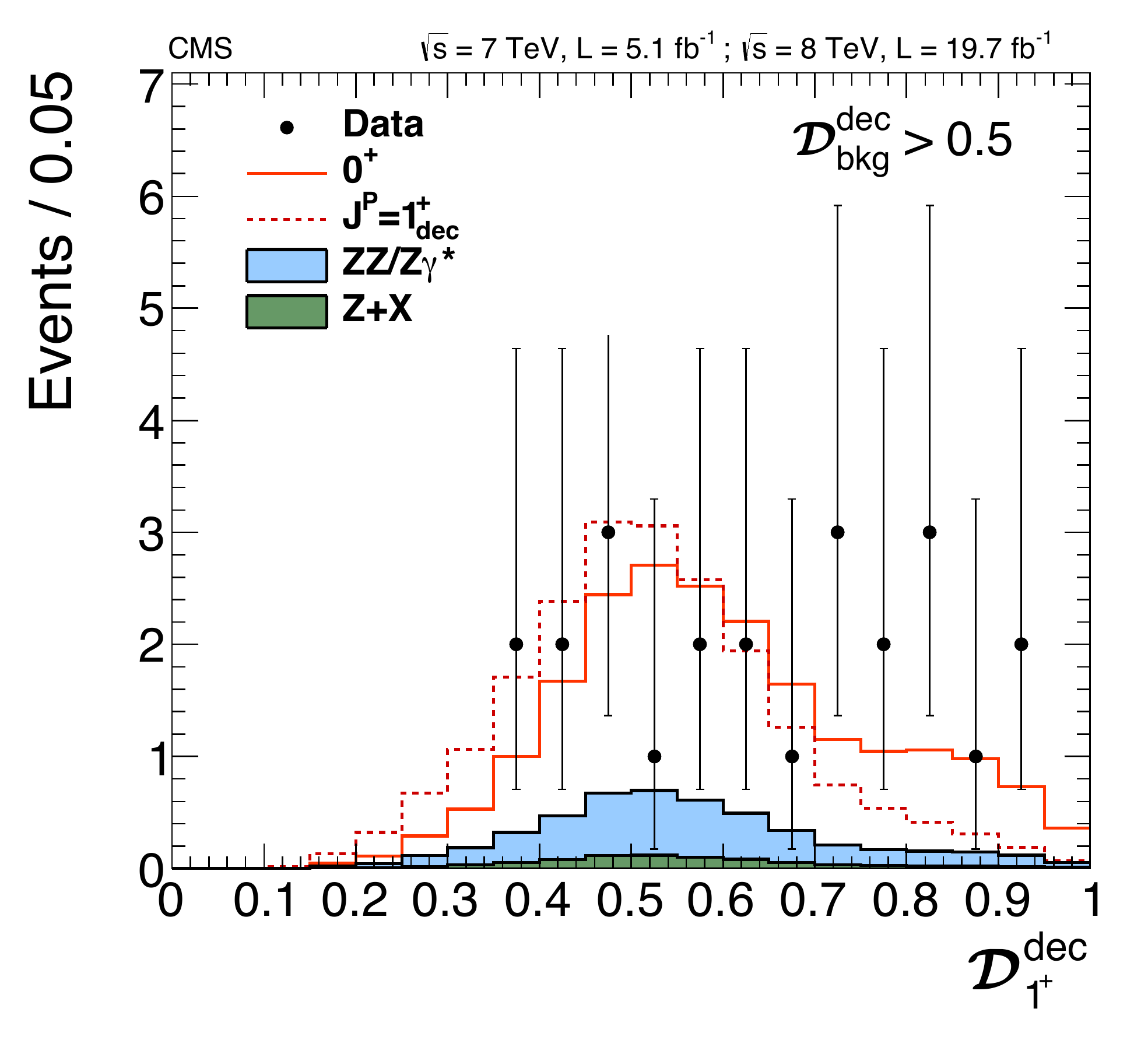}
    \caption{Distributions of $\spinKD$ with a requirement $\mathcal{D}_\text{bkg}^\text{(dec)} > 0.5$.  Distributions in data (points
      with error bars) and expectations for background and signal are
      shown: six alternative $J^P$ hypotheses are shown. $J^P=0^-$
      (upper left), $0_\mathrm{h}^+$ (upper middle), $1^-(\Pq\Paq)$ (upper right), $1^-$ (lower left),
      $1^+(\Pq\Paq)$ (lower middle), $1^+$ (lower right).  \label{fig:jp_kd1}}
  \end{center}
\end{figure*}

\begin{figure*}[th!]
  \begin{center}
    \includegraphics[width=0.32\linewidth]{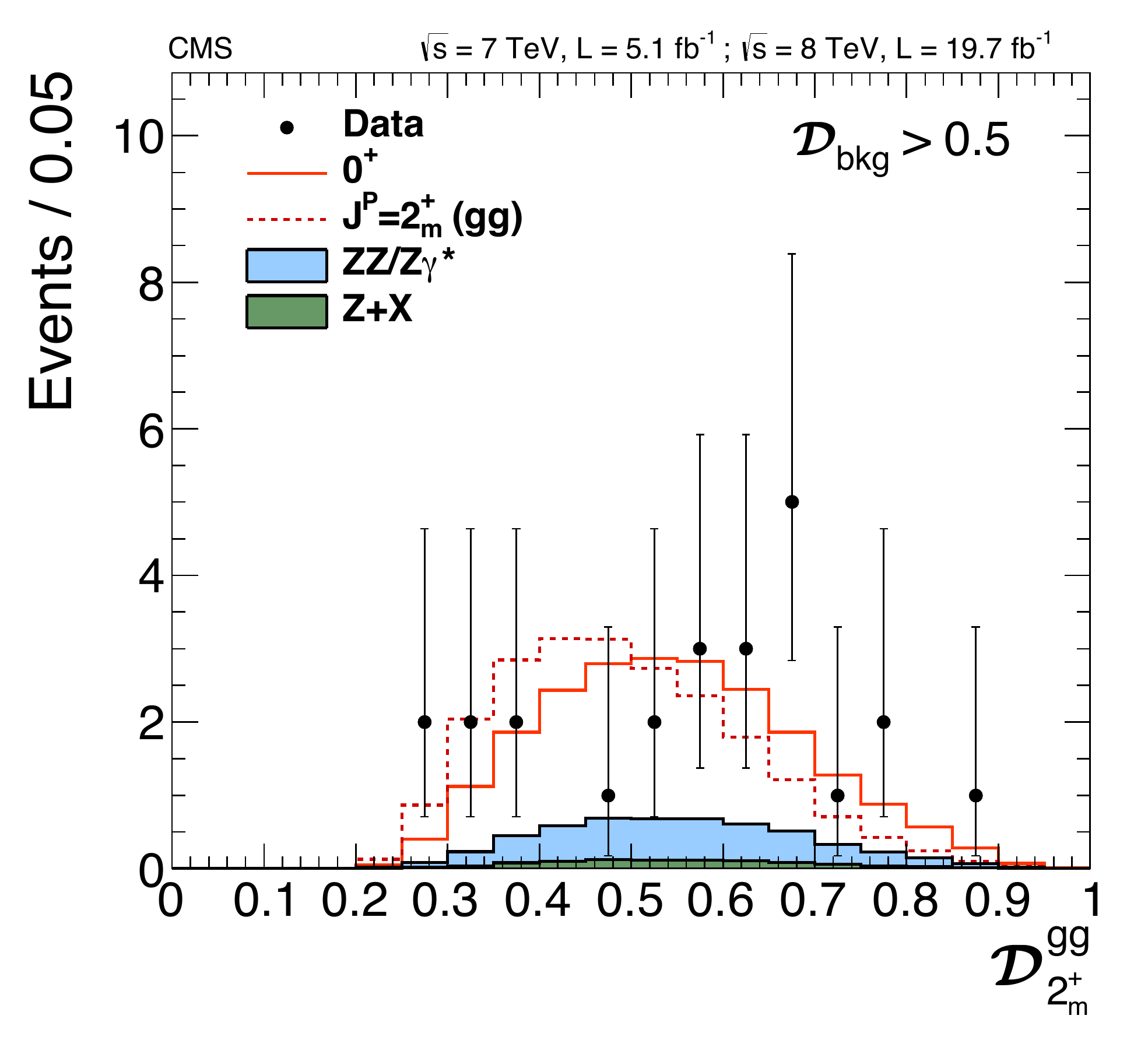}
    \includegraphics[width=0.32\linewidth]{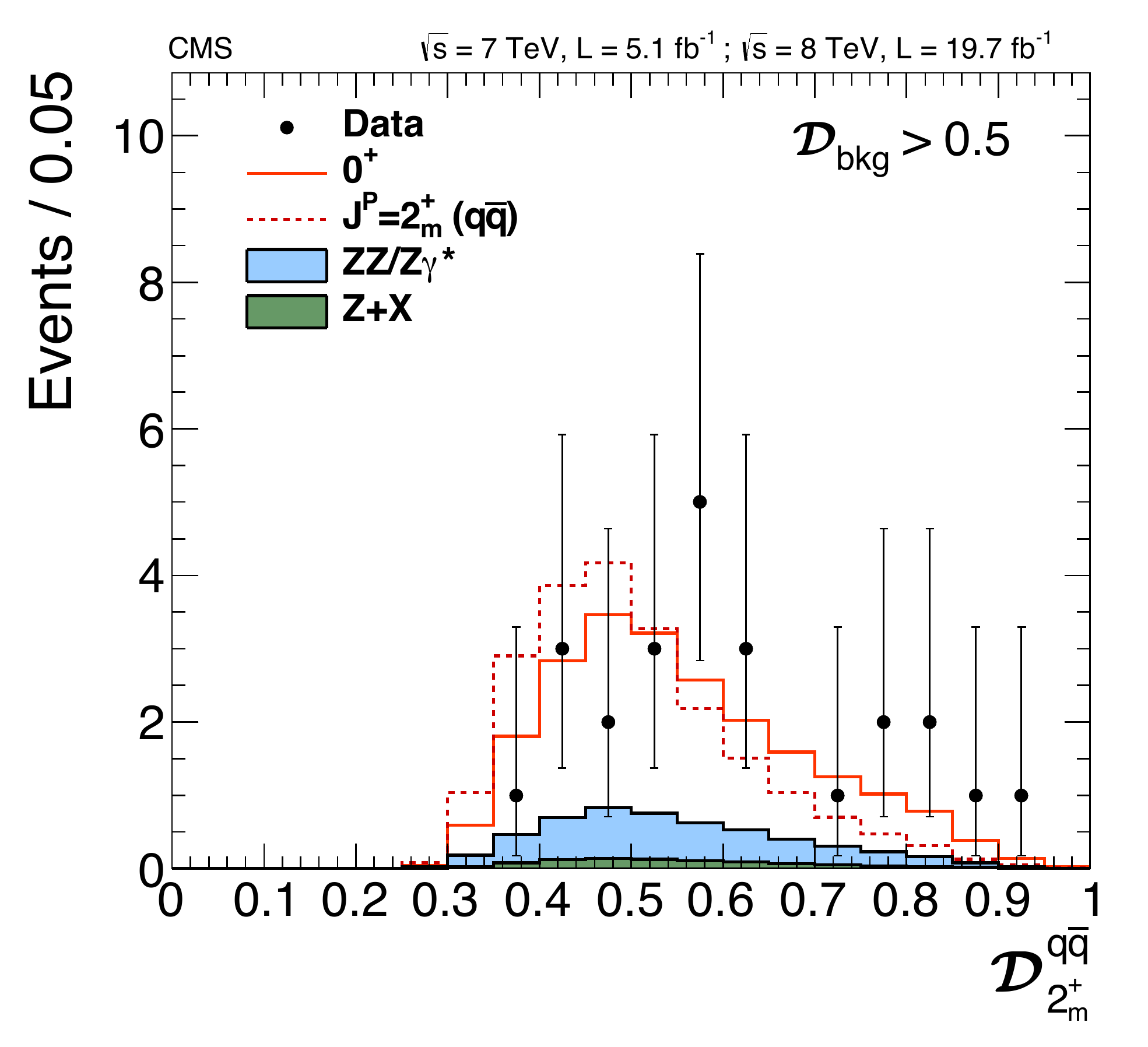}
    \includegraphics[width=0.32\linewidth]{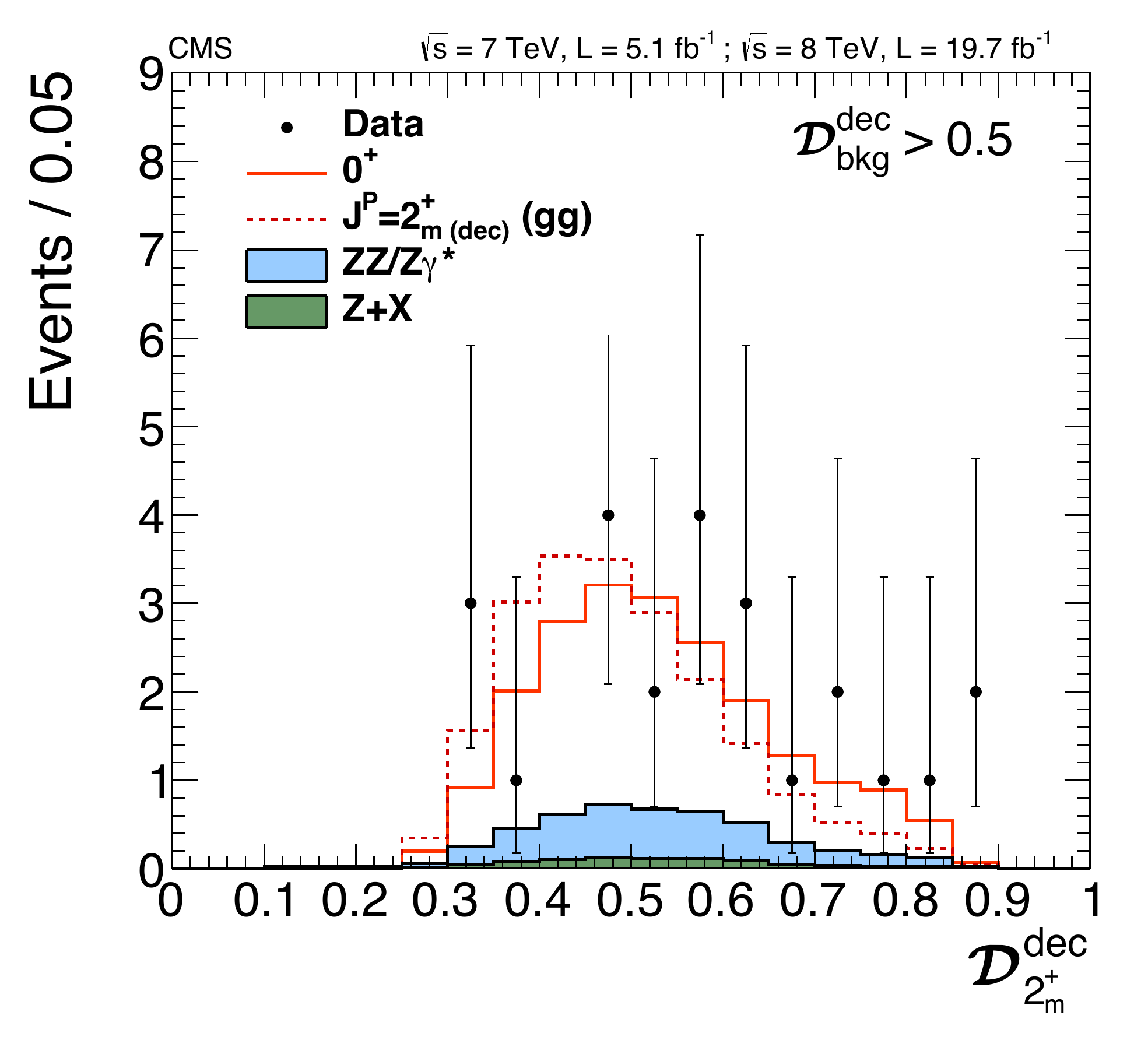}
    \includegraphics[width=0.32\linewidth]{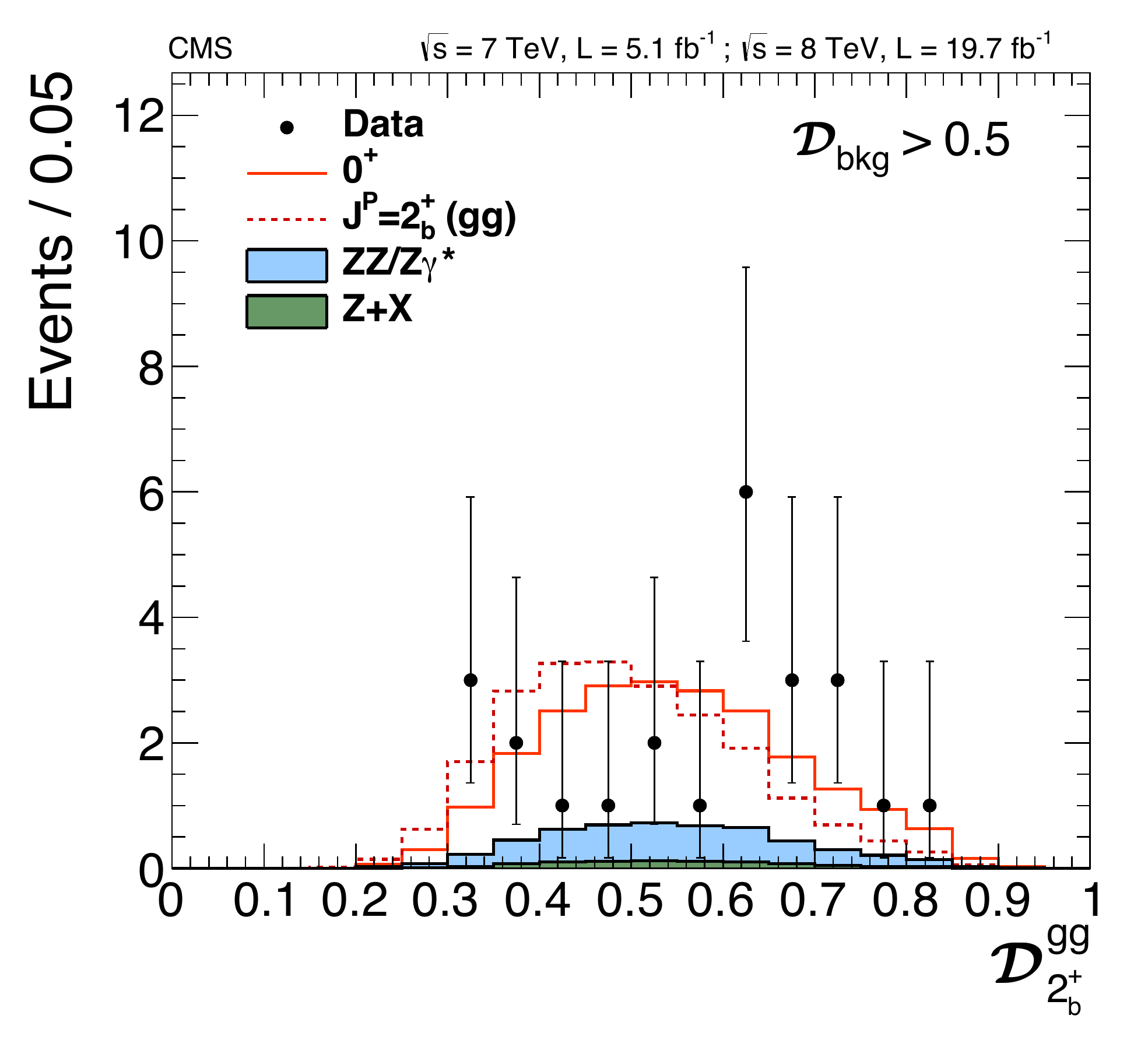}
    \includegraphics[width=0.32\linewidth]{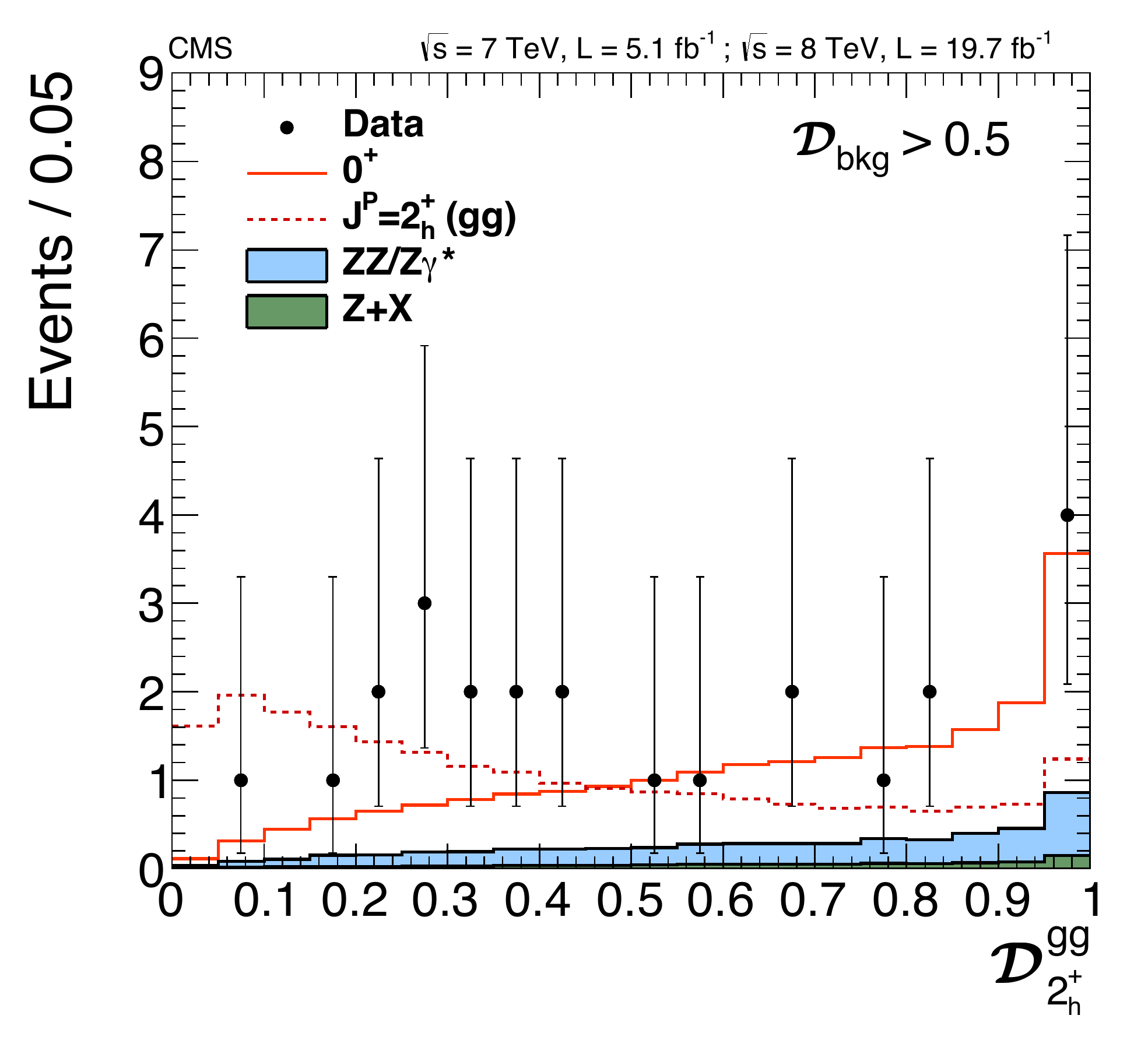}
    \includegraphics[width=0.32\linewidth]{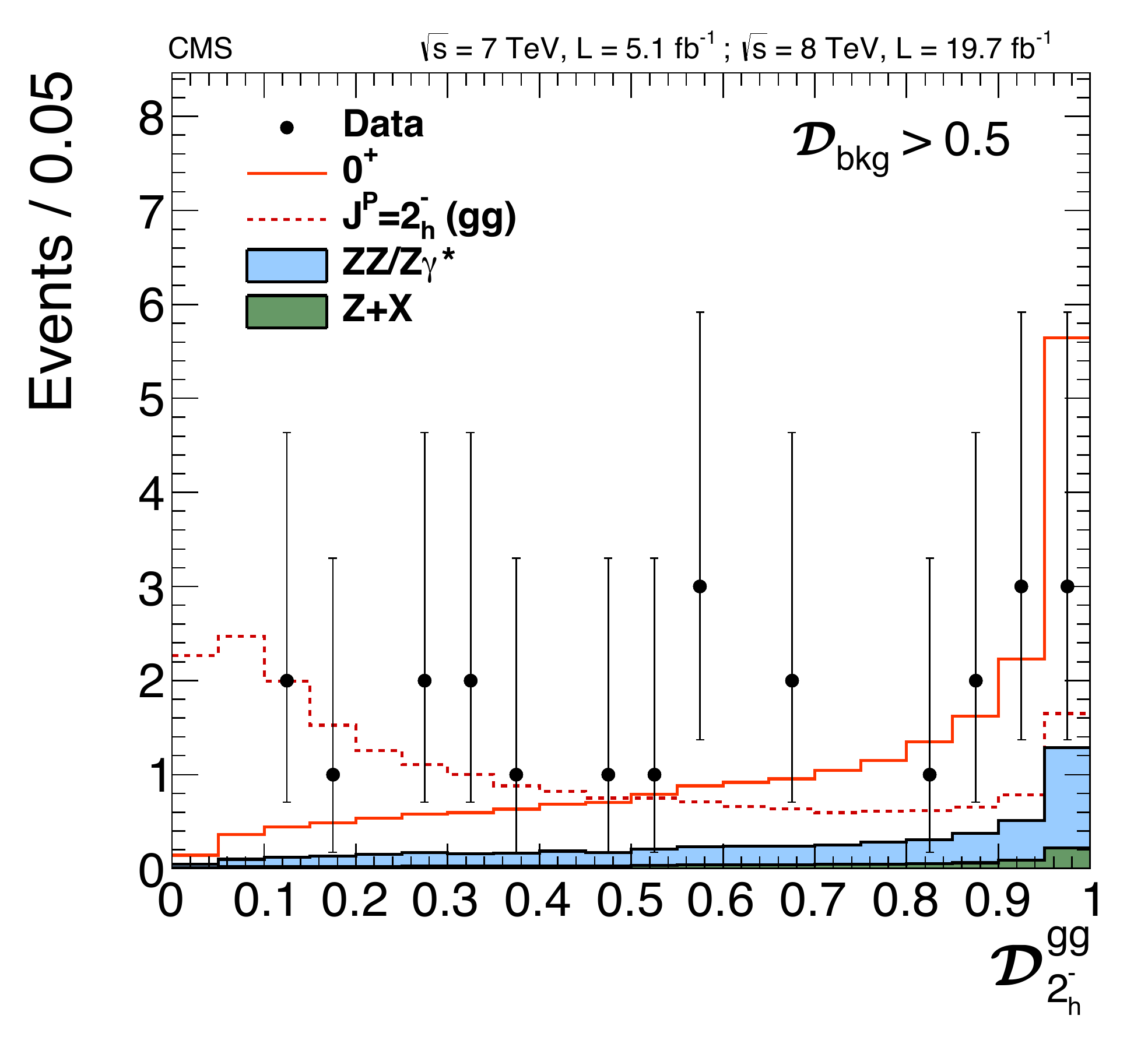}
    \caption{Distributions of $\spinKD$ with a requirement $\mathcal{D}_\text{bkg}^\text{(dec)} > 0.5$.  Distributions in data
      (points with error bars) and expectations for background and
      signal are shown: six alternative $J^P$ hypotheses are shown.
      $J^P=2^{+}_\mathrm{m}$ for gluon fusion (upper left), $2^{+}_\mathrm{m}$ for
      VBF (upper middle), $2^{+}_\mathrm{m}$ for the production-independent
      scenario (upper right), $2_\mathrm{b}^+(\Pg\Pg)$ (lower left), $2_\mathrm{h}^+(
        \Pg\Pg)$, (lower middle), $2_\mathrm{h}^-(\Pg\Pg)$ (lower right).  \label{fig:jp_kd2}}
  \end{center}
\end{figure*}

Figure~\ref{fig:supermela} shows expected and observed distributions
for the discriminants $\mathcal{D}_\text{bkg}$ and
$\mathcal{D}_\text{bkg}^{\text{dec}}$.  The distributions are very
similar for the SM and all alternative signal hypotheses but differ
significantly from the background.  Figures~\ref{fig:jp_kd1}
and~\ref{fig:jp_kd2} show distributions for the $\mathcal{D}_{J^P}$
observables for all tested signal hypotheses. Only one alternative
hypothesis is shown on each figure.  The distributions show events
with $\mathcal{D}_\text{bkg}^\text{ (dec)} > 0.5$ to enhance the
fraction of signal events for illustration purposes only. For the
hypothesis tests, the full range of the discriminant is used.

\begin{figure}[th!]
  \begin{center}
    \includegraphics[width=\cmsFigWidthStd]{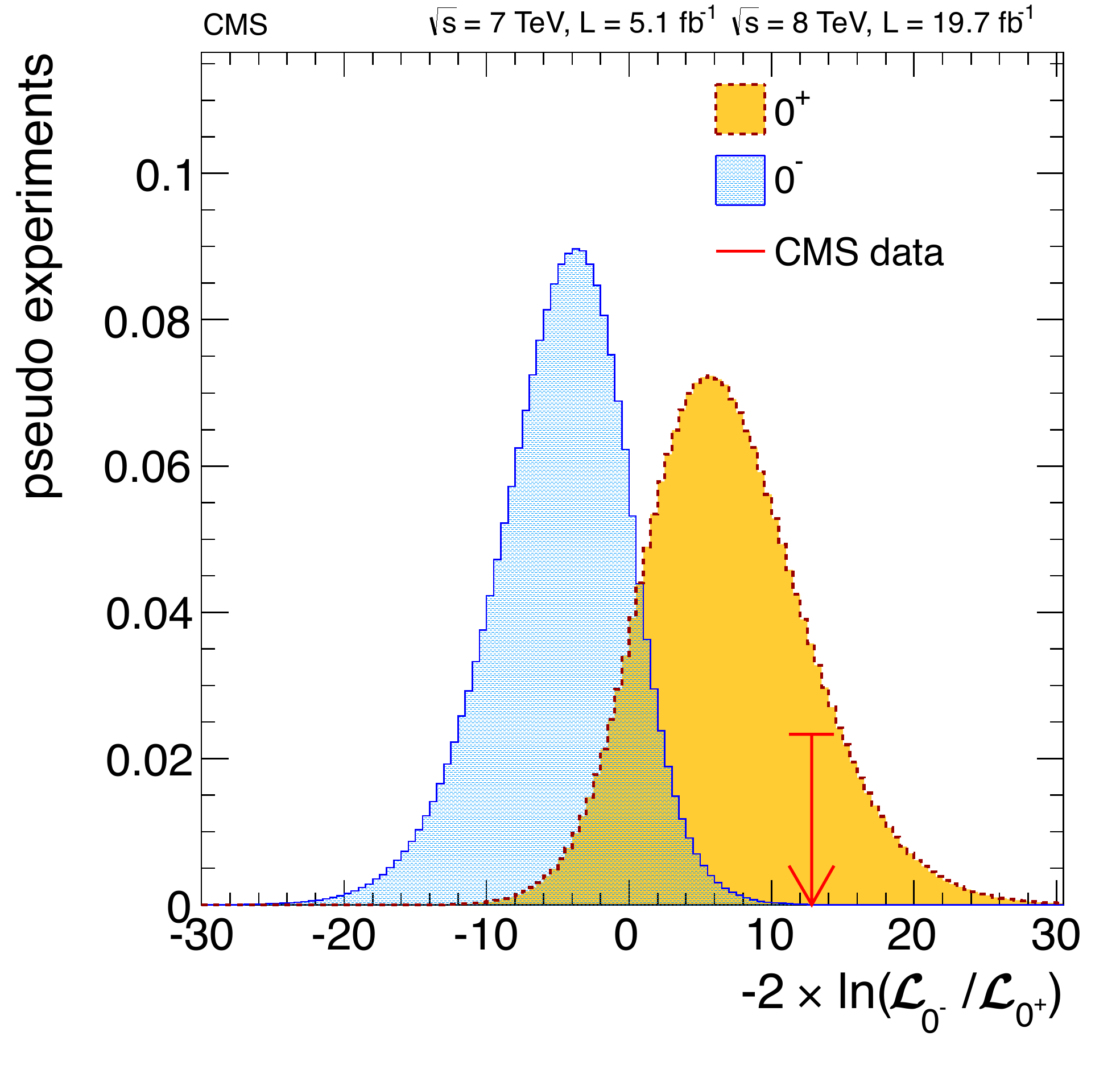}
    \includegraphics[width=\cmsFigWidthStd]{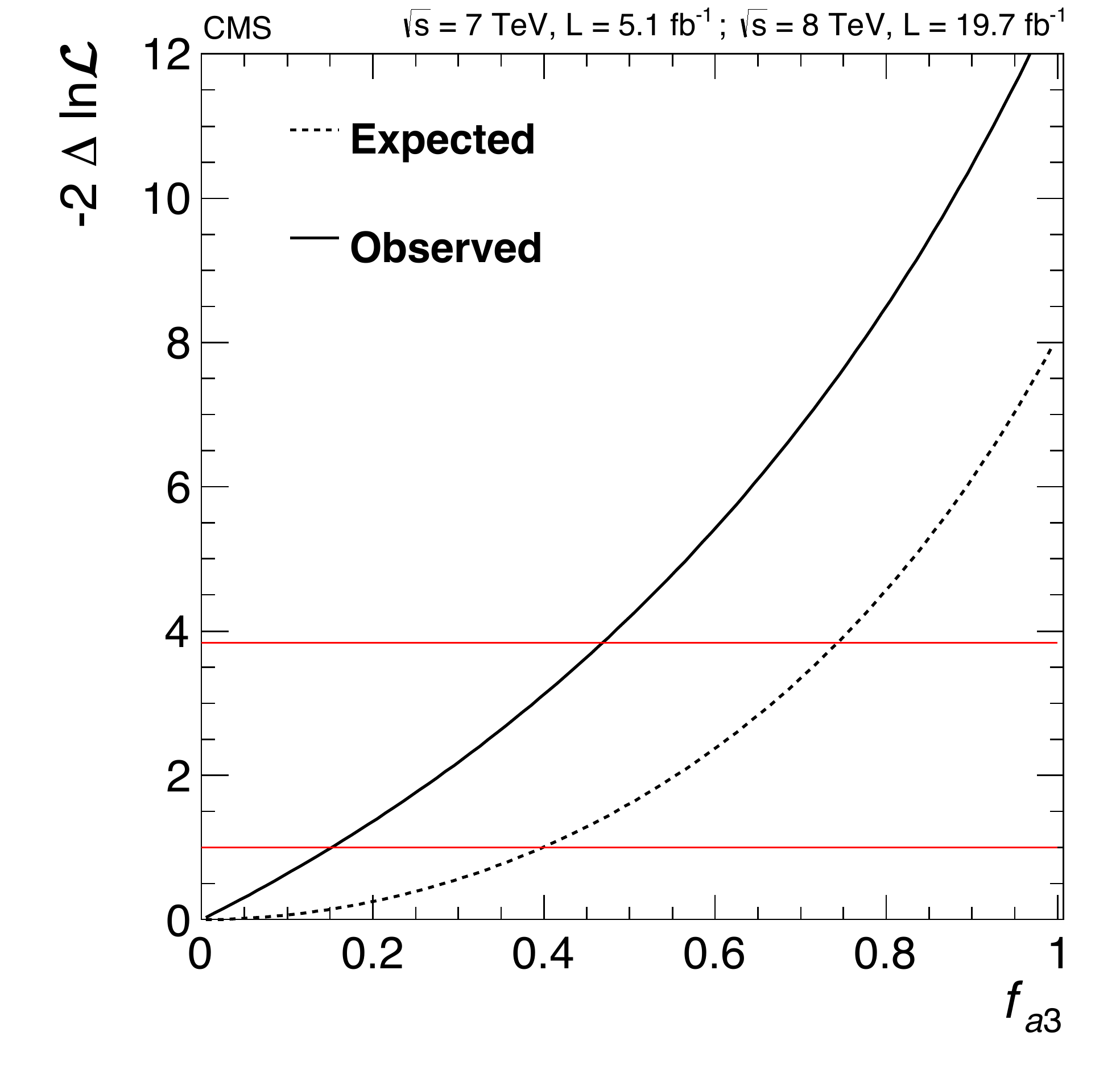}
    \caption{(\cmsLeft) Distribution of the test statistic
      $q=-2{\ln(\mathcal{L}_{0^-}/\mathcal{L}_{0^+})}$ of the
      pseudoscalar boson hypothesis tested against the SM Higgs boson
      hypothesis.  Distributions for the SM Higgs boson are
      represented by the yellow histogram, and those for the
      alternative $J^P$ hypotheses are represented by the blue
      histogram. The arrow indicates the observed value.  (\cmsRight)
      Average expected and observed distribution of $-2\Delta\ln L$ as
      a function of $f_{a3}$. The horizontal lines at
      $-2\Delta \ln \mathcal{L} = 1$ and 3.84 represent the 68\% and
      95\% \CL's, respectively. \label{fig:sep_fa3}} \end{center}
\end{figure}

The alternative signal models are defined by the tensor structure of
couplings; however, the absolute values of couplings, and hence, the
expected event yields are not uniquely defined.  The cross sections
for alternative signal hypotheses are left floating in the fit.  The
same approach is taken for the SM Higgs boson hypothesis: \ie, the
overall SM Higgs boson signal strength $\strength$ is the best-fit
value as it comes out from data. This way, the overall signal event
yield is not a part of the discrimination between alternative
hypotheses.  Consequently, for pairwise tests of alternative signal
hypotheses with respect to the SM Higgs boson, the test statistic is
defined using the ratio of signal plus background likelihoods for two
signal hypotheses $q=-2{\ln(\mathcal{L}_{J^P}/\mathcal{L}_{0^+})}$.
The expected distribution of $q$ for the pseudoscalar hypothesis (blue
histogram) and the SM Higgs boson (orange histogram) are shown in
Fig.~\ref{fig:sep_fa3}~(\cmsLeft). Similar distributions for the test
statistic $q$ are obtained for the other alternative hypotheses
considered.  The pseudoexperiments are generated using the nuisance
parameters fitted in data.

To quantify the consistency of the observed test statistics
$q_\text{obs}$ with respect to the SM Higgs boson hypothesis ($0^+$),
we assess the probability $p = P( q \leq q_\text{obs} \, | \, 0^+
+\text{bkg} )$ and convert it into a number of standard deviations $Z$
via the Gaussian one-sided tail integral:
\begin{equation}
\label{eq:significance}
p = \int^\infty_Z \! \frac{1}{\sqrt{2\pi}} \exp\left(-x^2/2\right) \, \mathrm{d}x.
\end{equation}
Similarly, the consistency of the observed data with alternative
signal hypotheses ($J^P$) is assessed from $P( q \geq q_\text{obs} \, |
\, J^P + \text{bkg} )$. The $\mathrm{CL_s}$ criterion, defined as $\mathrm{CL_s} = { P( q \geq q_\text{obs} \, | \, J^P + \text{bkg} ) } / { P( q
  \geq q_\text{obs} \, | \, 0^+ + \text{bkg} ) } < \alpha,$ is used for
the final inference of whether a particular alternative signal
hypotheses is excluded or not with a given confidence level $(1 -
\alpha)$.

\begin{figure}[th!]
  \begin{center}
    \includegraphics[width=\cmsFigWidth]{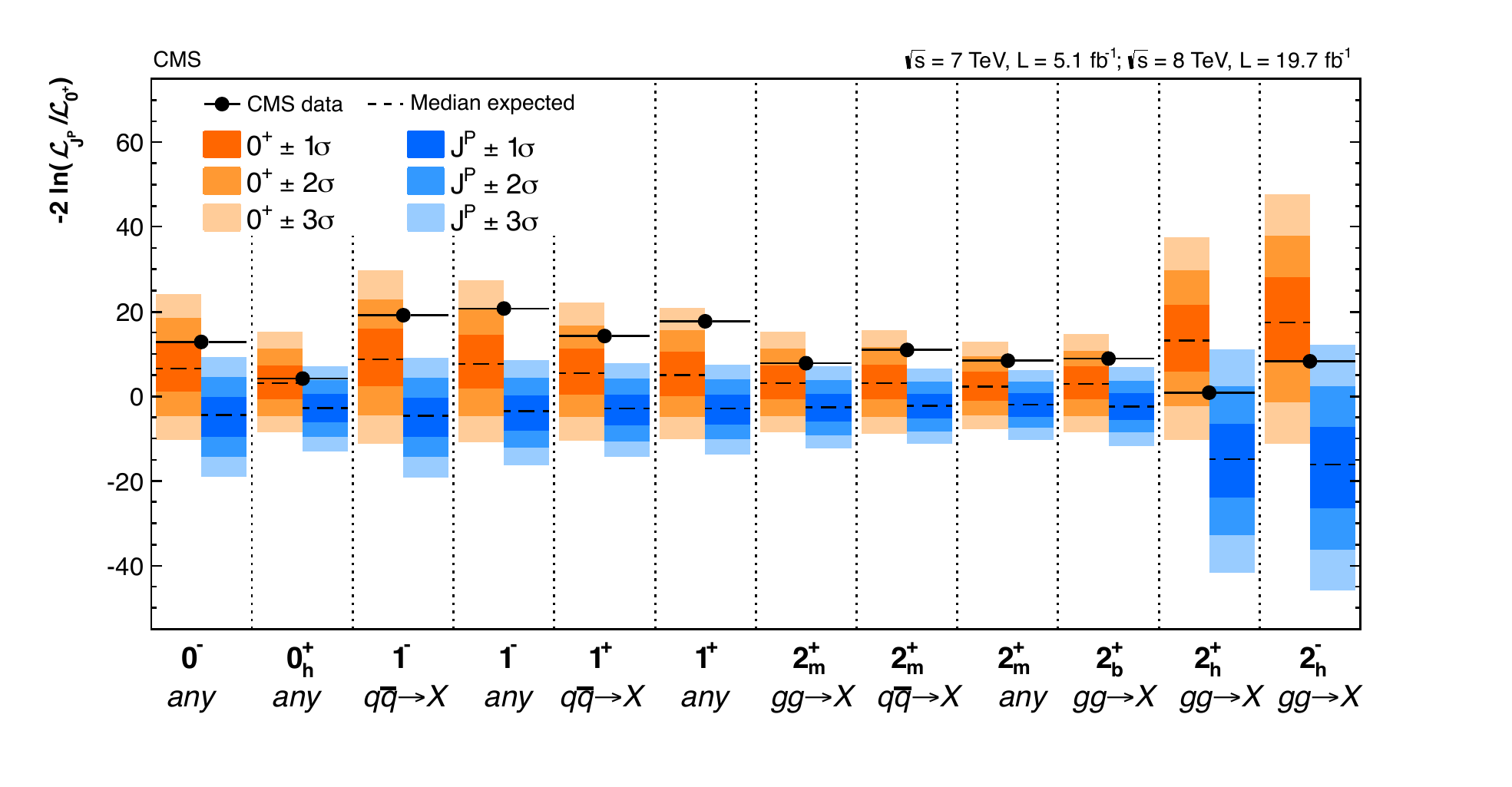}
    \caption{Summary of the expected and observed values for the
      test-statistic $q$ distributions for the twelve alternative
      hypotheses tested with respect to the SM Higgs boson.  The
      orange (blue) bands represent the 1$\sigma$, 2$\sigma$, and
      3$\sigma$ around the median expected value for the SM Higgs
      boson hypothesis (alternative hypothesis).  The black point
      represents the observed
      value.  \label{fig:jp_summary}} \end{center}
\end{figure}

The expected separations between alternative signal hypotheses are
quoted for two cases. In the first case, the expected SM Higgs boson
signal strength and the alternative signal cross sections are equal to
the ones obtained in the fit of the data. The second case assumes the
nominal SM Higgs boson signal strength ($\strength=1$, as indicated in
parentheses for expectations quoted in Table~\ref{tab:jpmodels}),
while the cross sections for the alternative signal hypotheses are
taken to be the same as for the SM Higgs boson (the $2\Pe2\Pgm$
channel is taken as a reference).  Since the observed signal strength
is very close to unity, the two results for the expected separations
are also similar.  The observed values of the test statistic in the
case of the SM Higgs boson versus a pseudoscalar boson are shown with
red arrows in Fig.~\ref{fig:sep_fa3}~(\cmsLeft).  Results obtained
from the test statistic distributions are summarized in
Table~\ref{tab:jpmodels} and in Fig.~\ref{fig:jp_summary}.

\begin{table*}[htbp]
  \centering
  \topcaption{ List of models used in the analysis of the spin and parity hypotheses
    corresponding to the pure states of the type noted.  The expected
    separation is quoted for two scenarios, where the signal strength
    for each hypothesis is predetermined from the fit to data and where
    events are generated with SM expectation for the signal cross
    section ($\mu=1$).  The observed separation quotes consistency of
    the observation with the $0^+$ model or $J^P$ model and corresponds
    to the scenario where the signal strength is floated in the fit to
    data.  The last column quotes the CL$_\mathrm{s}$ value for the $J^P$ model.
    \label{tab:jpmodels}}
  \begin{scotch}{cccccc}
    $J^P$ model & $J^P$ production & Expected ($\mu=1$) &  Obs. $0^+$  & Obs. $J^P$ & CL$_\mathrm{s}$  \\
    \hline
    $0^-$             & any              &  2.4$\sigma$ (2.7$\sigma$)  & $-$1.0$\sigma$  & +3.8$\sigma$  &  0.05\%  \\
    $0_\mathrm{h}^+$   & any              &  1.7$\sigma$ (1.9$\sigma$)  & $-$0.3$\sigma$  & +2.1$\sigma$  &  4.5\% \\
    $1^-$             & $\Pq\Paq\to \X$  &  2.7$\sigma$ (2.7$\sigma$)  & $-$1.4$\sigma$  & +4.7$\sigma$  &  0.002\% \\
    $1^-$             & any              &  2.5$\sigma$ (2.6$\sigma$)  & $-$1.8$\sigma$  & +4.9$\sigma$  &  0.001\% \\
    $1^+$             & $\Pq\Paq\to \X$  &  2.1$\sigma$ (2.3$\sigma$)  & $-$1.5$\sigma$  & +4.1$\sigma$  &  0.02\% \\
    $1^+$             & any              &  2.0$\sigma$ (2.1$\sigma$)  & $-$2.1$\sigma$  & +4.8$\sigma$  &  0.004\% \\
    $2_\mathrm{m}^+$     & $\Pg\Pg \to \X$       &  1.9$\sigma$ (1.8$\sigma$)  & $-$1.1$\sigma$  & +3.0$\sigma$  &  0.9\% \\
    $2_\mathrm{m}^+$     & $\Pq\Paq\to \X$       &  1.7$\sigma$ (1.7$\sigma$)  & $-$1.7$\sigma$  & +3.8$\sigma$  &  0.2\% \\
    $2_\mathrm{m}^+$     & any                   &  1.5$\sigma$ (1.5$\sigma$)  & $-$1.6$\sigma$  & +3.4$\sigma$  &  0.7\% \\
    $2_\mathrm{b}^+$     & $\Pg\Pg \to \X$       &  1.6$\sigma$ (1.8$\sigma$)  & $-$1.4$\sigma$  & +3.4$\sigma$  &  0.5\% \\
    $2_\mathrm{h}^+$     & $\Pg\Pg \to \X$       &  3.8$\sigma$ (4.0$\sigma$)  & +1.8$\sigma$    & +2.0$\sigma$  &  2.3\% \\
    $2_\mathrm{h}^-$     & $\Pg\Pg \to \X$       &  4.2$\sigma$ (4.5$\sigma$)  & +1.0$\sigma$    & +3.2$\sigma$  &  0.09\% \\
  \end{scotch}
\end{table*}

The observed value of the test statistic is larger than the median
expected for the SM Higgs boson. This happens for many distributions
because of strong kinematic correlations between different signal
hypotheses, most prominently seen in the $m_{\cPZ_2}$ distributions.
The pseudoscalar ($0^-$) and all spin-1 hypotheses tested are
excluded at the 99.9\% or higher \CL All tested spin-2 models are
excluded at the 95\% or higher \CL The $0^+_\mathrm{h}$ hypothesis is
disfavored, with a $\mathrm{CL_s}$ value of 4.5\%.

In addition to testing pure $J^P$ states against the SM Higgs boson
hypothesis, a measurement for a possible mixture of CP-even and CP-odd
states or other effects leading to anomalous couplings in the
$\PH\rightarrow\cPZ\cPZ$ decay amplitude in
Eq.~(\ref{eq:fullampl-spin0}) is performed.  The $\mathcal{D}_{0^-}$
discriminant is designed for the discrimination between the third and
the first amplitude contributions in Eq.~(\ref{eq:fullampl-spin0})
when the phase $\phi_{a3}$ between $a_3$ and $a_1$ couplings is not
determined from the data~\cite{Heinemeyer:2013tqa}.  For example, even
when restricting the coupling ratios to be real, there remains an
ambiguity where $\phi_{a3}=0$ or $\pi$.  The interference between the
two terms ($a_1$ and $a_3$) is found to have a negligible effect on
the discriminant distribution or the overall yield of events. The
parameter $f_{a3}$ is defined as
\begin{equation}
f_{a3} =  \frac{|a_{3}|^2\sigma_3}{| a_{1}|^2\sigma_1+ | a_{2}|^2\sigma_2+ |a_{3}|^2\sigma_3},
\label{eq:fractions}
\end{equation}
where $\sigma_i$ is the effective cross section
$\PH\rightarrow\cPZ\cPZ\rightarrow 2\Pe2\Pgm$ corresponding to
$a^{}_{i}=1, a_{j \ne i}=0$. The $4\Pe$ and $4\Pgm$ final states may
lead to either constructive or destructive interference of identical
leptons, and therefore to slightly different cross-section ratios.
When testing the CP-odd contribution, the second term in the amplitude
is assumed to be zero ($a_2=0$).  The measured value of $f_{a3}$ can
be used to extract the coupling constants in any parameterization.
For example, following Eq.~(\ref{eq:fullampl-spin0}), the couplings are
\begin{equation}
\frac{\abs{a_3}}{\abs{a_1}}=
\sqrt{\frac{f_{a3}}{(1-f_{a3})}} \times \sqrt{\frac{\sigma_1}{\sigma_3}},
\label{eq:frac2}
\end{equation}
where ${\sigma_1}/{\sigma_3}=6.36$ for a boson with mass 125.6\GeV.
The $f_{a3}$ parameter does not define the mixture of parity-even and
parity-odd states, because it would also depend on the relative
strength of their couplings to vector bosons.

Figure~\ref{fig:sep_fa3}~(\cmsRight) shows a likelihood scan of
$-2\ln\mathcal{L}$, where the likelihood for the event $i$,
$\mathcal{L}^i\equiv\mathcal{L}_{f_{a3}}^i \propto
(1-f_{a3})\mathcal{L}_{2D}^{i,0^+}+f_{a3}\mathcal{L}_{2D}^{i,0^-}$.
The normalization due to the acceptance is accounted for in
$\mathcal{L}_{2D}^{J^P}$, defined in Eq.~(\ref{eqn:likjcp}), and the
normalization of the likelihood $\mathcal{L}_{f_{a3}}^i$ depends on
$f_{a3}$. From the likelihood scan as a function of $f_{a3}$, the
fraction of a CP-odd amplitude contribution to the cross section
$f_{a3}=\valFaThree$, and a limit $f_{a3}<\ulFaThree$ at the 95\% \CL,
are inferred.  The limit on $f_{a3}$ can be converted into a limit on
amplitude constants using the convention of
Eq.~(\ref{eq:fullampl-spin0}): $\abs{a_3/a_1} < 2.4$ at the
95\% \CL~The statistical coverage of the results obtained in the
likelihood scan has also been tested with the Feldman-Cousins
approach~\cite{Feldman:1997qc} yielding a consistent result.


\section{Summary}
\label{sec:conclusions}

The observation and the measurements of the properties of a Higgs
boson candidate in the four-lepton decay channel have been presented.
The four-lepton invariant mass distributions are presented for
$m_{4\ell} > 70\GeV$ using data samples corresponding to integrated
luminosities of \usedLumiA at $\sqrt{s} = 7$\TeV and \usedLumiB at
$\sqrt{s}= 8$\TeV. For the measurements, the following experimental
observables are employed: the measured four-lepton mass, the mass
uncertainty, kinematic discriminants, and information sensitive to the
production mechanism, such as associated dijet characteristics and
transverse momentum of the four-lepton system.

The observation of the new
boson~\cite{Chatrchyan:2012ufa,Chatrchyan:2013lba,Chatrchyan:2012br}
is confirmed in the $4\ell$ final state, with a local significance of
$\obsSign$ standard deviations above the expected background. Upper
limits at the 95\% \CL exclude the SM-like Higgs boson in the mass
ranges 114.5--119.0\GeV and 129.5--832.0\GeV, for an expected
exclusion range for the background-only hypothesis of 115--740\GeV.
The measured mass of the new boson is $\valMass$\GeV.  The measured
width of this resonance is smaller than $\ulWidth$\GeV at the 95\%
\CL The production cross section of the new boson times the branching
fraction to four leptons is measured to be $\valMu$ times that
predicted by the standard model. Those associated with fermions and
vector bosons are $\muF=\valMuF$ and $\muV=\valMuV$, respectively,
consistent with the SM expectations.

The spin parity of the boson is studied, and the observation is
consistent with the pure scalar hypothesis when compared to several
other spin-parity hypotheses.  The fraction of a CP-odd contribution
to the decay amplitude, expressed through the fraction $f_{a3}$ of the
corresponding decay rate, is $f_{a3}=\valFaThree$, and thus consistent
with the expectation for the SM Higgs boson.  The hypotheses of a
pseudoscalar and all tested spin-1 boson hypotheses are excluded at
the 99\% \CL or higher. All tested spin-2 boson hypotheses are
excluded at the 95\% \CL or higher.

The production and decay properties of the observed new boson in the
four-lepton final state are consistent, within their uncertainties,
with the expectations for the SM Higgs boson.

\section*{Acknowledgements}
\hyphenation{Bundes-ministerium Forschungs-gemeinschaft Forschungs-zentren} We congratulate our colleagues in the CERN accelerator departments for the excellent performance of the LHC and thank the technical and administrative staffs at CERN and at other CMS institutes for their contributions to the success of the CMS effort. In addition, we gratefully acknowledge the computing centers and personnel of the Worldwide LHC Computing Grid for delivering so effectively the computing infrastructure essential to our analyses. Finally, we acknowledge the enduring support for the construction and operation of the LHC and the CMS detector provided by the following funding agencies: the Austrian Federal Ministry of Science and Research and the Austrian Science Fund; the Belgian Fonds de la Recherche Scientifique, and Fonds voor Wetenschappelijk Onderzoek; the Brazilian Funding Agencies (CNPq, CAPES, FAPERJ, and FAPESP); the Bulgarian Ministry of Education and Science; CERN; the Chinese Academy of Sciences, Ministry of Science and Technology, and National Natural Science Foundation of China; the Colombian Funding Agency (COLCIENCIAS); the Croatian Ministry of Science, Education and Sport, and the Croatian Science Foundation; the Research Promotion Foundation, Cyprus; the Ministry of Education and Research, Recurrent Financing Contract No. SF0690030s09 and European Regional Development Fund, Estonia; the Academy of Finland, Finnish Ministry of Education and Culture, and Helsinki Institute of Physics; the Institut National de Physique Nucl\'eaire et de Physique des Particules~/~CNRS and Commissariat \`a l'\'Energie Atomique et aux \'Energies Alternatives~/~CEA, France; the Bundesministerium f\"ur Bildung und Forschung, Deutsche Forschungsgemeinschaft, and Helmholtz-Gemeinschaft Deutscher Forschungszentren, Germany; the General Secretariat for Research and Technology, Greece; the National Scientific Research Foundation and National Innovation Office, Hungary; the Department of Atomic Energy and the Department of Science and Technology, India; the Institute for Studies in Theoretical Physics and Mathematics, Iran; the Science Foundation, Ireland; the Istituto Nazionale di Fisica Nucleare, Italy; the Korean Ministry of Education, Science and Technology and the World Class University program of NRF, Republic of Korea; the Lithuanian Academy of Sciences; the Mexican Funding Agencies (CINVESTAV, CONACYT, SEP, and UASLP-FAI); the Ministry of Business, Innovation and Employment, New Zealand; the Pakistan Atomic Energy Commission; the Ministry of Science and Higher Education and the National Science Centre, Poland; the Funda\c{c}\~ao para a Ci\^encia e a Tecnologia, Portugal; JINR, Dubna, the Ministry of Education and Science of the Russian Federation, the Federal Agency of Atomic Energy of the Russian Federation, Russian Academy of Sciences, and the Russian Foundation for Basic Research; the Ministry of Education, Science and Technological Development of Serbia; the Secretar\'{\i}a de Estado de Investigaci\'on, Desarrollo e Innovaci\'on and Programa Consolider-Ingenio 2010, Spain; the Swiss Funding Agencies (ETH Board, ETH Zurich, PSI, SNF, UniZH, Canton Zurich, and SER); the National Science Council, Taipei; the Thailand Center of Excellence in Physics, the Institute for the Promotion of Teaching Science and Technology of Thailand, Special Task Force for Activating Research and the National Science and Technology Development Agency of Thailand; the Scientific and Technical Research Council of Turkey and the Turkish Atomic Energy Authority; the Science and Technology Facilities Council, United Kingdom; the U.S. Department of Energy and the U.S. National Science Foundation.

Individuals have received support from the Marie-Curie program and the European Research Council and EPLANET (European Union); the Leventis Foundation; the A. P. Sloan Foundation; the Alexander von Humboldt Foundation; the Belgian Federal Science Policy Office; the Fonds pour la Formation \`a la Recherche dans l'Industrie et dans l'Agriculture (FRIA-Belgium); the Agentschap voor Innovatie door Wetenschap en Technologie (IWT-Belgium); the Ministry of Education, Youth and Sports (MEYS) of the Czech Republic; the Council of Science and Industrial Research, India; the Compagnia di San Paolo (Torino); the HOMING PLUS programme of Foundation for Polish Science, cofinanced by EU, Regional Development Fund; and the Thalis and Aristeia programmes cofinanced by EU-ESF and the Greek NSRF.

\bibliography{auto_generated}   

\providecommand{\href}[2]{#2}\begingroup\raggedright\begin{thebibliography}{100}%
\makeatletter
\providecommand{\hrefCMSnoop }[0]{\@secondoftwo}%
\makeatother
\providecommand{\doi}{\texttt{doi:}\begingroup \urlstyle{tt}\Url}

\bibitem{StandardModel67_1}
\hrefCMSnoop {} {S.~L. Glashow, ``{Partial-Symmetries of Weak Interactions}'',}
  \textit{ Nucl. Phys.} \textbf{ 22} (1961) 579,
\href{http://dx.doi.org/10.1016/0029-5582(61)90469-2}{\doi{10.1016/0029-5582(61)90469-2}}.

\bibitem{StandardModel67_2}
\hrefCMSnoop {} {S.~Weinberg, ``{A Model of Leptons}'',} \textit{ Phys. Rev.
  Lett.} \textbf{ 19} (1967) 1264,
\href{http://dx.doi.org/10.1103/PhysRevLett.19.1264}{\doi{10.1103/PhysRevLett.19.1264}}.

\bibitem{StandardModel67_3}
\hrefCMSnoop {} {A.~Salam, ``Weak and electromagnetic interactions'',} in
  \textit{ Elementary particle physics: relativistic groups and analyticity},
  N.~Svartholm, ed., p.~367.
\newblock Almqvist \& Wiksell, 1968.
\newblock Proceedings of the eighth Nobel symposium.

\bibitem{PhysRevLett.30.1343}
\hrefCMSnoop {} {D.~J. Gross and F.~Wilczek, ``Ultraviolet Behavior of
  Non-Abelian Gauge Theories'',} \textit{ Phys. Rev. Lett.} \textbf{ 30} (1973)
  1343,
  \href{http://dx.doi.org/10.1103/PhysRevLett.30.1343}{\doi{10.1103/PhysRevLett.30.1343}}.

\bibitem{PhysRevLett.30.1346}
\hrefCMSnoop {} {H.~D. Politzer, ``Reliable Perturbative Results for Strong
  Interactions?'',} \textit{ Phys. Rev. Lett.} \textbf{ 30} (1973) 1346,
  \href{http://dx.doi.org/10.1103/PhysRevLett.30.1346}{\doi{10.1103/PhysRevLett.30.1346}}.

\bibitem{Englert:1964et}
\hrefCMSnoop {} {F.~Englert and R.~Brout, ``{Broken Symmetry and the Mass of
  Gauge Vector Mesons}'',} \textit{ Phys. Rev. Lett.} \textbf{ 13} (1964) 321,
  \href{http://dx.doi.org/10.1103/PhysRevLett.13.321}{\doi{10.1103/PhysRevLett.13.321}}.

\bibitem{Higgs:1964ia}
\hrefCMSnoop {} {P.~W. Higgs, ``{Broken symmetries, massless particles and
  gauge fields}'',} \textit{ Phys. Lett.} \textbf{ 12} (1964) 132,
  \href{http://dx.doi.org/10.1016/0031-9163(64)91136-9}{\doi{10.1016/0031-9163(64)91136-9}}.

\bibitem{Higgs:1964pj}
\hrefCMSnoop {} {P.~W. Higgs, ``{Broken Symmetries and the Masses of Gauge
  Bosons}'',} \textit{ Phys. Rev. Lett.} \textbf{ 13} (1964) 508,
  \href{http://dx.doi.org/10.1103/PhysRevLett.13.508}{\doi{10.1103/PhysRevLett.13.508}}.

\bibitem{Guralnik:1964eu}
\hrefCMSnoop {} {G.~S. Guralnik, C.~R. Hagen, and T.~W.~B. Kibble, ``{Global
  Conservation Laws and Massless Particles}'',} \textit{ Phys. Rev. Lett.}
  \textbf{ 13} (1964) 585,
  \href{http://dx.doi.org/10.1103/PhysRevLett.13.585}{\doi{10.1103/PhysRevLett.13.585}}.

\bibitem{Higgs:1966ev}
\hrefCMSnoop {} {P.~W. Higgs, ``{Spontaneous Symmetry Breakdown without
  Massless Bosons}'',} \textit{ Phys. Rev.} \textbf{ 145} (1966) 1156,
  \href{http://dx.doi.org/10.1103/PhysRev.145.1156}{\doi{10.1103/PhysRev.145.1156}}.

\bibitem{Kibble:1967sv}
\hrefCMSnoop {} {T.~W.~B. Kibble, ``{Symmetry breaking in non-Abelian gauge
  theories}'',} \textit{ Phys. Rev.} \textbf{ 155} (1967) 1554,
  \href{http://dx.doi.org/10.1103/PhysRev.155.1554}{\doi{10.1103/PhysRev.155.1554}}.

\bibitem{PhysRev.122.345}
\hrefCMSnoop {} {Y.~Nambu and G.~Jona-Lasinio, ``Dynamical Model of Elementary
  Particles Based on an Analogy with Superconductivity. I'',} \textit{ Phys.
  Rev.} \textbf{ 122} (1961) 345,
  \href{http://dx.doi.org/10.1103/PhysRev.122.345}{\doi{10.1103/PhysRev.122.345}}.

\bibitem{GellMann:1960np}
\hrefCMSnoop {} {M.~Gell-Mann and M.~Levy, ``{The axial vector current in beta
  decay}'',} \textit{ Nuovo Cim.} \textbf{ 16} (1960) 705,
\href{http://dx.doi.org/10.1007/BF02859738}{\doi{10.1007/BF02859738}}.

\bibitem{Cornwall:1973tb}
\hrefCMSnoop {} {J.~M. Cornwall, D.~N. Levin, and G.~Tiktopoulos, ``{Uniqueness
  of spontaneously broken gauge theories}'',} \textit{ Phys. Rev. Lett.}
  \textbf{ 30} (1973) 1268,
\href{http://dx.doi.org/10.1103/PhysRevLett.30.1268}{\doi{10.1103/PhysRevLett.30.1268}}.

\bibitem{Cornwall:1974km}
\hrefCMSnoop {} {J.~M. Cornwall, D.~N. Levin, and G.~Tiktopoulos, ``{Derivation
  of Gauge Invariance from High-Energy Unitarity Bounds on the s Matrix}'',}
  \textit{ Phys. Rev. D} \textbf{ 10} (1974) 1145,
  \href{http://dx.doi.org/10.1103/PhysRevD.10.1145}{\doi{10.1103/PhysRevD.10.1145}}.
Erratum, \doi{10.1103/PhysRevD.11.972}.

\bibitem{LlewellynSmith:1973ey}
\hrefCMSnoop {} {C.~H. Llewellyn~Smith, ``{High-Energy Behavior and Gauge
  Symmetry}'',} \textit{ Phys. Lett. B} \textbf{ 46} (1973) 233,
\href{http://dx.doi.org/10.1016/0370-2693(73)90692-8}{\doi{10.1016/0370-2693(73)90692-8}}.

\bibitem{Lee:1977eg}
\hrefCMSnoop {} {B.~W. Lee, C.~Quigg, and H.~B. Thacker, ``{Weak Interactions
  at Very High-Energies: The Role of the Higgs Boson Mass}'',} \textit{ Phys.
  Rev. D} \textbf{ 16} (1977) 1519,
  \href{http://dx.doi.org/10.1103/PhysRevD.16.1519}{\doi{10.1103/PhysRevD.16.1519}}.

\bibitem{EWKlimits}
\href {http://cdsweb.cern.ch/record/1313716} {{ALEPH, CDF, D0, DELPHI, L3,
  OPAL, and SLD Collaborations, the LEP Electroweak Working Group, the Tevatron
  Electroweak Working Group, and the SLD Electroweak and Heavy Flavour Groups},
  ``Precision Electroweak Measurements and Constraints on the Standard
  Model'',} CERN PH-EP-2010-095, CERN,SLAC,Fermilab, 2010.
\newblock \href{http://www.arXiv.org/abs/1012.2367}{\texttt{ arXiv:1012.2367}},
  At this time, the most up-to-date Higgs boson mass constraints come from
  \url{lepewwg.web.cern.ch/LEPEWWG/plots/winter2012}.

\bibitem{Aad:2012tfa}
\hrefCMSnoop {} {{ ATLAS} Collaboration, ``{Observation of a new particle in
  the search for the Standard Model Higgs boson with the ATLAS detector at the
  LHC}'',} \textit{ Phys. Lett. B} \textbf{ 716} (2012) 1,
  \href{http://dx.doi.org/10.1016/j.physletb.2012.08.020}{\doi{10.1016/j.physletb.2012.08.020}},
\href{http://www.arXiv.org/abs/1207.7214}{\texttt{ arXiv:1207.7214}}.

\bibitem{Chatrchyan:2012ufa}
\hrefCMSnoop {} {{ CMS} Collaboration, ``{Observation of a new boson at a mass
  of 125 GeV with the CMS experiment at the LHC}'',} \textit{ Phys. Lett. B}
  \textbf{ 716} (2012) 30,
  \href{http://dx.doi.org/10.1016/j.physletb.2012.08.021}{\doi{10.1016/j.physletb.2012.08.021}},
\href{http://www.arXiv.org/abs/1207.7235}{\texttt{ arXiv:1207.7235}}.

\bibitem{Chatrchyan:2013lba}
\hrefCMSnoop {} {{ CMS} Collaboration, ``{Observation of a new boson with mass
  near 125 GeV in pp collisions at $\sqrt{s}$ = 7 and 8 TeV}'',} \textit{ JHEP}
  \textbf{ 06} (2013) 081,
  \href{http://dx.doi.org/10.1007/JHEP06(2013)081}{\doi{10.1007/JHEP06(2013)081}},
\href{http://www.arXiv.org/abs/1303.4571}{\texttt{ arXiv:1303.4571}}.

\bibitem{Barate:2003sz}
\hrefCMSnoop {} {{ALEPH, DELPHI, L3, OPAL Collaborations, and the LEP Working
  Group for Higgs boson searches}, ``{Search for the standard model Higgs boson
  at LEP}'',} \textit{ Phys. Lett. B} \textbf{ 565} (2003) 61,
  \href{http://dx.doi.org/10.1016/S0370-2693(03)00614-2}{\doi{10.1016/S0370-2693(03)00614-2}},
  \href{http://www.arXiv.org/abs/hep-ex/0306033}{\texttt{
  arXiv:hep-ex/0306033}}.

\bibitem{PhysRevLett.109.071804}
\hrefCMSnoop {} {{ CDF and D0} Collaboration, ``Evidence for a Particle
  Produced in Association with Weak Bosons and Decaying to a Bottom-Antibottom
  Quark Pair in Higgs Boson Searches at the Tevatron'',} \textit{ Phys. Rev.
  Lett.} \textbf{ 109} (2012) 071804,
  \href{http://dx.doi.org/10.1103/PhysRevLett.109.071804}{\doi{10.1103/PhysRevLett.109.071804}},
  \href{http://www.arXiv.org/abs/1207.6436}{\texttt{ arXiv:1207.6436}}.

\bibitem{Tevatron2013}
\hrefCMSnoop {} {{CDF and D0 Collaborations}, ``{Higgs boson studies at the
  Tevatron}'',} \textit{ Phys. Rev. D} \textbf{ 88} (2013) 052014,
  \href{http://dx.doi.org/10.1103/PhysRevD.88.052014}{\doi{10.1103/PhysRevD.88.052014}},
\href{http://www.arXiv.org/abs/1303.6346}{\texttt{ arXiv:1303.6346}}.

\bibitem{ATLAS:2012ac}
\hrefCMSnoop {} {{ ATLAS} Collaboration, ``{Search for the Standard Model Higgs
  boson in the decay channel $\PH \to$ ZZ(*) $\to 4 \ell$ with 4.8 fb$^{-1}$ of
  $\Pp\Pp$ collision data at $\sqrt{s}=7$ TeV with ATLAS}'',} \textit{ Phys.
  Lett. B} \textbf{ 710} (2012) 383,
  \href{http://dx.doi.org/10.1016/j.physletb.2012.03.005}{\doi{10.1016/j.physletb.2012.03.005}},
\href{http://www.arXiv.org/abs/1202.1415}{\texttt{ arXiv:1202.1415}}.

\bibitem{ATLAS:2012ae}
\hrefCMSnoop {} {{ ATLAS} Collaboration, ``{Combined search for the Standard
  Model Higgs boson using up to 4.9 fb$^{-1}$ of pp collision data at
  $\sqrt{s}=7$ TeV with the ATLAS detector at the LHC}'',} \textit{ Phys. Lett.
  B} \textbf{ 710} (2012) 49,
  \href{http://dx.doi.org/10.1016/j.physletb.2012.02.044}{\doi{10.1016/j.physletb.2012.02.044}},
\href{http://www.arXiv.org/abs/1202.1408}{\texttt{ arXiv:1202.1408}}.

\bibitem{atlas:20127tev}
\hrefCMSnoop {} {{ ATLAS} Collaboration, ``{Combined search for the Standard
  Model Higgs boson in pp collisions at $\sqrt{s} = 7$ TeV with the ATLAS
  detector}'',} \textit{ Phys. Rev. D} \textbf{ 86} (2012) 032003,
  \href{http://dx.doi.org/10.1103/PhysRevD.86.032003}{\doi{10.1103/PhysRevD.86.032003}},
\href{http://www.arXiv.org/abs/1207.0319}{\texttt{ arXiv:1207.0319}}.

\bibitem{Chatrchyan:2012dg}
\hrefCMSnoop {} {{ CMS} Collaboration, ``{Search for the standard model Higgs
  boson in the decay channel H to ZZ to 4 leptons in pp collisions at
  $\sqrt{s}$ = 7 TeV}'',} \textit{ Phys. Rev. Lett.} \textbf{ 108} (2012)
  111804,
  \href{http://dx.doi.org/10.1103/PhysRevLett.108.111804}{\doi{10.1103/PhysRevLett.108.111804}},
\href{http://www.arXiv.org/abs/1202.1997}{\texttt{ arXiv:1202.1997}}.

\bibitem{Chatrchyan:2012hr}
\hrefCMSnoop {} {{ CMS} Collaboration, ``{Search for the standard model Higgs
  boson in the $\PH$ to $\cPZ\cPZ$ to $\ell \ell \tau \tau$ decay channel in pp
  collisions at $\sqrt{s}=7$ TeV}'',} \textit{ JHEP} \textbf{ 03} (2012) 081,
  \href{http://dx.doi.org/10.1007/JHEP03(2012)081}{\doi{10.1007/JHEP03(2012)081}},
\href{http://www.arXiv.org/abs/1202.3617}{\texttt{ arXiv:1202.3617}}.

\bibitem{Chatrchyan:2012tx}
\hrefCMSnoop {} {{ CMS} Collaboration, ``{Combined results of searches for the
  standard model Higgs boson in pp collisions at $\sqrt{s} = 7$ TeV}'',}
  \textit{ Phys. Lett. B} \textbf{ 710} (2012) 26,
  \href{http://dx.doi.org/10.1016/j.physletb.2012.02.064}{\doi{10.1016/j.physletb.2012.02.064}},
\href{http://www.arXiv.org/abs/1202.1488}{\texttt{ arXiv:1202.1488}}.

\bibitem{Chatrchyan:2012br}
\hrefCMSnoop {} {{ CMS} Collaboration, ``{Study of the Mass and Spin-Parity of
  the Higgs Boson Candidate via Its Decays to Z Boson Pairs}'',} \textit{ Phys.
  Rev. Lett.} \textbf{ 110} (2013) 081803,
  \href{http://dx.doi.org/10.1103/PhysRevLett.110.081803}{\doi{10.1103/PhysRevLett.110.081803}},
\href{http://www.arXiv.org/abs/1212.6639}{\texttt{ arXiv:1212.6639}}.

\bibitem{Aad:2013wqa}
\hrefCMSnoop {} {{ ATLAS} Collaboration, ``{Measurements of Higgs boson
  production and couplings in diboson final states with the ATLAS detector at
  the LHC}'',} \textit{ Phys. Lett. B} \textbf{ 726} (2013) 88,
  \href{http://dx.doi.org/10.1016/j.physletb.2013.08.010}{\doi{10.1016/j.physletb.2013.08.010}},
\href{http://www.arXiv.org/abs/1307.1427}{\texttt{ arXiv:1307.1427}}.

\bibitem{Aad:2013xqa}
\hrefCMSnoop {} {{ ATLAS} Collaboration, ``{Evidence for the spin-0 nature of
  the Higgs boson using ATLAS data}'',} \textit{ Phys. Lett. B} \textbf{ 726}
  (2013) 120,
  \href{http://dx.doi.org/10.1016/j.physletb.2013.08.026}{\doi{10.1016/j.physletb.2013.08.026}},
\href{http://www.arXiv.org/abs/1307.1432}{\texttt{ arXiv:1307.1432}}.

\bibitem{cms:2008zzk}
\hrefCMSnoop {} {{ CMS} Collaboration, ``The {CMS} experiment at the {CERN}
  {LHC}'',} \textit{ JINST} \textbf{ 3} (2008) S08004,
\href{http://dx.doi.org/10.1088/1748-0221/3/08/S08004}{\doi{10.1088/1748-0221/3/08/S08004}}.

\bibitem{Chatrchyan:2009ad}
\hrefCMSnoop {} {{ CMS} Collaboration, ``{Commissioning and Performance of the
  CMS Silicon Strip Tracker with Cosmic Ray Muons}'',} \textit{ JINST} \textbf{
  5} (2010) T03008,
  \href{http://dx.doi.org/10.1088/1748-0221/5/03/T03008}{\doi{10.1088/1748-0221/5/03/T03008}},
\href{http://www.arXiv.org/abs/0911.4996}{\texttt{ arXiv:0911.4996}}.

\bibitem{Chatrchyan:2009sr}
\hrefCMSnoop {} {{ CMS} Collaboration, ``{Alignment of the CMS Silicon Tracker
  during Commissioning with Cosmic Rays}'',} \textit{ JINST} \textbf{ 5} (2010)
  T03009,
  \href{http://dx.doi.org/10.1088/1748-0221/5/03/T03009}{\doi{10.1088/1748-0221/5/03/T03009}},
\href{http://www.arXiv.org/abs/0910.2505}{\texttt{ arXiv:0910.2505}}.

\bibitem{Chatrchyan:2013dga}
\hrefCMSnoop {} {{ CMS} Collaboration, ``{Energy calibration and resolution of
  the CMS electromagnetic calorimeter in pp collisions at $\sqrt{s}$ = 7
  TeV}'',} \textit{ JINST} \textbf{ 8} (2013) P09009,
  \href{http://dx.doi.org/10.1088/1748-0221/8/09/P09009}{\doi{10.1088/1748-0221/8/09/P09009}},
\href{http://www.arXiv.org/abs/1306.2016}{\texttt{ arXiv:1306.2016}}.

\bibitem{powheg}
\hrefCMSnoop {} {S.~Frixione, P.~Nason, and C.~Oleari, ``{Matching NLO QCD
  computations with parton shower simulations: the POWHEG method}'',} \textit{
  JHEP} \textbf{ 11} (2007) 070,
  \href{http://dx.doi.org/10.1088/1126-6708/2007/11/070}{\doi{10.1088/1126-6708/2007/11/070}},
\href{http://www.arXiv.org/abs/0709.2092}{\texttt{ arXiv:0709.2092}}.

\bibitem{Bagnaschi:2011tu}
\hrefCMSnoop {} {E.~Bagnaschi, G.~Degrassi, P.~Slavich, and A.~Vicini, ``{Higgs
  production via gluon fusion in the POWHEG approach in the SM and in the
  MSSM}'',} \textit{ JHEP} \textbf{ 02} (2012) 088,
  \href{http://dx.doi.org/10.1007/JHEP02(2012)088}{\doi{10.1007/JHEP02(2012)088}},
\href{http://www.arXiv.org/abs/1111.2854}{\texttt{ arXiv:1111.2854}}.

\bibitem{Nason:2009ai}
\hrefCMSnoop {} {P.~Nason and C.~Oleari, ``{NLO Higgs boson production via
  vector-boson fusion matched with shower in POWHEG}'',} \textit{ JHEP}
  \textbf{ 02} (2010) 037,
  \href{http://dx.doi.org/10.1007/JHEP02(2010)037}{\doi{10.1007/JHEP02(2010)037}},
\href{http://www.arXiv.org/abs/0911.5299}{\texttt{ arXiv:0911.5299}}.

\bibitem{Gao:2010qx}
Y.~Gao\hrefCMSnoop {} { {et~al.}, ``{Spin determination of single-produced
  resonances at hadron colliders}'',} \textit{ Phys. Rev. D} \textbf{ 81}
  (2010) 075022,
  \href{http://dx.doi.org/10.1103/PhysRevD.81.075022}{\doi{10.1103/PhysRevD.81.075022}},
\href{http://www.arXiv.org/abs/1001.3396}{\texttt{ arXiv:1001.3396}}.

\bibitem{Bolognesi:2012}
S.~Bolognesi\hrefCMSnoop {} { {et~al.}, ``{On the spin and parity of a
  single-produced resonance at the LHC}'',} \textit{ Phys. Rev. D} \textbf{ 86}
  (2012) 095031,
  \href{http://dx.doi.org/10.1103/PhysRevD.86.095031}{\doi{10.1103/PhysRevD.86.095031}},
\href{http://www.arXiv.org/abs/1208.4018}{\texttt{ arXiv:1208.4018}}.

\bibitem{Anderson:2013fba}
I.~Anderson\hrefCMSnoop {} { {et~al.}, ``{Constraining anomalous HVV
  interactions at proton and lepton colliders}'',} (2013).
\href{http://www.arXiv.org/abs/1309.4819}{\texttt{ arXiv:1309.4819}}.

\bibitem{Sjostrand:2006za}
\hrefCMSnoop {} {T.~Sj{\"{o}}strand, S.~Mrenna, and P.~Z. Skands, ``{PYTHIA 6.4
  Physics and Manual}'',} \textit{ JHEP} \textbf{ 05} (2006) 026,
  \href{http://dx.doi.org/10.1088/1126-6708/2006/05/026}{\doi{10.1088/1126-6708/2006/05/026}},
\href{http://www.arXiv.org/abs/hep-ph/0603175}{\texttt{ arXiv:hep-ph/0603175}}.

\bibitem{Passarino:2010qk}
\hrefCMSnoop {} {G.~Passarino, C.~Sturm, and S.~Uccirati, ``{Higgs
  pseudo-observables, second Riemann sheet and all that}'',} \textit{ Nucl.
  Phys. B} \textbf{ 834} (2010) 77,
  \href{http://dx.doi.org/10.1016/j.nuclphysb.2010.03.013}{\doi{10.1016/j.nuclphysb.2010.03.013}},
\href{http://www.arXiv.org/abs/1001.3360}{\texttt{ arXiv:1001.3360}}.

\bibitem{Goria:2011wa}
\hrefCMSnoop {} {S.~Goria, G.~Passarino, and D.~Rosco, ``{The Higgs Boson
  Lineshape}'',} \textit{ Nucl. Phys. B} \textbf{ 864} (2012) 530,
  \href{http://dx.doi.org/10.1016/j.nuclphysb.2012.07.006}{\doi{10.1016/j.nuclphysb.2012.07.006}},
\href{http://www.arXiv.org/abs/1112.5517}{\texttt{ arXiv:1112.5517}}.

\bibitem{Kauer:2012hd}
\hrefCMSnoop {} {N.~Kauer and G.~Passarino, ``{Inadequacy of zero-width
  approximation for a light Higgs boson signal}'',} \textit{ JHEP} \textbf{ 08}
  (2012) 116,
  \href{http://dx.doi.org/10.1007/JHEP08(2012)116}{\doi{10.1007/JHEP08(2012)116}},
\href{http://www.arXiv.org/abs/1206.4803}{\texttt{ arXiv:1206.4803}}.

\bibitem{Heinemeyer:2013tqa}
\hrefCMSnoop {} {{LHC Higgs Cross Section Working Group}, ``{Handbook of LHC
  Higgs Cross Sections: 3. Higgs Properties}'',} CERN Report CERN-2013-004,
  2013.
\newblock
  \href{http://dx.doi.org/10.5170/CERN-2013-004}{\doi{10.5170/CERN-2013-004}},
  \href{http://www.arXiv.org/abs/1307.1347}{\texttt{ arXiv:1307.1347}}.

\bibitem{Passarino:2012ri}
\hrefCMSnoop {} {G.~Passarino, ``{Higgs Interference Effects in $\Pg \Pg \to
  \cPZ\cPZ$ and their Uncertainty}'',} \textit{ JHEP} \textbf{ 08} (2012) 146,
  \href{http://dx.doi.org/10.1007/JHEP08(2012)146}{\doi{10.1007/JHEP08(2012)146}},
\href{http://www.arXiv.org/abs/1206.3824}{\texttt{ arXiv:1206.3824}}.

\bibitem{Kauer:2012ma}
\href {http://pos.sissa.it/archive/conferences/145/027/RADCOR2011_027.pdf}
  {N.~Kauer, ``{Signal-background interference in $gg\to\PH\to{\rm VV}$}'',} in
  \textit{ 10th Int. Sym. on Radiative Corrections (RADCOR2011)}, p.~027.
\newblock 2011.
\newblock \href{http://www.arXiv.org/abs/1201.1667}{\texttt{ arXiv:1201.1667}}.
\newblock
PoS RADCOR2011.

\bibitem{deFlorian:2012mx}
\hrefCMSnoop {} {D.~de~Florian, G.~Ferrera, M.~Grazzini, and D.~Tommasini,
  ``{Higgs boson production at the LHC: transverse momentum resummation effects
  in the $\PH\to2\gamma$, $\PH\to\PW\PW\to2\ell2\nu$ and
  $\PH\to\cPZ\cPZ\to4\ell$ decay modes}'',} \textit{ JHEP} \textbf{ 06} (2012)
  132,
  \href{http://dx.doi.org/10.1007/JHEP06(2012)132}{\doi{10.1007/JHEP06(2012)132}},
\href{http://www.arXiv.org/abs/1203.6321}{\texttt{ arXiv:1203.6321}}.

\bibitem{Anastasiou:2008tj}
\hrefCMSnoop {} {C.~Anastasiou, R.~Boughezal, and F.~Petriello, ``{Mixed
  QCD-electroweak corrections to Higgs boson production in gluon fusion}'',}
  \textit{ JHEP} \textbf{ 04} (2009) 003,
  \href{http://dx.doi.org/10.1088/1126-6708/2009/04/003}{\doi{10.1088/1126-6708/2009/04/003}},
\href{http://www.arXiv.org/abs/0811.3458}{\texttt{ arXiv:0811.3458}}.

\bibitem{deFlorian:2009hc}
\hrefCMSnoop {} {D.~de~Florian and M.~Grazzini, ``{Higgs production through
  gluon fusion: updated cross sections at the Tevatron and the LHC}'',}
  \textit{ Phys. Lett. B} \textbf{ 674} (2009) 291,
  \href{http://dx.doi.org/10.1016/j.physletb.2009.03.033}{\doi{10.1016/j.physletb.2009.03.033}},
\href{http://www.arXiv.org/abs/0901.2427}{\texttt{ arXiv:0901.2427}}.

\bibitem{Baglio:2010ae}
\hrefCMSnoop {} {J.~Baglio and A.~Djouadi, ``{Higgs production at the LHC}'',}
  \textit{ JHEP} \textbf{ 03} (2011) 055,
  \href{http://dx.doi.org/10.1007/JHEP03(2011)055}{\doi{10.1007/JHEP03(2011)055}},
\href{http://www.arXiv.org/abs/1012.0530}{\texttt{ arXiv:1012.0530}}.

\bibitem{LHCHiggsCrossSectionWorkingGroup:2011ti}
\hrefCMSnoop {} {{LHC Higgs Cross Section Working Group}, ``{Handbook of LHC
  Higgs Cross Sections: 1. Inclusive Observables}'',} CERN Report
  CERN-2011-002, 2011.
\newblock
  \href{http://dx.doi.org/10.5170/CERN-2011-002}{\doi{10.5170/CERN-2011-002}},
  \href{http://www.arXiv.org/abs/1101.0593}{\texttt{ arXiv:1101.0593}}.

\bibitem{Djouadi:1991tka}
\hrefCMSnoop {} {A.~Djouadi, M.~Spira, and P.~M. Zerwas, ``{Production of Higgs
  bosons in proton colliders: QCD corrections}'',} \textit{ Phys. Lett. B}
  \textbf{ 264} (1991) 440,
\href{http://dx.doi.org/10.1016/0370-2693(91)90375-Z}{\doi{10.1016/0370-2693(91)90375-Z}}.

\bibitem{Dawson:1990zj}
\hrefCMSnoop {} {S.~Dawson, ``{Radiative corrections to Higgs boson
  production}'',} \textit{ Nucl. Phys. B} \textbf{ 359} (1991) 283,
\href{http://dx.doi.org/10.1016/0550-3213(91)90061-2}{\doi{10.1016/0550-3213(91)90061-2}}.

\bibitem{Spira:1995rr}
\hrefCMSnoop {} {M.~Spira, A.~Djouadi, D.~Graudenz, and R.~M. Zerwas, ``{Higgs
  boson production at the LHC}'',} \textit{ Nucl. Phys. B} \textbf{ 453} (1995)
  17,
  \href{http://dx.doi.org/10.1016/0550-3213(95)00379-7}{\doi{10.1016/0550-3213(95)00379-7}},
\href{http://www.arXiv.org/abs/hep-ph/9504378}{\texttt{ arXiv:hep-ph/9504378}}.

\bibitem{Harlander:2002wh}
\hrefCMSnoop {} {R.~V. Harlander and W.~B. Kilgore, ``{Next-to-next-to-leading
  order Higgs production at hadron colliders}'',} \textit{ Phys. Rev. Lett.}
  \textbf{ 88} (2002) 201801,
  \href{http://dx.doi.org/10.1103/PhysRevLett.88.201801}{\doi{10.1103/PhysRevLett.88.201801}},
\href{http://www.arXiv.org/abs/hep-ph/0201206}{\texttt{ arXiv:hep-ph/0201206}}.

\bibitem{Anastasiou:2002yz}
\hrefCMSnoop {} {C.~Anastasiou and K.~Melnikov, ``{Higgs boson production at
  hadron colliders in NNLO QCD}'',} \textit{ Nucl. Phys. B} \textbf{ 646}
  (2002) 220,
  \href{http://dx.doi.org/10.1016/S0550-3213(02)00837-4}{\doi{10.1016/S0550-3213(02)00837-4}},
\href{http://www.arXiv.org/abs/hep-ph/0207004}{\texttt{ arXiv:hep-ph/0207004}}.

\bibitem{Ravindran:2003um}
\hrefCMSnoop {} {V.~Ravindran, J.~Smith, and W.~L. van Neerven, ``{NNLO
  corrections to the total cross section for Higgs boson production in
  hadron-hadron collisions}'',} \textit{ Nucl. Phys. B} \textbf{ 665} (2003)
  325,
  \href{http://dx.doi.org/10.1016/S0550-3213(03)00457-7}{\doi{10.1016/S0550-3213(03)00457-7}},
\href{http://www.arXiv.org/abs/hep-ph/0302135}{\texttt{ arXiv:hep-ph/0302135}}.

\bibitem{Catani:2003zt}
\hrefCMSnoop {} {S.~Catani, D.~de~Florian, M.~Grazzini, and P.~Nason, ``{Soft
  gluon resummation for Higgs boson production at hadron colliders}'',}
  \textit{ JHEP} \textbf{ 07} (2003) 028,
  \href{http://dx.doi.org/10.1088/1126-6708/2003/07/028}{\doi{10.1088/1126-6708/2003/07/028}},
\href{http://www.arXiv.org/abs/hep-ph/0306211}{\texttt{ arXiv:hep-ph/0306211}}.

\bibitem{Actis:2008ug}
\hrefCMSnoop {} {S.~Actis, G.~Passarino, C.~Sturm, and S.~Uccirati, ``{NLO
  electroweak corrections to Higgs boson production at hadron colliders}'',}
  \textit{ Phys. Lett. B} \textbf{ 670} (2008) 12,
  \href{http://dx.doi.org/10.1016/j.physletb.2008.10.018}{\doi{10.1016/j.physletb.2008.10.018}},
\href{http://www.arXiv.org/abs/0809.1301}{\texttt{ arXiv:0809.1301}}.

\bibitem{Ciccolini:2007jr}
\hrefCMSnoop {} {M.~Ciccolini, A.~Denner, and S.~Dittmaier, ``{Strong and
  electroweak corrections to the production of Higgs + 2-jets via weak
  interactions at the LHC}'',} \textit{ Phys. Rev. Lett.} \textbf{ 99} (2007)
  161803,
  \href{http://dx.doi.org/10.1103/PhysRevLett.99.161803}{\doi{10.1103/PhysRevLett.99.161803}},
  \href{http://www.arXiv.org/abs/0707.0381}{\texttt{ arXiv:0707.0381}}.

\bibitem{Ciccolini:2007ec}
\hrefCMSnoop {} {M.~Ciccolini, A.~Denner, and S.~Dittmaier, ``{Electroweak and
  QCD corrections to Higgs production via vector-boson fusion at the LHC}'',}
  \textit{ Phys. Rev. D} \textbf{ 77} (2008) 013002,
  \href{http://dx.doi.org/10.1103/PhysRevD.77.013002}{\doi{10.1103/PhysRevD.77.013002}},
\href{http://www.arXiv.org/abs/0710.4749}{\texttt{ arXiv:0710.4749}}.

\bibitem{Figy:2003nv}
\hrefCMSnoop {} {T.~Figy, D.~Zeppenfeld, and C.~Oleari, ``{Next-to-leading
  order jet distributions for Higgs boson production via weak-boson fusion}'',}
  \textit{ Phys. Rev. D} \textbf{ 68} (2003) 073005,
  \href{http://dx.doi.org/10.1103/PhysRevD.68.073005}{\doi{10.1103/PhysRevD.68.073005}},
\href{http://www.arXiv.org/abs/hep-ph/0306109}{\texttt{ arXiv:hep-ph/0306109}}.

\bibitem{Arnold:2008rz}
\hrefCMSnoop {} {K.~Arnold {et~al.}, ``{VBFNLO: A parton level Monte Carlo for
  processes with electroweak bosons}'',} \textit{ Comput. Phys. Commun.}
  \textbf{ 180} (2009) 1661,
  \href{http://dx.doi.org/10.1016/j.cpc.2009.03.006}{\doi{10.1016/j.cpc.2009.03.006}},
\href{http://www.arXiv.org/abs/0811.4559}{\texttt{ arXiv:0811.4559}}.

\bibitem{Bolzoni:2010xr}
\hrefCMSnoop {} {P.~Bolzoni, F.~Maltoni, S.-O. Moch, and M.~Zaro, ``{Higgs
  Boson Production via Vector-Boson Fusion at NNLO at Next-to-Next-to-Leading
  Order in QCD}'',} \textit{ Phys. Rev. Lett.} \textbf{ 105} (2010) 011801,
  \href{http://dx.doi.org/10.1103/PhysRevLett.105.011801}{\doi{10.1103/PhysRevLett.105.011801}},
\href{http://www.arXiv.org/abs/1003.4451}{\texttt{ arXiv:1003.4451}}.

\bibitem{Melia:2011tj}
\hrefCMSnoop {} {T.~Melia, P.~Nason, R.~Rontsch, and G.~Zanderighi,
  ``{$\PW^+\PW^-$, $\PW\cPZ$ and $\cPZ\cPZ$ production in the POWHEG BOX}'',}
  \textit{ JHEP} \textbf{ 11} (2011) 078,
  \href{http://dx.doi.org/10.1007/JHEP11(2011)078}{\doi{10.1007/JHEP11(2011)078}},
\href{http://www.arXiv.org/abs/1107.5051}{\texttt{ arXiv:1107.5051}}.

\bibitem{Alwall:2007st}
J.~Alwall\hrefCMSnoop {} { {et~al.}, ``{MadGraph/MadEvent v4: the new web
  generation}'',} \textit{ JHEP} \textbf{ 09} (2007) 028,
  \href{http://dx.doi.org/10.1088/1126-6708/2007/09/028}{\doi{10.1088/1126-6708/2007/09/028}},
\href{http://www.arXiv.org/abs/0706.2334}{\texttt{ arXiv:0706.2334}}.

\bibitem{Binoth:2008pr}
\hrefCMSnoop {} {T.~Binoth, N.~Kauer, and P.~Mertsch, ``{Gluon-induced QCD
  corrections to pp $\to \cPZ\cPZ \to \ell\bar{\ell}\ell'\bar{\ell'}$}'',}
  (2008).
\href{http://www.arXiv.org/abs/0807.0024}{\texttt{ arXiv:0807.0024}}.

\bibitem{cteq66}
H.-L. Lai\hrefCMSnoop {} { {et~al.}, ``{Uncertainty induced by QCD coupling in
  the CTEQ global analysis of parton distributions}'',} \textit{ Phys. Rev. D}
  \textbf{ 82} (2010) 054021,
  \href{http://dx.doi.org/10.1103/PhysRevD.82.054021}{\doi{10.1103/PhysRevD.82.054021}},
\href{http://www.arXiv.org/abs/1004.4624}{\texttt{ arXiv:1004.4624}}.

\bibitem{Lai:2010vv}
H.-L. Lai\hrefCMSnoop {} { {et~al.}, ``{New parton distributions for collider
  physics}'',} \textit{ Phys. Rev. D} \textbf{ 82} (2010) 074024,
  \href{http://dx.doi.org/10.1103/PhysRevD.82.074024}{\doi{10.1103/PhysRevD.82.074024}},
\href{http://www.arXiv.org/abs/1007.2241}{\texttt{ arXiv:1007.2241}}.

\bibitem{Chatrchyan:2011id}
\hrefCMSnoop {} {{ CMS} Collaboration, ``{Measurement of the Underlying Event
  Activity at the LHC with $\sqrt{s}= 7$ TeV and Comparison with $\sqrt{s} =
  0.9$ TeV}'',} \textit{ JHEP} \textbf{ 09} (2011) 109,
  \href{http://dx.doi.org/10.1007/JHEP09(2011)109}{\doi{10.1007/JHEP09(2011)109}},
\href{http://www.arXiv.org/abs/1107.0330}{\texttt{ arXiv:1107.0330}}.

\bibitem{Agostinelli:2002hh}
\hrefCMSnoop {} {{ GEANT4} Collaboration, ``{GEANT4---a simulation toolkit}'',}
  \textit{ Nucl. Instrum. Meth. A} \textbf{ 506} (2003) 250,
\href{http://dx.doi.org/10.1016/S0168-9002(03)01368-8}{\doi{10.1016/S0168-9002(03)01368-8}}.

\bibitem{GEANT}
\hrefCMSnoop {} {J.~Allison {et~al.}, ``{Geant4 developments and
  applications}'',} \textit{ IEEE Trans. Nucl. Sci.} \textbf{ 53} (2006) 270,
\href{http://dx.doi.org/10.1109/TNS.2006.869826}{\doi{10.1109/TNS.2006.869826}}.

\bibitem{Dittmaier:2012vm}
\hrefCMSnoop {} {{LHC Higgs Cross Section Working Group}, ``{Handbook of LHC
  Higgs Cross Sections: 2. Differential Distributions}'',} CERN Report
  CERN-2012-002, 2012.
\newblock
  \href{http://dx.doi.org/10.5170/CERN-2012-002}{\doi{10.5170/CERN-2012-002}},
  \href{http://www.arXiv.org/abs/1201.3084}{\texttt{ arXiv:1201.3084}}.

\bibitem{CMS-PAS-PFT-09-001}
\href {http://cdsweb.cern.ch/record/1194487} {{ CMS} Collaboration,
  ``Particle-Flow Event Reconstruction in CMS and Performance for Jets, Taus,
  and MET'',} CMS Physics Analysis Summary CMS-PAS-PFT-09-001, 2009.

\bibitem{CMS-PAS-PFT-10-001}
\href {http://cdsweb.cern.ch/record/1247373} {{CMS Collaboration},
  ``Commissioning of the Particle-flow Event Reconstruction with the first
  {LHC} collisions recorded in the {CMS} detector'',} CMS Physics Analysis
  Summary CMS-PAS-PFT-10-001, 2010.

\bibitem{CMS-PAS-PFT-10-002}
\href {http://cms-physics.web.cern.ch/cms-physics/public/PFT-10-002-pas.pdf} {{
  CMS} Collaboration, ``Commissioning of the Particle--Flow Reconstruction in
  Minimum--Bias and Jet Events from pp Collisions at $7$~TeV'',} CMS Physics
  Analysis Summary CMS-PAS-PFT-10-002, 2010.

\bibitem{CMS-PAS-PFT-10-003}
\href {http://cms-physics.web.cern.ch/cms-physics/public/PFT-10-003-pas.pdf} {{
  CMS} Collaboration, ``Commissioning of the particle-flow event reconstruction
  with leptons from J/$\psi$ and W decays at $7$~TeV'',} CMS Physics Analysis
  Summary CMS-PAS-PFT-10-003, 2010.

\bibitem{Baffioni:2006cd}
S.~Baffioni\hrefCMSnoop {} { {et~al.}, ``{Electron reconstruction in CMS}'',}
  \textit{ Eur. Phys. J. C} \textbf{ 49} (2007) 1099,
\href{http://dx.doi.org/10.1140/epjc/s10052-006-0175-5}{\doi{10.1140/epjc/s10052-006-0175-5}}.

\bibitem{CMS-PAS-EGM-10-004}
\href {http://cdsweb.cern.ch/record/1299116} {{ CMS} Collaboration, ``Electron
  reconstruction and identification at $\sqrt{s} = 7$ TeV'',} CMS Physics
  Analysis Summary CMS-PAS-EGM-10-004, 2010.

\bibitem{CMS_DPS_2011-003}
\href {http://cdsweb.cern.ch/record/1360227} {{ CMS} Collaboration, ``Electron
  commissioning results at $\sqrt{s}=7$ TeV'',} CMS Detector Performance
  Summary CMS-DP-2011-003, 2011.

\bibitem{CMS_DP_2013-003}
\href {http://cdsweb.cern.ch/record/1523273} {{ CMS} Collaboration, ``Electron
  performance with 19.6~fb$^{-1}$ of data collected at $\sqrt{s}=8$ TeV with
  the CMS detector'',} CMS Detector Performance Summary CMS-DP-2013-003, 2013.

\bibitem{Adam:2003kg}
\hrefCMSnoop {} {W.~Adam, R.~Fr{\"u}hwirth, A.~Strandlie, and T.~Todorov,
  ``{Reconstruction of electrons with the Gaussian sum filter in the CMS
  tracker at LHC}'',} \textit{ J. Phys. G} \textbf{ 31} (2005) N9,
  \href{http://dx.doi.org/10.1088/0954-3899/31/9/N01}{\doi{10.1088/0954-3899/31/9/N01}},
\href{http://www.arXiv.org/abs/physics/0306087}{\texttt{
  arXiv:physics/0306087}}.

\bibitem{CMS_DPS_2012-015}
\href {http://cdsweb.cern.ch/record/1478128} {{ CMS} Collaboration, ``ECAL
  Performance Plots, 2012 Data'',} CMS Detector Performance Summary
  CMS-DP-2012-015, 2012.

\bibitem{tmva}
\href {http://pos.sissa.it/archive/conferences/050/040/ACAT_040.pdf} {H.~Voss,
  A.~H{\"o}cker, J.~Stelzer, and F.~Tegenfeldt, ``{TMVA: Toolkit for
  Multivariate Data Analysis with ROOT}'',} in \textit{ XIth International
  Workshop on Advanced Computing and Analysis Techniques in Physics Research
  (ACAT)}, p.~040.
\newblock 2007.
\newblock
\href{http://www.arXiv.org/abs/physics/0703039}{\texttt{
  arXiv:physics/0703039}}.
\newblock

\bibitem{Chatrchyan:2009qm}
\hrefCMSnoop {} {{ CMS} Collaboration, ``{Performance and Operation of the CMS
  Electromagnetic Calorimeter}'',} \textit{ JINST} \textbf{ 5} (2010) T03010,
  \href{http://dx.doi.org/10.1088/1748-0221/5/03/T03010}{\doi{10.1088/1748-0221/5/03/T03010}},
\href{http://www.arXiv.org/abs/0910.3423}{\texttt{ arXiv:0910.3423}}.

\bibitem{CrystalBall}
\href {http://www.slac.stanford.edu/pubs/slacreports/slac-r-236.html}
  {M.~Oreglia, ``{A study of the reactions $\psi^\prime \to \gamma \gamma
  \psi$}''}.
\newblock PhD thesis, Stanford University, 1980.
\newblock {{SLAC} Report {SLAC-R-236}}.

\bibitem{Chatrchyan:2012xi}
\hrefCMSnoop {} {{ CMS} Collaboration, ``{Performance of CMS muon
  reconstruction in pp collision events at $\sqrt{s} = 7$ TeV}'',} \textit{
  JINST} \textbf{ 7} (2012) P10002,
  \href{http://dx.doi.org/10.1088/1748-0221/7/10/P10002}{\doi{10.1088/1748-0221/7/10/P10002}},
\href{http://www.arXiv.org/abs/1206.4071}{\texttt{ arXiv:1206.4071}}.

\bibitem{Chatrchyan:2014wfa}
\hrefCMSnoop {} {{ CMS Collaboration} Collaboration, ``{Alignment of the CMS
  tracker with LHC and cosmic ray data}'',} (2014).
  \href{http://www.arXiv.org/abs/1403.2286}{\texttt{ arXiv:1403.2286}}.
Submitted to JINST.

\bibitem{Cacciari:2007fd}
\hrefCMSnoop {} {M.~Cacciari and G.~P. Salam, ``{Pileup subtraction using jet
  areas}'',} \textit{ Phys. Lett. B} \textbf{ 659} (2008) 119,
  \href{http://dx.doi.org/10.1016/j.physletb.2007.09.077}{\doi{10.1016/j.physletb.2007.09.077}},
\href{http://www.arXiv.org/abs/0707.1378}{\texttt{ arXiv:0707.1378}}.

\bibitem{Cacciari:2008gn}
\hrefCMSnoop {} {M.~Cacciari, G.~P. Salam, and G.~Soyez, ``{The Catchment Area
  of Jets}'',} \textit{ JHEP} \textbf{ 04} (2008) 005,
  \href{http://dx.doi.org/10.1088/1126-6708/2008/04/005}{\doi{10.1088/1126-6708/2008/04/005}},
\href{http://www.arXiv.org/abs/0802.1188}{\texttt{ arXiv:0802.1188}}.

\bibitem{Cacciari:2011ma}
\hrefCMSnoop {} {M.~Cacciari, G.~P. Salam, and G.~Soyez, ``{FastJet user manual
  (for version 3.0.2)}'',} \textit{ Eur. Phys. J. C} \textbf{ 72} (2012) 1896,
  \href{http://dx.doi.org/10.1140/epjc/s10052-012-1896-2}{\doi{10.1140/epjc/s10052-012-1896-2}},
\href{http://www.arXiv.org/abs/1111.6097}{\texttt{ arXiv:1111.6097}}.

\bibitem{Catani:kt}
\hrefCMSnoop {} {S.~Catani, Y.~L. Dokshitzer, M.~Seymour, and B.~Webber,
  ``{Longitudinally invariant $k_T$ clustering algorithms for hadron hadron
  collisions}'',} \textit{ Nucl. Phys. B} \textbf{ 406} (1993) 187,
\href{http://dx.doi.org/10.1016/0550-3213(93)90166-M}{\doi{10.1016/0550-3213(93)90166-M}}.

\bibitem{Ellis:kt}
\hrefCMSnoop {} {S.~D. Ellis and D.~E. Soper, ``{Successive combination jet
  algorithm for hadron collisions}'',} \textit{ Phys. Rev. D} \textbf{ 48}
  (1993) 3160,
  \href{http://dx.doi.org/10.1103/PhysRevD.48.3160}{\doi{10.1103/PhysRevD.48.3160}},
\href{http://www.arXiv.org/abs/hep-ph/9305266}{\texttt{ arXiv:hep-ph/9305266}}.

\bibitem{Baffioni:2006sk}
S.~Baffioni\hrefCMSnoop {} { {et~al.}, ``{Discovery potential for the SM Higgs
  boson in the $\PH \to \cPZ\cPZ^{(*)}\to \Pep\Pem\Pep\Pem$ decay channel}'',}
  \textit{ J. Phys. G} \textbf{ 34} (2007) N23,
\href{http://dx.doi.org/10.1088/0954-3899/34/2/N01}{\doi{10.1088/0954-3899/34/2/N01}}.

\bibitem{CMS:2011aa}
\hrefCMSnoop {} {{ CMS} Collaboration, ``{Measurement of the Inclusive $W$ and
  $Z$ Production Cross Sections in pp Collisions at $\sqrt{s}=7$ TeV}'',}
  \textit{ JHEP} \textbf{ 10} (2011) 132,
  \href{http://dx.doi.org/10.1007/JHEP10(2011)132}{\doi{10.1007/JHEP10(2011)132}},
\href{http://www.arXiv.org/abs/1107.4789}{\texttt{ arXiv:1107.4789}}.

\bibitem{antikt}
\hrefCMSnoop {} {{M. Cacciari and G. P. Salam and G. Soyez}, ``{The anti-$k_T$
  jet clustering algorithm}'',} \textit{ JHEP} \textbf{ 04} (2008) 063,
  \href{http://dx.doi.org/10.1088/1126-6708/2008/04/063}{\doi{10.1088/1126-6708/2008/04/063}},
  \href{http://www.arXiv.org/abs/0802.1189}{\texttt{ arXiv:0802.1189}}.

\bibitem{Cacciari:fastjet2}
\hrefCMSnoop {} {{M. Cacciari, G. P. Salam}, ``{Dispelling the $N^{3}$ myth for
  the ${k_T}$ jet-finder}'',} \textit{ Phys. Lett. B} \textbf{ 641} (2006) 57,
  \href{http://dx.doi.org/10.1016/j.physletb.2006.08.037}{\doi{10.1016/j.physletb.2006.08.037}},
  \href{http://www.arXiv.org/abs/hep-ph/0512210}{\texttt{
  arXiv:hep-ph/0512210}}.

\bibitem{cmsJEC}
\hrefCMSnoop {} {{ CMS} Collaboration, ``{Determination of Jet Energy
  Calibration and Transverse Momentum Resolution in CMS}'',} \textit{ JINST}
  \textbf{ 6} (2011) P11002,
  \href{http://dx.doi.org/10.1088/1748-0221/6/11/P11002}{\doi{10.1088/1748-0221/6/11/P11002}},
\href{http://www.arXiv.org/abs/1107.4277}{\texttt{ arXiv:1107.4277}}.

\bibitem{jetIdPAS}
\href {http://cdsweb.cern.ch/record/1581583} {{ CMS} Collaboration, ``Pileup
  Jet Identification'',} CMS Physics Analysis Summary CMS-PAS-JME-13-005, 2013.

\bibitem{MCFM}
\hrefCMSnoop {} {J.~M. Campbell and R.~K. Ellis, ``{MCFM for the Tevatron and
  the LHC}'',} \textit{ Nucl. Phys. Proc. Suppl.} \textbf{ 205} (2010) 10,
  \href{http://dx.doi.org/10.1016/j.nuclphysbps.2010.08.011}{\doi{10.1016/j.nuclphysbps.2010.08.011}},
\href{http://www.arXiv.org/abs/1007.3492}{\texttt{ arXiv:1007.3492}}.

\bibitem{Campbell:1999ah}
\hrefCMSnoop {} {J.~M. Campbell and R.~K. Ellis, ``{An update on vector boson
  pair production at hadron colliders}'',} \textit{ Phys. Rev. D} \textbf{ 60}
  (1999) 113006,
  \href{http://dx.doi.org/10.1103/PhysRevD.60.113006}{\doi{10.1103/PhysRevD.60.113006}},
  \href{http://www.arXiv.org/abs/hep-ph/9905386}{\texttt{
  arXiv:hep-ph/9905386}}.

\bibitem{Campbell:2011bn}
\hrefCMSnoop {} {J.~M. Campbell, R.~K. Ellis, and C.~Williams, ``{Vector boson
  pair production at the LHC}'',} \textit{ JHEP} \textbf{ 07} (2011) 018,
  \href{http://dx.doi.org/10.1007/JHEP07(2011)018}{\doi{10.1007/JHEP07(2011)018}},
\href{http://www.arXiv.org/abs/1105.0020}{\texttt{ arXiv:1105.0020}}.

\bibitem{Aad:2013bjm}
\hrefCMSnoop {} {{ ATLAS} Collaboration, ``{Measurement of hard double-parton
  interactions in $\PW(\to\ell\nu)$ + 2 jet events at $\sqrt{s}$ = 7 TeV with
  the ATLAS detector}'',} \textit{ New J. Phys.} \textbf{ 15} (2013) 033038,
  \href{http://dx.doi.org/10.1088/1367-2630/15/3/033038}{\doi{10.1088/1367-2630/15/3/033038}},
\href{http://www.arXiv.org/abs/1301.6872}{\texttt{ arXiv:1301.6872}}.

\bibitem{Landau:1944if}
\hrefCMSnoop {} {L.~Landau, ``{On the energy loss of fast particles by
  ionization}'',} \textit{ J. Phys. (USSR)} \textbf{ 8} (1944)
201.

\bibitem{Choi:2002jk}
\hrefCMSnoop {} {S.~Y. Choi, D.~J. Miller, M.~M. {M\"uhlleitner}, and P.~M.
  Zerwas, ``{Identifying the Higgs spin and parity in decays to Z pairs}'',}
  \textit{ Phys. Lett. B} \textbf{ 553} (2003) 61,
  \href{http://dx.doi.org/10.1016/S0370-2693(02)03191-X}{\doi{10.1016/S0370-2693(02)03191-X}},
\href{http://www.arXiv.org/abs/hep-ph/0210077}{\texttt{ arXiv:hep-ph/0210077}}.

\bibitem{Soni:1993jc}
\hrefCMSnoop {} {A.~Soni and R.~M. Xu, ``{Probing CP violation via Higgs decays
  to four leptons}'',} \textit{ Phys. Rev. D} \textbf{ 48} (1993) 5259,
  \href{http://dx.doi.org/10.1103/PhysRevD.48.5259}{\doi{10.1103/PhysRevD.48.5259}},
\href{http://www.arXiv.org/abs/hep-ph/9301225}{\texttt{ arXiv:hep-ph/9301225}}.

\bibitem{Barger:1993wt}
V.~D. Barger\hrefCMSnoop {} { {et~al.}, ``{Higgs bosons: Intermediate mass
  range at $\Pep\Pem$ colliders}'',} \textit{ Phys. Rev. D} \textbf{ 49} (1994)
  79,
  \href{http://dx.doi.org/10.1103/PhysRevD.49.79}{\doi{10.1103/PhysRevD.49.79}},
\href{http://www.arXiv.org/abs/hep-ph/9306270}{\texttt{ arXiv:hep-ph/9306270}}.

\bibitem{Allanach:2002gn}
B.~C. Allanach\hrefCMSnoop {} { {et~al.}, ``{Exploring small extra dimensions
  at the Large Hadron Collider}'',} \textit{ JHEP} \textbf{ 12} (2002) 039,
  \href{http://dx.doi.org/10.1088/1126-6708/2002/12/039}{\doi{10.1088/1126-6708/2002/12/039}},
\href{http://www.arXiv.org/abs/hep-ph/0211205}{\texttt{ arXiv:hep-ph/0211205}}.

\bibitem{Buszello:2002uu}
\hrefCMSnoop {} {C.~P. Buszello, I.~Fleck, P.~Marquard, and J.~J. van~der Bij,
  ``{Prospective analysis of spin- and CP-sensitive variables in $H \to Z Z \to
  \ell^+_1 \ell^-_1 \ell^+_2 \ell^-_2$ at the LHC}'',} \textit{ Eur. Phys. J.
  C} \textbf{ 32} (2004) 209,
  \href{http://dx.doi.org/10.1140/epjc/s2003-01392-0}{\doi{10.1140/epjc/s2003-01392-0}},
\href{http://www.arXiv.org/abs/hep-ph/0212396}{\texttt{ arXiv:hep-ph/0212396}}.

\bibitem{Godbole:2007cn}
\hrefCMSnoop {} {R.~M. Godbole, D.~J. Miller, and M.~M. {M\"uhlleitner},
  ``{Aspects of CP violation in the $\PH\cPZ\cPZ$ coupling at the LHC}'',}
  \textit{ JHEP} \textbf{ 12} (2007) 031,
  \href{http://dx.doi.org/10.1088/1126-6708/2007/12/031}{\doi{10.1088/1126-6708/2007/12/031}},
\href{http://www.arXiv.org/abs/0708.0458}{\texttt{ arXiv:0708.0458}}.

\bibitem{Keung:2008ve}
\hrefCMSnoop {} {W.-Y. Keung, I.~Low, and J.~Shu, ``{Landau-Yang Theorem and
  Decays of a Z' Boson into Two Z Bosons}'',} \textit{ Phys. Rev. Lett.}
  \textbf{ 101} (2008) 091802,
  \href{http://dx.doi.org/10.1103/PhysRevLett.101.091802}{\doi{10.1103/PhysRevLett.101.091802}},
\href{http://www.arXiv.org/abs/0806.2864}{\texttt{ arXiv:0806.2864}}.

\bibitem{Antipin:2008hj}
\hrefCMSnoop {} {O.~Antipin and A.~Soni, ``{Towards establishing the spin of
  warped gravitons}'',} \textit{ JHEP} \textbf{ 10} (2008) 018,
  \href{http://dx.doi.org/10.1088/1126-6708/2008/10/018}{\doi{10.1088/1126-6708/2008/10/018}},
\href{http://www.arXiv.org/abs/0806.3427}{\texttt{ arXiv:0806.3427}}.

\bibitem{Hagiwara:2009wt}
\hrefCMSnoop {} {K.~Hagiwara, Q.~Li, and K.~Mawatari, ``{Jet angular
  correlation in vector-boson fusion processes at hadron colliders}'',}
  \textit{ JHEP} \textbf{ 07} (2009) 101,
  \href{http://dx.doi.org/10.1088/1126-6708/2009/07/101}{\doi{10.1088/1126-6708/2009/07/101}},
\href{http://www.arXiv.org/abs/0905.4314}{\texttt{ arXiv:0905.4314}}.

\bibitem{DeRujula:2010ys}
A.~De~Rujula\hrefCMSnoop {} { {et~al.}, ``{Higgs look-alikes at the LHC}'',}
  \textit{ Phys. Rev. D} \textbf{ 82} (2010) 013003,
  \href{http://dx.doi.org/10.1103/PhysRevD.82.013003}{\doi{10.1103/PhysRevD.82.013003}},
\href{http://www.arXiv.org/abs/1001.5300}{\texttt{ arXiv:1001.5300}}.

\bibitem{Gainer:2011xz}
\hrefCMSnoop {} {J.~S. Gainer, K.~Kumar, I.~Low, and R.~Vega-Morales,
  ``{Improving the sensitivity of Higgs boson searches in the golden
  channel}'',} \textit{ JHEP} \textbf{ 11} (2011) 027,
  \href{http://dx.doi.org/10.1007/JHEP11(2011)027}{\doi{10.1007/JHEP11(2011)027}},
\href{http://www.arXiv.org/abs/1108.2274}{\texttt{ arXiv:1108.2274}}.

\bibitem{Chen:2012jy}
\hrefCMSnoop {} {Y.~Chen, N.~Tran, and R.~Vega-Morales, ``{Scrutinizing the
  Higgs Signal and Background in the $2\Pe2\Pgm$ Golden Channel}'',} \textit{
  JHEP} \textbf{ 01} (2013) 182,
  \href{http://dx.doi.org/10.1007/JHEP01(2013)182}{\doi{10.1007/JHEP01(2013)182}},
\href{http://www.arXiv.org/abs/1211.1959}{\texttt{ arXiv:1211.1959}}.

\bibitem{Avery:2012um}
P.~Avery\hrefCMSnoop {} { {et~al.}, ``{Precision studies of the Higgs boson
  decay channel $\PH \to \cPZ\cPZ^{*} \to 4\ell$ with MEKD}'',} \textit{ Phys.
  Rev. D} \textbf{ 87} (2013) 055006,
  \href{http://dx.doi.org/10.1103/PhysRevD.87.055006}{\doi{10.1103/PhysRevD.87.055006}},
\href{http://www.arXiv.org/abs/1210.0896}{\texttt{ arXiv:1210.0896}}.

\bibitem{Artoisenet:2013puc}
P.~Artoisenet\hrefCMSnoop {} { {et~al.}, ``{A framework for Higgs
  characterisation}'',} \textit{ J. High Energy Phys.} \textbf{ 11} (2013)
043.

\bibitem{PhysRev.168.1926}
\hrefCMSnoop {} {N.~Cabibbo and A.~Maksymowicz, ``Angular Correlations in
  ${K}_{e4}$ Decays and Determination of Low-Energy $\pi-\pi$ Phase Shifts'',}
  \textit{ Phys. Rev.} \textbf{ 168} (1968) 1926,
  \href{http://dx.doi.org/10.1103/PhysRev.168.1926}{\doi{10.1103/PhysRev.168.1926}}.

\bibitem{Landau:1948xy}
\hrefCMSnoop {} {L.~Landau, ``On the angular momentum of a two-photon
  system'',} \textit{ Dokl. Akad. Nawk. Ser. Fiz. (USSR)} \textbf{ 60} (1948)
207.

\bibitem{Yang:1950rg}
\hrefCMSnoop {} {C.-N. Yang, ``{Selection Rules for the Dematerialization of a
  Particle into Two Photons}'',} \textit{ Phys. Rev.} \textbf{ 77} (1950) 242,
\href{http://dx.doi.org/10.1103/PhysRev.77.242}{\doi{10.1103/PhysRev.77.242}}.

\bibitem{Randall:1999vf}
\hrefCMSnoop {} {L.~Randall and R.~Sundrum, ``{An Alternative to
  Compactification}'',} \textit{ Phys. Rev. Lett.} \textbf{ 83} (1999) 4690,
  \href{http://dx.doi.org/10.1103/PhysRevLett.83.4690}{\doi{10.1103/PhysRevLett.83.4690}},
\href{http://www.arXiv.org/abs/hep-th/9906064}{\texttt{ arXiv:hep-th/9906064}}.

\bibitem{Randall:1999ee}
\hrefCMSnoop {} {L.~Randall and R.~Sundrum, ``{Large Mass Hierarchy from a
  Small Extra Dimension}'',} \textit{ Phys. Rev. Lett.} \textbf{ 83} (1999)
  3370,
  \href{http://dx.doi.org/10.1103/PhysRevLett.83.3370}{\doi{10.1103/PhysRevLett.83.3370}},
\href{http://www.arXiv.org/abs/hep-ph/9905221}{\texttt{ arXiv:hep-ph/9905221}}.

\bibitem{Agashe:2007zd}
\hrefCMSnoop {} {K.~Agashe, H.~Davoudiasl, G.~Perez, and A.~Soni, ``{Warped
  Gravitons at the LHC and Beyond}'',} \textit{ Phys. Rev. D} \textbf{ 76}
  (2007) 036006,
  \href{http://dx.doi.org/10.1103/PhysRevD.76.036006}{\doi{10.1103/PhysRevD.76.036006}},
\href{http://www.arXiv.org/abs/hep-ph/0701186}{\texttt{ arXiv:hep-ph/0701186}}.

\bibitem{Christensen:2008py}
\hrefCMSnoop {} {N.~D. Christensen and C.~Duhr, ``{FeynRules - Feynman rules
  made easy}'',} \textit{ Comput. Phys. Commun.} \textbf{ 180} (2009) 1614,
  \href{http://dx.doi.org/10.1016/j.cpc.2009.02.018}{\doi{10.1016/j.cpc.2009.02.018}},
\href{http://www.arXiv.org/abs/0806.4194}{\texttt{ arXiv:0806.4194}}.

\bibitem{BNN1}
R.~M. Neal, ``Bayesian learning for neural networks''.
\newblock Springer-Verlag, 1996.
\newblock Lecture Notes in Statistics, Vol. 118.

\bibitem{BNN2}
\href {http://www.sciencedirect.com/science/article/pii/S0893608000000988}
  {J.~Lampinen and A.~Vehtari, ``Bayesian approach for neural networks - review
  and case studies'',} \textit{ Neural Networks} \textbf{ 14} (2001) 257.

\bibitem{bdt1}
L.~Breiman, J.~H. Friedman, R.~A. Olshen, and C.~J. Stone, ``{Classification
  and Regression Trees}''.
\newblock Wadsworth, Monterey, CA, 1984.

\bibitem{bdt2}
\hrefCMSnoop {} {J.~H. Friedman, ``{Stochastic gradient boosting}'',} \textit{
  Comput. Stat. Data Anal.} \textbf{ 38} (2002) 367,
\href{http://dx.doi.org/10.1016/S0167-9473(01)00065-2}{\doi{10.1016/S0167-9473(01)00065-2}}.

\bibitem{Cousins:1994yw}
\hrefCMSnoop {} {R.~D. Cousins, ``{Why isn't every physicist a Bayesian?}'',}
  \textit{ Am. J. Phys.} \textbf{ 63} (1995) 398,
\href{http://dx.doi.org/10.1119/1.17901}{\doi{10.1119/1.17901}}.

\bibitem{Pivk:2004ty}
\hrefCMSnoop {} {M.~Pivk and F.~R. Le~Diberder, ``{SPlot: A Statistical tool to
  unfold data distributions}'',} \textit{ Nucl. Instrum. Meth. A} \textbf{ 555}
  (2005) 356,
  \href{http://dx.doi.org/10.1016/j.nima.2005.08.106}{\doi{10.1016/j.nima.2005.08.106}},
\href{http://www.arXiv.org/abs/physics/0402083}{\texttt{
  arXiv:physics/0402083}}.

\bibitem{LHC-HCG}
\href {http://cdsweb.cern.ch/record/1379837} {{ATLAS and CMS Collaborations,
  LHC Higgs Combination Group}, ``Procedure for the LHC Higgs boson search
  combination in Summer 2011'',} ATL-PHYS-PUB 2011-11/CMS NOTE 2011/005, 2011.

\bibitem{Bozzi:2005wk}
\hrefCMSnoop {} {G.~Bozzi, S.~Catani, D.~de~Florian, and M.~Grazzini,
  ``{Transverse-momentum resummation and the spectrum of the Higgs boson at the
  LHC}'',} \textit{ Nucl. Phys. B} \textbf{ 737} (2006) 73,
  \href{http://dx.doi.org/10.1016/j.nuclphysb.2005.12.022}{\doi{10.1016/j.nuclphysb.2005.12.022}},
\href{http://www.arXiv.org/abs/hep-ph/0508068}{\texttt{ arXiv:hep-ph/0508068}}.

\bibitem{deFlorian:2011xf}
\hrefCMSnoop {} {D.~de~Florian, G.~Ferrera, M.~Grazzini, and D.~Tommasini,
  ``{Transverse-momentum resummation: Higgs boson production at the Tevatron
  and the LHC}'',} \textit{ JHEP} \textbf{ 11} (2011) 064,
  \href{http://dx.doi.org/10.1007/JHEP11(2011)064}{\doi{10.1007/JHEP11(2011)064}},
\href{http://www.arXiv.org/abs/1109.2109}{\texttt{ arXiv:1109.2109}}.

\bibitem{Bozzi:2003jy}
\hrefCMSnoop {} {G.~Bozzia, S.~Catani, D.~de~Florian, and M.~Grazzini, ``{The
  q(T) spectrum of the Higgs boson at the LHC in QCD perturbation theory}'',}
  \textit{ Phys. Lett. B} \textbf{ 564} (2003) 65,
  \href{http://dx.doi.org/10.1016/S0370-2693(03)00656-7}{\doi{10.1016/S0370-2693(03)00656-7}},
\href{http://www.arXiv.org/abs/hep-ph/0302104}{\texttt{ arXiv:hep-ph/0302104}}.

\bibitem{Alekhin:2011sk}
\hrefCMSnoop {} {S.~Alekhin {et~al.}, ``{The PDF4LHC Working Group Interim
  Report}'',} (2011).
\href{http://www.arXiv.org/abs/1101.0536}{\texttt{ arXiv:1101.0536}}.

\bibitem{Botje:2011sn}
\hrefCMSnoop {} {M.~Botje {et~al.}, ``{The PDF4LHC Working Group Interim
  Recommendations}'',} (2011).
\href{http://www.arXiv.org/abs/1101.0538}{\texttt{ arXiv:1101.0538}}.

\bibitem{Martin:2009iq}
\hrefCMSnoop {} {A.~D. Martin, W.~J. Stirling, R.~S. Thorne, and G.~Watt,
  ``{Parton distributions for the LHC}'',} \textit{ Eur. Phys. J. C} \textbf{
  63} (2009) 189,
  \href{http://dx.doi.org/10.1140/epjc/s10052-009-1072-5}{\doi{10.1140/epjc/s10052-009-1072-5}},
\href{http://www.arXiv.org/abs/0901.0002}{\texttt{ arXiv:0901.0002}}.

\bibitem{Ball:2011mu}
\hrefCMSnoop {} {{ NNPDF} Collaboration, ``{Impact of Heavy Quark Masses on
  Parton Distributions and LHC Phenomenology}'',} \textit{ Nucl. Phys. B}
  \textbf{ 849} (2011) 296,
  \href{http://dx.doi.org/10.1016/j.nuclphysb.2011.03.021}{\doi{10.1016/j.nuclphysb.2011.03.021}},
\href{http://www.arXiv.org/abs/1101.1300}{\texttt{ arXiv:1101.1300}}.

\bibitem{lumiPAS}
\href {http://cdsweb.cern.ch/record/1376102} {{ CMS} Collaboration, ``Absolute
  Calibration of the CMS Luminosity Measurement: Summer 2011 Update'',} CMS
  Physics Analysis Summary CMS-PAS-EWK-11-001, 2011.

\bibitem{CMS:2013gfa}
\href {http://cds.cern.ch/record/1598864} {{ CMS} Collaboration, ``{CMS
  Luminosity Based on Pixel Cluster Counting - Summer 2013 Update}'',} CMS
  Physics Analysis Summary CMS-PAS-LUM-13-001, 2013.

\bibitem{Denner:2011mq}
A.~Denner\hrefCMSnoop {} { {et~al.}, ``{Standard Model Higgs-Boson Branching
  Ratios with Uncertainties}'',} \textit{ Eur. Phys. J. C} \textbf{ 71} (2011)
  1753,
  \href{http://dx.doi.org/10.1140/epjc/s10052-011-1753-8}{\doi{10.1140/epjc/s10052-011-1753-8}},
\href{http://www.arXiv.org/abs/1107.5909}{\texttt{ arXiv:1107.5909}}.

\bibitem{Junk}
\hrefCMSnoop {} {T.~Junk, ``Confidence level computation for combining searches
  with small statistics'',} \textit{ Nucl. Instrum. Meth. A} \textbf{ 434}
  (1999) 435,
  \href{http://dx.doi.org/10.1016/S0168-9002(99)00498-2}{\doi{10.1016/S0168-9002(99)00498-2}},
  \href{http://www.arXiv.org/abs/hep-ex/9902006}{\texttt{
  arXiv:hep-ex/9902006}}.

\bibitem{Read:2002hq}
\hrefCMSnoop {} {A.~L. Read, ``{Presentation of search results: The CL(s)
  technique}'',} \textit{ J. Phys. G} \textbf{ 28} (2002) 2693,
\href{http://dx.doi.org/10.1088/0954-3899/28/10/313}{\doi{10.1088/0954-3899/28/10/313}}.

\bibitem{PDG}
\hrefCMSnoop {} {{ Particle Data Group} Collaboration, ``{Review of Particle
  Physics}'',} \textit{ Phys. Rev. D} \textbf{ 86} (2012) 010001,
\href{http://dx.doi.org/10.1103/PhysRevD.86.010001}{\doi{10.1103/PhysRevD.86.010001}}.

\bibitem{Cowan:2010st}
\hrefCMSnoop {} {G.~Cowan, K.~Cranmer, E.~Gross, and O.~Vitells, ``Asymptotic
  formulae for likelihood-based tests of new physics'',} \textit{ Eur. Phys. J.
  C} \textbf{ 71} (2011) 1,
  \href{http://dx.doi.org/10.1140/epjc/s10052-011-1554-0}{\doi{10.1140/epjc/s10052-011-1554-0}},
  \href{http://www.arXiv.org/abs/1007.1727}{\texttt{ arXiv:1007.1727}}.

\bibitem{Gross:2010qma}
\hrefCMSnoop {} {E.~Gross and O.~Vitells, ``{Trial factors or the look
  elsewhere effect in high energy physics}'',} \textit{ Eur. Phys. J. C}
  \textbf{ 70} (2010) 525,
  \href{http://dx.doi.org/10.1140/epjc/s10052-010-1470-8}{\doi{10.1140/epjc/s10052-010-1470-8}},
\href{http://www.arXiv.org/abs/1005.1891}{\texttt{ arXiv:1005.1891}}.

\bibitem{CMS:2012bw}
\hrefCMSnoop {} {{ CMS} Collaboration, ``{Observation of Z decays to four
  leptons with the CMS detector at the LHC}'',} \textit{ JHEP} \textbf{ 12}
  (2012) 034,
  \href{http://dx.doi.org/10.1007/JHEP12(2012)034}{\doi{10.1007/JHEP12(2012)034}},
\href{http://www.arXiv.org/abs/1210.3844}{\texttt{ arXiv:1210.3844}}.

\bibitem{Feldman:1997qc}
\hrefCMSnoop {} {G.~J. Feldman and R.~D. Cousins, ``{A unified approach to the
  classical statistical analysis of small signals}'',} \textit{ Phys. Rev. D}
  \textbf{ 57} (1998) 3873,
  \href{http://dx.doi.org/10.1103/PhysRevD.57.3873}{\doi{10.1103/PhysRevD.57.3873}},
\href{http://www.arXiv.org/abs/physics/9711021}{\texttt{
  arXiv:physics/9711021}}.

\end{thebibliography}\endgroup

\cleardoublepage \appendix\section{The CMS Collaboration \label{app:collab}}\begin{sloppypar}\hyphenpenalty=5000\widowpenalty=500\clubpenalty=5000\textbf{Yerevan Physics Institute,  Yerevan,  Armenia}\\*[0pt]
S.~Chatrchyan, V.~Khachatryan, A.M.~Sirunyan, A.~Tumasyan
\vskip\cmsinstskip
\textbf{Institut f\"{u}r Hochenergiephysik der OeAW,  Wien,  Austria}\\*[0pt]
W.~Adam, T.~Bergauer, M.~Dragicevic, J.~Er\"{o}, C.~Fabjan\cmsAuthorMark{1}, M.~Friedl, R.~Fr\"{u}hwirth\cmsAuthorMark{1}, V.M.~Ghete, C.~Hartl, N.~H\"{o}rmann, J.~Hrubec, M.~Jeitler\cmsAuthorMark{1}, W.~Kiesenhofer, V.~Kn\"{u}nz, M.~Krammer\cmsAuthorMark{1}, I.~Kr\"{a}tschmer, D.~Liko, I.~Mikulec, D.~Rabady\cmsAuthorMark{2}, B.~Rahbaran, H.~Rohringer, R.~Sch\"{o}fbeck, J.~Strauss, A.~Taurok, W.~Treberer-Treberspurg, W.~Waltenberger, C.-E.~Wulz\cmsAuthorMark{1}
% ...
\vskip\cmsinstskip
% ...
1:~~Also at Vienna University of Technology, Vienna, Austria\\
2:~~Also at CERN, European Organization for Nuclear Research, Geneva, Switzerland\\
% ...

\end{sloppypar}
\end{document}